\documentclass[preprint]{aastex631}

\usepackage{graphicx}
\usepackage{CJKutf8}
\usepackage{multirow}

\newcommand{\RObs}{Rubin Observatory}

\newcommand{\anns}{{\textquotesingle Ayl\'{o}\textquotesingle chaxnim}}

\newcommand{\rh}{\ensuremath{r_{\mathrm{h}}}}

\newcommand{\update}[1]{\textcolor{black}{#1}}

\shorttitle{Tuning LSST for Solar System Science}
\shortauthors{Schwamb et al.}

\graphicspath{{./}{figures/}{movies/}}

\begin{document}
\title{ Tuning the Legacy Survey of Space and Time (LSST) Observing Strategy for Solar System Science }



\author[0000-0003-4365-1455]{Megan E. Schwamb}
\correspondingauthor{Megan E. Schwamb}
\email{m.schwamb@qub.ac.uk}
\affiliation{Astrophysics Research Centre, School of Mathematics and Physics, Queen's University Belfast, Belfast BT7 1NN, UK}

\author[0000-0001-5916-0031]{R. Lynne Jones}
\affiliation{Rubin Observatory, 950 N. Cherry Ave., Tucson, AZ 85719, USA}
\affiliation{Aerotek, Suite 150, 4321 Still Creek Drive, Burnaby, BC V5C6S, Canada}
\author[00000-0003-2874-6464]{Peter Yoachim}
\affiliation{Department of Astronomy, University of Washington, 3910 15th Ave NE, Seattle, WA 98195, USA}

\author[0000-0001-8736-236X]{Kathryn Volk}
\affiliation{Lunar and Planetary Laboratory, The University of Arizona, 1629 E University Blvd, Tucson, AZ 85721, USA}
\affiliation{Planetary Science Institute, 11700 East Fort Lowell, Suite 106, Tucson, AZ 85719, USA}

\author[0000-0002-8910-1021]{Rosemary C. Dorsey}
\affiliation{School of Physical and Chemical Sciences | Te Kura Mat\={u}, University of Canterbury,  Private Bag 4800, Christchurch 8140, New Zealand}

\author[0000-0002-9298-7484]{Cyrielle Opitom}
\affiliation{Institute for Astronomy, University of Edinburgh, Royal Observatory, Edinburgh, EH9 3HJ, UK}

\author[0000-0002-4439-1539]{Sarah Greenstreet}
\affiliation{DiRAC Institute and the Department of Astronomy, University of Washington, 3910 15th Ave NE, Seattle, WA 98195, USA}

\author[0000-0002-3818-7769]{Tim Lister}
\affiliation{Las Cumbres Observatory, 6740 Cortona Drive Suite 102, Goleta, CA 93117, USA}

\author[0000-0001-9328-2905]{Colin Snodgrass}
\affiliation{Institute for Astronomy, University of Edinburgh, Royal Observatory, Edinburgh, EH9 3HJ, UK}

\author[0000-0002-4950-6323]{Bryce T. Bolin}
\altaffiliation{NASA Postdoctoral Program Fellow}
\affiliation{Goddard Space Flight Center, 8800 Greenbelt Road, Greenbelt, MD 20771, USA}
\affiliation{Division of Physics, Mathematics and Astronomy, California Institute of Technology, Pasadena, CA 91125, USA}
\affiliation{Infrared Processing and Analysis Center, California Institute of Technology, Pasadena, CA 91125, USA}

\author[0000-0002-0271-2664]{Laura Inno}
\affiliation{Science and Technology Department, Parthenope University of Naples, CDN IC4, 80143, Naples, Italy}
\affiliation{INAF-Osservatorio Astronomico di Capodimonte, Salita Moraliello, 16, 80131, Naples, Italy}

\author[0000-0003-3257-4490]{Michele T. Bannister}
\affiliation{School of Physical and Chemical Sciences | Te Kura Mat\={u}, University of Canterbury,  Private Bag 4800, Christchurch 8140, New Zealand}

\author[0000-0002-1398-6302]{Siegfried Eggl}
\affiliation{Department of Aerospace Engineering,
University of Illinois at Urbana-Champaign,
Urbana, IL 61801, USA}
\affiliation{Department of Astronomy,
University of Illinois at Urbana-Champaign,
Urbana, IL 61801, USA}
\affiliation{National Center for Supercomputing Applications,
University of Illinois at Urbana-Champaign,
Urbana, IL 61801, USA}
\affiliation{IMCCE, Paris Observatory,
77 Avenue Denfert-Rochereau, 75014 Paris, France}

\author[0000-0002-1701-8974]{Michael Solontoi}
\affiliation{Department of Physics and Engineering, Monmouth College, 700 E Broadway, Monmouth, IL 61462, USA}

\author[0000-0002-6702-7676]{Michael S. P. Kelley}
\affil{Department of Astronomy, University of Maryland, College Park, MD 20742-0001, USA}

\author[0000-0003-1996-9252]{Mario Juri\'c}
\affiliation{DiRAC Institute and the Department of Astronomy, University of Washington, 3910 15th Ave NE, Seattle, WA 98195, USA}

\author[0000-0001-7737-6784]{Hsing~Wen~Lin (\begin{CJK*}{UTF8}{gbsn}林省文\end{CJK*})}
\affiliation{Department of Physics, University of Michigan, Ann Arbor, MI 48109, USA}

\author[0000-0003-1080-9770]{Darin Ragozzine}
  \affiliation{Brigham Young University, Department of Physics and Astronomy, N283 ESC, Provo, UT 84602, USA}

\author[0000-0003-0743-9422]{Pedro H. Bernardinelli} 
\affiliation{DiRAC Institute and the Department of Astronomy, University of Washington, 3910 15th Ave NE, Seattle, WA 98195, USA} 

\author[0000-0003-3240-6497]{Steven R. Chesley}
\affiliation{Jet Propulsion Laboratory, California Institute of Technology, 4800 Oak Grove Dr., Pasadena, CA 91109, USA}

\author[0000-0002-6939-9211]{Tansu Daylan}
\affiliation{Department of Astrophysical Sciences, Princeton University, 4 Ivy Lane, Princeton, NJ 08544, USA }
\affiliation{LSSTC Catalyst Fellow}

\author[0000-0003-4914-3646]{Josef \v{D}urech}
\affiliation{Astronomical Institute, Faculty of Mathematics and Physics, Charles University, V Hole\v{s}ovi\v{c}k\'ach 2, 180 00, Prague 8,Czech Republic}

\author[0000-0001-6680-6558]{Wesley C. Fraser}
\affiliation{Herzberg Astronomy and Astrophysics Research Centre, National Research Council, 5071 W. Saanich Rd. Victoria BC, V9E 2E7, Canada}

\author[0000-0002-5624-1888]{Mikael Granvik}
\affiliation{Department of Physics, P.O.~Box 64, 00014 University of Helsinki, Finland}
\affiliation{Asteroid Engineering Laboratory, Lule\aa{} University of Technology, Box 848, SE-981 28 Kiruna, Sweden}

\author[0000-0003-2781-6897]{Matthew M. Knight}
  \affiliation{Physics Department, United States Naval Academy, 572C Holloway Rd, Annapolis, MD 21402, USA} 
  
\author[0000-0002-9548-1526]{Carey M. Lisse}
  \affiliation{Space Exploration Sector, Johns Hopkins University Applied Physics Laboratory, 11100 Johns Hopkins Road, Laurel, MD 20723, USA}

\author[0000-0002-1226-3305]{Renu Malhotra}
  \affiliation{Lunar and Planetary Laboratory, The University of Arizona, 1629 E University Blvd, Tucson, AZ 85721, USA}

\author[0000-0001-5750-4953]{William J. Oldroyd}
  \affiliation{Department of Astronomy \& Planetary Science, Northern Arizona University, P.O. Box 6010, Flagstaff, AZ 86011, USA}
  
  \author[0000-0002-1506-4248]{Audrey Thirouin}
  \affiliation{Lowell Observatory, 1400 W. Mars Hill Road, Flagstaff, AZ 86001, USA}
  
\author[0000-0002-4838-7676]{Quanzhi Ye (\begin{CJK*}{UTF8}{gbsn}叶泉志\end{CJK*})}
\affil{Department of Astronomy, University of Maryland, College Park, MD 20742-0001, USA}
\affiliation{Center for Space Physics, Boston University, 725 Commonwealth Ave, Boston, MA 02215, USA}

\begin{abstract}
The Vera C. Rubin Observatory is expected to start the Legacy Survey of Space and Time (LSST) in \update{early to mid-2025}. This multi-band wide-field synoptic survey will transform our view of the solar system, with the discovery and monitoring of over 5 million small bodies. The final survey strategy chosen for LSST has direct implications on the discoverability and characterization of solar system minor planets and passing interstellar objects. Creating an inventory of the solar system is one of the four main LSST science drivers. The LSST observing cadence is a complex optimization problem that must balance the priorities and needs of all the key LSST science areas. To design the best LSST survey strategy, a series of operation simulations using the Rubin Observatory scheduler have been generated to explore the various options for tuning observing parameters and prioritizations. We explore the impact of the various simulated LSST observing strategies on studying the solar system's small body reservoirs. We examine what are the best observing scenarios and review what are the important considerations for maximizing LSST solar system science. \update{In general, most of the LSST cadence simulations produce $\pm5\%$ or less variations in our chosen key metrics, but a subset of the simulations significantly hinder science returns with much larger losses in the discovery and light curve metrics.} 
\end{abstract}

\section{Introduction} 
\label{sec:intro}

The Vera C. Rubin Observatory is currently under construction on Cerro Pach\'on in Chile. When completed, the observatory will house the 8.36-m Simonyi Survey Telescope equipped with the Rubin Observatory LSST Camera (LSSTCam) which covers a 9.6 deg$^2$ circular field-of-view (FOV). This provides the unique depth and temporal sky coverage that will enable Rubin Observatory's planned 10-year Legacy Survey of Space and Time \citep[LSST; ][]{2019ApJ...873..111I,2022ApJS..258....1B} to be an unprecedented discovery machine for solar system small bodies. With survey operations currently expected to begin in \update{early to mid-2025}, current predictions estimate that Rubin Observatory will detect over 5 million new solar system objects. In each of the solar system's small body reservoirs, an order of magnitude more objects will be discovered during the LSST than cataloged to date in the Minor Planet Center (MPC) \footnote{\url{https://www.minorplanetcenter.net/}} \citep{2009EM&P..105..101J, 2009arXiv0912.0201L, 2010Icar..205..605S, 2015MNRAS.446.2059S, 2016AJ....152..103S, 2016AJ....151..172G, 2017AJ....154...13V, 2018Icar..303..181J, 2018arXiv180201783S, 2019ApJ...873..111I, 2020Icar..33813517F}. In addition, the survey is expected to discover at least several interstellar objects (ISOs) passing through the solar system \citep{2009ApJ...704..733M, 2016ApJ...825...51C,2017AJ....153..133E, 2017ApJ...850L..38T,2018AJ....155..217S,2021ApJ...922...39L,2022PSJ.....3...71H}. Beyond discovery, the dawn of Rubin Observatory will also usher in a revolution for time domain planetary astronomy. The LSST will monitor most of its 5+ million small body discoveries over a ten-year period, with likely hundreds of observations per object split across 6 broad-band ($ugrizy$) filters \citep{2009arXiv0912.0201L, 2019ApJ...873..111I}. This will enable an unparalleled probe of activity within various regions of the solar system, including cometary outgassing/sublimation, cometary outbursts,  rotational breakup events, and asteroid collisions \citep{2009EM&P..105..101J, 2009arXiv0912.0201L, 2018arXiv180201783S, 2021RNAAS...5..143S}. The large number of observations per object will also provide opportunities to study rotational light curves, phase curves, and photometric colors which probe the shape, size, rotation rate, and surface composition of these small bodies \citep{2009EM&P..105..101J, 2009arXiv0912.0201L, 2018arXiv180201783S}.

The LSST will be a collection of surveys operating in tandem. The main component of the LSST is the Wide-Fast-Deep (WFD), a wide-field survey covering $\sim$18,000 deg$^2$ of the sky with a universal observing strategy. Although there is tuning to the implementation of the WFD that is possible, the main requirements for the WFD are outlined in the LSST Science Requirements Document \citep[SRD; ][]{lsstSRD}. The SRD defines the WFD as $\sim$18,000 deg$^2$ of sky uniformly covered to a median total of 825 $\sim$30 s exposures divided across the six filters over a ten year period. Approximately 80-90$\%$ of the LSST's on-sky observing time will be devoted to the WFD. The remaining time is expected to be used for non-WFD observing and will likely be split between observing other portions of the sky in mini-surveys (taking up more than a few percent of the observing time) with different cadences, micro-surveys (observing strategies that require $\sim$1\% of the observing time), and approximately 5$\%$ of the on-sky time dedicated to Deep Drilling Fields (DDFs; a small number of dedicated pointings that will receive intensive observing at a higher cadence than the WFD). For a full description of the various components of the LSST and the requirements set by the SRD, readers are directed to \cite{lsstSRD}, \cite{2019ApJ...873..111I}, \cite{2022ApJS..258....1B}, and references therein. 
 
How exactly Rubin Observatory will scan the night sky is not fully settled. The Rubin Observatory Project and Operations teams have engaged with the wider user community to optimize the LSST observing strategy \update{in order to maximize the future science returns from the resulting dataset and facilitate the best science with the survey \citep{2022ApJS..258....1B}}. Partitioning out the non-WFD LSST observing time and fine tuning the WFD observing cadence can be likened to cutting a cake and dividing it out to attendees at a birthday party. There are many ways to cut and serve the slices of cake, but the various slicing/serving options may result in very different outcomes. For example, cutting even slices such that everyone gets the same portion size of cake is much more equitable and will likely result in a much happier crowd than cutting half the cake for the first person served and dividing the other half of the cake amongst the rest of the attendees. LSST has four key science drivers: probing dark energy and dark matter, exploring the transient optical sky, inventorying the solar system, and mapping the Milky Way \citep{2009arXiv0912.0201L, 2019ApJ...873..111I}. What may be beneficial for one science driver in a proposed LSST observing cadence may negatively impact the returns from another. Optimizing the LSST strategy is thus a fine balance to tune the cadence parameters to obtain the best science from each of the LSST's key drivers while evenly distributing the ``unhappiness" such that no science area is overly impacted by the finalized cadence decisions.

As highlighted in \cite{2022ApJS..258....1B}, optimizing the LSST cadence is a multivariate problem. In order to facilitate exploring the various options for modifying the LSST survey strategy and the resulting impacts on the survey's main science drivers, the Rubin Observatory LSST Scheduler Team has developed a suite of cadence simulations \citep{2014SPIE.9150E..14C, 2014SPIE.9150E..15D, 2017arXiv170804058L, jones_r_lynne_2020_4048838} using the Rubin Observatory scheduler \citep[\tt{rubin$\_$sim/OpSim};][]{2019AJ....157..151N} and the python-based LSST Metrics Analysis Framework \citep[MAF,][]{2014SPIE.9149E..0BJ}. The Rubin Observatory Survey Cadence Optimization Committee (SCOC) has been synthesizing the feedback from the LSST user community and the output from the MAF metrics to produce a formal recommendation on how to optimize the LSST survey strategy \update{\citep{SCOC_Report_1, SCOC_Report_2}}. The SCOC is expected to finish its \update{main} deliberations \update{by} the end of 202\update{3}. The SCOC may request some additional fine-tuning of the observatory strategy and recommend changes for the first year of the survey based on the knowledge gained during commissioning and benchmarking of the telescope-camera system \citep{2022ApJS..258....1B}. Once Rubin Observatory science operations start, it is expected that the SCOC will periodically review the performance of the LSST observing cadence and subsequently recommend modifications as needed. 

\update{This paper is a contribution to the Astrophysical Journal Supplement Series focus issue on \emph{Rubin LSST Survey Strategy Optimization}\footnote{\url{https://iopscience.iop.org/journal/0067-0049/page/rubin\_cadence}}. The focus issue aims to capture the knowledge learned during the process of selecting and finalizing the LSST initial cadence and identifying what observing strategies are or are not suitable for each of the key LSST science areas. We refer the reader to the opening paper by \cite{2022ApJS..258....1B} for a more detailed introduction to the focus issue.} The work presented in this paper stems from the Rubin Observatory LSST Solar System Science Collaboration's (SSSC) efforts to provide feedback to the SCOC. \update{The LSST SRD does not set performance requirements based on detecting a certain number of solar system objects in the various small body populations. Instead, the SRD outlines the minimum requirements and stretch goals for the observing specifications of the LSST such as single exposure depth, sky coverage, number of visits, astrometric precision, and co-added 10-year depths that would enable science in all the four main survey science drivers. What it means to maximize the returns on LSST Solar System science in the context of survey cadence decisions is up for interpretation by the Rubin data rights community and the SCOC. The baseline  survey strategy that was being simulated at the start of the cadence optimization process, showed  an order of magnitude increase in Solar System discoveries across each of the minor planet populations \citep{2009arXiv0912.0201L, 2017AJ....154...13V, 2018Icar..303..181J}. Determining that the LSST needs to discover N objects of class X to measure Y at the Z confidence level in order provide the next leap forward in our understanding of the Solar System is extremely challening to do. Many of the science questions that the LSST will address are not necessarily well understood (or even formulated) yet.  In most cases, it is very difficult to take existing models of the solar system, its formation and evolution, and transform that into the the total number of particular kinds of objects needed and the photometric precision required to distinguish between the available models. The analysis presented in this work is the SSSC's attempt based on the collaboration's science priorities \citep{2018arXiv180201783S} to find quantitative proxies that can be calculated within MAF and use these outputted metrics to identify which potential LSST observing strategies are the best and worst at enabling solar system science. }

\update{In this paper, we} review the LSST cadence simulations and MAF metrics focusing on the impact on the detection and monitoring of solar system minor planets and ISOs.  In Section \ref{sec:sim_moving_objects}, we provide an overview of how LSST moving object discoveries are simulated and how the relevant MAF metrics are calculated. Section \ref{sec:sims} briefly describes the LSST cadence simulations utilized in this work. In Section \ref{sec:analysis}, we examine the impact of various survey strategy choices and identify tension points with moving object detection and characterization. In Section \ref{sec:not_simulated}, we discuss additional factors that are currently not explored in the cadence simulations which have the potential for hindering or enhancing solar system science with the LSST. Finally, we draw together in Section \ref{sec:conclusions} conclusions and recommendations for tuning the LSST survey strategy in order to maximize solar system science as well as identify areas for future work. Given the length of this paper, we have included a table of acronyms and their expansions in Appendix \ref{appendix:acronyms}.

\section{Simulating LSST Solar System Detections}
\label{sec:sim_moving_objects}

Simulating observations of solar system objects requires considerations beyond those commonly used for most other astrophysical sources. Of foremost importance are their non-sidereal motions and the fact that a common rest frame cannot simultaneously approximate all of them. Solar system object proper motions range from $\lesssim1$\arcsec~hr$^{-1}$ for distant trans-Neptunian objects (TNOs) to $\gtrsim1$\degr~hr$^{-1}$ for closely approaching and impacting near-Earth objects (NEOs). Next, their brightnesses may greatly vary depending on their orbits around the Sun and how closely they approach the Earth. Furthermore, cometary activity (i.e., sublimation-driven mass loss) can enhance or even dominate the intrinsic brightness of active objects in response to solar insolation, and make them extended objects. Finally, their brightnesses also vary with the phase (Sun-target-observer) angle. Other brightness variations, e.g., due to the rotational light curve, or outbursts of activity, can be treated in ways similar to any other astrophysical source.

To partially illustrate the added complexities of modeling solar system objects, take, for example, a 1-km radius object in a parabolic orbit observed at solar opposition.  Such an object would have an apparent magnitude of
\begin{equation}
    m = H(1,1,0) + 5 \log_{10}(\rh) + 5 \log_{10}(\Delta) + 2.5 \log_{10}(\Phi),
\end{equation}
where $H(1,1,0)$ (or more simply, $H$) is the absolute magnitude\footnote{Defined as the apparent magnitude of the target as seen by the Sun at a distance of 1~au (i.e., \rh=1~au, $\Delta$=1~au, $\phi$=0\degr)}, $\rh{}$ is the heliocentric distance in units of au, $\Delta$ is the observer-target distance in units of au, and $\Phi$ is the phase function evaluated at phase angle $\phi$.  Let the 1-km object have a geometric albedo\footnote{Ratio of the brightness at 0\degr{} phase to that of a white disk with the same geometric cross section.} of 4\%, then $H\simeq17.6$~mag.  The apparent brightness of this target would range from 25th magnitude at 6~au to 17th magnitude at 1.5~au from the Sun, within Rubin Observatory's nominal capabilities.  If the 1-km object was active, the coma contribution to small-aperture photometry may be estimated as
\begin{equation}
    m_c = H_y + 2.5 (2 - k) \log_{10}(\rh) + 2.5 \log_{10}(\Delta) + 2.5 \log_{10}(\Phi_c),
\end{equation}
where $H_y$ is the cometary absolute magnitude, $k$ is the heliocentric distance power-law slope for activity, and $\Phi_c$ is the phase function of the coma.  Here, the apparent magnitude varies as $\Delta$, rather than $\Delta^2$ in order to account for the spatial extendedness of the coma and fixed-angular photometric apertures where the aperture is smaller than the apparent size of the coma.  Let $k=-4$ and the comet with $m$=25~mag at 6~au may brighten to $m$=13~mag at 1.5~au.  Move our hypothetical 1~km object to an inner-Earth orbit, and the LSST may not even observe it if survey operations never allow for low solar elongation (low-SE) observations. Thus, solar system objects have the potential to be undetected at some epochs during LSST operations, saturate during others, or be missed altogether.

In order to address the above challenges when simulating observations of individual objects, a survey simulator must have knowledge of a target's orbit and its activity state (e.g., cometary or inactive). Furthermore, to assess a survey's ability to detect, discover, and characterize solar system object populations, distributions of representative orbits that account for the variety of orbits are also needed.  Model distributions of solar system small bodies' orbits (and their physical properties) are desirable.  Such models are generally derived from the known solar system populations but debiased to account for discovery efficiencies.  The survey simulator and solar system object orbital distributions are described in Sections \ref{sec:simulator} and \ref{sec:simulating-ssos}, respectively.  The metrics used to analyze the simulated observations are described in Section \ref{sec:metrics}.

\subsection{Rubin Observatory Scheduler and Operations Simulator}
\label{sec:simulator}

Various aspects of the current and previous iterations of the Rubin Observatory scheduler and operations simulator (OpSim) are described in \cite{2014SPIE.9150E..14C}, \cite{2014SPIE.9150E..15D}, \cite{2016SPIE.9910E..13D}, \cite{Yoachim2016},  \cite{2017arXiv170804058L},  \cite{2018Icar..303..181J}, \cite{2019AJ....157..151N}, \cite{jones_r_lynne_2020_4048838}, \cite{2022ApJS..258....1B} and references therein. We provide a brief overview here. The Rubin Observatory scheduling software is part of {\tt rubin\_sim} \update{\citep{peter_yoachim_2022_7374619}}, an open-source Rubin-developed python package. The {\tt rubin\_sim} \update{package} contains the primary \update{LSST} scheduling algorithm that will be used to choose pointings for the telescope based on real-time telemetry, goal target maps, and configurable survey parameters.
At the top level, the scheduler uses a decision tree to generate observations in real-time. The decision tree steps through the potential observing modes of 1) DDFs, 2) Paired observations in a large contiguous area 3) Paired observations in twilight and 4) single observations selected using a greedy algorithm. The DDFs are pre-scheduled for optimal times, all the other observing modes use a modified Markov Decision Process (MDP) similar to the one presented in \citet{2019AJ....157..151N} to generate lists of desired observations. The MDP typically considers slew time, image depth, and desired footprint coverage when selecting potential observations.
The scheduling algorithm is paired with a model observatory to simulate the full 10-year LSST for these investigations. The model observatory includes a kinematic model of the telescope along with realistic weather logs, scheduled and unscheduled downtime, and a sky brightness model \citep{Yoachim2016}.
Various survey strategy experiments are performed by either modifying the scheduler decision tree (e.g., inserting a new observing mode for taking high airmass observations in twilight), or by altering the MDP algorithm (e.g., adding a new basis function).

\subsection{Simulating Small Body Populations}
\label{sec:simulating-ssos}

The {\tt movingObjects} module in {\tt rubin\_sim} generates the observations of a model small body population as the objects would be seen in a particular simulated survey, taking into account their motion and expected changes in brightness. As a first step, ephemerides are generated from the sample orbits using {\tt OpenOrb} \citep{2009MPS...44.1853G}; the precise camera footprint is applied to determine which detections could be acquired based on their positions. Then trailing losses and color terms between the reference band and the observed filter are added to each record, to be used later when combined with a (potentially modified) H value to calculate apparent magnitudes for each observation. 

We use a typical sample size of 5,000 orbits per population. This generally provides enough statistical accuracy across the orbital distribution to reach accuracies of a few percent at the 50\% completeness level for discovery and characterization metrics, while keeping compute requirements for a few hundred simulations reasonable. We then clone the potential observations of these orbits over a range of H values (a simple linear array, chosen with appropriate values for each individual population), in order to be able to measure discovery and characterization metrics across the expected range of observable values for each population. The cloning takes place as part of metric calculation, within the {\tt MAF} (Metrics Analysis Framework) module of {\tt rubin\_sim}. At the metric calculation stage, the measured apparent magnitude is generated for each observation provided by the {\tt movingObjects} module, taking into account each individual H value within the range used for cloning, as well as the effects of phase angle, distance from Earth and Sun, trailing losses and filter color terms. Using this apparent magnitude and the expected $5\sigma$ depth of each visit, the signal-to-noise (SNR) ratio of the object in each visit is reported. In addition, the probability of detection is also reported; this is close to requiring a SNR=5 for detection, but adds statistical scatter which has the effect of smoothing the cutoff at $5\sigma$, allowing occasional detection of fainter objects or occasional losses of slightly brighter objects. 

The process of generating simulated small body populations is described in more detail, specifically for an NEO population, in \citet{2018Icar..303..181J}. For survey strategy evaluations,  we include a range of sample populations from inner solar system objects like NEOs, through mid-system objects like main belt asteroids and Jovian Trojans, all the way to outer solar system bodies like TNOs and comets. These include: 
\begin{itemize}
    \item NEOs based on a sample of orbits from \citet{2018Icar..312..181G}. \update{A random set of 5,000 orbits are drawn from the full 802,000 synthetic NEO sample instantiated by Granvik\footnote{The full 802,000 object Granvik sample is available for download from \url{https://www.mv.helsinki.fi/home/mgranvik/data/Granvik+_2018_Icarus/}; the subset is selected as described in detail at \url{https://github.com/lsst-sssc/SSSC_test_populations_gitlfs/blob/master/MAF_TEST/granvik/Granvik\%20NEO\%20Model.ipynb}.} and used for general NEO evaluation. In addition, Earth MOID (Minimum Orbit Intersection Distance) values were calculated for the full Granvik sample, and a subset of 5,000 orbits with MOID values $<0.05$ au were randomly selected to represent the Potentially Hazardous Asteroid (PHA) population. }
    \item \update{An \textquotesingle Ayl\'{o}\textquotesingle chaxnim population\footnote{Previously this population was referred to as the Vatira population or Vatiras \citep{2012Icar..217..355G} before the discovery of the first known object \textquotesingle Ayl\'{o}\textquotesingle chaxnim \citep{2020MPEC....A...99B, 2020MNRAS.493L.129G, 2020MNRAS.494L...6D, 2020MNRAS.496.3572P, Ip_2022, BolinAN2022}.}, consisting of 10,000 objects with orbits inside the orbit of Venus, was created via rejection sampling of the probability distribution for orbital elements given in the Granvik NEO model   \citep{2018Icar..312..181G}. Orbital elements were drawn from the Granvik distribution and rejected unless they were compatible with the definition of Vatiras given in the same reference, i.e. objects between the the apocenter distance of Mercury ($Q_M=0.307$ au) and the pericenter distance of Venus ($q_V=0.718$ au). Angular orbital elements not provided through the Granvik model were sampled from uniform distributions. To achieve reasonable statistical signal, this population is simulated with a larger sample size as each individual orbit is inherently unlikely to be observed.}
    
    \item Main Belt Asteroids (MBAs) and Jupiter Trojans, based on a random sample of 5,000 main belt asteroids and 5,000 Jupiter Trojan asteroids (respectively) from the Panoramic Survey Telescope and Rapid Response System (Pan-STARRS) Synthetic Solar System Model (S3M) \citep{2011PASP..123..423G}.
    \item Trans-Neptunian Objects (TNOs), based on a random sample of 5,000 objects from the L7 model from CFEPS (Canada-France Ecliptic Plane Survey)  \citep{2011AJ....142..131P}. 
    \item Oort Cloud Comet (OCC) populations, created from the long period comet model of \citet{2019AJ....157..181V}. Two different samples of 5,000 comets are created, one with a maximum perihelion distance of 5~au and another with a maximum perihelion distance of 20~au.
\end{itemize}
For the comet populations, we include a cometary brightening function, based on the Af$\rho$ quantity of \citet{1984AJ.....89..579A}, using a translation from $H$ to cometary nuclei radii, and parameters appropriate for long-period comets.  Intrinsic light curves due to variations in shape of the objects or surface albedo or color variations are not included for any population but would be useful to include in the future. These populations do not include every population across the Solar System but serve as a representative sample covering a wide range of apparent velocities, sky coverage, and orbital parameters for the purposes of evaluating the impacts of changes in survey strategy.

A variety of solar system reflectance spectra are assigned to the members of these populations, in order to determine color terms for the LSST filters. The general simple rule of thumb is that Bus-DeMeo \citep{2009Icar..202..160D} SEDs (spectral energy distributions) are assigned to objects depending on their semimajor axes; orbits with semimajor axes smaller than 2~au are assigned to S types, orbits with semimajor axes larger than 4~au are assigned to C types, and between 2--4~au orbits are assigned randomly to S vs. C with a linear increase in probability as a function of semimajor axis, in accordance with \citet{2001AJ....122.2749I}. This means \textquotesingle Ayl\'{o}\textquotesingle chaxnims are entirely S type, the Trojans are entirely C types, while PHAs, NEOs, and MBAs are a mix of S and C types. The TNOs are assigned a significantly redder, TNO-specific SED, appropriate for the typical colors of a red dynamically excited TNO or a bluer object from the red cold classicals. The OCC populations are assigned D type SEDs, as a reasonable approximation for the colors of the cometary nuclei. Colors for these populations are shown in Table~\ref{tab:colors}. In reality, objects in each of these populations show a range of colors, so this is a simplification but is sufficient for survey strategy evaluation purposes \update{as the same $H$-orbit-color distributions are applied to all the LSST cadence simulations used in this work}. 

\begin{deluxetable}{ccccc}
\tablecaption{Rubin colors for these model SEDs. LSST catalogs will contain measurements reported as `top-of-atmosphere' AB magnitudes. \label{tab:colors}}
\tablewidth{0pt}
\tablehead{
\colhead{Color (mag)} &  \colhead{S} & \colhead{D} &   \colhead{C}  & \colhead{TNO} \\
}
\startdata
$u-r$  &   2.13 &   1.90 &   1.72 &     2.55 \\
$g-r$  &   0.65 &   0.58 &   0.48 &     0.92 \\
$i-r$  &  -0.19 &  -0.21 &  -0.11 &    -0.38 \\
$z-r$ &  -0.14 &  -0.30 &  -0.12 &    -0.59 \\
$y-r$  &  -0.14 &  -0.39 &  -0.12 &    -0.70 \\
\enddata
\end{deluxetable}

To illustrate this process more concretely, for each orbit in the test population: 
\begin{enumerate}
    \item The rough positions of the object at each night throughout the survey lifetime are calculated using OpenOrb.
    \item If the rough positions are within a tolerance value of any visit in a simulated survey, a more precise position at the time of that visit is calculated, along with the expected $V$ band magnitude as calculated by OpenOrb for the $H$ value recorded with the database (typically a fiducial placeholder value of $H$=20 mag).
    \item If the position at that time lands within the camera footprint aligned with the boresight and rotation angle of the visit, the position is recorded as a potential observation.
    \item The trailing loss and color term for that particular visit are recorded (depending on the seeing of the visit, velocity of the object, the color of the object, and the filter used for the visit). \update{Solar system objects will be moving during LSSTCam exposures. Depending on the object's velocity and the observation's exposure time, a solar system object's point spread function (PSF) can appear extended along the direction of motion. Compared to a point source of the same apparent magnitude, a trailed source will have a lower SNR because the photons are spread across more pixels on the detector. As a result, the Rubin Observatory's detection algorithm is not as sensitive to trailed sources. The algorithm uses a stellar PSF-like matched filter to find sources in the LSST images that are at or above the $5\sigma$ SNR detection limit. The trailing loss calculated in this step accounts for both the decrease in SNR and drop in detection efficiency compared to stationary point sources. We refer the reader to Section 5.1.4 of \cite{2018Icar..303..181J} for further details. }
    \item The series of potential observations are evaluated for an array of $H$ values. For example, the NEO population is evaluated for $H$ values ranging from 16 to 28 magnitudes, at steps of 0.2 magnitudes. At $H$=16 mag, the apparent magnitude of the object  \update{that will be measured by the Rubin Observatory source detection pipeline} in each visit is calculated by combining the ephemeris $V$ magnitude, the trailing losses, the color terms, and an offset between the fiducial $H$ value and the current `clone' value of $H$=16 mag (for cometary populations, there is also a calculation of the cometary brightening). At $H$=22 mag, the same process is repeated but more of the potential observations of the object will fall below the $5\sigma$ SNR limit, so fewer observations will be considered (and at a lower SNR) for the calculation of each metric for each object\update{.}
    \item The result is a series of values for each metric, corresponding to the combination of the positions resulting from each orbit with the apparent magnitudes resulting from a range of $H$ values. 
\end{enumerate}
This is repeated over all of the orbits in the test population.

\subsection{\update{LSST} Solar System Science Metrics}
\label{sec:metrics}

With the {\tt movingObjects} and {\tt MAF} modules of {\tt rubin\_sim}, the calculation of any arbitrary quantity per object is straightforward. The MAF software identifies the observations of a given object (or more specifically, orbit and $H$ value, in the case of cloning) and passes these to the MAF {\tt  Metric}, where the value can be calculated based on the acquired observations, and then saved. Summary values across the entire population, such as fraction of objects with light curve inversion potential or ``discoverable'' objects, can be calculated from the results. 

Generally speaking, our current solar system science metrics can be split into two categories: discovery metrics and characterization metrics. Discovery metrics relate to which objects could be discovered in the survey, while characterization metrics cover a broad range of science areas such as likelihood of detecting activity on the surface of an object or likelihood of acquiring a color measurement. For each metric, the value per orbit-$H$ magnitude combination is calculated and recorded, and then a `summary value' is evaluated across the entire population. For discovery metrics, this summary value is the fraction of the population that can be linked by Rubin Observatory's Solar System Processing (SSP) pipelines \citep{lsstMOPS,lsstSSP} -- the discovery completeness. For characterization metrics, the summary value is typically the fraction of the population that meets a given threshold -- the fraction of the population which is likely to meet light curve inversion requirements, for example -- although it can also be the mean or median or maximum (etc.) value of the metric across the population. These summary values are reported at either a particular $H$ value or cumulatively, for objects with $H$ less than or equal to a given $H$ value. These summary values at both a bright $H$ (large size) and a fainter $H$ (smaller size) are pulled out for each population for comparison of multiple simulations. The particular $H$ values used are dependent on the population; typically the bright $H$ value is where the metric results reach their highest value and then remain constant with decreasing $H$. The fainter $H$ values are typically set close to where the baseline survey strategy reaches about 50\% for that metric result. \update{The discovery and characterization metrics used in this paper are listed in Tables \ref{tab:metrics} and \ref{tab:secondary_metrics}. The details of how these metrics are calculated is described below in Sections \ref{sec:discovery-metrics}, \ref{sec:lightcurve_metrics}, and \ref{sec:color-light-curve}.}

\begin{deluxetable}{ll}
\tablecaption{Key solar system MAF metrics used in this analysis.  \label{tab:metrics}}
\tablehead{
\colhead{Population } & \colhead{Main Metrics}
}
\tabletypesize{\scriptsize}
\startdata
 \multicolumn{2}{c}{Discovery Metrics} \\
 \hline
 \hline
  \multirow{2}{*}{\textquotesingle Ayl\'{o}\textquotesingle chaxnims$^{a,b}$} &  \update{3 nightly pairs in 15 nights discovery completeness for $H\leq$ 16.0} \\
 & \update{3 nightly pairs in 15 nights discovery completeness for $H\leq$ 20.5} \\
  &   \update{4 detections in 1 night discovery completeness for $H\leq$ 16.0$^{d}$} \\
  & \update{4 detections in 1 night discovery completeness for $H\leq$ 20.5$^{d}$} \\
 \hline
 \multirow{2}{*}{PHAs} &  \update{3 nightly pairs in 15 nights discovery completeness for $H\leq$ 16.0}  \\
 & \update{3 nightly pairs in 15 nights discovery completeness for $H\leq$ 22.0}  \\
 \hline
  \multirow{2}{*}{NEOs} & \update{3 nightly pairs in 15 nights discovery completeness for $H\leq$ 16.0}  \\
 & \update{3 nightly pairs in 15 nights discovery completeness for $H\leq$ 22.0}  \\
  \hline
   \multirow{2}{*}{MBAs} & \update{3 nightly pairs in 15 nights discovery completeness for $H\leq$ 16.0}  \\
 & \update{3 nightly pairs in 15 nights discovery completeness for $H\leq$ 21.0}  \\
  \hline
    \multirow{2}{*}{Jupiter Trojans} & \update{3 nightly pairs in 15 nights discovery completeness for $H\leq$ 14.0}  \\
 & \update{3 nightly pairs in 15 nights discovery completeness for $H\leq$ 18.0}  \\
  \hline
     \multirow{2}{*}{TNOs} & \update{3 nightly pairs in 15 nights discovery completeness for $H\leq$ 6.0}  \\
 & \update{3 nightly pairs in 15 nights discovery completeness for $H\leq$ 8.0}  \\
  \hline
      \multirow{2}{*}{\update{OCCs$^c$ with $q\leq5$ au}}    &  \update{3 nightly pairs in 15 nights discovery completeness for $H\leq$ 8.0 }\\
      & \update{3 nightly pairs in 15 nights discovery completeness for $H\leq$ 17.0} \\
  \hline
       \multirow{2}{*}{\update{OCCs$^c$  with  $q\leq$ 20 au}}    &  \update{3 nightly pairs in 15 nights discovery completeness for $H\leq$ 8.0} \\
& \update{3 nightly pairs in 15 nights discovery completeness  for $H\leq$ 12.0} \\ 
\hline
  \multicolumn{2}{c}{Light Curve Metrics} \\
  \hline
  \hline
 \multirow{2}{*}{PHAs} &  Fraction of $H$ = 16.0 with sufficient observations for light curve inversion \\
 &  Fraction of $H$ = 19.0 with sufficient observations for light curve inversion \\
  \hline
   \multirow{2}{*}{NEOs} & Fraction of $H$ = 16.0 with sufficient observations for  light curve inversion \\
 &  Fraction of $H$ = 19.0 with sufficient observations for light curve inversion \\
  \hline
    \multirow{2}{*}{MBAs} &  Fraction of $H$ = 16.0 with sufficient observations for light curve inversion \\
 &  Fraction of  $H$ = 18.0  with sufficient observations for light curve inversion \\
  \hline
     \multirow{2}{*}{Jupiter Trojans} &  Fraction of $H$ = 14.0  with sufficient observations for light curve inversion \\
 &  Fraction of $H$ = 15.0 with sufficient observations for light curve inversion \\
\enddata
\tablenotetext{^a}{Previously referred to in the literature as Vatiras.}
\tablenotetext{^b}{Metrics for the \textquotesingle Ayl\'{o}\textquotesingle chaxnims are only analyzed for simulation families that include low solar elongation (low-SE) twilight observations.}
\tablenotetext{^c}{Metrics for OCCs are only calculated since the v2.0 simulations.}
\tablenotetext{^d}{\update{Only assessed for simulations that take 4 observations per pointing during twilight.}}

\tablecomments{In the figures presented in this work, these metrics are normalized and compared to the baseline simulation for a range of cadence simulation families by varying a different survey strategy parameter.}
\end{deluxetable}

\begin{deluxetable}{ll}
\tablecaption{Secondary solar system MAF metrics used in this analysis.  \label{tab:secondary_metrics}}
\tablehead{
\colhead{Population } & \colhead{Secondary Metrics}
}
\tabletypesize{\scriptsize}
\startdata
  \multicolumn{2}{c}{Color-Light Curve Metrics} \\
  \hline
  \hline
 \multirow{2}{*}{PHAs} &  \update{Fraction of $H$ = 16.0 with the equivalent of 40 SNR=5 detections or 10 SNR=20 detections per filter in $grizy$ } \\
 & \update{ Fraction of $H$ = 19.0 with the equivalent of 40 SNR=5 detections or 10 SNR=20 detections per filter in $grizy$ } \\
  \hline
   \multirow{2}{*}{NEOs} & \update{Fraction of $H$ = 16.0 with the equivalent of 40 SNR=5 detections or 10 SNR=20 detections per filter in $grizy$ } \\
 &  \update{Fraction of $H$ = 19.0 with the equivalent of 40 SNR=5 detections or 10 SNR=20 detections per filter in $grizy$ } \\
  \hline
    \multirow{2}{*}{MBAs} &  \update{Fraction of $H$ = 16.0 with the equivalent of 40 SNR=5 detections or 10 SNR=20 detections per filter in $grizy$ } \\
 &  \update{Fraction of $H$ = 18.0 with the equivalent of 40 SNR=5 detections or 10 SNR=20 detections per filter in $grizy$ }  \\
  \hline
     \multirow{2}{*}{Jupiter Trojans} &  \update{Fraction of $H$ = 14.0 with the equivalent of 40 SNR=5 detections or 10 SNR=20 detections per filter in $grizy$ }\\
 &  \update{Fraction of $H$ = 15.0 with the equivalent of 40 SNR=5 detections or 10 SNR=20 detections per filter in $grizy$ }\\
 \hline
      \multirow{2}{*}{TNOs} & \update{Fraction of $H=$ 6.0 with at least 30 SNR $>5$ observations in 1 filter and 20 observations in 3 other filters } \\
 &  \update{Fraction of $H=$ 8.0 with at least 30 SNR $>5$ observations in 1 filter and 20 observations in 3 other filters } \\
  \hline
      \multirow{2}{*}{\update{OCCs$^a$ with $q\leq5$ au}}      &  \update{Fraction of $H=$ 8.0 with at least 30 SNR $>5$ observations in 1 filter and 20 observations in 3 other filters  } \\
      & \update{Fraction of $H=$ 17.0 with at least 30 SNR $>5$ observations in 1 filter and 20 observations in 3 other filters } \\
\hline
      \multirow{2}{*}{\update{OCCs$^a$  with  $q\leq$ 20 au}}  &  \update{Fraction of $H=$ 8.0 with at least 30 SNR $>5$ observations in 1 filter and 20 observations in 3 other filters} \\
& \update{Fraction of $H=$ 12.0 with at least 30 SNR $>5$ observations in 1 filter and 20 observations in 3 other filters}  \\ 
\enddata
\tablenotetext{^a}{Metrics for OCCs are only calculated since the v2.0 simulations.}

\tablecomments{In the figures presented in this work, these metrics are normalized and compared to the baseline simulation for a range of cadence simulation families by varying a different survey strategy parameter.}
\end{deluxetable}

\subsubsection{Discovery Metrics}
\label{sec:discovery-metrics}

 \update{The SSP pipelines \citep{lsstMOPS,lsstSSP} will link transient sources from the nightly visits into ``tracklets" (potential linkages in the same night using linear extrapolation). SSP will identify new moving objects by attempting to link together 3 tracklets from within a 15 day window onto a heliocentric orbit.} The current baseline LSST object discovery guidelines require pairs of observations on three separate nights, within a window of 15 days as the design goal and 30 days as the stretch goal; the 15 day requirement is a confident lower limit, but a 30 day window is a reasonable extension that is useful to also consider. Thus the basic discovery metric searches for precisely this: pairs of observations on at least three different nights within 15 or 30 days, using the probabilistic detection value to determine what is visible or not\footnote{The probabilistic detection likelihood depends on the expected 5\update{$\sigma$} point source depth, determined by sky brightness, airmass, and seeing alone; it does not take into account potential crowding in the field.}. The metric allows for setting the minimum and maximum time separation between the individual visits in each pair; in the default configuration, the minimum time separation was set to 0~minutes\footnote{The minimum time separation for pairs of visits was set to 0 minutes during the metric runs analyzed in this paper. In the future, we will be using 5 minutes as the minimum separation time. However, we do not anticipate there being a significant drop in metric performance, as the overwhelming number of pairs of visits are acquired at very close to the goal time separation, around 33 minutes.}, and the maximum time separation was set to 90 minutes, corresponding to the approximate limits suggested by early expectations for configuration for the solar system processing pipelines and very widely bracketing the typical expected separation of visits. The overwhelming majority of visits in the survey are acquired in pairs with a separation of 22 to 30 minutes (depending on the details of the survey configuration), the pairs of visits are usually acquired in `adjoining' filters (i.e. $g$ and $r$ or $r$ and $i$ visits for a pair), and coupled with the large field of view of Rubin, most although not all observations of an object are followed up by a second observation in the same night.  \update{It is also helpful to consider objects that could be discovered via more traditional methods of identifying four observations on the same night (i.e. `quad detections`). This is particularly useful when considering observations of near or interior to Earth asteroids within the special near-sun twilight micro-survey, where observations are purposefully obtained in quads in order to secure identifications of these rapidly disappearing asteroids.} If the observations of a given orbit-$H$ combination meet the required criteria at least once, the object is considered `discovered'; to compare the results across different simulations of survey strategy, the discovery completeness is reported at both a bright and faint $H$ value for each population. More details about the discovery metrics are presented in \citet{2018Icar..303..181J}, including more background about the potential for false positive discoveries. In short, we do not expect a significant number of false positive detections, regardless of survey strategy choices, with the criteria of 3 pairs of detections over a window of either 15 or 30 nights; this is due to a range of factors, including the low fraction of false positive detections coming from difference imaging and the low likelihood of pairs of detections on three separate nights aligning within expected residuals for initial orbit determination.  

The various survey strategy simulations and populations expose some basic trends:
\begin{itemize}
    \item Discovery completeness for slow moving populations, such as TNOs, depends strongly on the total area included in the survey. Because these objects move so slowly year over year and are relatively `easy' to discover via linking, the footprint itself is the most important consideration of the survey strategy, particularly for the brighter TNOs. The completeness for the fainter TNOs can also vary slightly depending on which filters are paired together in visits and whether the most sensitive filters are used often enough within the window. 
    \item Discovery completeness for fast-moving populations, such as NEOs, depends more strongly on the number of visits per pointing. Since NEOs travel across much more of the sky on the timescale of the survey, the footprint isn't as much of a constraint as for TNOs. However, the total number of visits in the survey is relatively constant with different survey strategies, and so the footprint influences the number of visits per pointing and thus the typical cadence of those visits. Fainter NEOs in particular may only be visible for a short period of time, thus more visits per pointing results in a higher likelihood of an object having observations suitable for discovery, and so a higher discovery completeness. For the brightest NEOs, the footprint weighs in as well, as covering more sky results in discovering more NEOs. 
    \item Intermediate populations, such as MBAs, fall in between these extremes. In general, we find a threshold number of visits per pointing results in good completeness for a given population, and this threshold increases as the $H$ value being evaluated gets \update{larger} and/or the population includes more small semimajor axis or high inclination or high eccentricity orbits. 
    \item The Jupiter Trojans show stronger variability with some kinds of survey strategy changes that include changes in the timing of observations. Some survey strategies focus visits on particular regions of the sky in particular years, such as in the rolling cadence. These variations can result in a higher or lower sampling of the Jupiter Trojan population depending on the timing of visits, as these asteroids are both more spatially constrained and moving across the sky.
    \item More relaxed discovery criteria result in more discoveries, but with similar trends. For example, 30 day windows perform about 2--5\% better than 15 day windows for fainter objects, depending on the population (brighter objects show little difference). However, these different criteria follow similar trends between survey strategies, meaning that evaluating 15 day windows shows similar preferences in survey strategy as evaluating 30 day windows. 
\end{itemize}


\subsubsection{Light Curve Metrics}
\label{sec:lightcurve_metrics}

Inner solar system objects have the potential to be subjects for sparse light curve inversion, inferring the shape of the asteroid from photometric measurements over a wide range of viewing geometries as suggested in \citet{2009arXiv0912.0201L} ({\it e.g.} \citet{2020A&A...642A.138M, 2016A&A...587A..48D, 2011A&A...530A.134H}). The Light Curve Inversion Metric evaluates the suitability of a set of observations for this process. The evaluation is based on the phase curve and ecliptic longitude coverage provided by the observations, as well as the overall number and SNR of each observation, considering observations in a single filter at a time. The ecliptic longitude range of the observations must be \update{more than} 90$^{\circ}$ ecliptic longitude and cover more than 5$^{\circ}$ of phase angle, as a proxy of the required range of viewing geometries. Further, there must be more than a threshold value of SNR-weighted observations, equivalent to about 50 SNR=100 observations or 250 SNR=20 observations, all in the same filter, in order to provide enough photometric measurements. Like all other metrics within MAF, the rotation of the asteroid and its impact on the photometric measurements is not considered; presumably this would be part of the light curve inversion process.  If all conditions are met, then light curve inversion is at least potentially likely, thus this metric provides a likely upper limit on the fraction of the population for which light curve inversion may be possible. This is evaluated per orbit-$H$ combination, and then the fraction of the population (at a bright and fainter $H$ value) is reported.  Outer solar system objects never achieve the required range of viewing geometries, and objects where the nuclei is shrouded with coma such as active comets are also not good candidates; this metric is not evaluated for these populations.
 
This metric is very sensitive to the number of observations per pointing, but also to the cadence of those observations. Generally, we find a trend across the simulations that the light curve inversion results track in a similar sense for all of the inner solar system populations, with NEOs being least sensitive to survey strategy variations, followed by PHAs, then MBAs, and finally Trojan asteroids showing the most variation in metric results as survey strategies change.

\subsubsection{Color-Light Curve Metrics}
\label{sec:color-light-curve}

There are several metrics relating to determining colors for the small body population members, tailored for inner solar system or outer solar system objects. As the LSST will not obtain instantaneous colors, each of these metrics also includes some requirement on measuring a light curve. 

For the inner solar system, the Color-Light Curve metric evaluates the number of SNR-weighted observations per bandpass to evaluate if the color could be determined in that bandpass. Essentially, this could be translated to fitting the light curve in each bandpass alone, then combining these light curves to evaluate the color. The equivalent of 40 SNR=5 detections or 10 SNR=20 detections per filter are required, but the more extensive requirements that relate to achieving a range of viewing geometries for light curve inversion are not, and no limitation is set on when the observations are acquired. The specific number of detections needed is based on an estimate of the amount of data that would be sufficient to measure basic light curve, color, and phase-curve parameters with scientifically meaningful uncertainties. Although work is still needed to use the sparse LSST-like cadence to determine these parameters, a preliminary assessment suggests that 20-40 observations per color should be sufficient. While phase curves are also necessary for this analysis, we elected to keep the metric simple by not requiring a particular spread in phase angles. In practice, almost any cadence will produce sufficient constraint on the phase curve to allow for colors to be determined for the vast majority of objects.  The summary values reported are the fraction of the population (at a given $H$ value) for which either 2 specific colors ($g-r$ or $g-i$ plus $g-z$ or $g-y$), 3 specific colors ($g-r$, $r-i$, $i-z$ or $r-i$, $i-z$, $z-y$), 4 colors ($g-r$, $r-i$, $i-z$, $z-y$), or all 5 colors (adding $u-g$ to the 4 color set) are potentially determined.

For the outer solar system, a slightly different Color-Light Curve metric evaluates the number of observations reaching a minimum threshold (SNR$\approx5$). This metric requires at least 30 observations in a `primary' bandpass and then 20 observations in the additional bandpass(es). This is equivalent to assuming a light curve fit in the primary bandpass with additional observations in the secondary bandpass serving to help fit the light curve and color, possibly simultaneously (such as would be possible with multi-band Lomb-Scargle fitting). The summary values reported are the fraction of the population (at a given $H$ value) for which 1, 2 or more colors can be fitted, without restrictions on which bandpasses are used.

\subsubsection{Metric Limitations}
\label{sec:metric_limitations}

As described above in Section \ref{sec:simulating-ssos}, the most accessible and up to date orbital and absolute magnitude distributions have been used to model the expected LSST solar system detections. The physical and orbital properties of the modeled synthetic small bodies are driven by observational data, but the LSST cadence simulations do have to make some assumptions about these small body populations. This is particularly true on the smallest size scales that have not been very well probed by past wide-field surveys. The distribution of different surface types applied to the various simulated small body reservoirs will also impact the apparent magnitude of the synthetic objects in the various optical filters. Additionally, we have to make simplifying assumptions about active objects. We assume all comets will generate dust coma with the same relation applied to calculate the observed apparent magnitude, and the effects of cometary outbursts are ignored. Also, rotational brightness variations due to shape or uneven surface albedo are not accounted for in these simulations. Thus, the exact number of solar system minor planets found by LSST will differ from that ``discovered'' in the simulations explored in this paper because of these choices. 

The smaller Solar System minor planet populations, such as the Main Belt Comets (MBCs), Jupiter Family Comets (JFCs), sungrazing comets, Neptune Trojans, and Centaurs, have not been simulated for this work. For the MBCs, sungrazing comets, and JFCs this is partly due to having to develop a representative activity model. We can instead use the populations that are simulated in the \texttt{rubin$\_$sim} simulations as proxies to help inform what the impact of various cadences might be. Simulated survey strategies that will improve the metrics for MBAs and NEOs will also likely enhance the discovery and monitoring of MBCs and JFCs. Cadences that improve the chances of finding near Sun \textquotesingle Ayl\'{o}\textquotesingle chaxnims will likely increase the LSST discovery rate of sungrazing comets, like the Kruetz family. As the Centaurs reside in the middle Solar System, the impacts on the Centaurs can be extrapolated using the TNO and MBA simulation metrics.

No ISOs were simulated for this work. With only two ISOs known to date \citep{2017Natur.552..378M,MPEC2019-R106}, the characteristics of this population are currently unconstrained. Long period comets are distributed across the sky with a much larger range of ecliptic latitudes compared to the MBAs and TNOs, due to the effects of passing stars and  galactic tides that shape the Oort Cloud into a shell rather than a flared disk shape \citep[][and references therein]{1967AJ.....72..716E, 1997Icar..129..106F, 2005ApJ...635.1348F, 2007AJ....134.1693H, 2010A&A...516A..72B, 2015SSRv..197..191D, 2019AJ....157..181V, 2020AJ....160..134H}.  \update{Recent predictions by \cite{2017AJ....153..133E}, \cite{2018AJ....155..217S}, and \cite{2022PSJ.....3...71H} suggest that LSST ISO discoveries will cover a wide range of ecliptic latitudes and heliocentric distances  similar to long period comets. Thus, we assume that the trends seen for the simulated LSST OCC discoveries can provide some broad guidance for how the cadence decisions will impact LSST ISO discoveries.  Like 1I/`Oumuamua which was discovered at 0.22 au \citep{2017Natur.552..378M} moving at 6.2$^\circ$ per day, a subset of ISOs discovered close to Earth will on short-timescales ($\lesssim$10 days) look similar to NEOs \citep[e.g.,][]{2016ApJ...825...51C}. Therefore, the NEO metrics are also insightful for gauging the potential impacts to the ISO discovery rate.}

The solar system MAF metrics assume equal detection efficiency across all areas of the survey footprint (even near the plane of the Galaxy where stellar crowding may be a factor). Rubin Observatory's data pipelines will detect solar system bodies using difference imaging. Templates representing the static sky will be subtracted from the nightly images and what remains will be a variable, transient, or moving source. This will help significantly in detecting solar system objects in regions of high stellar density, but stellar crowding will likely cause some decrease in the efficiency of Rubin Observatory's Difference Image Analysis (DIA) and SSP pipelines. The MAF solar system metrics are likely overly optimistic near and in the Galactic plane where stellar crowding is the highest. This should be kept in mind when examining the cadence simulations modifying the LSST Galactic plane observing strategy. 

The Rubin scheduler aims to take image pairs, each night per pointing, to facilitate the identification of moving solar system objects \citep{lsstSRD}. The time between these repeat observations is a tunable survey parameter. \update{The Rubin SSP pipelines \citep{lsstMOPS,lsstSSP} require motion within a single night for initial discovery. Transient sources that appear stationary between the two images taken on the same night will not be included in the daily tracklets that the SSP algorithm will try to link with tracklets from previous nights. The Rubin SSP pipelines as currently planned will not be able to detect objects bodies beyond $\sim$100-150 au (see Section \ref{sec:nightly_sep} for a detailed estimate), but other search algorithms will likely be developed by the wider community to search for very slow moving objects in the LSST data. The MAF solar system discovery metric does not account for SSP's slow motion limit.} The only metrics really impacted by this are the estimated TNO discoveries. As long as the median separation between the observations is similar for a set of cadence simulations, then the output from the discovery metrics can be compared. We note that some care must be taken when considering the impact of varying the time separation between repeat observations, and we refer the reader to Section \ref{sec:nightly_sep} for further discussion. 

There are currently no MAF metrics that measure how precise small body orbital predictions and characterization will be based on Rubin Observatory observations. The accuracy of the orbits of moving objects is primarily driven by the observational arc length. There were no reasons to consider the observational arc length separately with a dedicated MAF metric, because all of the observing strategy options currently being considered as part of the LSST cadence optimization exercise (see Section \ref{sec:sims}) have repeat coverage of the entire LSST footprint over several years. This should be sufficient for the needs of the majority of astrometric and dynamical solar system science cases. We also note that, if \update{a} cadence option not covering the entire sky over the majority of the 10-year time span was evaluated, it would be undesirable for other science cases such as proper motion measurements.

\update{The likelihood of having satellite streaks and glints present in LSST images is increasing with every satellite constellation launch (e.g. Starlink, Project Kuiper and OneWeb). The impact of future satellite constellations is not currently taken into account by the metrics. We discuss the potential impacts of the ongoing industrialization of the near-Earth environment in Section \ref{sec:satcons}.} 

Keeping these caveats in mind, the LSST cadence simulations and the MAF metrics can be used to explore the impact of various changes to the LSST observing strategy. Some care is required in examining certain families of simulations. Overall, by adopting the same synthetic small body populations for each of the cadence simulations and focusing on the relative change in the MAF metrics compared to the baseline survey, we can still gain a good understanding of the impact caused by tuning various LSST observing parameters. 

\section{Overview of the LSST Cadence Simulations (Versions 1.5-2.2)}
\label{sec:sims}

Over the past several years, a variety of LSST cadence simulations have been generated \citep[e.g., ][]{2009arXiv0912.0201L, 2017arXiv170804058L, 2019ApJ...873..111I, jones_r_lynne_2020_4048838, SMTN-017} exploring various avenues for optimizing the WFD survey and exploring different scenarios for what to do with the remaining $\sim$10$-$20$\%$ of survey time. We examine the LSST cadence simulations produced after the implementation of the Feature Based Scheduler system \citep{2019AJ....157..151N}, as this iteration of the Rubin scheduler is closest to the version that will be in place during survey operations, starting with the version 1.5 simulation release. At the time of this paper's submission, additional families of simulations have been released up to version 2.2.  The v1.5 simulations were released in May 2020, version 1.6 in August 2020, v1.7 in  January 2021, and v1.7.1 in April 2021. These simulations cover a wide range of variations of the survey strategy that informed the first round of the SCOC's review.  After assessing the feedback from the Rubin user community, the SCOC recommended a new round of simulations (v2.0) to inform their final deliberations \citep{SCOC_Report_1, 2022ApJS..258....1B}. The v2.0 cadence simulations were made available in November 2021. Two additional smaller  sets of simulations were released in April and June 2022 (v2.1 and v2.2) that clarify/explore some limited options identified after community review of the 2.0 cadence simulations, including DDF observing options, new parameters for implementing the twilight low-SE solar system observations, and revised scenarios for Galactic plane observing. \update{The v2.2 simulation runs were redone with an update in the scheduler configuration after the submission of this paper due to an issue with the sky distribution of $u$-band observations. We use the updated v2.2 simulations in our analysis.} The v1.5-v2.2 cadence simulations are described in detail in \cite{jones_r_lynne_2020_4048838}, and \cite{SMTN-017}. Short descriptions of the simulations are also available in online Jupyter notebooks\footnote{\url{https://github.com/lsst-pst/survey_strategy/blob/main/fbs_1.7/SummaryInfo.ipynb} and \url{https://github.com/lsst-pst/survey_strategy/blob/main/fbs_2.0/SummaryInfo_v2.1.ipynb}}. The resulting MAF metrics derived from these simulations are available in online CSV (comma-separated values) files\footnote{\url{https://github.com/lsst-pst/survey_strategy/tree/main/fbs_1.7} and  \url{https://github.com/lsst-pst/survey_strategy/tree/main/fbs_2.0}}.

We focus in this paper on the key survey strategy parameters that drive significant changes in the detectability and characterization of solar system objects or would lead to unique planetary astronomy datasets that only Rubin Observatory could provide. Several of the simulation families were repeated in later versions with improvements to the Rubin scheduler, changes to the prescription used in the scheduler, or modifications to the planned survey footprint. For this work, if a simulation family was repeated in later releases, we only review the latest version. We also note that the OCC orbital distributions were only incorporated as MAF metrics in release 2.0 and onward. We include the OCC metrics when available. The v2.1 simulations include a range of families that explore the final details of the DDF observing strategy.  No solar system metrics were run against these v2.1 DDF families as very small numbers of solar system objects will be discovered in these fields compared to the rest of the survey footprint due to the fact that the DDFs take 5$\%$ of the observing time at locations high off the ecliptic. The main lever arm for solar system science in relation to the DDFs is the fraction of total observing time spent on the DDFs which is explored in Section \ref{sec:DDFs}. Simulations covering rotational and positional dithers between repeat survey pointings are also not explored here because of the negligible impact on the solar system metrics.

Appendix \ref{appendix:sims} gives a brief overview of the LSST cadence simulations evaluated in this paper. The LSST cadence simulations can be divided into several broad categories or families exploring different modifications to the survey footprint,  filter distribution, intra-night visits, DDF observing strategy, visit exposure times, rolling cadence strategies, and micro-surveys. Each simulation family explores changing one parameter in the LSST observing strategy. The footprint families explore the shape and location of the WFD on-sky footprint as well as the possible adoption of a variety of mini-surveys,  strategies surveying the sky outside the WFD footprint or with a different cadence to the WFD that require a few percent or more of the total available LSST observing time. One such example of a mini-survey is observing the northern ecliptic region. \update{Micro-surveys} are small observing campaigns requesting $\sim$0.3$\%$- 3$\%$ of the total observing budget. Rolling cadence in this context focuses on prioritizing observing some parts of the sky over others in order to acquire more photometric data points in a given observing season. This enables faster and better identification of supernovae, kilonovae, and other astrophysical transients \citep{2017arXiv170804058L, RollingCadence}.

\section{Impact of Survey Strategy Choices}
\label{sec:analysis}
How to evaluate whether a specific LSST survey strategy is ``good" or ``bad" for solar system science has a complex answer. How does one weigh a significant improvement in NEO discoveries to a large loss in the number of TNOs found? It depends on which population one is interested in studying and on the science goals one wants to achieve. We choose a unified approach when evaluating the various LSST cadence simulations. We equally consider the impact on the main solar system populations probed by LSST:  NEOs, PHAs, TNOs, MBAs, ISOs, and OCCs. Secondary consideration is given to the smaller populations such as giant planet Trojans and Inner-Earth objects (IEOs; objects on orbits interior to Earth's orbit). \update{Although an ISO population is not simulated for this work, we use the OCCs and NEOs metrics where appropriate to examine the impact on the ISOs in the various cadence simulations (see Section \ref{sec:metric_limitations}).} The SSSC Science Roadmap \citep{2018arXiv180201783S} sets out the collaboration's science priorities with LSST data. The document was designed specifically to guide future cadence decisions and ranks the key solar system research themes for investigation with LSST. Based on the SSSC Science Roadmap, for each LSST cadence simulation, we evaluate in priority order the impact on 1) discovery/orbital characterization, 2) color measurements, and 3) rotational light curves. 

We have found that per small body population, the light curve inversion and discovery metrics sufficiently encapsulate the requirements for obtaining reliable broadband colors, such that the majority of cadence simulation families are evaluated using these two metrics alone. The main metrics used in our analysis and the parameters used in their calculation are listed in Table \ref{tab:metrics}. We focus our analysis on the discovery metric that best matches \update{the} SSP discovery requirements (3 tracklets detected within 15 nights) as other variations of the discovery metrics require bespoke community-developed software tools. In a small number of instances reviewing the color-light curve metrics calculated for 4 colors was also useful for interpretation (see Table \ref{tab:secondary_metrics} for input parameters), but we will primarily focus on the discovery and light curve inversions for this work. When examining a given cadence experiment, we normalize all the metric values calculated to the relevant baseline cadence or reference simulation that we consider the default scheduler parameter setting or configuration for this cadence experiment. See Figure \ref{fig:v1.5_footprint} for an example where the resulting solar system metrics for discovery (top) and light curve inversion (bottom) are presented. We note that the Jupiter Trojans have the most variable metrics due to their smaller numbers and constrained positions on the sky. Metric results for the most recent baseline survey simulation at the time of submission (\texttt{baseline\_v2.1\_10yrs}) are shown in Figure \ref{fig:baseline_metric_values} and listed in Appendix~\ref{appendix:BaselineMetrics}.

\begin{figure}
\begin{center}
\includegraphics[width=0.93\columnwidth]{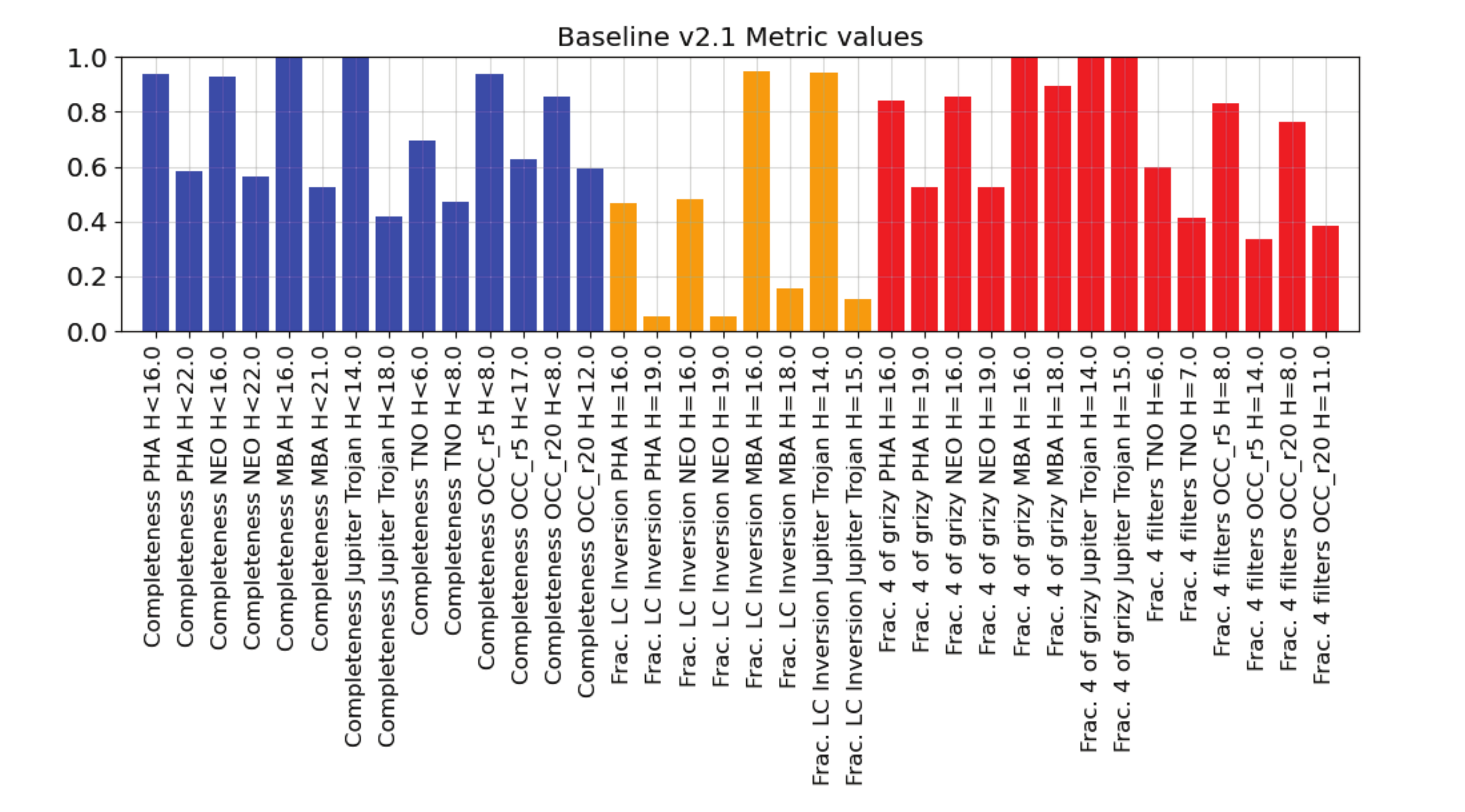}
\caption{\update{Metric values for the primary metrics under consideration for \texttt{baseline\_v2.1\_10yrs} from the latest version of {\tt rubin\_sim}. These values represent the fraction of the simulated population which would "pass" the metric requirements. `Completeness' refers to the discovery completeness for each sample population at the indicated $H$ value, while `Fraction LC Inversion' refers to the fraction of each population which would have observations which meet the metric requirements implying that object would be a good subject for lightcurve inversion. Likewise for 'Fraction 4 filters', showing the fraction of each population which would be likely to obtain colors in four filters. Full descriptions of the metrics are listed in Tables \ref{tab:metrics} and \ref{tab:secondary_metrics}. Numerical values are  provided in Appendix~\ref{appendix:BaselineMetrics}.}
\label{fig:baseline_metric_values}}
\end{center}
\end{figure}

We deem reductions in the relevant metrics larger than $\sim$5$\%$ unsuitable. The small body science goals set out in the SSSC Science Roadmap \citep{2018arXiv180201783S} are derived from increasing sample sizes by an order of magnitude. This $\sim$5$\%$ threshold prevents a ``death by a thousand cuts" scenario where all the tuned cadence parameters produce individually small impacts on the metrics but when combined cause a significant reduction in solar system science. This constraint also buffers against any future unexpected small observing time losses. We have provided written feedback to the SCOC identifying which cadence simulations pass or fail our criteria (are ``good" or ``bad" for solar system science).  In this paper, we will not identify every simulation that fails our thresholds as this can be readily identified using the relevant figures within the following sections and the MAF output. Instead, we focus on examining the trends in the solar system metrics as each knob is turned and providing recommendations based on this analysis.

\subsection{Survey Footprint}
\label{sec:footprint}

\begin{figure}
\begin{center}
\includegraphics[width=0.82\columnwidth]{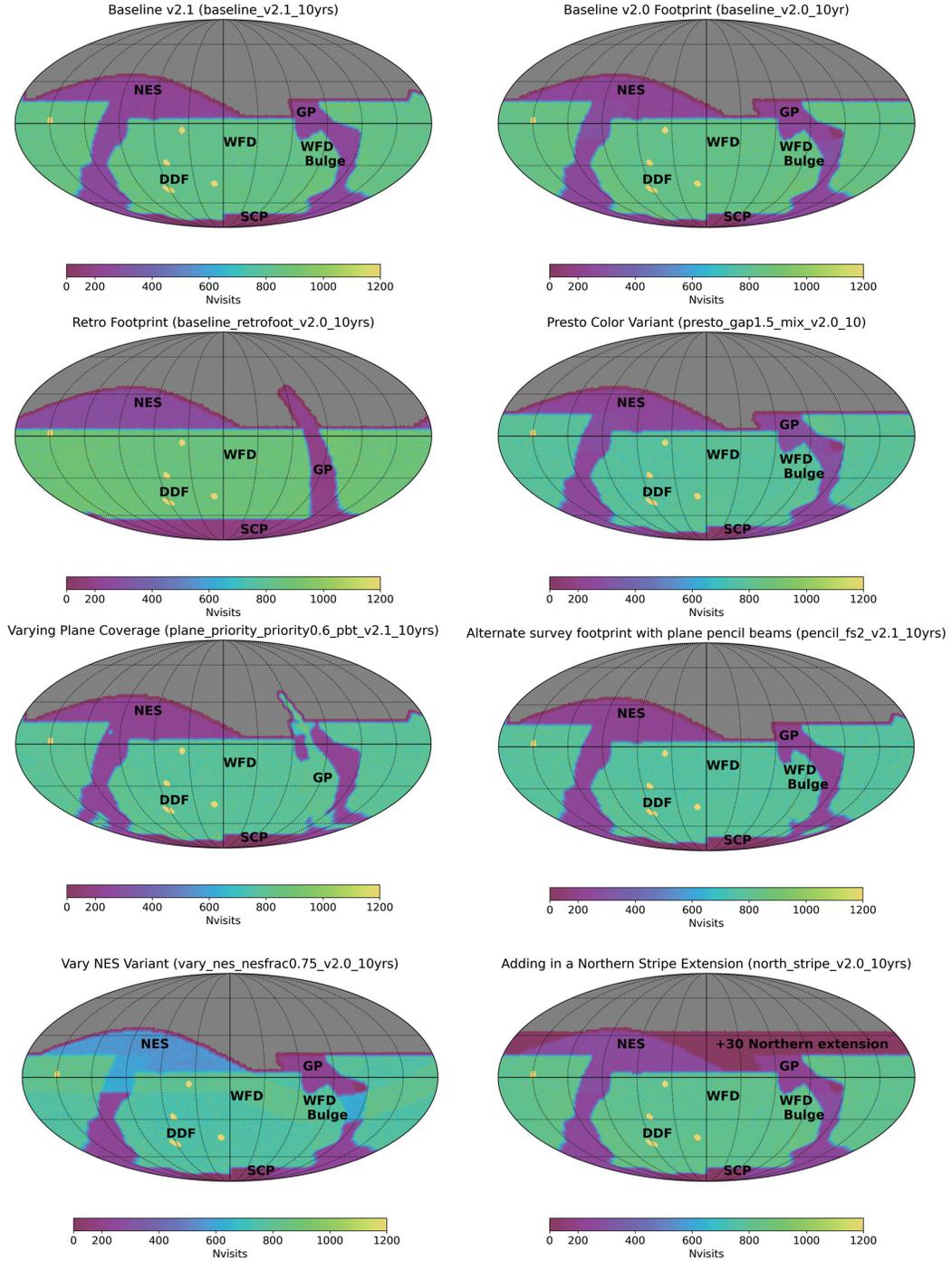}
\caption{The total number of visits in all filters after 10 years for a representative sample of LSST simulations. In several of these scenarios the effective on-sky footprint of the Wide-Fast-Deep (WFD) survey and other observing areas, including the  NES (Northern Ecliptic Spur), Galactic Plane (GP), and South Celestial Pole (SCP) regions, change depending on how the observing time on-sky is divided. The Deep Drilling Fields (DDFs) are also visible as a collection of single \update{fields} receiving a higher number of observations than a WFD pointing, with each DDF receiving approximately $1\%$ of the total LSST observing time. \update{The plots are centered on $\alpha$=0 and $\delta$=0. Right ascension and declination lines are marked every 30$^\circ$.}  \label{fig:footprints}}
\end{center}
\end{figure}

The LSST footprint determines what sky is observed over the ten year survey and how the total number of on-sky visits gets \update{apportioned across} the major components of the LSST. Examples can be seen in Figure \ref{fig:footprints} which depicts a representative set of footprints explored in the v2.0-v2.1 simulations. In this section, we focus on the arguments for incorporating the northern ecliptic region into the LSST with the Northern Ecliptic Spur (NES) mini-survey. We also examine the amount of \update{observing time that} should be divided between the Galactic plane and NES mini-surveys and options for the shape and extent of the WFD footprint. Later sections will discuss variations on how these visits are executed such as how they are distributed over time (Sections \ref{sec:visits} and \ref{sec:rolling}) and by filter (Section \ref{sec:filter_dist}). Small modifications to the footprint using much less than a few percent of the observing time are presented in the micro-surveys discussion in Section \ref{sec:all_micro-surveys}. 

\subsubsection{Northern Ecliptic Spur (NES)}
\label{sec:NES}

\begin{figure}
\begin{center}
\includegraphics[width=0.6\columnwidth]{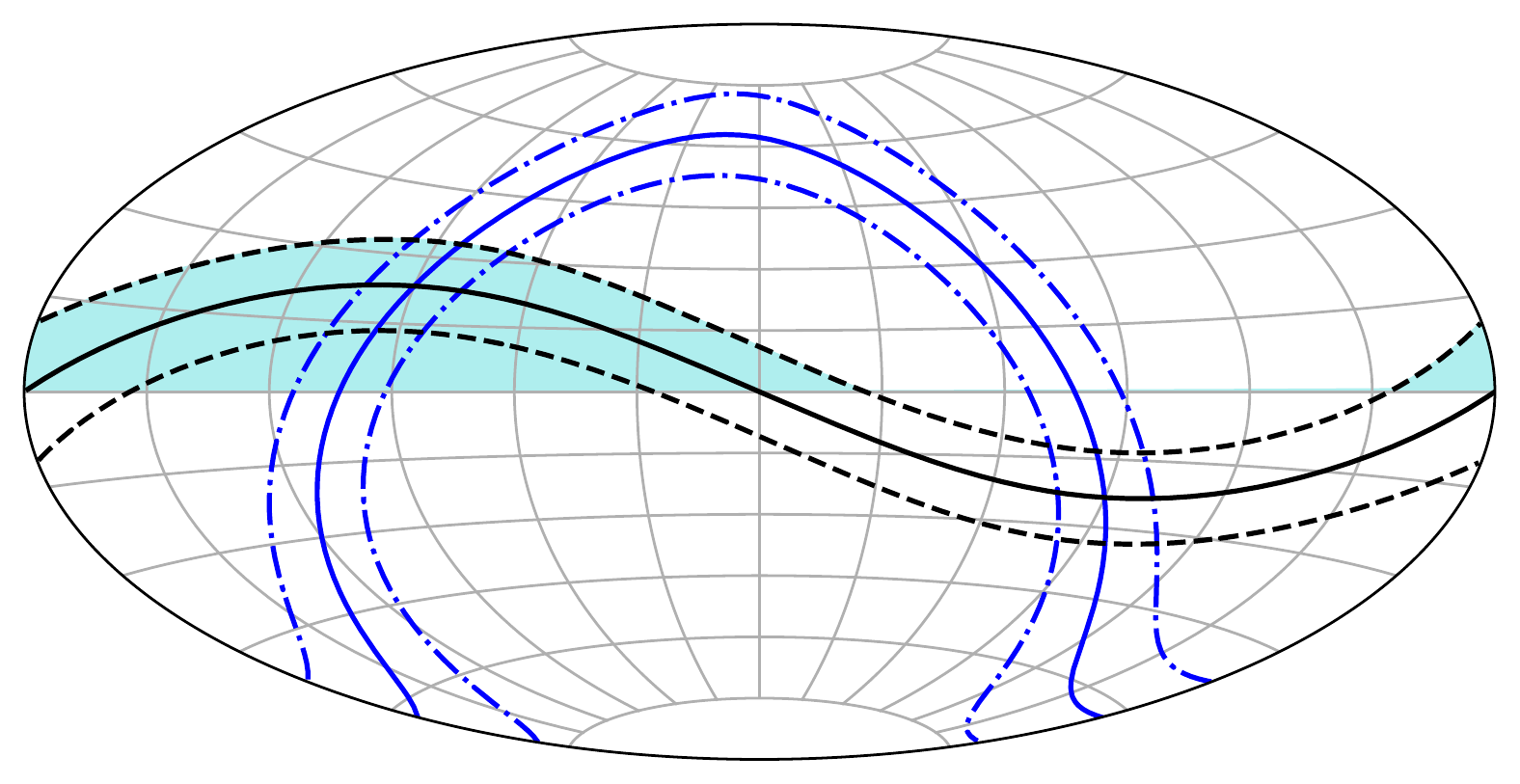}
\caption{The footprint of the Northern Ecliptic Spur (NES) in equatorial coordinates. The light blue represents the pointings requested as part of the NES. The solid black line represents the ecliptic. The dashed black lines represent $\pm$ 10$^{\circ}$ ecliptic latitude. The solid blue line plots the center of the Galactic plane. The dashed blue lines mark $\pm$ 10$^{\circ}$ galactic latitude. \update{The plot is centered on $\alpha$=0 and $\delta$=0. Right ascension is marked every 30$^\circ$ and declination lines are visible every 15$^\circ$ up to and including $\pm$75$^\circ$.} \label{fig:NES}}
\end{center}
\end{figure}

The WFD by its design requirements is meant to cover the majority of the sky in the southern celestial hemisphere below 0$^{\circ}$ declination \citep{lsstSRD}, but the Simonyi Survey Telescope is capable of observing the entire ecliptic. The extent of the WFD has evolved over time (as shown in Figure \ref{fig:footprints} and later discussed in Section \ref{sec:WFD_north}), but no matter what the proposed variations to the WFD sky coverage are, the full extent of the ecliptic plane will not be incorporated into the WFD footprint. The NES mini-survey aims to remedy this situation by ensuring higher airmass observations of the northern ecliptic are taken as part of the LSST \citep{2009arXiv0912.0201L, 2017arXiv170804058L, 2018arXiv181201149S, 2019ApJ...873..111I,2022ApJS..258....1B}. The NES region, shown in Figure \ref{fig:NES}, is \update{comprised} of $\sim$604 pointings covering in total $\sim$5800 deg$^2$ spanning from 0$^{\circ}$ declination to +10$^{\circ}$ ecliptic latitude. In order to make this goal achievable with the non-WFD time, the NES mini-survey has typically been implemented in the cadence simulations to receive a smaller number of visits per field ($\sim$250; as shown in Figure \ref{fig:footprints}) compared to the WFD. The NES mini-survey includes observations taken in a combination of the $griz$ filters only, where solar system objects are typically the brightest. The observing time dedicated to the NES is explored in Section \ref{sec:no-wfd}; here we focus on the impact of including or excluding the NES mini-survey from the LSST.  

\begin{figure}
\begin{center}
\includegraphics[width=0.94\columnwidth]{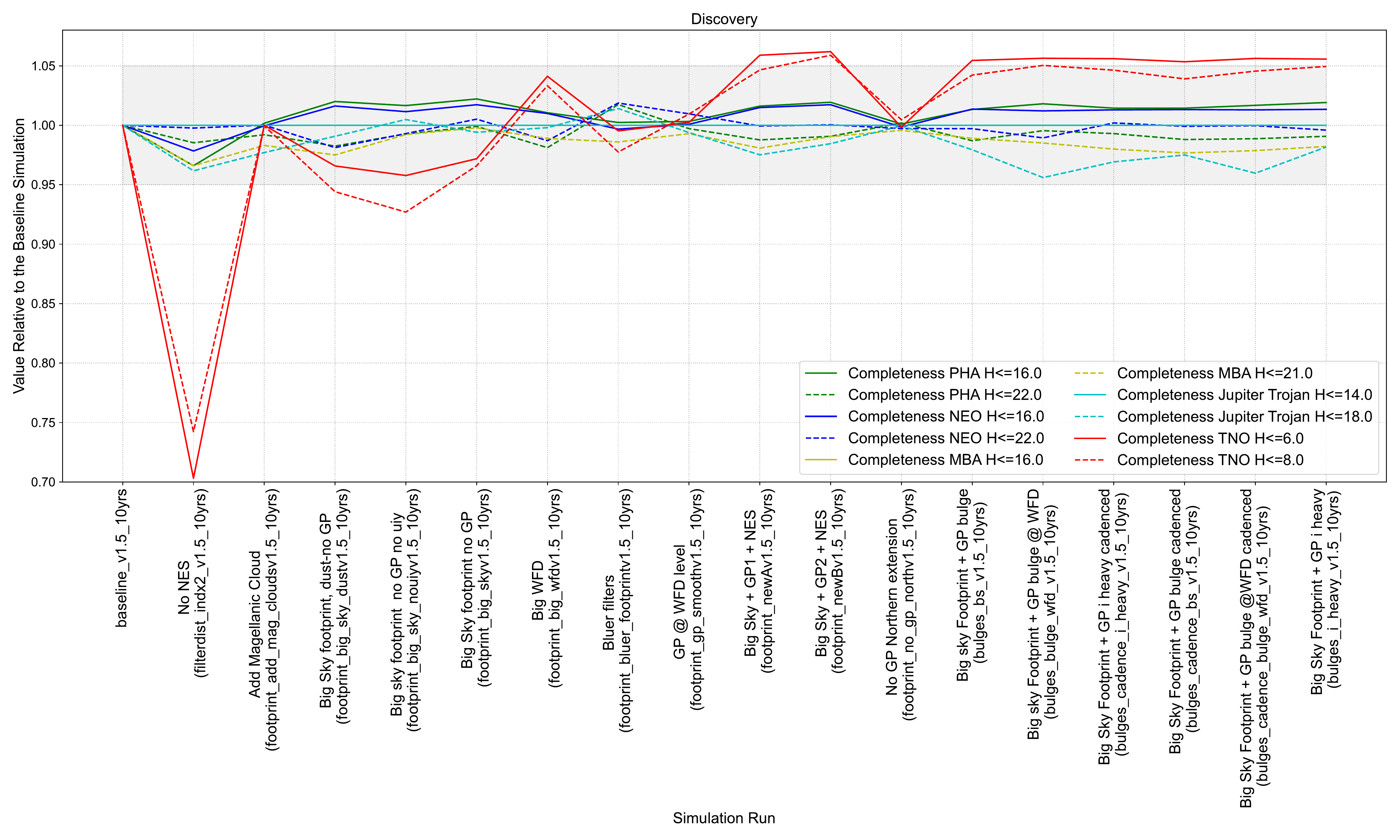}
\includegraphics[width=0.94\columnwidth]{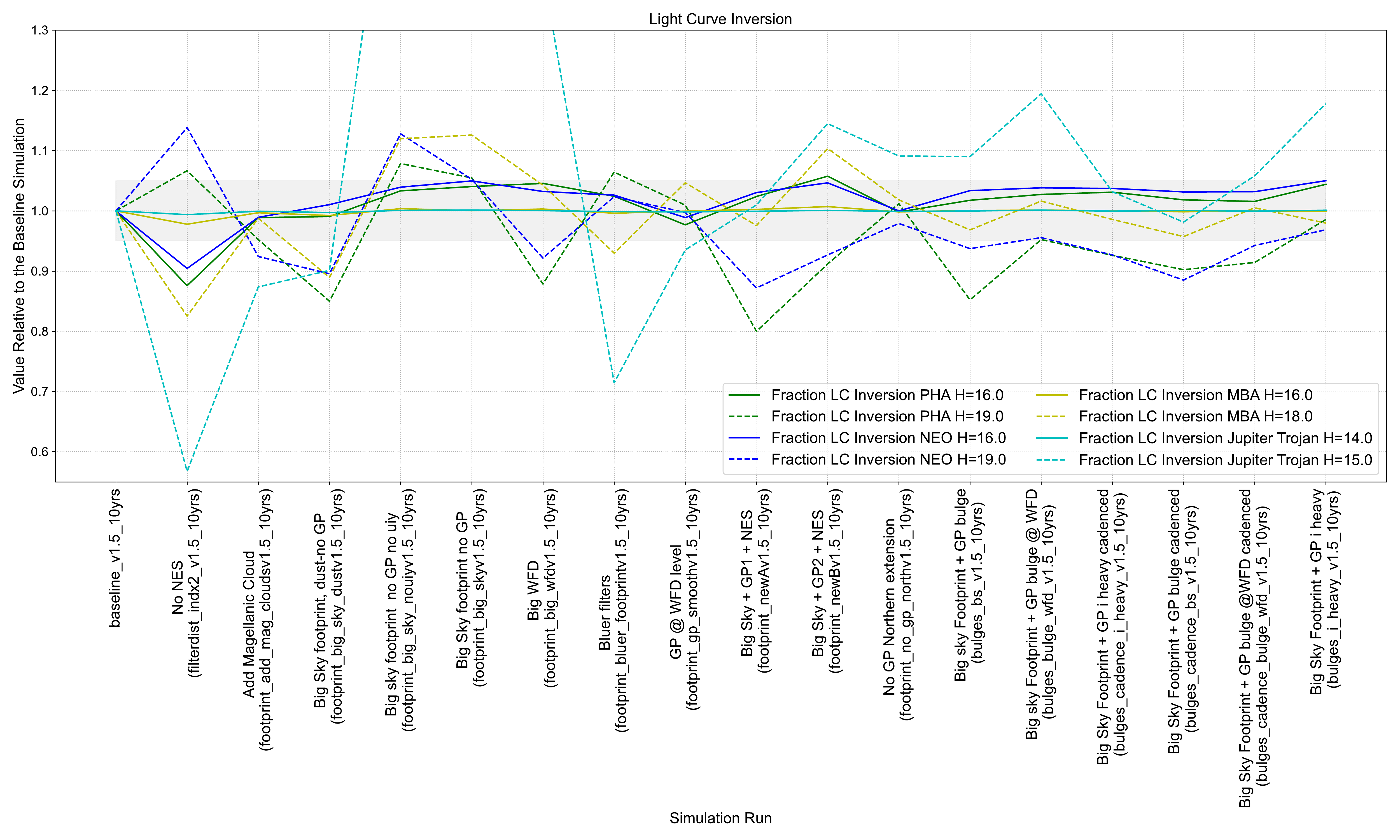}
\caption{Possible tuning options for the LSST footprint from the v1.5 experiments. \update{The baseline (reference) simulation with the default scheduler configuration} for this cadence experiment is the first entry on the left. All values have been normalized by this simulation's output. The gray shading outlines changes that are within $\pm5\%$ of the baseline simulation. Top: Discovery Metrics. Bottom: Light Curve Inversion Metrics. We have truncated the bottom plot's y-axis for visibility. The change in the $H$=15 Jupiter Trojan detections in some of the runs extends well above 1.2. \label{fig:v1.5_footprint}}
\end{center}
\end{figure}

The NES mini-survey is crucial for inventorying the outer Solar System. About half of the ecliptic plane is covered within the WFD footprint. Over the 10 year span of the LSST, inner solar system populations like the MBAs will complete full orbits. This means that MBAs in the northern hemisphere at the start of the survey will be in favorable positions to be detected within the WFD during the later years of the survey. This is not true for outer solar system objects whose orbital periods are well beyond $\sim$10 years. Outer solar system bodies will only have a small fraction of their orbital periods covered by the LSST.  Thus, the vast majority of TNOs located in the NES at the start of the survey will remain in the northern hemisphere, missing the WFD footprint.  This is reflected in Figure \ref{fig:v1.5_footprint}, where the first two simulations plotted are \texttt{baseline\_v1.5\_10yrs}, which includes the NES mini-survey, and  \texttt{filterdist\_indx2\_v1.5\_10yrs} which excludes the NES. TNO discoveries suffer nearly a 30$\%$ loss with the exclusion of the NES mini-survey while there is only a very small drop for the inner solar system populations. Although not simulated at the time in this cadence experiment, populations that are more uniformly distributed on the sky (such as OCCs and ISOs) also benefit from the inclusion of the NES which provides additional sky coverage and therefore more chances for discovery.

Figure \ref{fig:v1.5_footprint} also shows that the light curve metrics for small MBAs suffer a bit more than a 15$\%$ loss when the NES mini-survey is not executed. Discovery relies on the object being above the 5$\sigma$ limiting magnitude on 3 nights, but to perform light curve inversion requires many more observations. The NES provides additional opportunities to observe those faint objects close to the LSST limiting magnitude, giving additional chances for the small MBAs to be observed in conditions where they might have sufficient SNR to contribute to shape modeling. The opposite effect is observed for the small PHAs and NEOs which benefit in simulations without the NES mini-survey (about a 6-10$\%$ increase in the light curve inversion metrics). These objects are typically detected close to Earth and so quickly become too faint to be detected. Thus, pushing the time used for the NES mini-survey into additional WFD visits enables more observations where these small PHAs and NEOs are detectable and can have light curves measured. The opposite effect can be seen for the larger PHAs and NEOs. The larger PHAs and NEOs suffer a $\sim$10$\%$ drop in the light curve metrics when the NES fields are excluded. Because large PHAs/NEOs are more likely to be above the limiting magnitude in a LSST image, surveying the NES creates new opportunities to monitor the brightness of large PHAs/NEOs as they pass by Earth. The Jupiter Trojans also take a significant hit when the NES is not included. This is likely due to their constrained positions on the sky.

\begin{figure}
\begin{center}
\includegraphics[width=0.87\columnwidth]{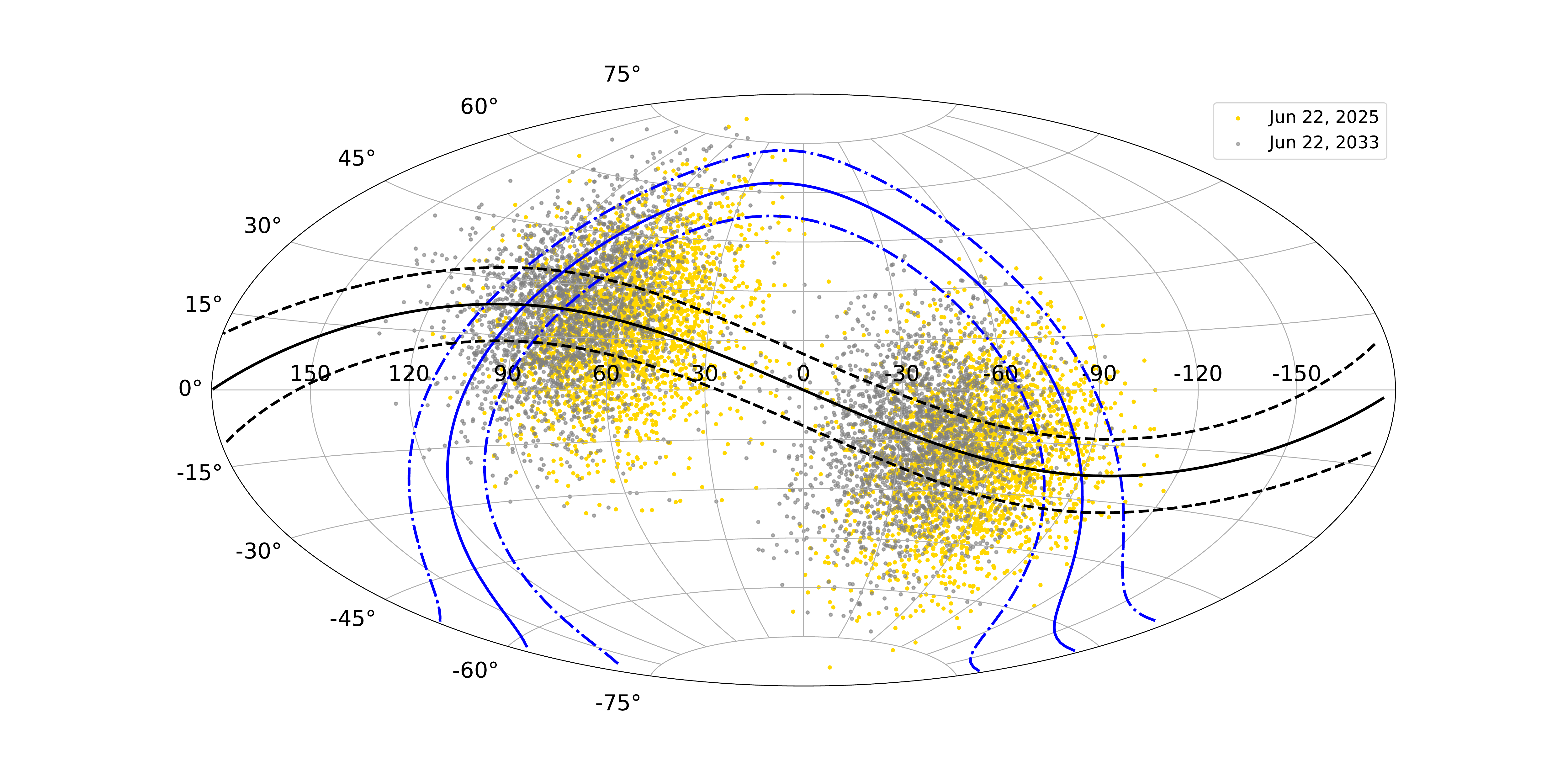}
\caption{\label{fig:Neptune_Trojans} The sky positions of the Neptune Trojan population from the \cite{2021Icar..36114391L} model on 2025 June 22 (gold) and 2033 June 22 (gray). The two epochs represent early and late times in the LSST, respectively. The leading L4 cloud is at north, left-hand side of the plot. The dashed black lines represent $\pm$ 10$^{\circ}$ ecliptic latitude. The solid blue line plots the center of the Galactic plane. The dashed blue lines delineate $\pm$ 10$^{\circ}$ galactic latitude. Probing the low inclination L4 Neptune Trojans requires the inclusion of the NES into the LSST footprint.}
\end{center}
\end{figure}

Not captured in the MAF metrics are the benefits that the NES mini-survey provides to small body populations that are distributed asymmetrically across the sky. They would only be partially sampled without the NES observations, as discussed in \cite{2018arXiv181201149S}. Two such cases are the Neptune Trojans and the resonant TNO populations. Over the ten-year period, the vast majority of the leading Neptune Trojan L4 cloud is accessible only via observations of the NES as shown in Figure~\ref{fig:Neptune_Trojans}. \cite{2019Icar..321..426L} find evidence for potential differences in the color distributions of the L4 (leading) and L5 (trailing) Neptune Trojans. The WFD and NES mini-survey combined are capable of sampling both clouds with sufficiently large numbers to test this further. Including the NES region in the LSST footprint enables characterization of the libration islands for the various mean motion resonances (MMR) with Neptune \citep{2018arXiv181201149S}. Only observing half the ecliptic with just the WFD and Galactic plane mini-survey would impact the study of the resonant TNO populations which preferentially come to perihelion at certain locations on the sky \citep[e.g., ][]{2012AJ....144...23G, 2021ARA&A..59..203G}. 

\begin{figure}
\begin{center}
\includegraphics[width=0.50\columnwidth]{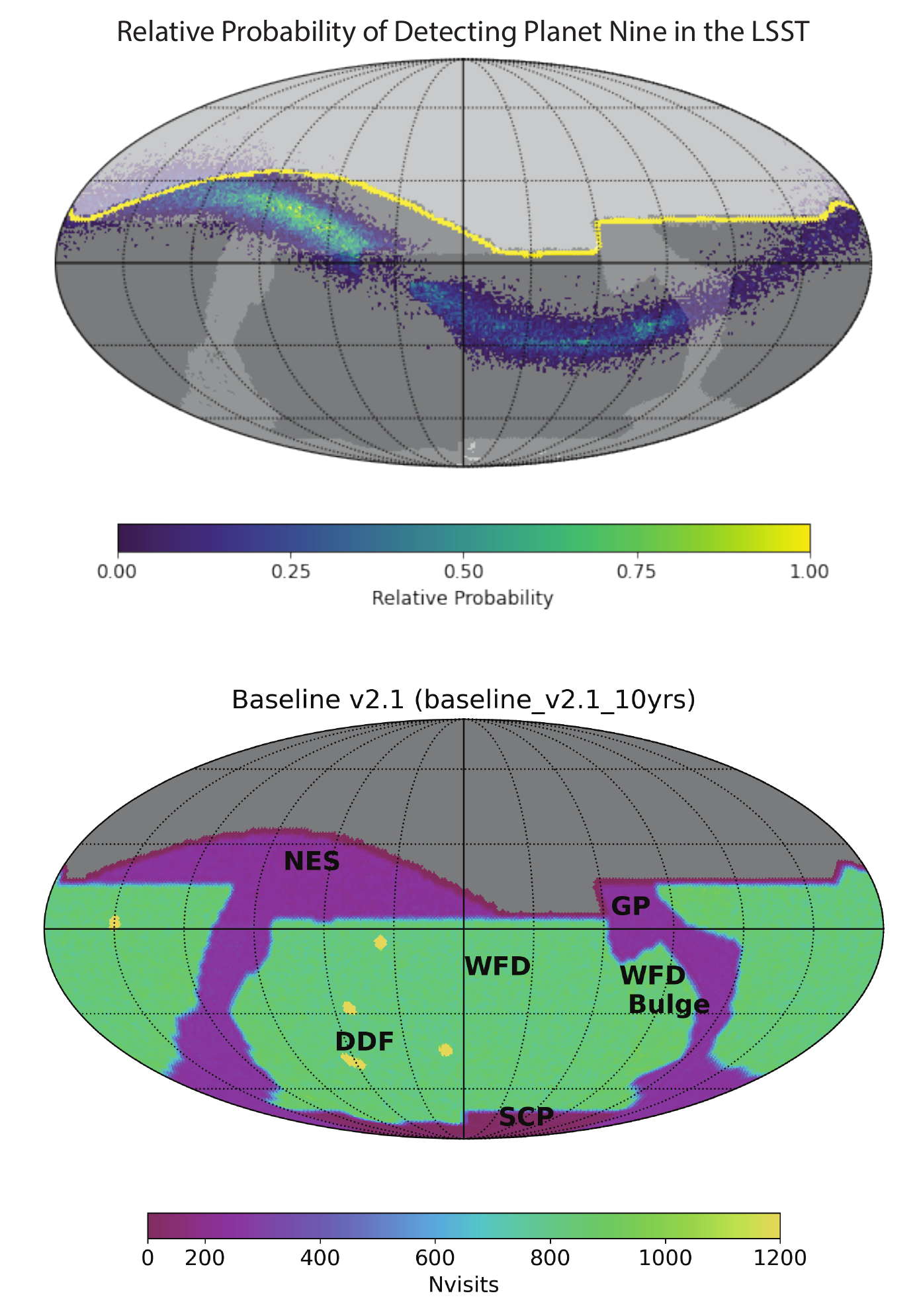}
\caption{The simulated probability of \update{Planet} Nine (top) compared to the possible number of visits in the LSST footprint from the \texttt{baseline\_v2.1\_10yrs} simulation (bottom). The Planet Nine probability density is taken from \cite{Brown_2022} which is based on 100,000 synthetic orbits and physical properties of Planet Nine (including on-sky locations and $V$-band apparent magnitudes) drawn from the distributions developed in \cite{2021AJ....162..219B}, where we have removed the ones that are flagged as being ruled out by constraints from the Zwicky Transient Facility (ZTF) \citep{2022AJ....163..102B} and the Dark Energy Survey (DES) \citep{2022ApJS..258...41B, 2022AJ....163..216B}.  The top figure has the LSST footprint shaded by the number of observations that reach 5$\sigma$ limiting magnitude of 24 in any filter. The most probable locations of Planet Nine are within the NES region, but the full LSST footprint is required to search and probe the majority of the \cite{2021AJ....162..219B} predicted Planet Nine parameter space.  \update{The plots are centered on $\alpha$=0 and $\delta$=0. Right ascension and declination lines are marked every 30$^\circ$.}\label{fig:P9_orbits}}
\end{center}
\end{figure}

\begin{figure}
\begin{center}
\includegraphics[width=0.9\columnwidth]{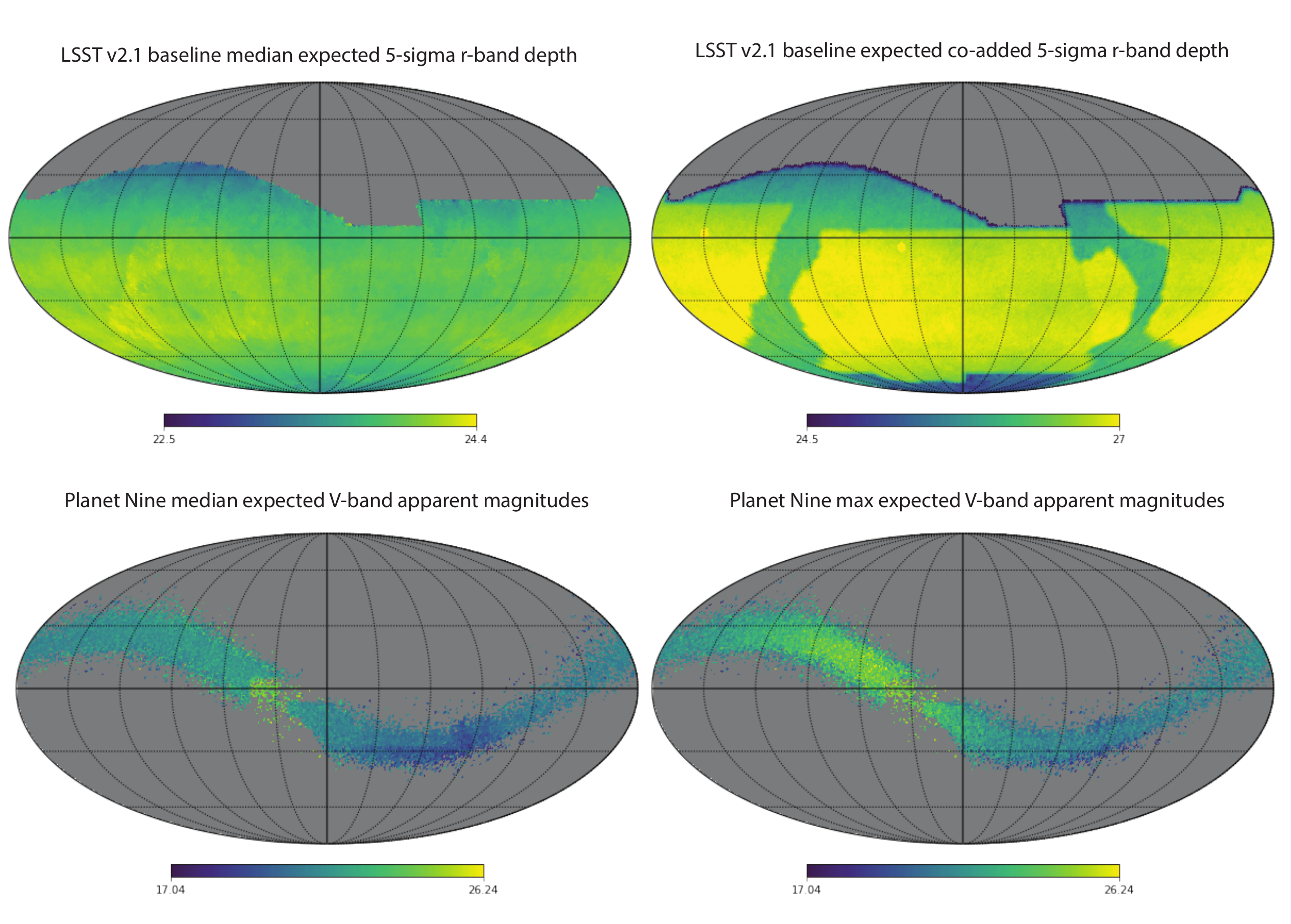}
 \caption{The sky-magnitude parameter space for a simulated  Planet Nine (bottom) compared to the possible LSST sky coverage and limiting magnitudes from the \texttt{baseline\_v2.1\_10yrs} simulation (top). \update{ The bottom left shows the median expected V-band apparent magnitude, and the bottom right has the maximum expected V-band magnitude from the distribution of  Planet Nines, as described in Figure \ref{fig:P9_orbits}.} The LSST individual exposure median (top left) and 10-year coadded (top right) $r$ magnitude depths per pointing in the survey footprint. Since the optical color of the potential Planet Nine is not constrained, we do not apply any $V-r$ color to allow for multiple comparisons depending \update{on} the reflectance model preferred by the reader. Observing the NES with Rubin Observatory is crucial for testing and constraining the Planet Nine parameter space. \update{The plots are centered on $\alpha$=0 and $\delta$=0. Right ascension and declination lines are marked every 30$^\circ$.}  \label{fig:P9_mags}}
\end{center}
\end{figure}

The NES mini-survey is crucially important for searching for additional distant planets in the Solar System and testing the apparent orbital clustering of Sedna-like  Inner Oort Cloud objects (IOCs; $q > $ 50 au and $a > $250 au) and extreme TNOs (ETNOs; objects on orbits with $q >$ 42 au and $a > $150 au). It has been proposed that a giant planet (``Planet Nine") is gravitationally shepherding the distant planetesimals onto similar orbits with aligned orbital poles and longitudes of perihelion \citep{2014Natur.507..471T, 2016AJ....151...22B,2016AJ....152..221S, 2019AJ....157...62B, 2019PhR...805....1B, 2021AJ....162..219B, 2021AJ....162...39O}. Recent modeling by \cite{2021AJ....162..219B} combined with constraints from the Zwicky Transient Facility (ZTF) \citep{2022AJ....163..102B} and the Dark Energy Survey (DES) \citep{2022ApJS..258...41B, 2022AJ....163..216B} predict Planet Nine to be residing at a semimajor axis of 700 au or higher. Although the current predictions made available in \cite{Brown_2022} do have Planet Nine distributed over a wide range of ecliptic longitudes, the most likely location of Planet Nine is close to the region where the Galactic plane intersects the northern ecliptic (see Figure \ref{fig:P9_orbits}). The bulk of the predicted Planet Nine sky locations are within the LSST footprint as implemented in the \texttt{baseline\_v2.1\_10yrs} simulation which includes the NES mini-survey. Figure \ref{fig:P9_mags} presents the estimated on-sky $V$-band apparent magnitude distribution for Planet Nine from \cite{Brown_2022} and the predicted LSST $r$-band limiting magnitudes from the \texttt{baseline\_v2.1\_10yrs} run. Over a wide range of possible $V-r$ colors, Planet Nine could potentially be bright enough to be visible with LSST images. If Planet Nine is bright enough to be imaged in single LSST exposures, the length of the survey in combination with the repeated coverage of the survey footprint effectively eliminates the possibility of missing Planet Nine in high galactic latitude fields due to coincidental overlap with another source. Closer to the galactic plane, stellar crowding will be significant and identifying sources will be difficult. This may require community-optimized search algorithms to look for Planet Nine in these observations. Rubin Observatory is also exploring additional options to enhance source extraction near the Galactic plane (see \cite{dmtn-129}). \update{If current searches fail to find Planet Nine,} Rubin Observatory will put the best observational constraints on the existence of Planet Nine over the next decade, and \update{will be} the facility with the best chance of directly \update{imaging }it \citep{2018AJ....155..243T}. \update{As noted in Section \ref{sec:metric_limitations}, Rubin Observatory's SSP pipelines are only sensitive to moving objects at heliocentric distances $\lesssim$ 100-150 au. We fully expect that there will be several community-led efforts to find very slow moving distant objects in the LSST transient catalogs to search for Planet Nine and explore the IOCs and ETNOs. Therefore, it is still important to consider this science case for LSST footprint considerations.}

Even if Planet Nine is not \update{visible in the LSST images}, the LSST would potentially be able to reveal its presence if the orbital alignment holds with the increased LSST sample of ETNOs and IOCs and matches the Planet Nine predictions. Whether or not the Planet Nine theory is correct, the distant IOCs and ETNOs are an important probe for studying the origin and evolution of the very distant outer solar system and testing alternatives to the Planet Nine theory \citep{2004AJ....128.2564M, 2006Icar..184...59B, 2006ApJ...643L.135G, 2011Icar..215..491K, 2012Icar..217....1B, 2020AJ....160...50Z, 2022A&A...662L...4E,2022ApJ...938L..23H}. Observing across the ecliptic will be crucial for creating a large enough sample to alleviate the challenging observational biases currently dealt with when combining the multiple datasets previously used to identify and test the apparent orbital clustering  \citep{2016ApJ...824L..23B, 2019AJ....157...62B, 2017AJ....154...50S, 2020PSJ.....1...28B, 2021PSJ.....2...59N}. 

\subsubsection{Extending the Wide-Fast-Deep (WFD) Footprint Northward}
\label{sec:WFD_north}

The NES mini-survey (as described in Section \ref{sec:NES}) was proposed when the northern limit of the WFD footprint was initially set to be +2$^{\circ}$ declination (see the \texttt{baseline\_retrofoot} simulation in Figure \ref{fig:footprints}). The originally planned WFD sky coverage used a simple cut in Galactic coordinates to identify the boundary of the WFD with the Galactic plane/bulge observing region \citep{2017arXiv170804058L, 2019ApJ...873..111I,jones_r_lynne_2020_4048838}. Combining this boundary with the sky coverage requirements and visit constraints for the WFD set the original northern declination limit. What sky is included within the WFD is a changeable LSST survey parameter, as long as the SRD requirements for the WFD survey area are met; at least 18,000 deg$^2$ with a median of 825 visits per field \citep{lsstSRD}. For extragalactic science, including cosmology and galaxy studies, and galactic science, such as the study of the Milky Way's structure, there have been proposals from the community requesting more of the WFD to be shifted to low-extinction and less crowded sky \citep{2018arXiv181200515L,2018arXiv181202204O,2022ApJS..259...58L}. Other arguments have also been raised for shifting the WFD footprint further northward, including overlap with future DESI \citep[Dark Energy Spectroscopic Instrument;][]{2022arXiv220510939A} and \emph{Nancy Grace Roman Space Telescope} observations \citep{2018arXiv181202204O,2019arXiv190410438C}. As this is a zero-sum game, visits are taken from near the Galactic plane with high stellar crowding and dust extinction and redistributed northward above 0$^{\circ}$ declination. This results in a  fraction of the WFD survey now covering the NES region as seen in the \texttt{Baseline v2.0} and  \texttt{v2.1} footprints shown in Figure \ref{fig:footprints}.  The SCOC has made the recommendation to use this new footprint as shown in Figure \ref{fig:footprints} extending the WFD northward although the final declination limits of the WFD and the exact boundary of the galactic/high dust extinction region can still be fine-tuned \citep{SCOC_Report_1}. As implemented in \texttt{baseline\_v2.0\_10yrs} simulation, the revised WFD footprint has two declination boundaries spanning from -72$^{\circ}$ to +12$^{\circ}$ declination with an interstellar dust extinction cutoff at approximately $E(B-V)$ = 0.2 mag or $A(V)=0.6$ mag \citep{SCOC_Report_1,SMTN-017}, where $E(B-V)$ is \update{the} dust reddening in magnitudes and $A(V)$ is the total $V$-band extinction. The northern boundary of the WFD varies with \update{right ascension} in this revised northward footprint; this is partly due to other additional constraints with the scheduler. We note that \texttt{baseline\_v2.1\_10yrs} simulation uses the same footprint as v2.0 but incorporates the Virgo cluster \update{($\alpha$=12 hrs, $\delta=$+12$^\circ$)}into the WFD \citep{SMTN-017}. 

Expanding the WFD footprint northward will cover part of the NES for ``free" with the time charged to the WFD time allocation, but part of the redistributed pointings in the 2$^{\circ}$ to 12$^{\circ}$ declination band is at high ecliptic latitude because part of the ecliptic plane crosses the Galactic plane in the southern hemisphere. Transferring WFD visits from the Galactic bulge region will reduce the number of photometric data points available for generating rotational light curves for some MBAs within the bulge, but how significant the impact will depend on the exact shape of the footprint. The OCCs, NEOs, and PHAs are distributed across a wide range of ecliptic latitudes, so observations at higher ecliptic latitudes will still find small bodies in these populations. \update{The} same arguments that hold for outer solar system objects in Section \ref{sec:NES} also apply in this case. \update{Assuming a 14 February 2025 start date, Neptune's on-sky position will have changed by about 1 hr in right ascension and 8$^\circ$ in declination by the end of LSST observations. Objects beyond 30 au will be moving slower than Neptune. Most of the TNOs and IOCs located in the NES at the start of the survey will remain in the NES throughout the duration of the LSST. As these distant objects do not move very far on-sky during the 10-year survey, any observations of the NES are beneficial for discovery as long as not too much time is taken away from near ecliptic pointings in the southern hemisphere.}

\begin{figure}
\begin{center}
\includegraphics[width=0.96\columnwidth]{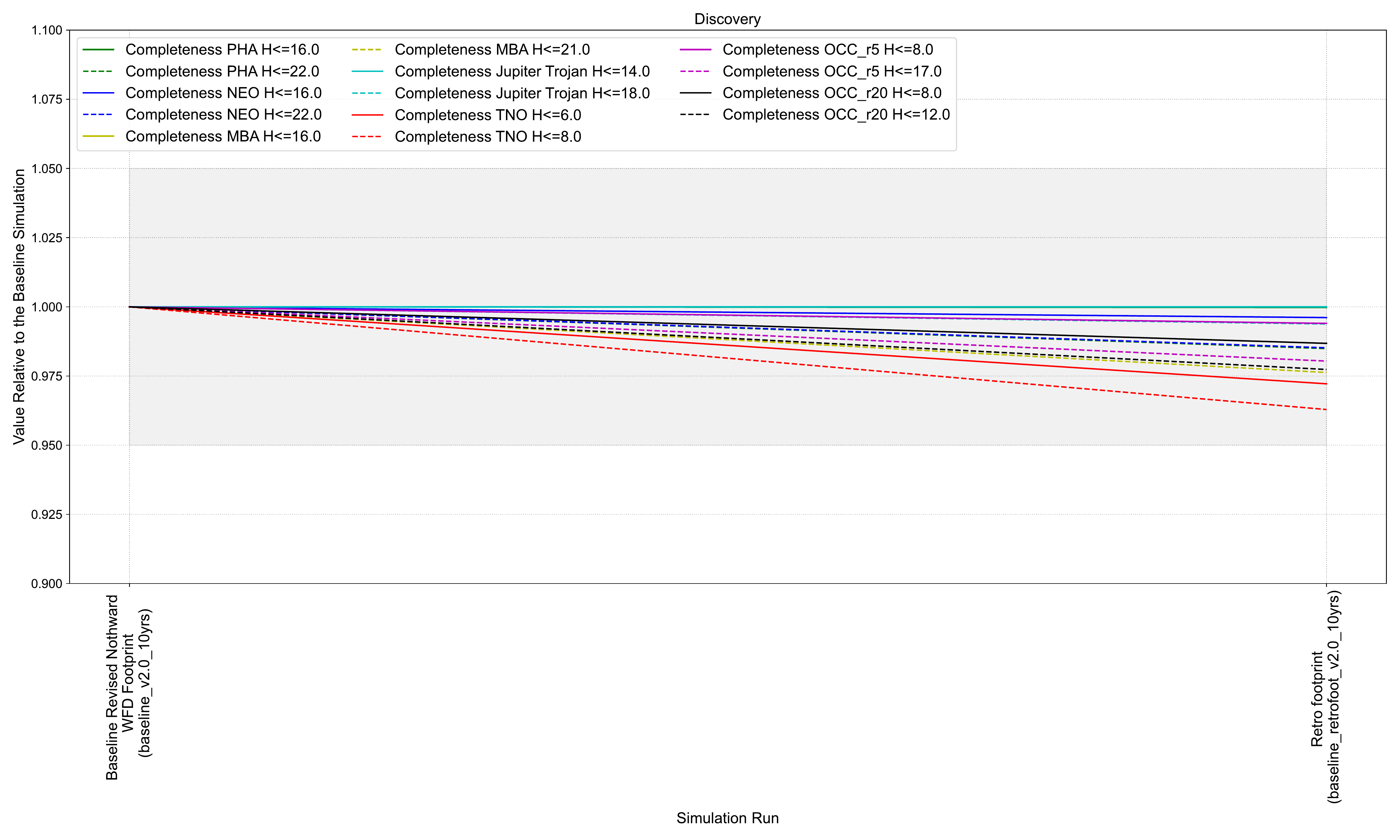}
\includegraphics[width=0.96\columnwidth]{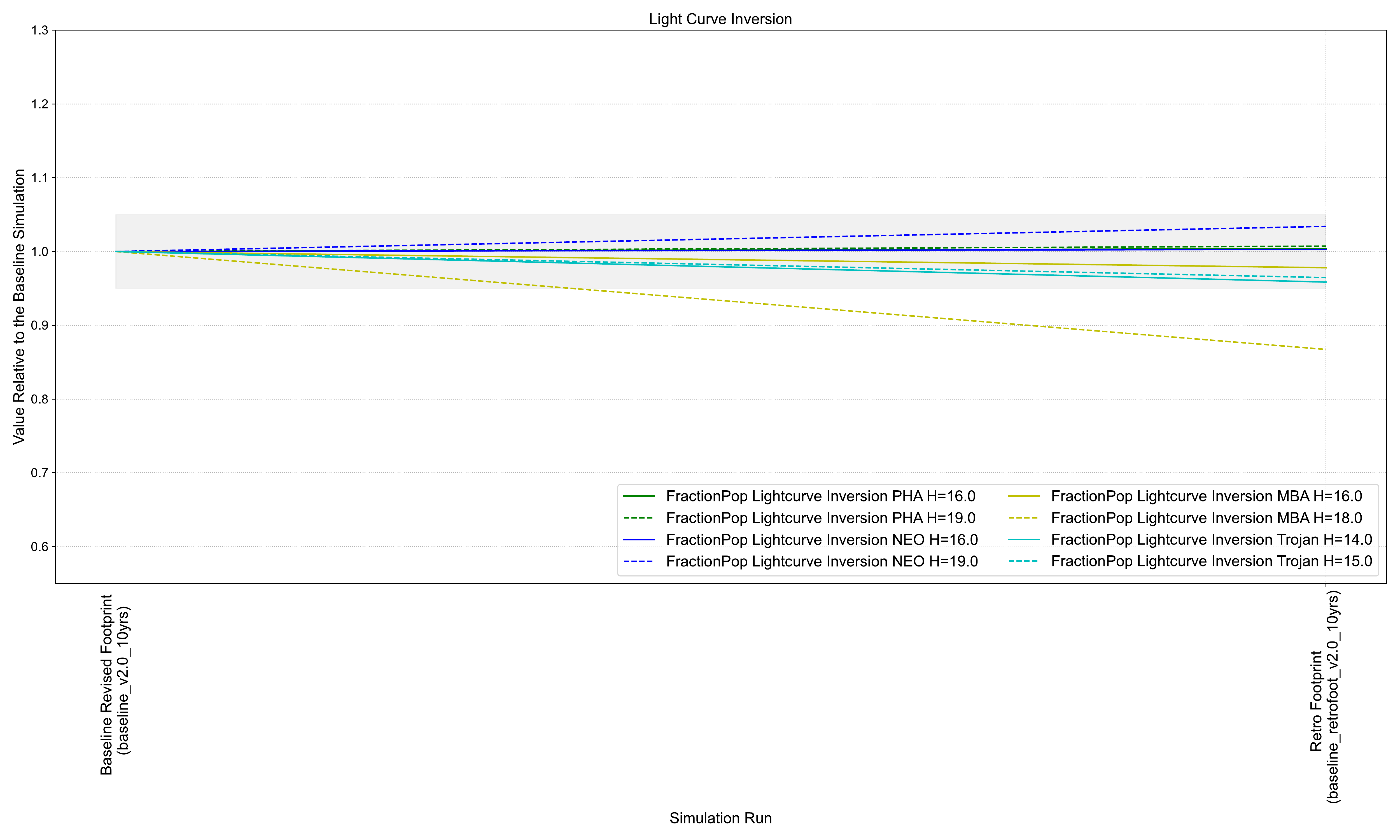}
\caption{Impact of the revised v2.0 LSST footprint with the northward and dust-extinction limited WFD. \update{The baseline (reference) simulation with the default scheduler configuration} for this cadence experiment is the first entry on the left. All values have been normalized by this simulation's output. The gray shading outlines changes that are within $\pm5\%$ of the baseline simulation. Top: Discovery Metrics. Bottom: Light Curve Inversion Metrics.\label{fig:v2.1_2.0_footprint}}
\end{center}
\end{figure}

In Figure \ref{fig:v2.1_2.0_footprint}, we evaluate the impact of the new northward WFD sky coverage, comparing \texttt{baseline\_v2.0\_10yrs} and \texttt{baseline\_retrofoot\_v2.0\_10yrs} simulations. All simulations predating the v2.0 simulations start from a variation of WFD with the old +2$^{\circ}$ declination limit. The v2.0 baseline also has additional changes to the observing cadence including tweaks to the rolling cadence implemented and the exposure time for $u$-band visits. For an apples to apples comparison, the v2.0 release includes \texttt{baseline\_retrofoot\_v2.0\_10yrs} which uses the original v1.5-1.7 baseline WFD+NES footprint leaving the other cadence parameters the same as the v2.0 simulation. The  northward WFD footprint produces a slight increase in discovery metrics (all less than $5\%$ change) for all populations except for the large Jupiter Trojans. There are also slight improvements in the light curve metrics, with the smallest MBAs seeing more than a $10\%$ increase with the extended WFD footprint. These increases may be more significant than represented in the MAF metrics if stellar crowding was taken into account. Although this comparison is to the v2.0 baseline, it will still \update{hold true} for the v2.1 baseline (\texttt{baseline\_v2.1\_10yrs}) that goes slightly more northward. We note that the addition of the Virgo Cluster is a minuscule change in area, and there are negligible impacts to any of the solar system metrics compared to \texttt{baseline\_v2.0\_10yrs} (as discussed in Section \ref{sec:other_micro-surveys}).  

\begin{figure}
\begin{center}
\includegraphics[width=0.93\columnwidth]{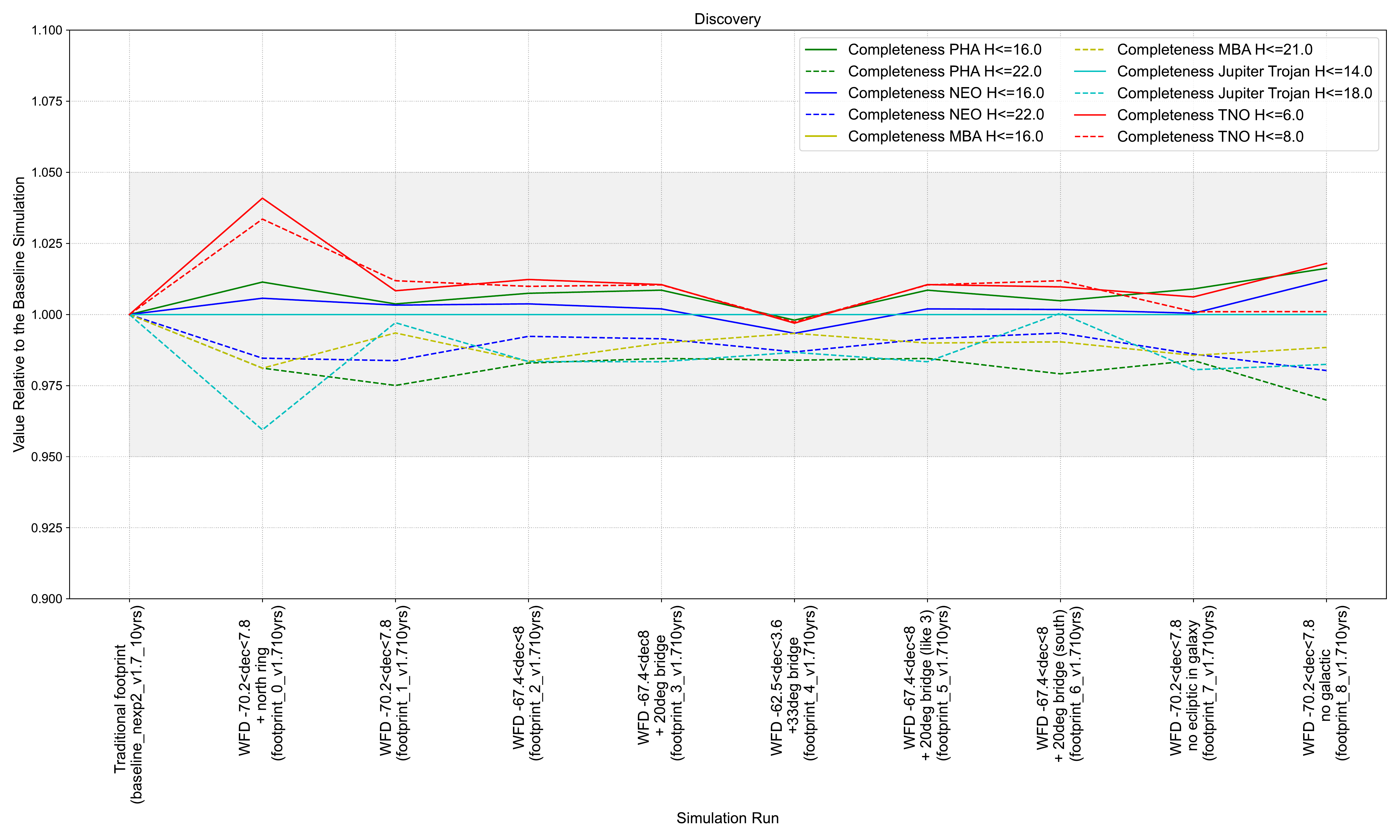}
\includegraphics[width=0.93\columnwidth]{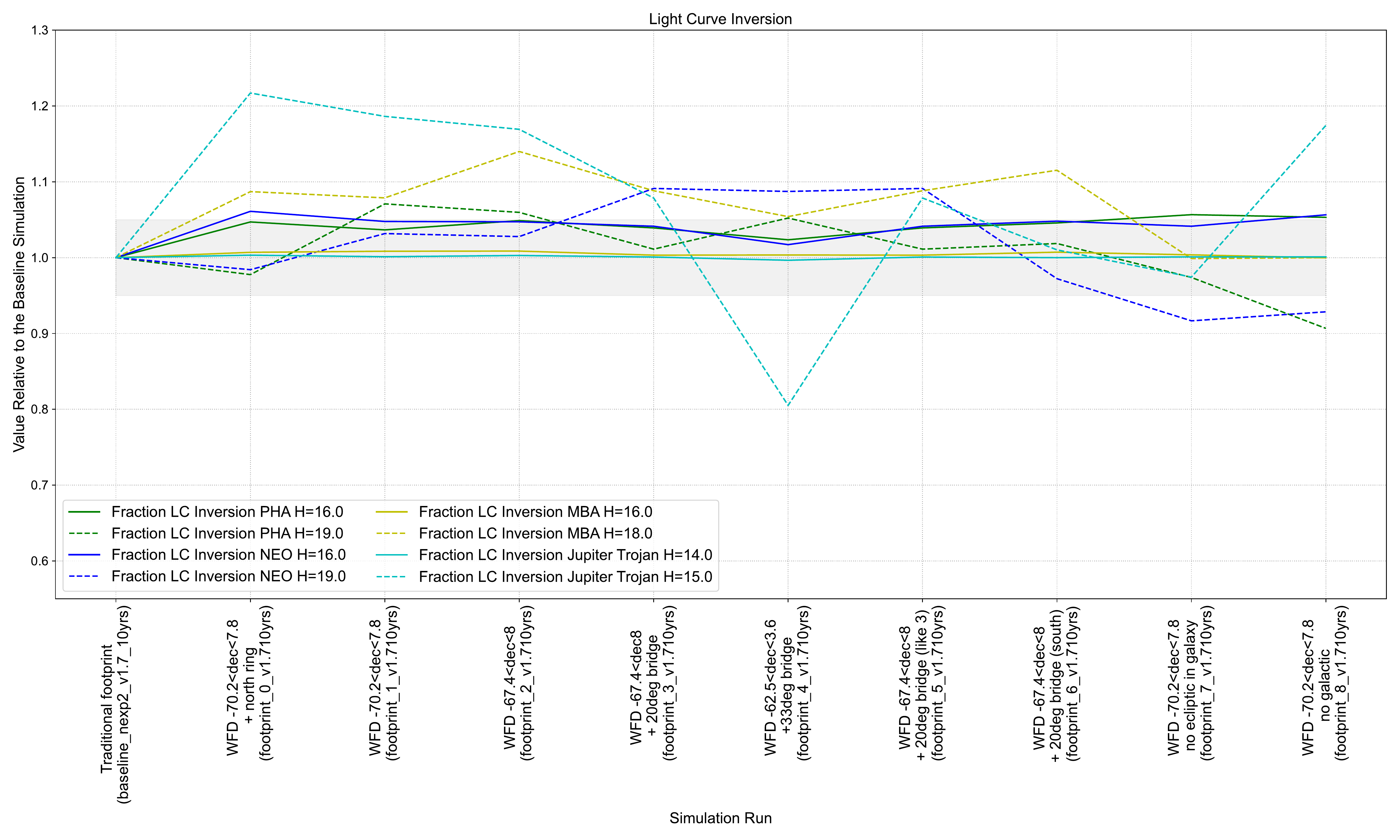}
\caption{Possible tuning options for the WFD (Wide-Fast-Deep) Survey footprint from the v1.7 experiments. As visits are taken away from the Galactic plane and bulge region, they are redistributed northward and southward to less dust extinction and less stellar crowded regions. \update{The baseline (reference) simulation with the default scheduler configuration} for this cadence experiment is the first entry on the left. All values have been normalized by this simulation's output. The gray shading outlines changes that are within $\pm5\%$ of the baseline simulation. Top: Discovery Metrics. Bottom: Light Curve Inversion Metrics.\label{fig:v1.7_footprint}}
\end{center}
\end{figure}

\begin{figure}
\begin{center}
\includegraphics[width=0.75\columnwidth]{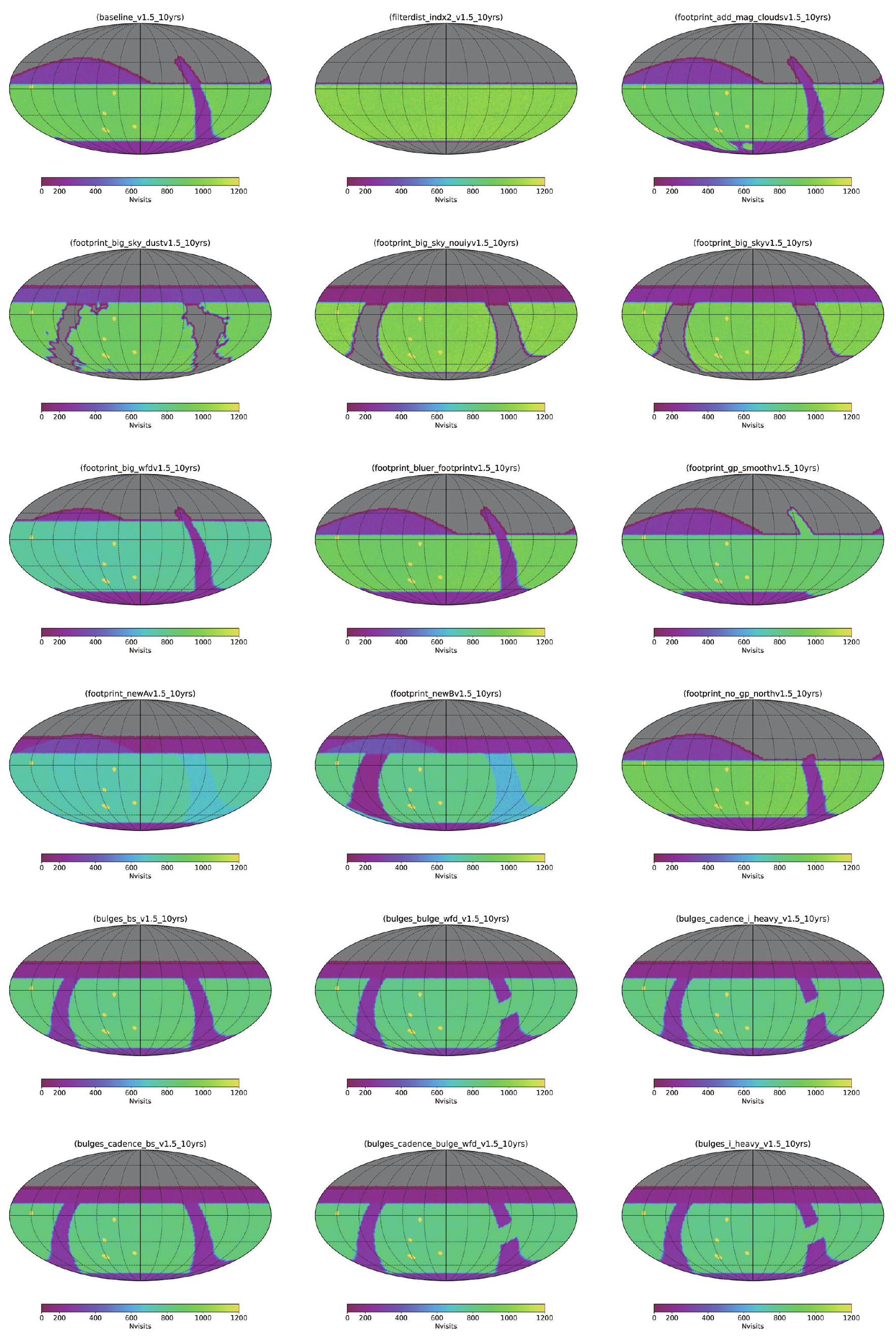}
\caption{The total number of visits in all filters after 10 years for the v1.5 footprint experiment simulations. The Deep Drilling Fields (DDFs) are also visible as \update{a collection of single fields} receiving \update{a higher} number of observations than a WFD pointing. \update{Each} DDF \update{receives} approximately $1\%$ of the total LSST observing time. The \texttt{filterdist\_indx2\_v1.5\_10yrs} run does not include DDFs.  \update{The plots are centered on $\alpha$=0 and $\delta$=0. Right ascension and declination lines are marked every 30$^\circ$.}  \label{fig:v1.5footprints}}
\end{center}
\end{figure}

\begin{figure}
\begin{center}
\includegraphics[width=0.67\columnwidth]{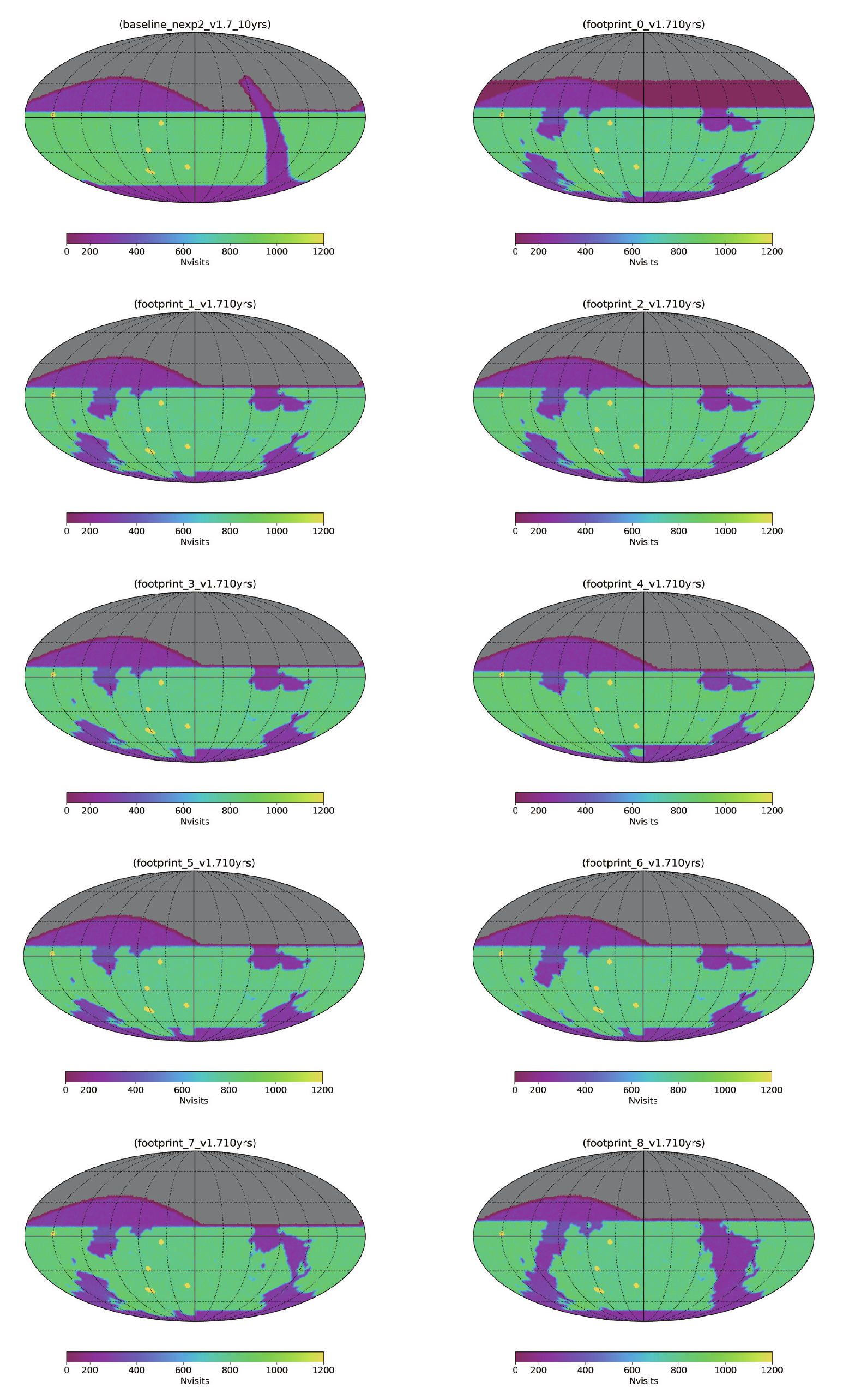}
\caption{The total number of visits in all filters after 10 years for the v1.7 Wide-Fast-Deep (WFD) survey footprint exploration simulations. As visits are taken away from the Galactic plane and bulge region, they are redistributed northward to pointings with less stellar crowding and dust extinction. The Deep Drilling Fields (DDFs) are also visible as a collection of \update{single fields} receiving a higher number of observations than a WFD pointing, with each DDF receiving approximately $1\%$ of the total LSST observing time. \update{The plots are centered on $\alpha$=0 and $\delta$=0. Right ascension and declination lines are marked every 30$^\circ$.}\label{fig:v1.7WFDfootprints}}
\end{center}
\end{figure}

For completeness, we briefly discuss the footprint experiments performed in the v1.5 and v1.7 releases that led to the revised northward WFD incorporated into the v2.0 and onward LSST cadence simulations. The discovery and light curve metrics are shown in Figures \ref{fig:v1.5_footprint} and \ref{fig:v1.7_footprint}. The  v1.5 footprint simulations are set up with different overheads than the v1.7/2.0/2.1 WFD footprint simulations that allow for a larger number of visits to be distributed across the sky. The v1.5 footprint family uses these additional visits to explore the impact of typically adding more northern visits in various configurations, modifying the number of visits in the Galactic plane and NES (sometimes in differing filters), and some changes to the extent of the WFD footprint. The sky map showing the total numbers of visits per pointing in these v1.5 simulations is shown in Figure \ref{fig:v1.5footprints}. In general, adding visits northward enhances TNO discovery statistics and in most cases, there are only small impacts on the ability to obtain light curves and produce shape inversion models of inner solar system objects. Instead of adding a small number of visits, the v1.7 WFD \texttt{footprint} experiments explore WFD variations on a dust-extinction limited footprint with variable North/South declination limits. The total numbers of visits in these v1.7 WFD \texttt{footprint} experiments are shown in Figure \ref{fig:v1.7WFDfootprints}. Overall, TNOs and outer solar system discoveries benefit the most, with the inner solar system object discoveries taking only a few percent loss in discoveries. The light curve metrics for the most part see 5-10$\%$ boosts in the various configurations of the more dust-free WFD, but they start to decrease more significantly for the smaller sized MBAs, NEOs, and PHAs as less of the ecliptic that intersects with the Galactic plane in the southern hemisphere is included in the WFD and the number of visits to those regions drops. Some caution needs to be taken in interpreting this result as this loss may be less than what is shown by the metrics as detection efficiency of solar system objects (by extension the ability to measure their light curves) decreases in crowded fields. The Jupiter Trojans are constrained in set locations on the sky; with small numbers of detections in the simulations, this is likely contributing to the variation observed in the light curve metrics. Overall, these v1.5 and 1.7 footprint experiments show that moving visits northward is an improvement and paved the way for the optimized v2.0/v2.1 WFD footprint and full LSST footprint.

\subsubsection{Varying the Fraction of Time Spent in the Non-WFD Regions}
\label{sec:no-wfd}


Three sets of simulations were done in which the visits to the NES and the Galactic plane were varied relative to the WFD coverage. 
In these simulations, varying numbers of extra visits to the NES or the galactic plane are added at the expense of removing that observing time from the WFD.
The simulations discussed below show how covering these specific areas with visits ranging from 1\% to 100\% of the WFD cadence affect metrics for solar system populations.

\begin{figure}
\begin{center}
\includegraphics[width=0.94\columnwidth]{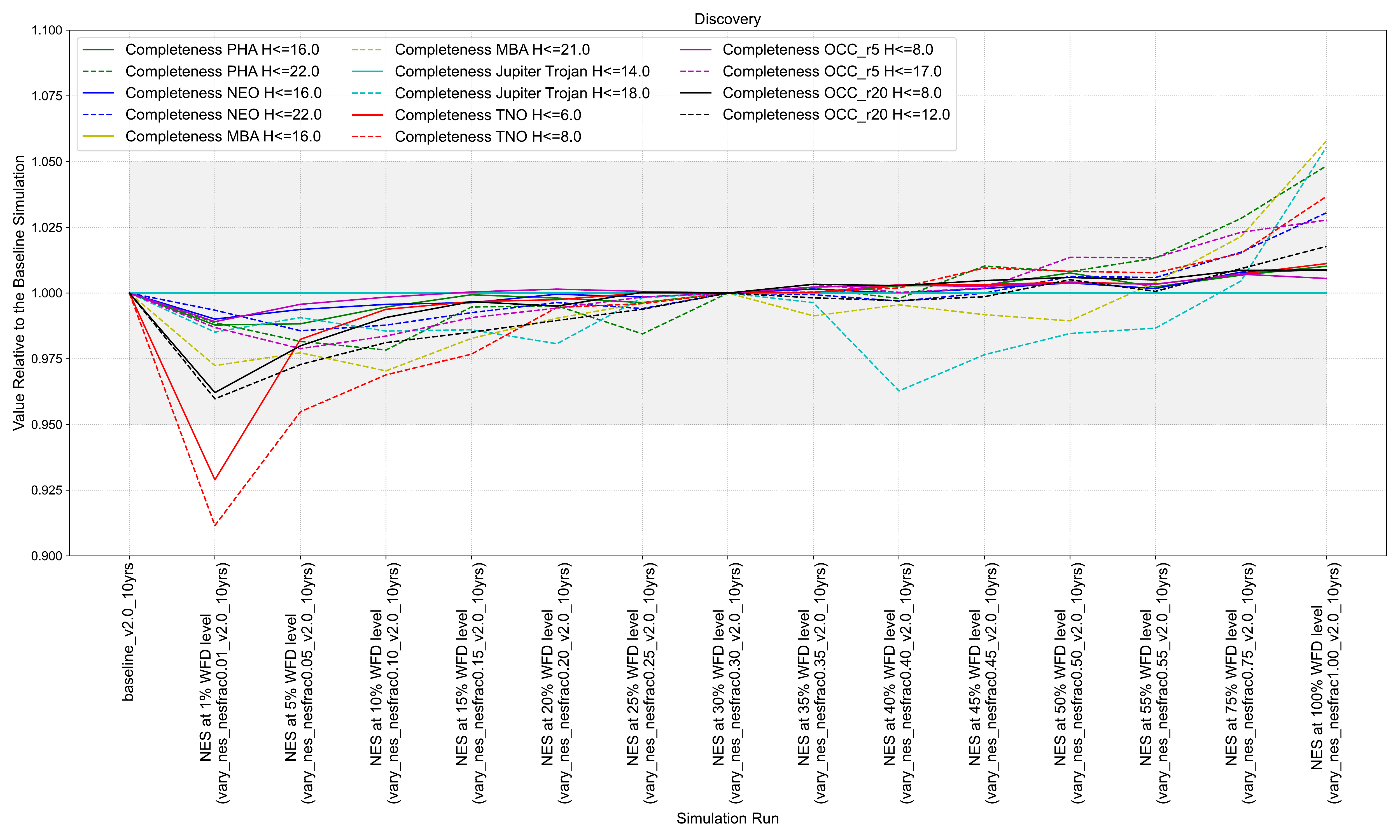}
\includegraphics[width=0.94\columnwidth]{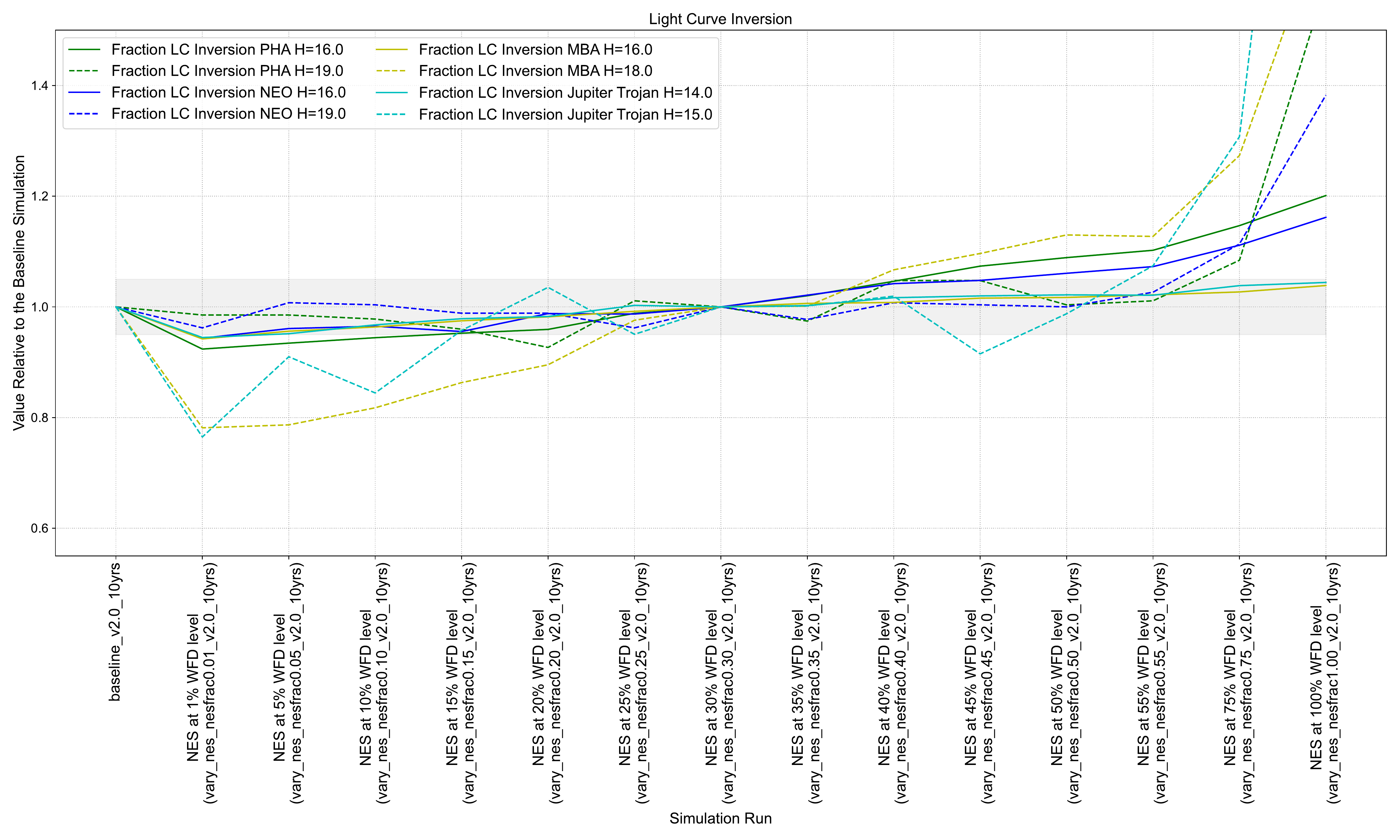}
\caption{Varying the time spent on the Northern Ecliptic Spur (NES). \update{The baseline (reference) simulation with the default scheduler configuration} for this cadence experiment is the first entry on the left. All values have been normalized by this simulation's output. The gray shading outlines changes that are within $\pm5\%$ of the baseline simulation. Top: Discovery Metrics. Bottom: Light Curve Inversion Metrics Note: The light curve inversion plot has been truncated for clarity. The MBAs and Jupiter Trojans extend beyond the plot for the 1.0 NES.
fraction. \label{fig:v2.1_2.0_vary_NES}}
\end{center}
\end{figure}

\begin{figure}
\begin{center}
\includegraphics[width=0.97\columnwidth]{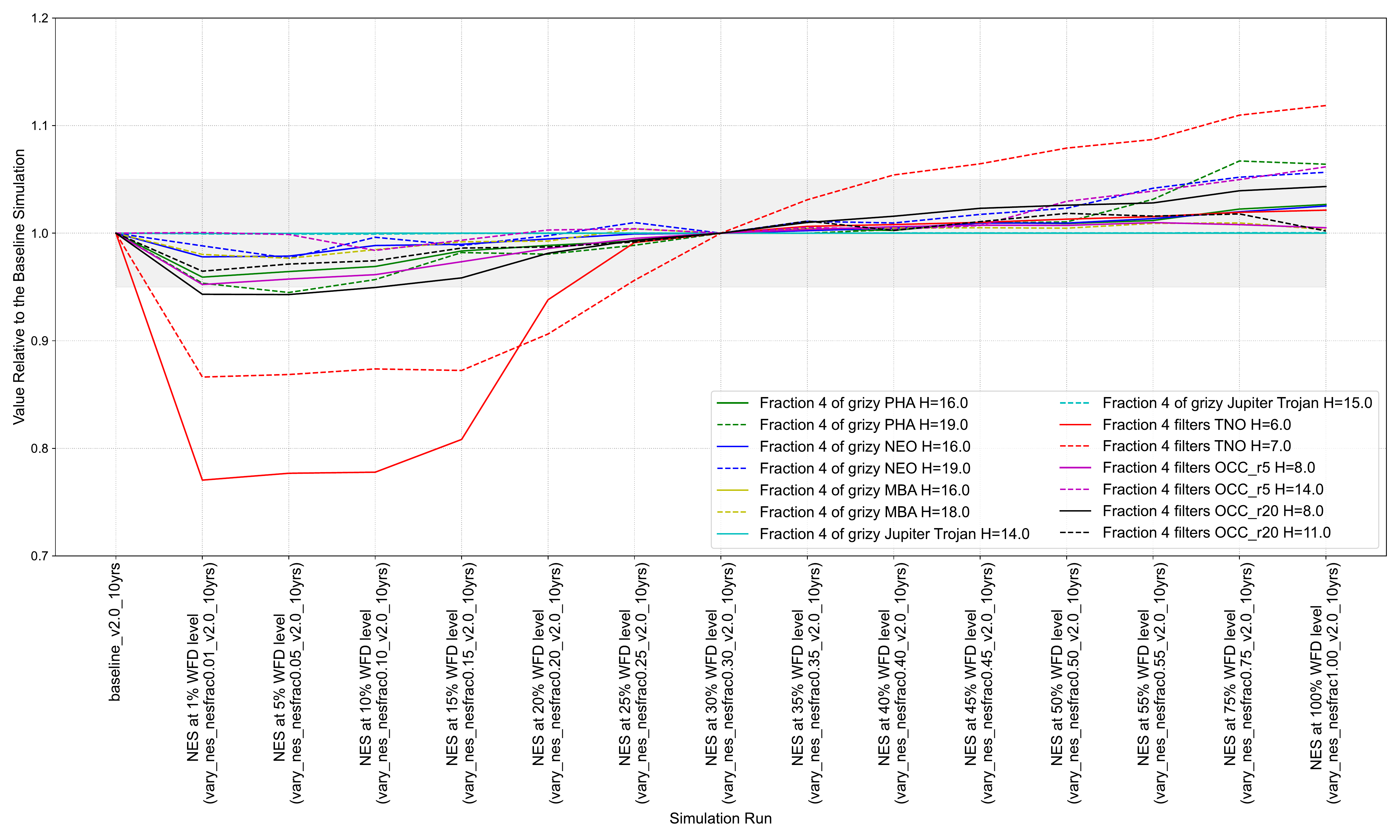}
\caption{Color-light curve metrics with varying time spent on the Northern Ecliptic Spur (NES) from the v2.0 simulations. \update{The baseline (reference) simulation with the default scheduler configuration} for this cadence experiment is the first entry on the left. All values have been normalized by this simulation's output.The gray shading outlines changes that are within $\pm 5\%$ of the baseline simulation. \label{fig:v2.1_2.0_vary_NES_4filters}}
\end{center}
\end{figure}

The \texttt{vary\_NES} family of simulations included coverage of the fields in the NES at 1\% of the WFD level, at 5-55\% of the WFD level in 5\% increments, and at 75\% and 100\% of the WFD level; the baseline simulation has the NES fields at 30\% of the WFD. 
The top panel of Figure~\ref{fig:v2.1_2.0_vary_NES} shows the discovery metrics for various solar system populations for these simulations.
At low coverage levels for the NES ($<$ 10\%), the discovery metrics for the TNO populations are reduced by more than 5\% relative to baseline, and most populations show increasing discoveries as the NES coverage increases.
Figure~\ref{fig:v2.1_2.0_vary_NES_4filters} shows the fraction of each solar system population (relative to baseline) that is observed in at least four of the $grizy$ filters as a function of the NES coverage. 
If the NES is covered at $\lesssim$ 15\% of the WFD, the fraction of TNOs (down to $H=6$) that are observed in at least four filters is reduced by more than 20\% compared to baseline; the NES must cover at least 25\% of WFD to not reduce this metric by more than 5\%. 
The TNO populations are the most affected in both discovery and color-light curve metrics when the NES is not covered to at least 25\% of the WFD level because they move slowly on-sky compared to closer-in solar system populations. 
Most of them will not move enough over the 10 year LSST time span to move from NES fields to WFD fields.
Covering the NES at $<$ 25\% of WFD also significantly decreases the number of faint MBAs and faint Jupiter Trojans that are expected to have light curve measurements (bottom panel of Figure~\ref{fig:v2.1_2.0_vary_NES}); all the populations generally improve in both the color-light curve and light curve metrics as NES coverage increases.

\begin{figure}
\begin{center}
\includegraphics[width=0.96\columnwidth]{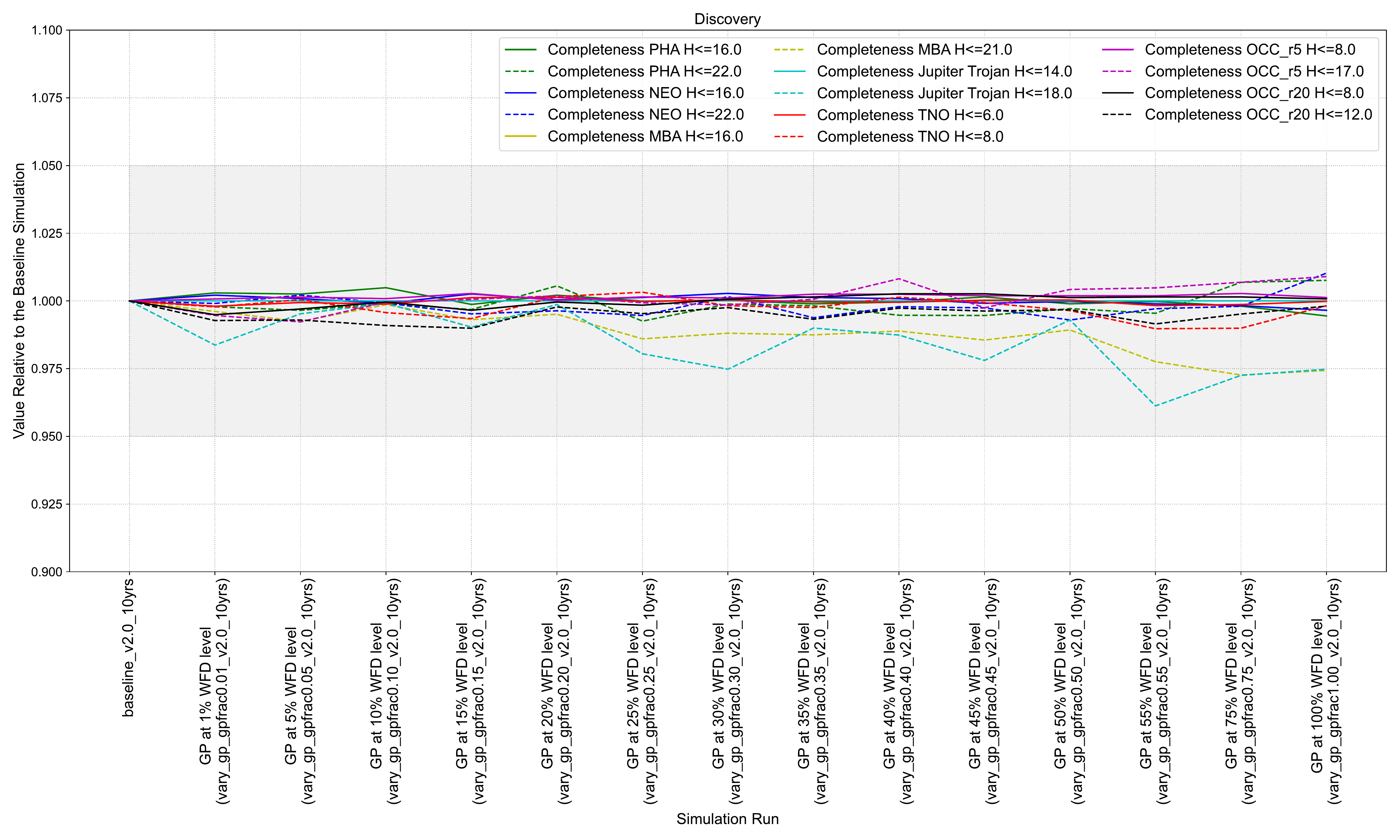}
\includegraphics[width=0.96\columnwidth]{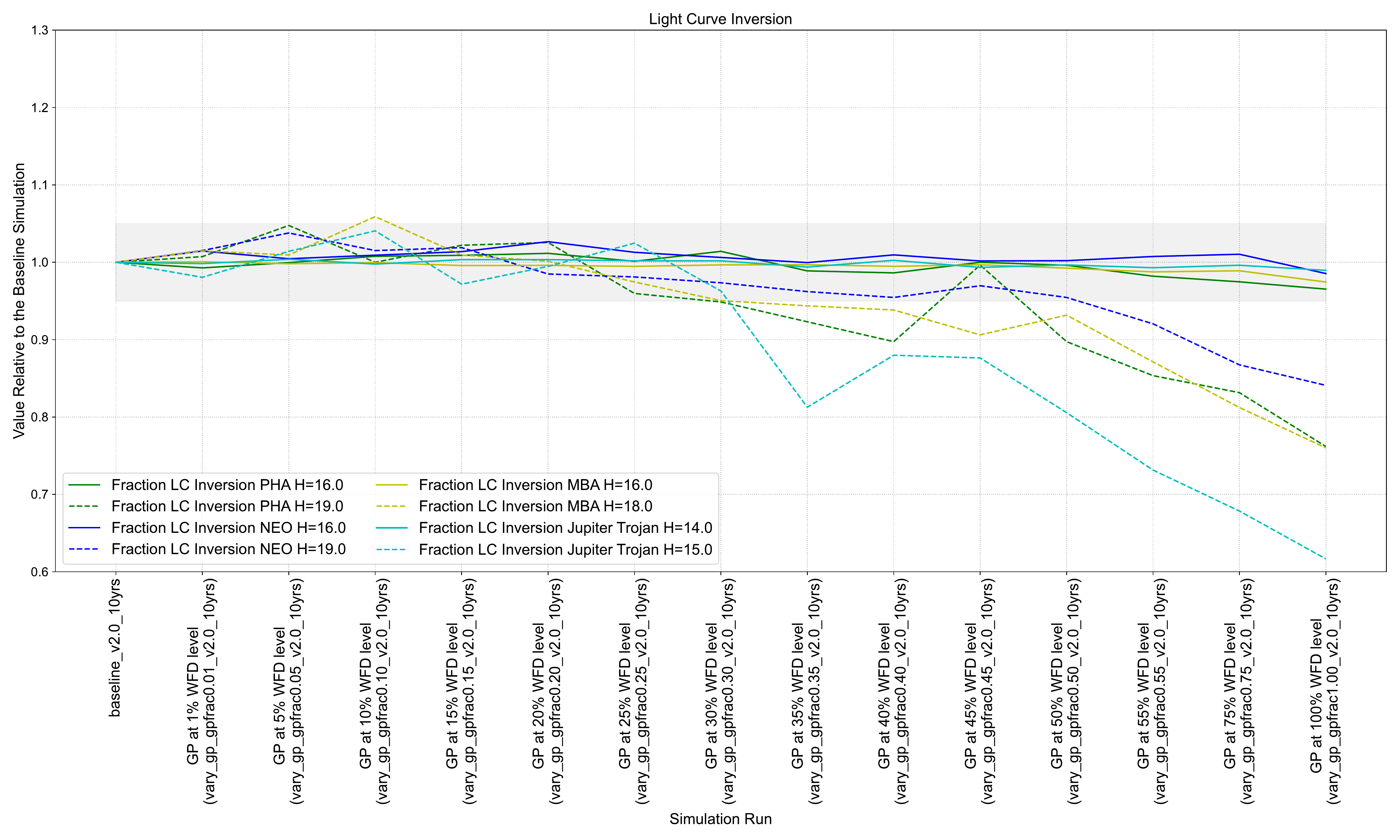}
\caption{Varying the time spent on the Galactic plane (v2.0 simulations).  \update{The baseline (reference) simulation with the default scheduler configuration} for this cadence experiment is the first entry on the left. All values have been normalized by this simulation's output. The gray shading outlines changes that are within $\pm5\%$ of the baseline simulation. Top: Discovery Metrics. Bottom: Light Curve Inversion Metrics. \label{fig:v2.1_2.0_vary_GP}}
\end{center}
\end{figure}

The \texttt{vary\_GP} family of simulations included coverage of the fields in the Galactic plane at 1\% of the WFD level, at 5-55\% of the WFD level in 5\% increments, and at 75\% and 100\% of the WFD level. 
Figure~\ref{fig:v2.1_2.0_vary_GP} shows the resulting solar system metrics for discovery and light curve inversion. 
The discovery metrics for different solar system populations are all within 5\% of baseline for these simulations, and the fraction of each population that has observations in multiple filters is also relatively unaffected. 
However, when the Galactic plane is covered at $>$ 30\% of the WFD level, the fraction of faint MBAs, Jupiter Trojans, and PHAs with light curve inversions all drop by 5\% or more (increasing losses with increasing Galactic plane coverage) compared to the baseline simulations.
This is likely simply a result of shifting time away from the WFD fields, decreasing the odds that the fainter solar system objects are above detection thresholds multiple times in the reduced number of visits to their fields. 

\begin{figure}
\begin{center}
\includegraphics[width=0.96\columnwidth]{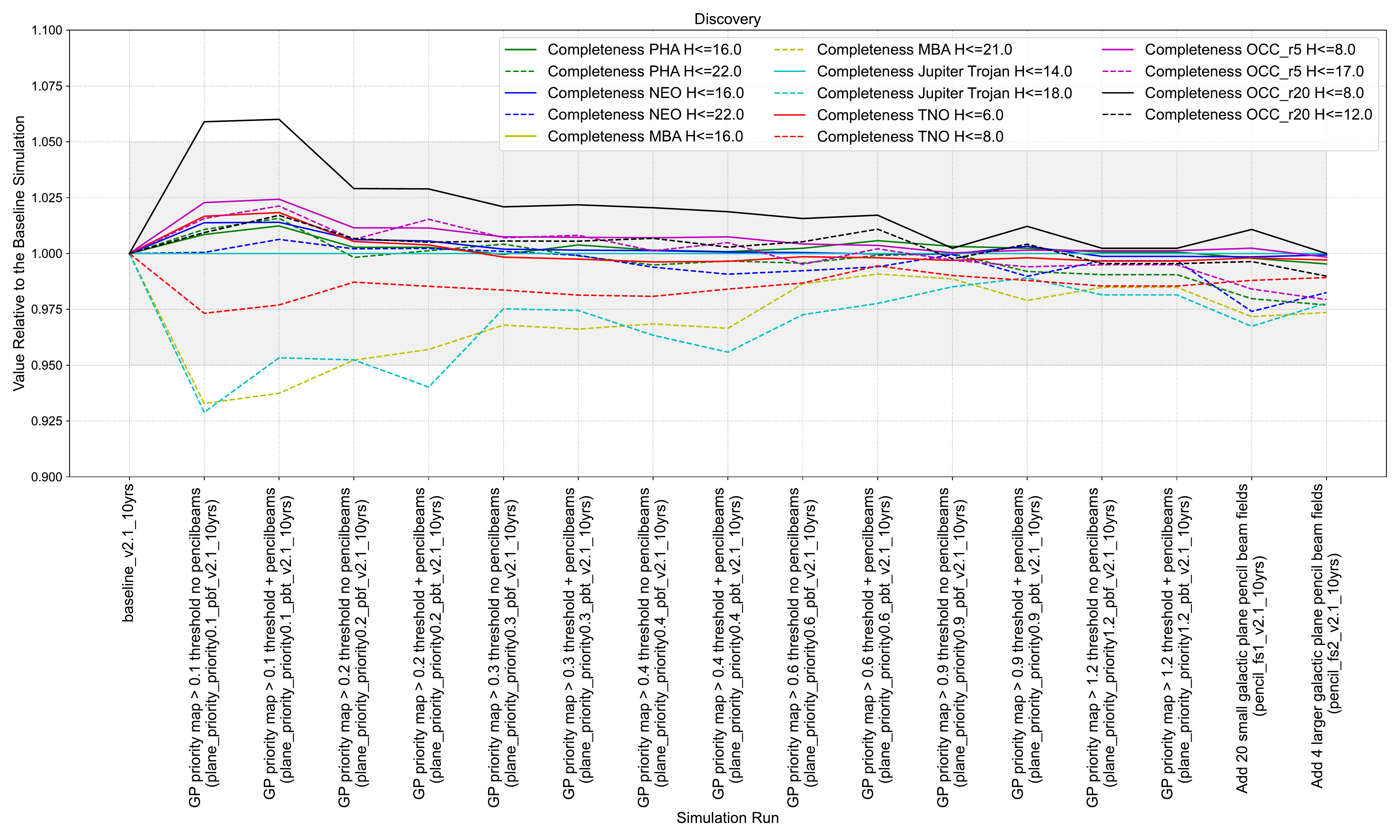}
\includegraphics[width=0.96\columnwidth]{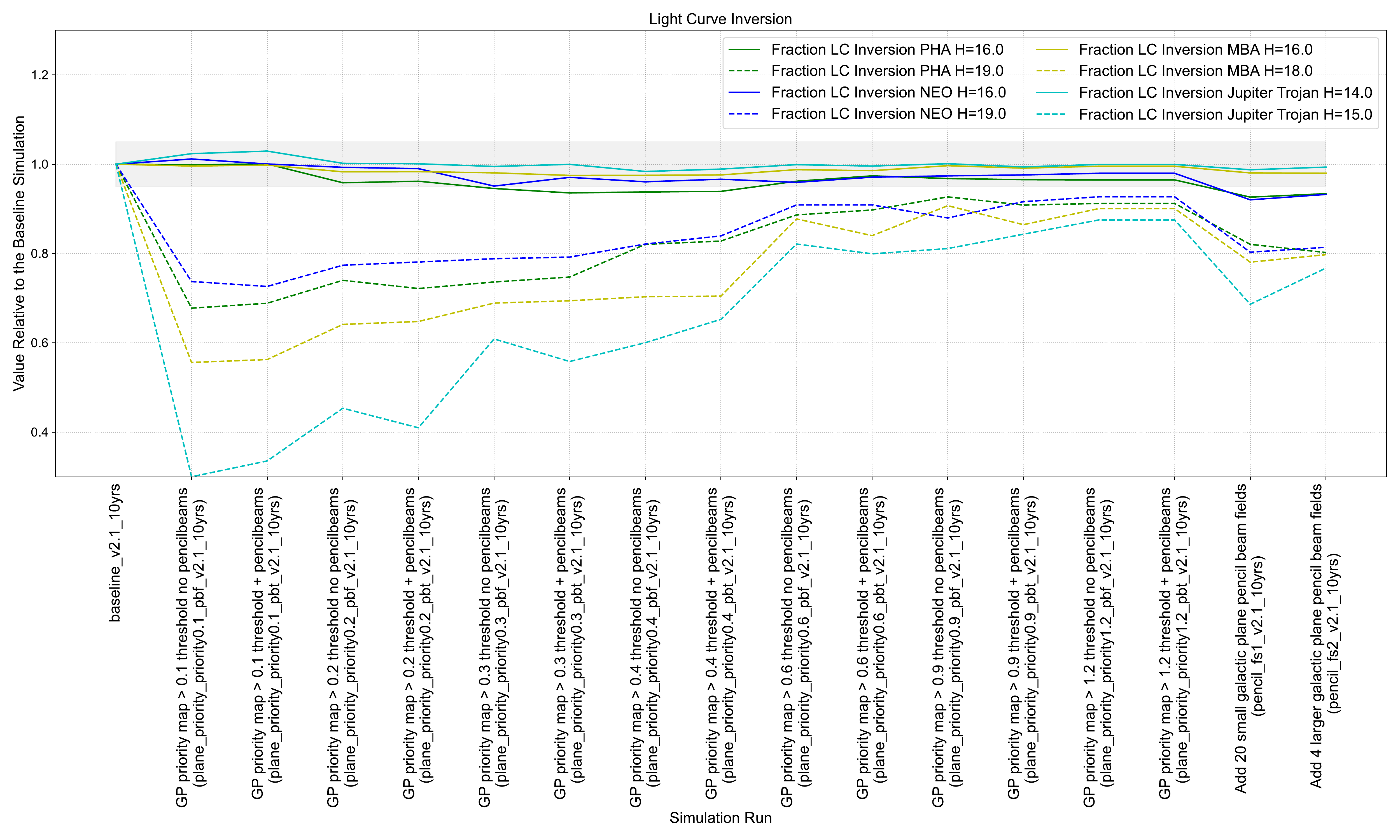}
\caption{Varying the time spent on the Galactic plane (v2.1 simulations). \update{The baseline (reference) simulation with the default scheduler configuration} for this cadence experiment is the first entry on the left. All values have been normalized by this simulation's output. The gray shading outlines changes that are within $\pm5\%$ of the baseline simulation. Top: Discovery Metrics. Bottom: Light Curve Inversion Metrics. \label{fig:v2.1_GP_priority}}
\end{center}
\end{figure}

\begin{figure}
\begin{center}
\includegraphics[width=0.97\columnwidth]{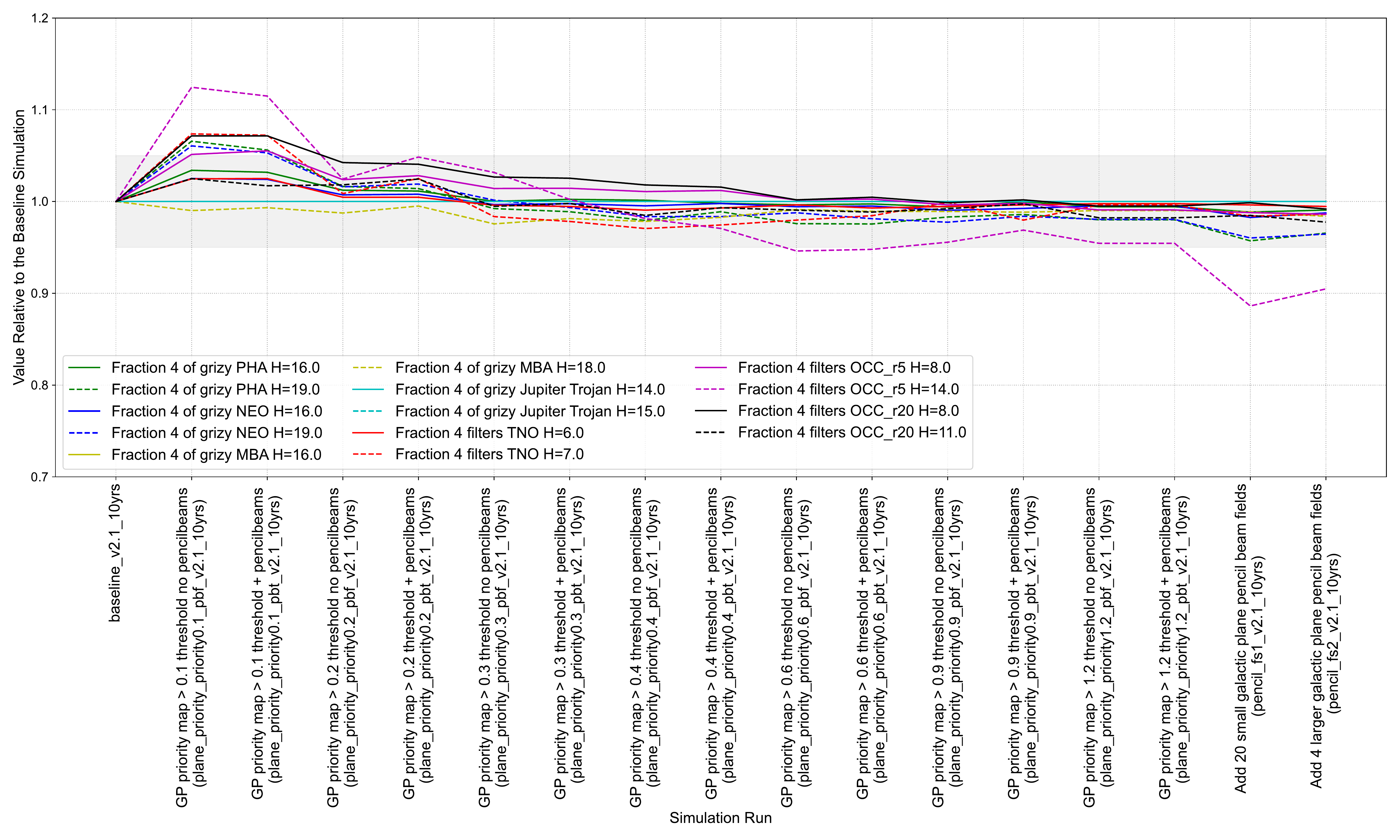}
\caption{Color-light curve metrics with varying time spent on the Galactic plane (v2.1 simulations). \update{The baseline (reference) simulation with the default scheduler configuration} for this cadence experiment is the first entry on the left. All values have been normalized by this simulation's output.The gray shading outlines changes that are within $\pm 5\%$ of the baseline simulation. \label{fig:v2.1_GP_priority_4filters}}
\end{center}
\end{figure}

The \texttt{plane\_priority} family of simulations varies how different regions of the Galactic plane are covered based on a priority map of the Galactic plane from the Rubin Observatory LSST SMLV (Stars, Milky Way, and Local Volume) and TVS (Transients and Variable Stars) science collaborations. 
Some of these simulations also have pencil beam fields in areas of the Galactic plane that the WFD is not planned to cover.  
These targeted pencil beam fields would be visited at the same level as the WFD. 
The GP \texttt{plane\_priority} simulations were completed with and without pencil beam fields, and two additional simulations were done with just four larger or 20 smaller GP pencil beam fields (the \texttt{pencil\_fs} simulations). 
The discovery metrics for different solar system populations are almost all within 5\% of baseline for these simulations, with only faint MBAs and faint Jupiter Trojans dropping slightly below those thresholds for the priority threshold at 0.1-0.2 (top panel of Figure~\ref{fig:v2.1_GP_priority}).
The color-light curve metrics (see Figure \ref{fig:v2.1_GP_priority_4filters}) are also generally similar across the family of simulations, with losses in the fraction of faint populations observed in four of the $grizy$ filters that hover around 5\% for the simulations with higher threshold values (above $\sim$0.4), with or without pencil beams; the simulations with the lowest thresholds show an enhancement in the color-light curve metric.
However for this entire family of simulations, the fraction of faint MBAs, Jupiter Trojans, NEOs, and PHAs with light curve inversions all suffer $>$ 5\% losses compared to the baseline simulation (bottom panel of Figure~\ref{fig:v2.1_GP_priority}); again, this is likely due to additional time shifted away from the WFD fields.
The set of simulations that cover the priority map at $>$ 0.6-1.2 threshold with or without pencil beams generally keep the light curve inversion losses for these populations to between 10\% and 20\% compared to baseline. 
The simulations with 4 larger or 20 smaller Galactic plane pencil beam fields added in addition to the plane priority maps have worse light curve inversion metrics for all solar system populations than simulations with just the priority maps.

\subsection{Exposure Times and Snaps}
\label{sec:exposures_and_snaps}
In this section we explore the various options for the total exposure time per visit and the number of observations (``snaps") taken at each visit. Both parameters directly impact the amount of open shutter time available and therefore how many exposures can be taken on any given night and in total by the survey per filter. The visit exposure time also impacts the individual image depth, increasing or decreasing the resulting image's 5$\sigma$ limiting magnitude.

\subsubsection{Snaps}
\label{sec:snaps}

The LSST cadence is currently planned with two exposures of equal length dubbed ``snaps", nominally 15 s each, to be taken back-to-back at each visit to an on-sky pointing, except in the case of $u$-band observing (see Section \ref{sec:u_band}). \update{The original plan was for the Rubin Observatory data management pipelines to compare the two snaps in order to identify and flag pixel-level artifacts (e.g. cosmic rays). Source detection would be performed on the image resulting from coadding the two exposures \citep{lsstSRD, 2019ApJ...873..111I}. \update{We note that the Rubin SSP pipeline\update{s'} discovery algorithm is agnostic to the number of snaps per visit, as it uses the transient sources detected in the coadded snaps image as input \citep{lsstMOPS,lsstSSP}}.}

\update{There now exist many algorithms published in the literature for identifying cosmic rays in single astronomical images \citep[e.g.][]{2000PASP..112..703R, 2001PASP..113.1420V, 2005AN....326..428S, 2018zndo...1482019M}. If these algorithms work well on LSSTCam images, there may be no strong reason for taking two snaps at each visit.} The decision on whether to implement one snap or two snaps per visit will be made during commissioning of the LSSTCam and the Rubin data management pipelines \citep{2019ApJ...873..111I} when the feasibility of single exposure cosmic ray rejection can be tested and the impact from satellite constellation steaks can be properly assessed (see Section \ref{sec:satcons}). With two snaps, each planned visit has two camera readouts and two camera shutter openings and closings. Although the readout time and the movement of the camera shutter are relatively quick (well less than a minute), the summed time lost to these overheads over the entire 10-year survey is non-negligible in the case of the two snaps observing cadence. As can be seen in Table \ref{tab:snaps} for the  v1.7 family of simulations (the most recent LSST cadence simulations exploring the number of snaps), switching to one snap generates an $\sim$8$\%$ gain in on-sky visits and an increase in the sky coverage reaching the WFD goal of 825 visits per pointing. 

\begin{figure}
\begin{center}
\includegraphics[width=0.96\columnwidth]{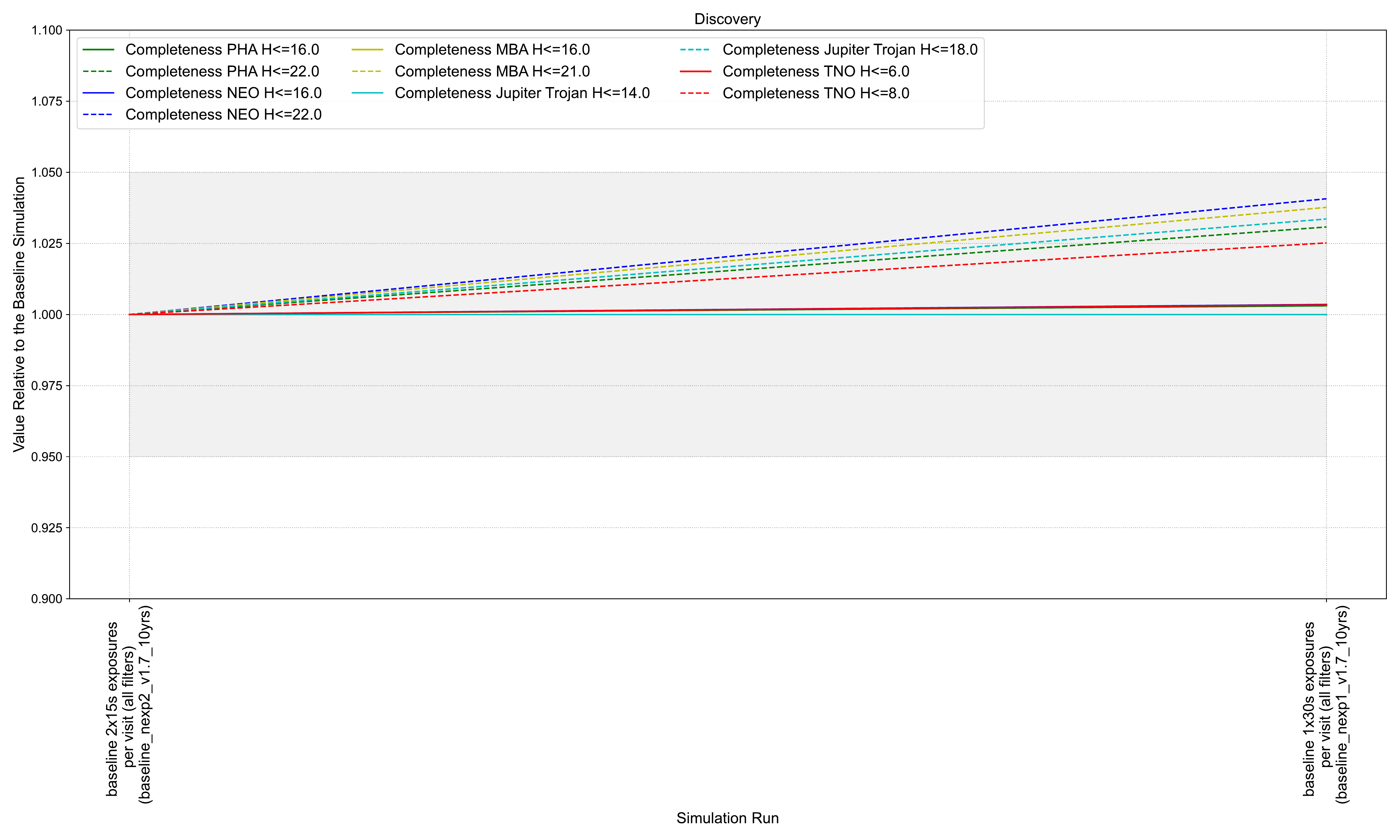}
\includegraphics[width=0.96\columnwidth]{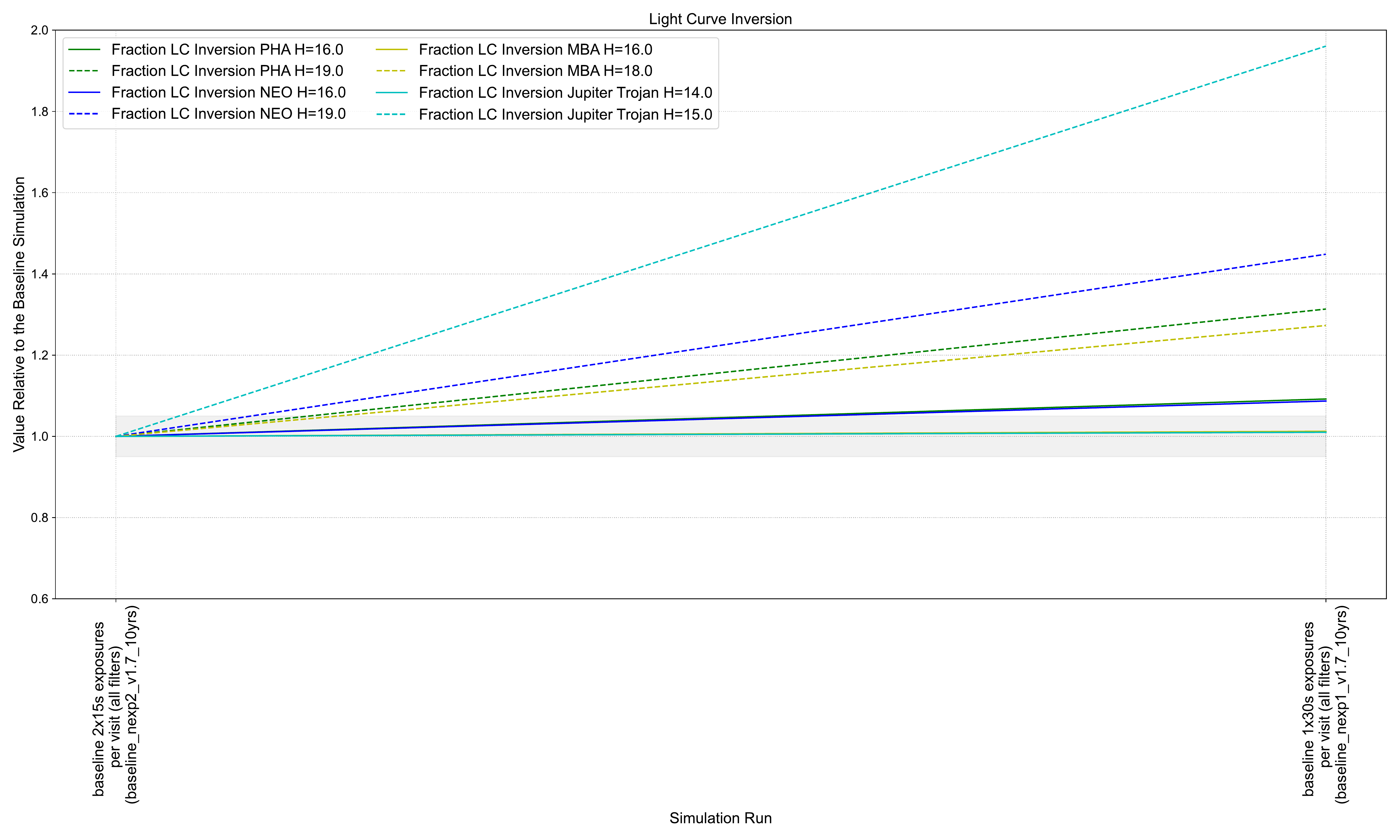}
\caption{Varying  the number of snaps per visit (v1.7 simulations). The reference simulation with 2 snaps case  is the first entry on the left. All values have been normalized by this simulation's output. The gray shading outlines changes that are within $\pm5\%$ of the baseline simulation. Top: Discovery Metrics. Bottom: Light Curve Inversion Metrics. \label{fig:v1.7_snaps}}
\end{center}
\end{figure}

\update{The impact of switching to a single snap cadence will depend on what the added exposures are used for. The SCOC has not yet made any decision about snaps and how to partition out the extra visits. If a significant portion of the gained visits can be distributed across the entire LSST footprint or WFD and NES, the increase in exposures will add either sky coverage and/or temporal coverage that will help with detection and monitoring of small bodies while enabling other science.} In the v1.7 simulations, the extra visits gained in the one snap case were divided out evenly between the WFD and other parts of the simulation's survey footprint. This produces an increase in both the detection and the light curve metrics (see Figure \ref{fig:v1.7_snaps}). The detection metrics for the small size end increase by a few percent. The extra visits provide additional chances for those objects near the survey brightness limit to get above the image 5$\sigma$ limiting magnitude and be detected. The largest bodies see only a very slight increase because the majority of times when they land within an exposure, they are already brighter than the limiting magnitude. The largest enhancement is seen with the light curve inversion \update{metrics, especially for the small end of the size distribution }where we find a $>$20$\%$ boost across the MBAs, PHAs, NEOs, and Jupiter Trojans. Like the case for discovery, the extra observations provide more opportunities for better temporal coverage to \update{probe rotations} and perform shape inversion. Since only 6 detections are required for discovery, it is the light curve inversion metric that shows the true benefits for color and light curve measurements from the extra on-sky observing. 

\update{As noted in \cite{2018arXiv181201149S}, there is some extra information that is potentially gained with two snaps per visit. Although the SSP pipelines are not currently planned to use any information from the individual snaps,  bespoke community software could be developed to take advantage of the two exposures per visit.} For those small body populations moving fast enough to be significantly trailed in the LSST images, such as NEOs and PHAs, the sequential snaps allow for 1) the on-sky direction of motion to be measured from the two streaks and 2) brightness variations (on the order of seconds) to be extracted from the streaks for ultra-fast rotators. \update{Only a very tiny fraction of the asteroids discovered will be rotating fast enough that sub-30 s resolution will be useful \citep{2000Icar..148...12P, 2009Icar..202..134W, 2009Icar..204..145M, 2011Icar..214..194H, 2014ApJ...788...17C, 2019ApJS..241....6C,2021pdss.data...10W, 2021PSJ.....2..191C,  2022ApJ...932L...5C}.  These are very limited benefits compared to the gains to all solar system populations from an extra $8\%$ of on-sky observing time. Therefore, we recommend incorporating single snap visits into the LSST cadence, if feasible.}

\begin{deluxetable}{clcc}
\tablecaption{Varying Snaps \label{tab:snaps}}
\tablewidth{0pt}
\tablehead{
\colhead{Exposures} &  \colhead{LSST Cadence} & \colhead{Total $\#$} &   \colhead{Area with} \\
\colhead{Per Visit} & \colhead{Simulation Name} & \colhead{of On-Sky} & \colhead{ $>$ 825  visits}   \\
 &   &  \colhead{Visits}  & \colhead{(degrees$^2$)}  
}
\startdata
2 $\times$ 15 s  (current SRD requirement) & \texttt{baseline$\_$nexp2$\_$v1.7$\_$10yrs} & 2045493 & 17982.71	 \\
1 $\times$ 30 s & \texttt{baseline$\_$nexp1$\_$v1.7$\_$10yrs} & 2208619 &18190.85 \\
\enddata
\end{deluxetable}

\subsubsection{Long u-band Observations}
\label{sec:u_band}

\begin{figure}
\begin{center}
\includegraphics[width=0.96\columnwidth]{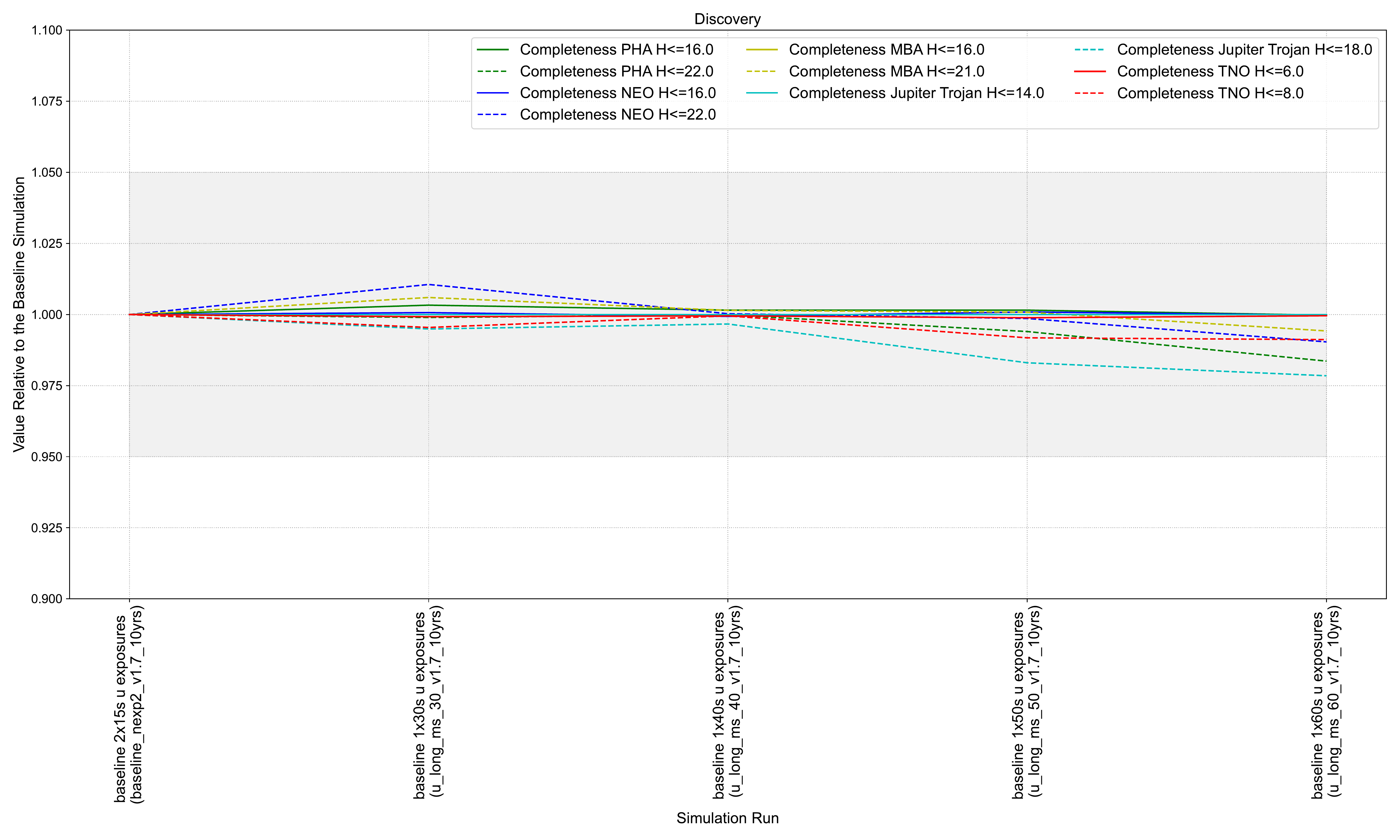}
\includegraphics[width=0.96\columnwidth]{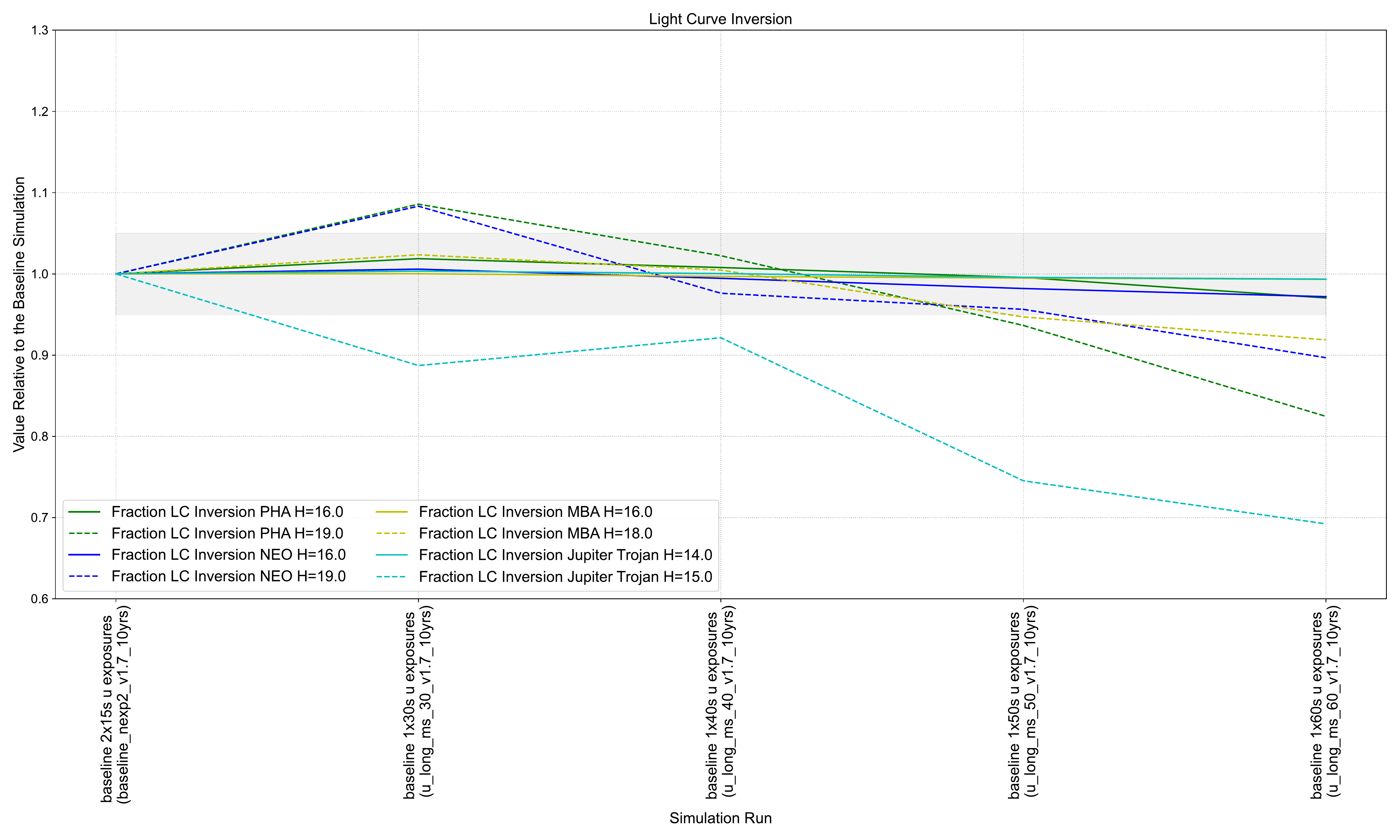}
\caption{Changing the length of the $u$-band exposures in the v1.7 simulations. \update{The baseline (reference) simulation with the default scheduler configuration} for this cadence experiment is the first entry on the left. All values have been normalized by this simulation's output. The gray shading outlines changes that are within $\pm 5\%$ of the baseline simulation. Top: Discovery Metrics. Bottom: Light Curve Inversion Metrics. \label{fig:v1.7_long_u}}
\end{center}
\end{figure}

\begin{figure}
\begin{center}
\includegraphics[width=0.94\columnwidth]{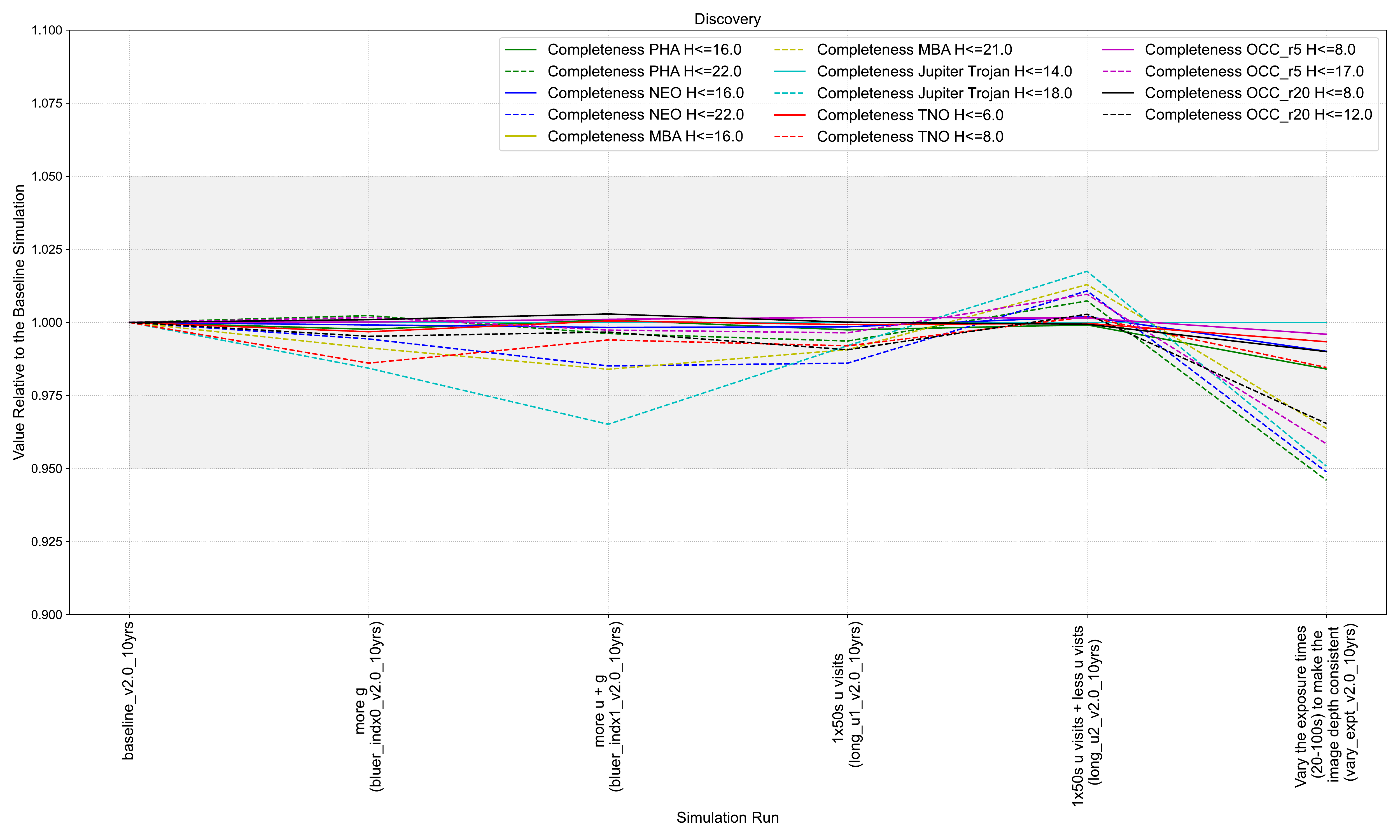}
\includegraphics[width=0.94\columnwidth]{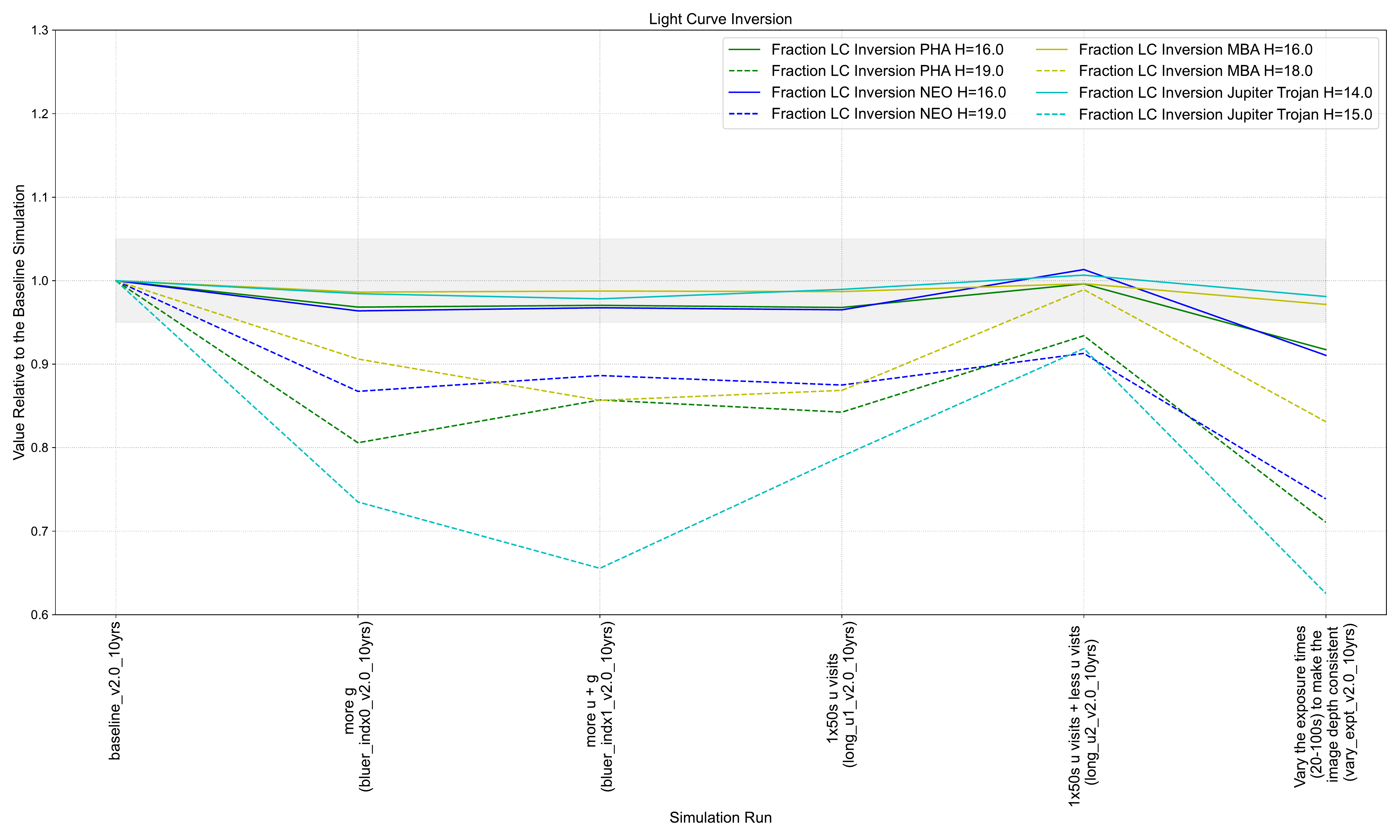}
\caption{\update{The impact of skewing the filter distribution bluer, increasing the exposure time of the $u$-band observations,} and the effect of varying exposure time per visit (\texttt{vary\_expt\_v2.0\_10yrs}). \update{All the simulations presented in this figure are from the v2.0 runs.} \update{The baseline (reference) simulation with the default scheduler configuration} for this cadence experiment is the first entry on the left. All values have been normalized by this simulation's output. The gray shading outlines changes that are within $\pm 5\%$ of the baseline simulation. Top: Discovery Metrics. Bottom: Light Curve Inversion Metrics. \label{fig:v2.0_blue_long_u}}
\end{center}
\end{figure}

The \texttt{long\_uX} and \texttt{u\_long} simulations explore the impact of using a single longer $u$-band exposure versus two shorter (15 s) exposures in the baseline. This was investigated as, because of the low level of sky background in u-band, read-out noise has a larger impact than in redder bands, and a single longer exposure allows to significantly improve the u-band depth, see e.g. \cite{jones_r_lynne_2020_4048838}. The \texttt{u\_long} family varies the duration of the single $u$-band exposure (30, 40, 50, or 60 s). It is expected that longer $u$-band exposures (40 or 50 s) will be advantageous for the detection of faint activity around solar system objects. Since the u-band also encompasses emission from the CN radical around 388 nm (and to a lesser extent from the NH radical), there might be slight gains for active comets inside 3 au, increasing with decreasing heliocentric distances, but this hasn't been modeled in detail yet. However, longer $u$-band exposure times (starting marginally with 50 s but more strongly for 60 s) result in a lower number of observations being performed in other filters and thus decrease the number of faint Jupiter Trojans and PHAs detected as well as the number of faint objects for which we can perform light curve inversion, as illustrated in Figure \ref{fig:v1.7_long_u}. The \texttt{long\_uX} uses a 50 s exposure, either keeping the same number of visits (\texttt{long\_u1}) or reducing it (\texttt{long\_u2}). Both of these tend to be worse than the baseline for solar system objects in terms of light curve inversion in particular for faint objects, as illustrated in Figures \ref{fig:v2.0_blue_long_u} and \ref{fig:v1.7_long_u}. This results from the fact that light curve inversion requires a certain number of observations above a certain SNR threshold, which might not be met for some objects in bluer filters where most solar system bodies are fainter. The  \texttt{long\_u2} family performs better for both detection and light curve inversion metrics and was identified as a good compromise as long as it is not done together with any of the \texttt{bluer\_indxXX} options mentioned in Section \ref{sec:filter_dist}, which is shifting more visits to blue filters over redder filters.

\subsubsection{Other Variations of Exposure Times}
\label{sec:other_exp_variations}

\begin{figure}
\begin{center}
\includegraphics[width=0.94\columnwidth]{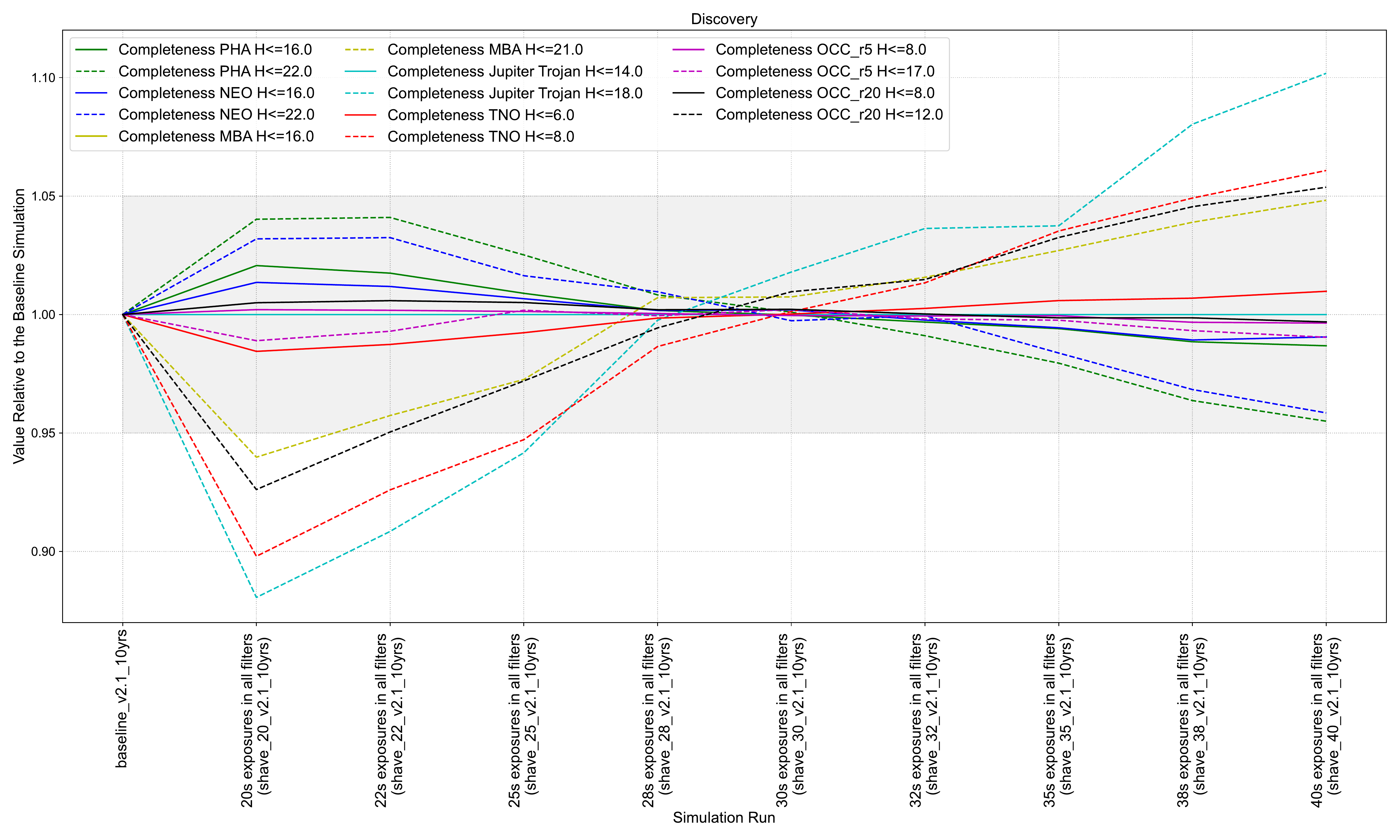}
\includegraphics[width=0.94\columnwidth]{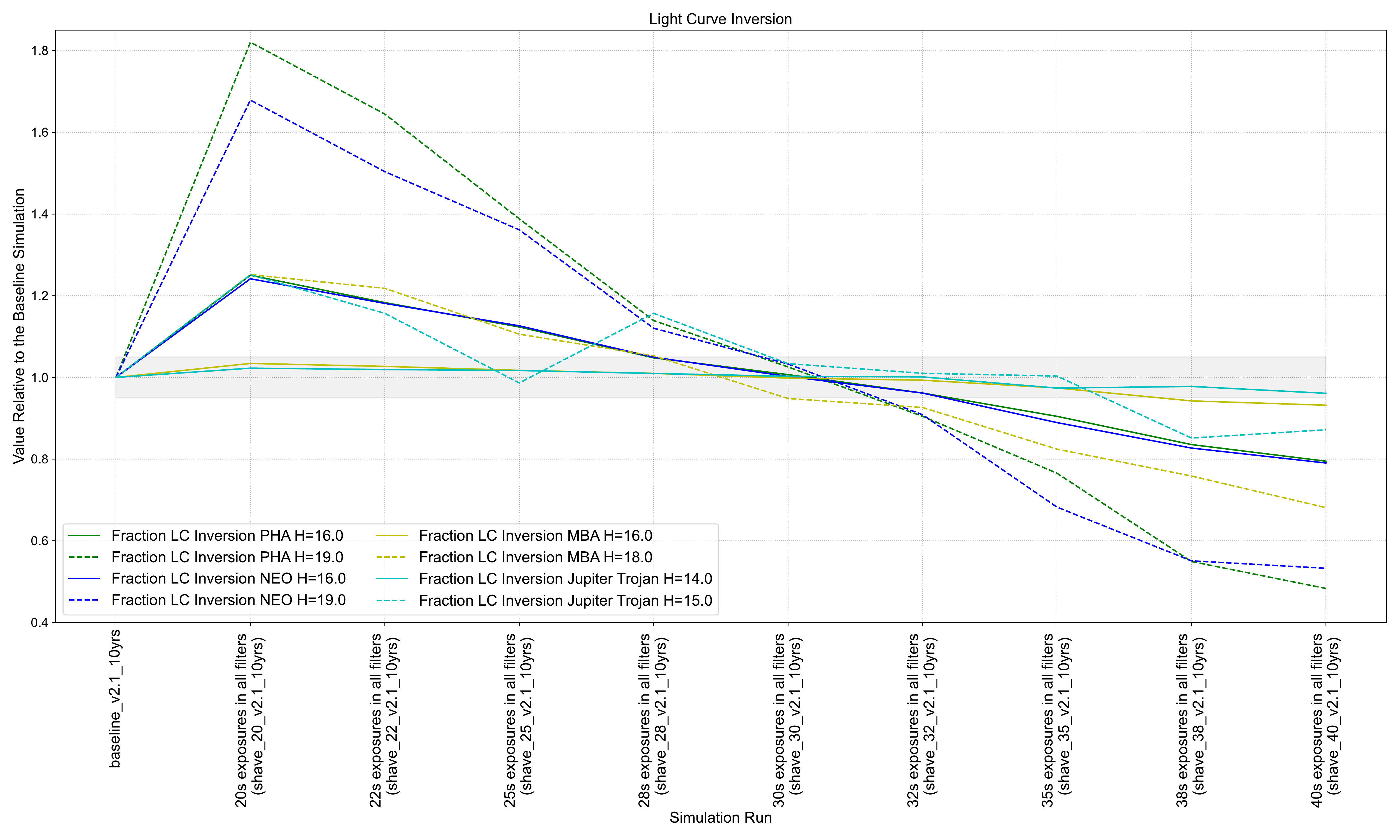}
\caption{Variations on the effective exposure time per visit in the v2.1 cadence simulations. \update{The baseline (reference) simulation with the default scheduler configuration} for this cadence experiment is the first entry on the left. All values have been normalized by this simulation's output. The gray shading outlines changes that are within $\pm 5\%$ of the baseline simulation. \update{ Except for the baseline simulation which has 2 $\times$ 15 s vists in $grizy$ and 1 $\times 30s$ in $u$, all other simulations in this run had single exposure visits per pointing.} Top: Discovery Metrics. Bottom: Light Curve Inversion Metrics. \label{fig:v2.1_vary_exposure_time}}
\end{center}
\end{figure}

\update{The visits in the v$\ge$2.0 survey simulations are typically set to 2 $\times$ 15 s exposures in the $grizy$ filters while the $u$ band has 1 $\times$ 30 s exposures.} A series of simulations (v2.1 \texttt{shave}) has been run to explore the impact of different exposure times on the survey metrics compared to the family's baseline simulation. As seen in the top panel of Figure~\ref{fig:v2.1_vary_exposure_time}, the relative effect on the discovery rate of TNOs and faint OCCs, and MBAs diminishes significantly with shorter exposure times compared to the baseline exposure time configuration as the 5$\sigma$ limiting magnitude decreases with exposure time. Shorter exposure times have a greater effect on fainter absolute magnitude TNOs dropping the discovery metrics by more than $\sim$5$\%$ for TNOs with $6<H<8$ compared to TNOs with $H<6$. The effect is similar for OCCs with the discovery metrics decreasing by $\sim$5$\%$ for OCCs with $8<H<12$ compared to OCCs with $H<8$. The discovery of NEOs and PHAs does see a small improvement with the shorter exposure cadence\update{s} owing to increased sky coverage resulting from the shorter exposure times allowing for more exposures to be taken \citep[e.g., ][]{Jedicke2016}. \update{This improvement in the discovery metrics is greater for fainter PHA and NEOs with $16<H<22$ than for bodies with $H<16$.} The enhanced discovery caused by the greater coverage is due to the fact that these closer-in objects tend to be detected at viewing geometries when they are closer to the Earth and are thus brighter to compensate for their smaller size. Examples of faint, close in asteroids whose discoveries are favored by the shorter cadence when they approach the Earth include asteroids on Earth-similar orbits and meteoroids \citep[e.g.,][]{Kwiatkowski2009,Granvik2013b,2014Icar..241..280B,Jedicke2018,Shober2019,delaFuenteMarcos2020,Bolin2020,Fedorets2020,Naidu2021,delaFuenteMarcos2022}.

When exposure times are increased to more than 30 s, more distant objects have improved discoveries but the discovery of PHAs and NEOs diminishes. The decreased discovery of PHAs could be due to the decreased sky coverage in the longer exposure scheme compared to the shorter exposure scheme. The degradation in the number of PHAs and NEOs in the longer exposures could also be due to trailing losses from their higher rate of motion \citep[e.g., ][]{Shao2014,2019PASP..131g8002Y}. One additional factor to consider in the longer exposure times is that they will be more susceptible to images being compromised from satellite trails which is more likely in longer exposures \citep[][]{Tyson:2020}, see Section \ref{sec:satcons} for a detailed discussion.

The effect of shortening the exposure time improves the light curve inversion metrics for all dynamical groups of objects included in the v2.1 cadence simulations as seen in the bottom panel of Figure~\ref{fig:v2.1_vary_exposure_time}. Shorter exposure times generally improve the light curve inversion metrics due to the improved coverage and improved density of detections enabled by shorter exposure cadences. The magnitude of improvement varies by dynamical class. For faint PHAs and NEOs with $H$ = 19, the light curve metric is almost doubled with a 20 and 22 s \update{exposures} compared to the baseline cadence. Larger PHAs and NEOs with $H$ = 16 see a moderate improvement as well with the shorter exposures. The higher density of coverage will also be useful for the study of the rotation states of Jupiter Trojans and asteroid family members in the Main Belt \cite[e.g., ]{Hanus2018a} as shown by the increase in the light curve metrics for Trojans and MBAs. The benefits of wider and more frequent coverage of the sky to light curve inversion may also extend to the monitoring and detection of \update{activity within the asteroid belt} \citep[e.g., ][]{Moreno2017}. The improvement for more close in objects may be explained by their higher sky-plane motion placing this in a wider range of possible areas of sky positions that is more easily covered with a shorter cadence. A good compromise exposure time for obtaining favorable discoveries for inner and outer Solar System objects as well as dense light curves seems to be the 30 s exposure cadence.

An additional simulation, \texttt{vary$\_$expt$\_$v2.0$\_$10yrs}, was designed to test the results of varying the exposure times between 20-100 s in the $ugrizy$ filters to provide consistency in the image depth in different filters. As seen in the top panel of Fig.~\ref{fig:v2.0_blue_long_u}, varying the exposure time between 20-100 s results in poorer discovery metrics relative to the baseline simulation for all classes of solar system objects used in the simulations. This is due to the fact that the longer exposures result in an overall decrease in survey coverage and a decreased chance to detect moving objects. As seen in the bottom panel of Fig.~\ref{fig:v2.0_blue_long_u}, the effect on light curve inversion metric is also worse for all classes of solar system objects. Therefore, varying the exposure times to achieve uniform visit depth is not recommended.

\subsection{Filter Cadence and Filter Distribution}
\label{sec:filter_dist}

This section explores \update{decisions focused around the choice of filter} i.e.  changing the distribution of observations across filters (to increase the \update{total} number of observations in bluer filters, for example) \update {or modifying how observations in different filters are interspersed within a night or throughout a lunation}. First we examine the effects of increasing the number of observations in $u$ and $g$ bands compared to the baseline. Next we explore the consequence of imposing that a certain number of observations are performed in various combinations of filters each year. Lastly we investigate changing the cadence of observations in $g$ band, taking advantage of bright time to schedule extra \update{visits} and reduce the gap between successive \update{observations} of a given field in $g$-band. 

\begin{figure}
\begin{center}
\includegraphics[width=0.95\columnwidth]{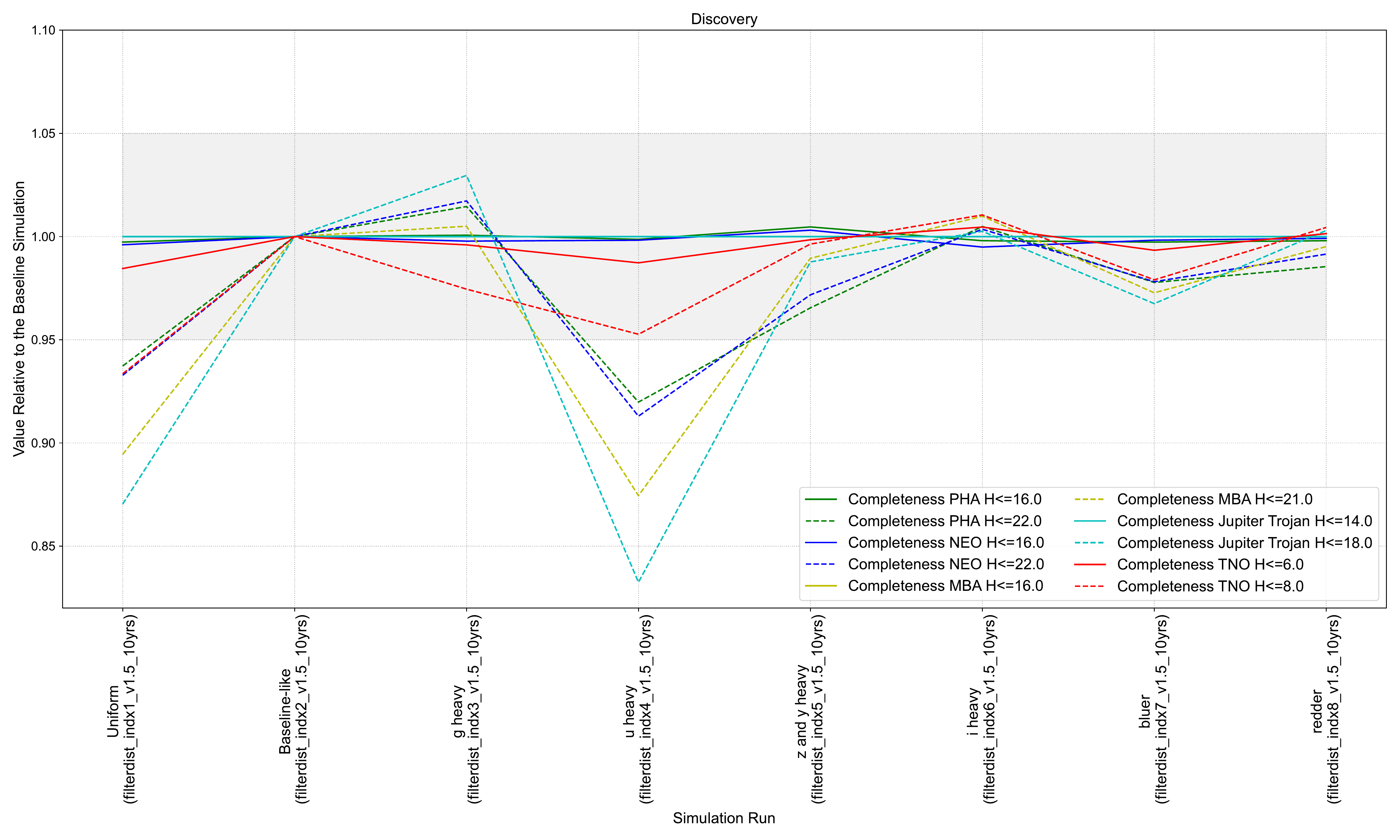}
\includegraphics[width=0.95\columnwidth]{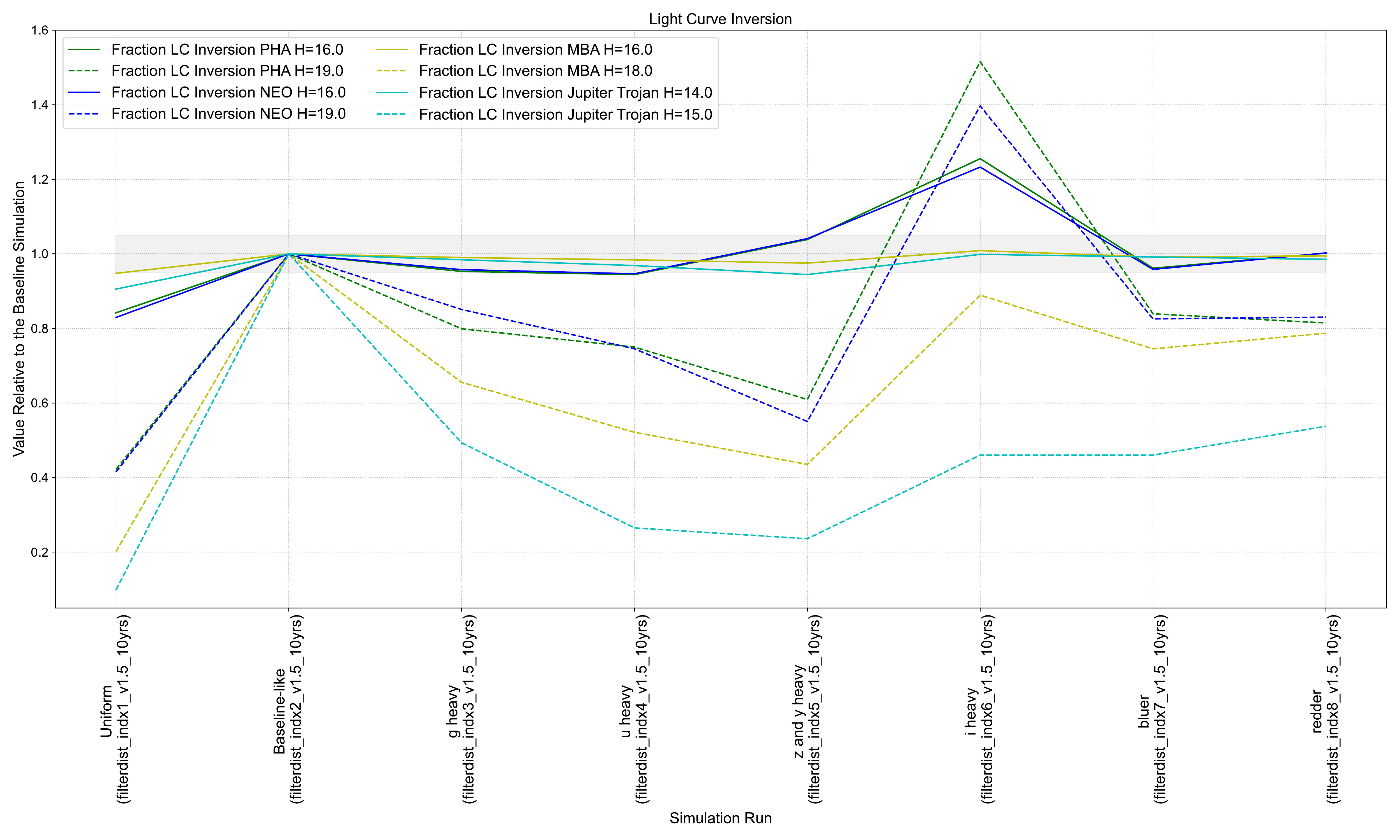}
\caption{Additional options for tuning the filter distribution (v1.5 simulations). \update{The baseline (reference) simulation with the default scheduler configuration} for this cadence experiment is the first entry on the left. All values have been normalized by this simulation's output. The gray shading outlines changes that are within $\pm 5\%$ of the baseline simulation. Top: Discovery Metrics. Bottom: Light Curve Inversion Metrics.\label{fig:v1.5_filters}}
\end{center}
\end{figure}
The \texttt{bluer\_indxXX} family of simulations have a bluer filter distribution, increasing the number of exposures in $g$, or $u$ and $g$ filters compared to the baseline (the filter balance in the baseline is `$u$': 0.07, `$g$': 0.09, `$r$': 0.22, `$i$': 0.22, `$z$': 0.20, `$y$': 0.20). This is done by removing visits in redder filters to redistribute them between $u$ and $g$. Similarly to the families discussed above, increasing the number of exposures in $u$ or $g$-band results in a severe decrease of the number of faint objects for which light curve inversion is possible. Even though this hasn't been modelled yet, active objects close to the Sun might benefit from increased $u$ and $g$ coverage, as these filter encompass emissions from CN and C$_2$ radicals respectively. 
As discussed above and illustrated in Figure \ref{fig:v1.5_filters}, a $u$-heavy distribution induces a significant decrease in the detection of faint solar system objects that are fainter in $u$-band, as illustrated in Table \ref{tab:colors}.

\begin{figure}
\begin{center}
\includegraphics[width=0.94\columnwidth]{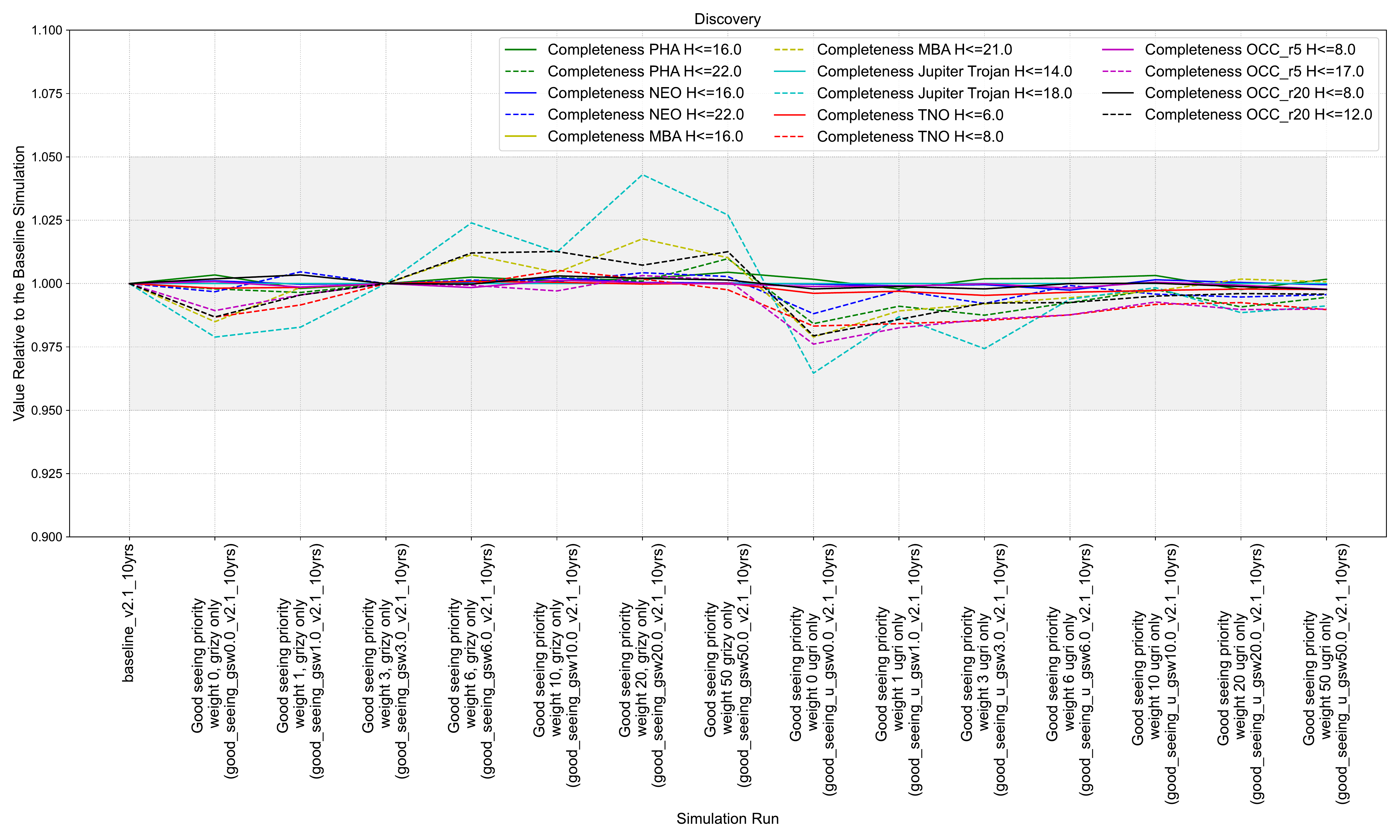}
\includegraphics[width=0.94\columnwidth]{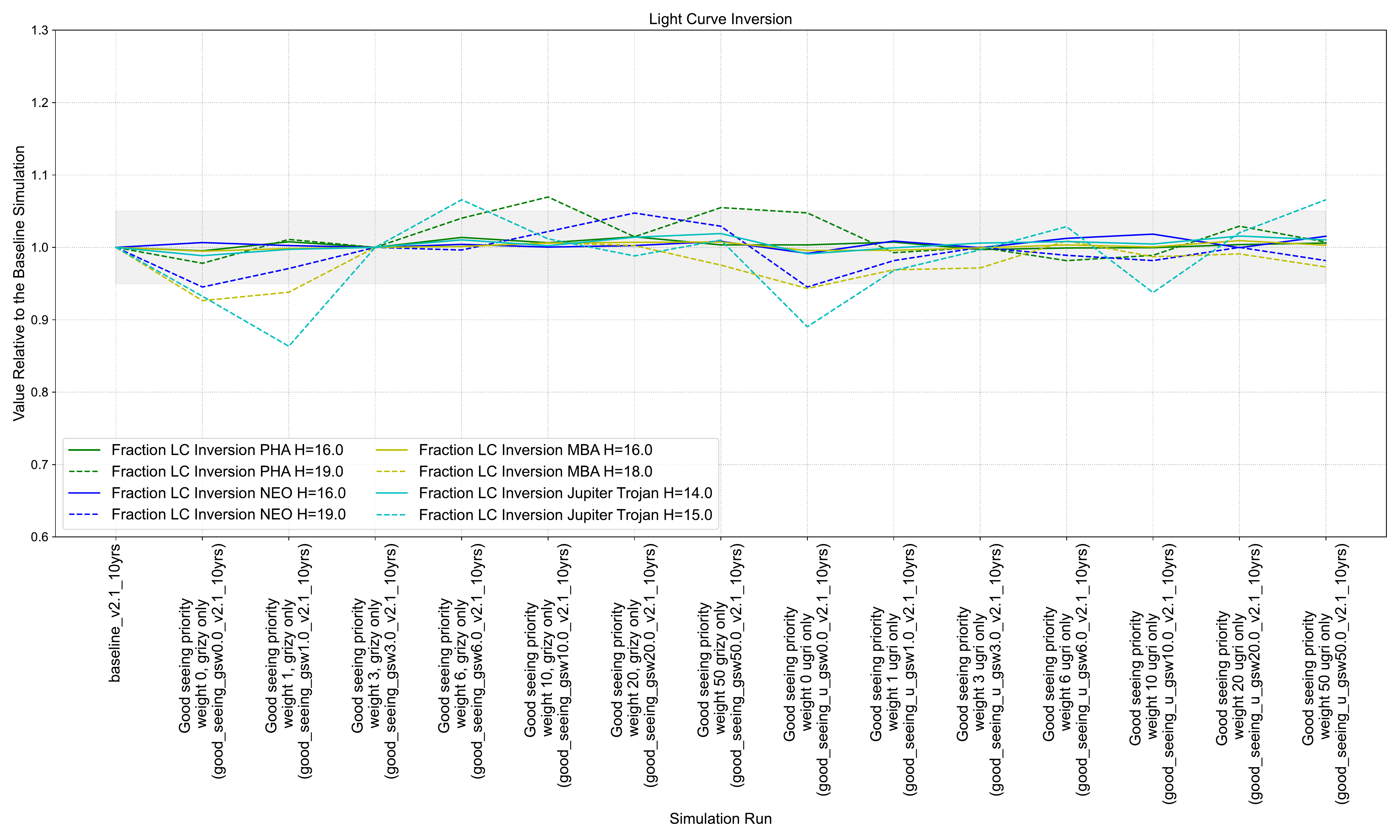}
\caption{The impact of adding  a requirement for 3 ``good seeing" (seeing $<$ 0.8$^{\prime\prime}$) images per year in various bandpasses (v2.0 simulations). \update{The baseline (reference) simulation with the default scheduler configuration} for this cadence experiment is the first entry on the left. All values have been normalized by this simulation's output. The \texttt{baseline\_v2.1\_10yrs} includes the good seeing requirement for $r$ and $i$ bands as the default. The gray shading outlines changes that are within $\pm 5\%$ of the baseline simulation. Top: Discovery Metrics. Bottom: Light Curve Inversion Metrics. \label{fig:v2.1_good_seeing}}
\end{center}
\end{figure}

Figure \ref{fig:v2.1_good_seeing} presents a set of \update{simulations} where emphasis is put on obtaining a handful of exposures with a seeing $<$ 0.8$^{\prime\prime}$ each year, varying the weight put on that constraint and the combinations of filters for which it has to be met (whether $i$ and $y$ are included or not). \update{These simulations come as a request for extragalactic science cases. Ensuring there are yearly good seeing images in several filters enhances strong lensing detection \citep{2019arXiv190205141V} and  galaxy studies \cite{Document-37637}.} In general, requirements of having a minimum number of good seeing images per year in various bandpasses do not impact strongly our discovery or light curve metrics (except for the \texttt{good\_seeing\_gsw1.0\_v2.1\_10yrs} and \texttt{good\_seeing\_u\_gsw0.0\_v2.1\_10yrs}).

\begin{figure}
\begin{center}
\includegraphics[width=0.95\columnwidth]{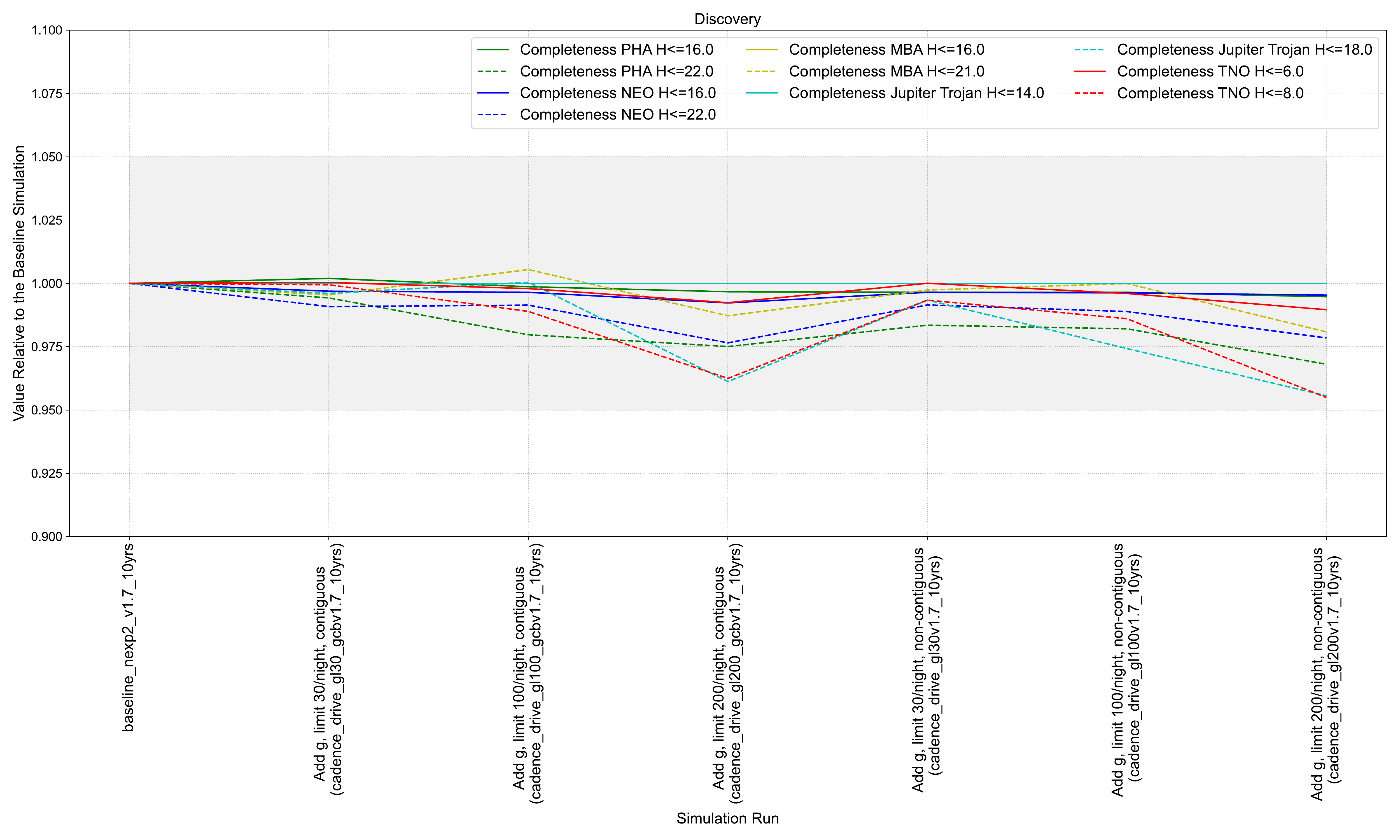}
\includegraphics[width=0.95\columnwidth]{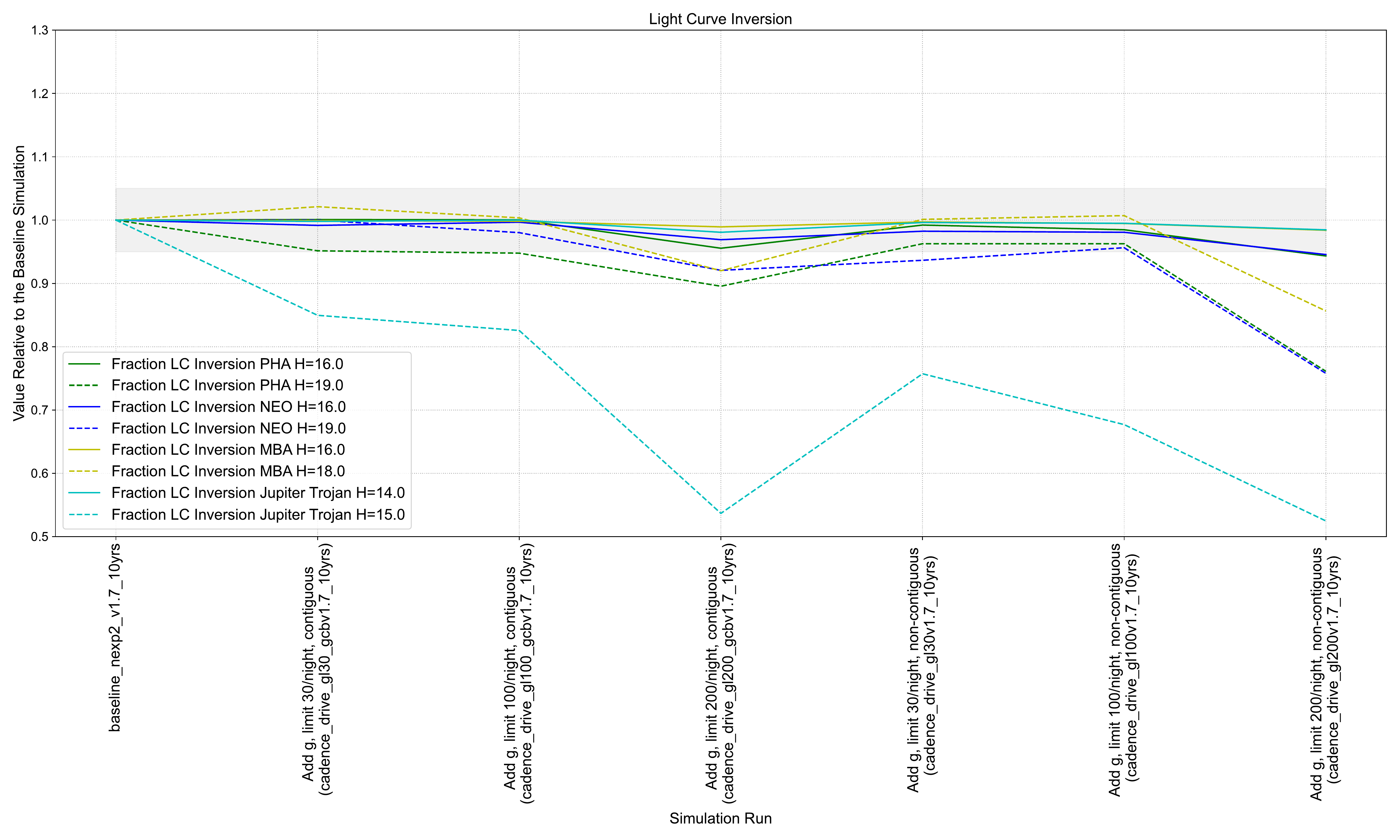}
\caption{Investigating ways of reducing the gaps between $g$ band visits over a month (v1.7 simulations). \update{The baseline (reference) simulation with the default scheduler configuration} for this cadence experiment is the first entry on the left. All values have been normalized by this simulation's output. The gray shading outlines changes that are within $\pm 5\%$ of the baseline simulation. Top: Discovery Metrics. Bottom: Light Curve Inversion Metrics. \label{fig:v1.7_cadence_drive}}
\end{center}
\end{figure}

The \texttt{cadence\_drive} family of simulations investigates reducing long gaps between $g$-band visits over a month by requiring a certain number of fill-in visits each night during bright time. Adding $g$-band visits during full moon time (and consequently reducing the number of visits in redder bands) is generally detrimental for solar system objects, and in particular for light curve inversion of faint Jupiter Trojans as illustrated in Figure \ref{fig:v1.7_cadence_drive}. A small number (30) of contiguous visits might be acceptable but in general the lowest possible number of $g$-band fill-in visits is preferable.

\subsection{Visits Within a Night}
\label{sec:visits}

The Rubin SSP pipelines will search nightly image pairs for new moving sources. Once the orbit of a solar system object is known sufficiently well, SSP will be able to predict the orbit and identify previously known small bodies in single LSST observations, but throughout the entire ten years new solar system discoveries will be made \citep{lsstMOPS,lsstSSP}. The majority of the TNOs and MBAs will be picked up within the first two years of the survey, but new comets, NEOs, and ISOs will continue to be discovered across the duration of the LSST \update{\citep{dmtn-109}}. Thus, it is important that nightly pairs be taken over the full span of the LSST. 

The LSST SRD \citep{lsstSRD} requires at least two observations per night at each observed pointing in order to facilitate accurate removal of the solar system ``cruft" that will pollute the millions of transient astrophysical LSST alerts sent out. A transient only seen in one but not in a repeat observation on the same night will most likely be due to a previously undiscovered moving small body. Multiple observations in the same night also help differentiate inner solar system objects from outer solar system bodies. Additionally, these repeat visits provide temporal and color information that can be used to probe the evolution of astrophysical transients \citep[e.g.,][]{2019MNRAS.485.4260S,2019PASPpresto,2022ApJS..258....2L, 2022ApJS..258....5A, 2022ApJS..259...58L} and minor planets. Below we explore several proposed options on the number, time separation, and filter choices of these intra-night visits. 

\subsubsection{Separation Between Nightly Pairs}
\label{sec:nightly_sep}

\begin{figure}
\begin{center}
\includegraphics[width=1.0\columnwidth]{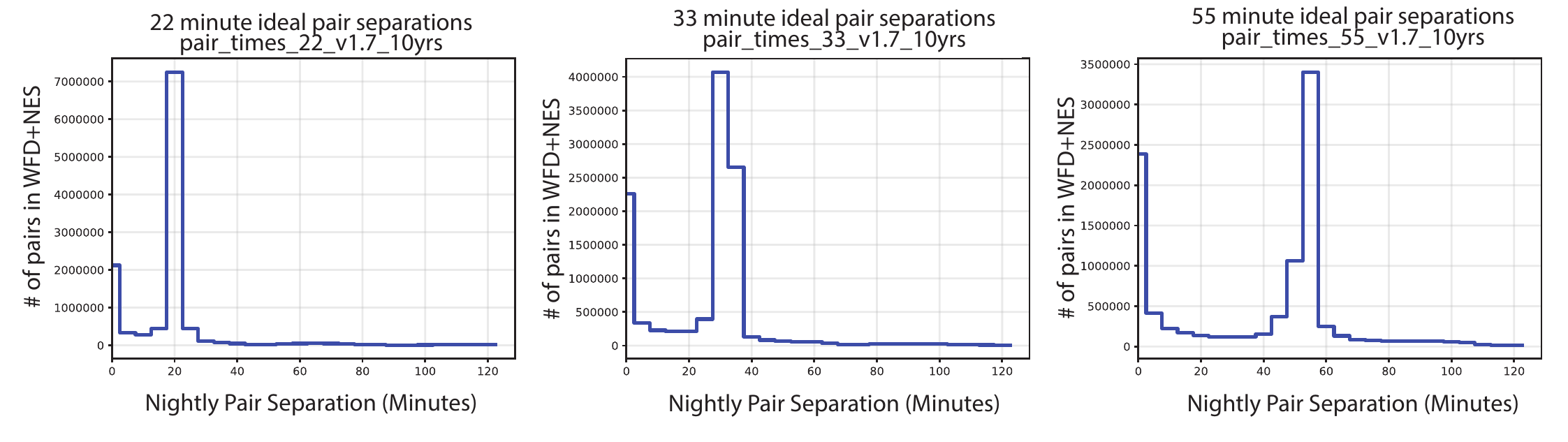}
\caption{Distribution of nightly pair separations across the Wide-Fast-Deep (WFD) and the Northern Ecliptic Spur (NES) from three simulations that make up the v1.7 \texttt{pair\_times} family. The histograms are truncated at 120 minutes.  \label{fig:pair_sep_hist}}
\end{center}
\end{figure}

Nightly pairs in combinations of the $g$, $r$, and $i$ filters are the most conducive to finding solar system objects. We explore the intra-night separations in these combinations of filters. \update{As discussed in Section \ref{sec:metric_limitations}, the SSP pipelines} require that motion be seen between the two exposures \citep{lsstMOPS,lsstSSP}. If a solar system body has not moved sufficiently for it to be identified as a new transient source in the next visit, SSP will not be able to spot that moving object. The separation between nightly pairs sets the furthest distance at which SSP can detect moving sources. The time between repeat visits also directly impacts the number of pairs observed per night, and thus the total area searchable for solar system objects. We aim to find the best pair time separation that increases the distances that SSP is sensitive to without making a significant trade off in observing efficiency. This would allow the Rubin SSP to detect more distant TNOs while not compromising the discovery and characterization of the more inward solar system populations. We note that the tunable parameter here is the Rubin scheduler's goal for spacing the repeat visits in a given night. In reality, this will be a distribution centered about the ideal value the Rubin scheduler is aiming for. This is shown in Figure \ref{fig:pair_sep_hist} for three examples from the v1.7 \texttt{pair\_times} simulations  which take mixed filters with ideal separations between 11, 22, 33, 44, and 55 minutes.

\begin{figure}
\begin{center}
\includegraphics[width=0.55\columnwidth]{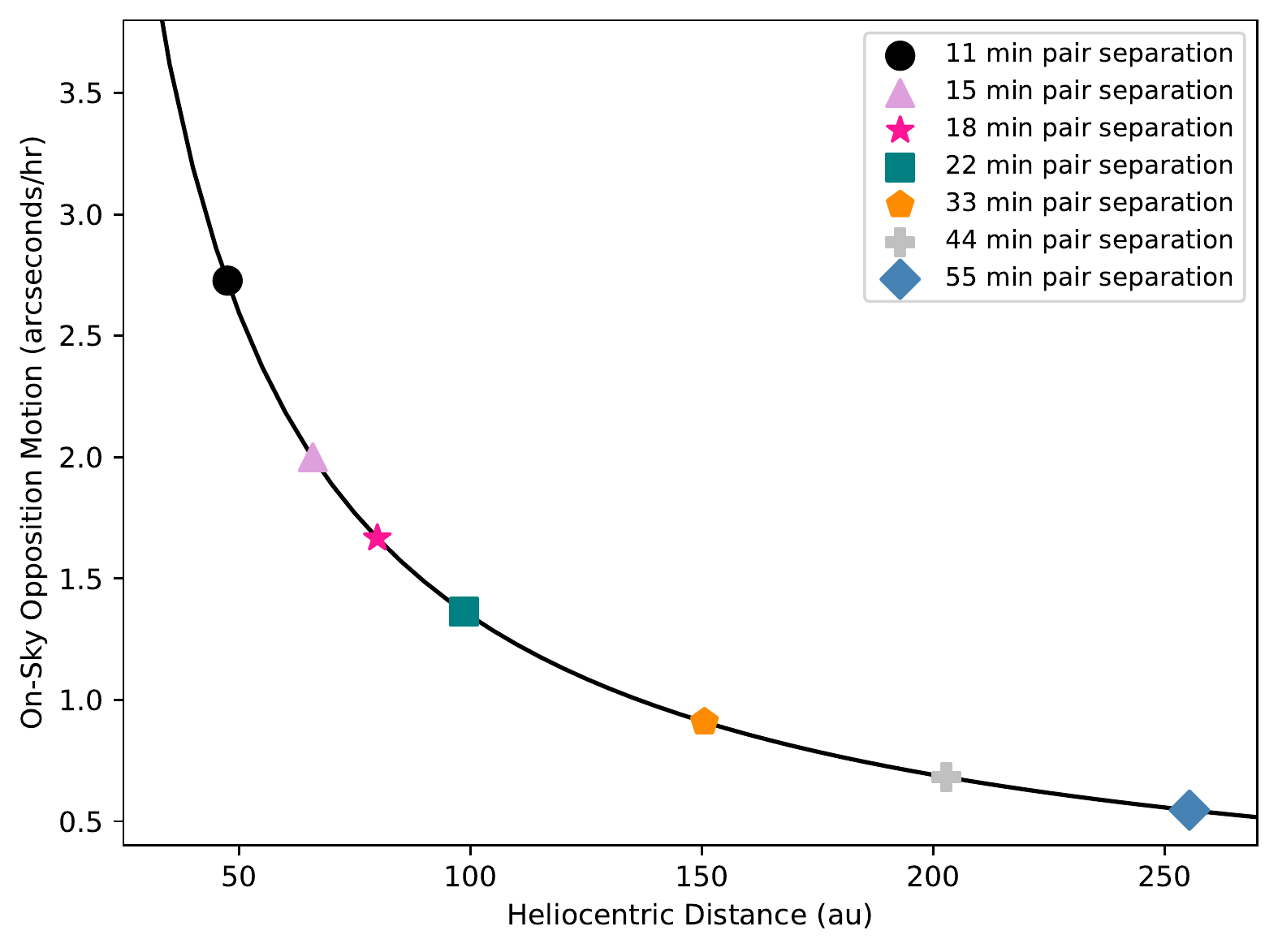}
\caption{The opposition on-sky motion observed  on Earth as a function of different heliocentric distance (solid line). The colored points represent the calculated slowest motion/distance detectable by the Rubin solar system Processing (SSP) pipeline for a range of nightly pair spacings. \label{fig:on_sky_motion}}
\end{center}
\end{figure}

Solar system objects appear to move fastest on-sky when they are at opposition, where the apparent motion is dominated by the parallax induced by the  Earth's movement. The on-sky rate of motion at opposition for a body exterior to Earth's orbit on a circular and  coplanar orbit can be defined as:
\begin{equation}
\frac{d\theta}{dt} = 148\left({\frac{1-\rh^{-0.5}}{\rh-1}}\right)
\label{eq:onsky}
\end{equation}
where $\rh$ is the body's heliocentric distance in au and $\frac{d\theta}{dt}$ is the apparent motion at opposition in arcseconds per hour \citep{1988AJ.....95.1256L}. We assume 140 mas as a conservative estimate for the astrometric uncertainty for sources near the LSST detection limit and a 3$\sigma$ positional shift for the SSP pipelines to successfully identify the moving object as a new source in the second observation (Private Communication, Mario Juri\'c 2022). This translates to solar system bodies having to move at least 0.5 arcseconds between the visits in order to become detectable by SSP, setting a minimum speed limit.  In Figure \ref{fig:on_sky_motion}, we estimate SSP's  motion limit  for the range of pair separations including those explored in the \texttt{pair\_times} simulations. The solid line represents the opposition on-sky rate of motion as calculated from Equation \ref{eq:onsky}. 

\begin{figure}
\begin{center}
\includegraphics[width=0.95\columnwidth]{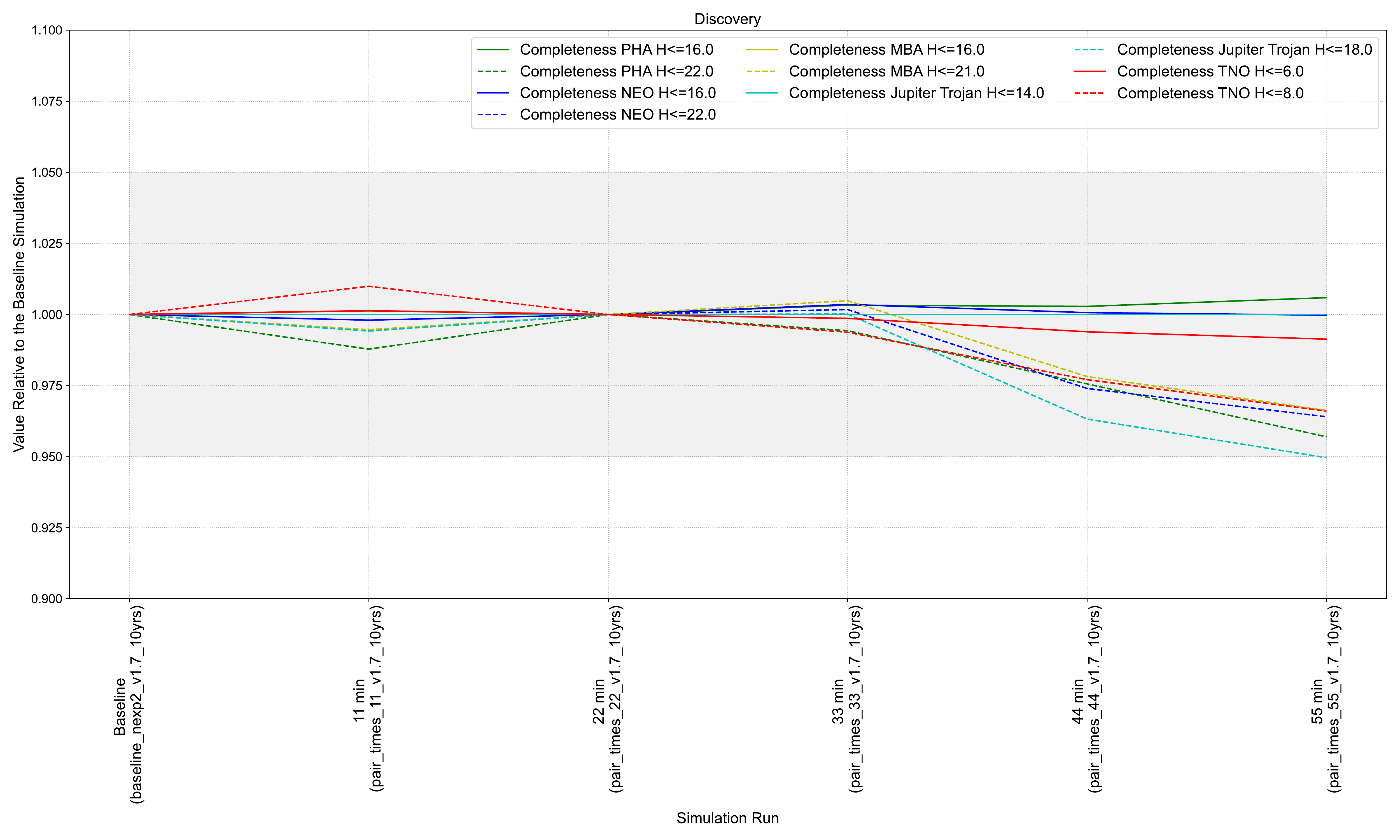}
\includegraphics[width=0.95\columnwidth]{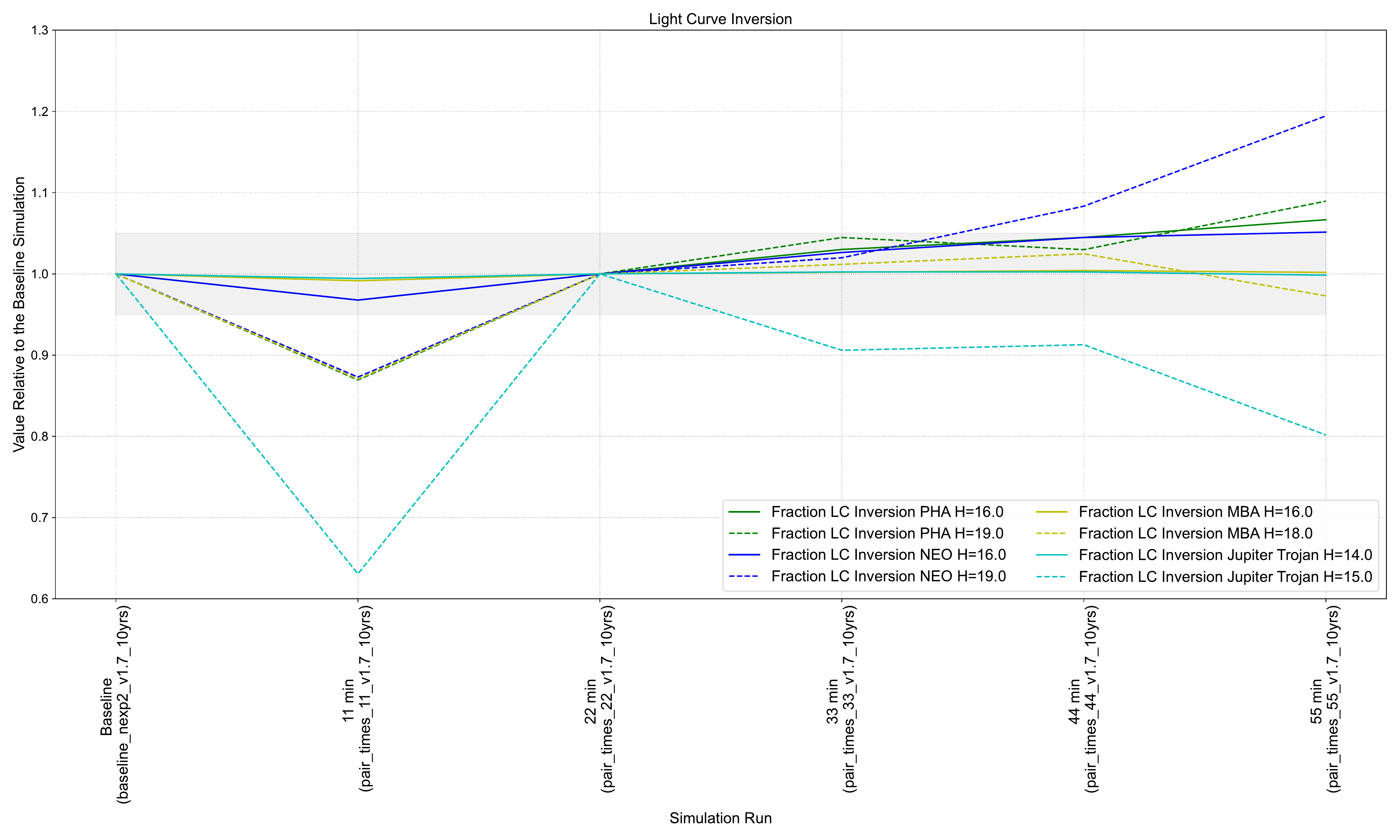}
\caption{Changing the ideal nightly pair separation (v1.7 simulations). \update{The baseline (reference) simulation with the default scheduler configuration} for this cadence experiment is the first entry on the left. All values have been normalized by this simulation's output. The gray shading outlines changes that are within $\pm 5\%$ of the baseline simulation. Top: Discovery Metrics. Bottom: Light Curve Inversion Metrics. \label{fig:v1.7_pair_times}}
\end{center}
\end{figure}

The bulk of the classical Kuiper belt extends from $\sim$42 au to 47.7 au, the  2:1 Mean Motion Resonance (MMR) with Neptune, but the Kuiper belt's scattered/scattering disk and detached/high-perihelion TNO population (with perihelion at $\sim$50-80 au) do extend well beyond that \citep{2001ApJ...554L..95T,2011AJ....142..131P, 2014AJ....148...55A, 2018ApJS..236...18B, 2022ApJS..258...41B}. Separations longer than 18 minutes are needed to search for objects beyond 80 au. Separations longer than 33 minutes start to slightly negatively affect the discovery metrics and significantly enhance the light curve metrics, as plotted in Figure \ref{fig:v1.7_pair_times}.  The loss of discovery at fainter absolute magnitudes is less than 5$\%$ even at 55 minute spacings. As the time gap gets longer, the pairs are more vulnerable to interruptions, mostly from weather. The fraction of $gri$ pairs peaks at 22 minutes, but the total visits and the on-sky area reaching 850 visits both increase with longer pair separations. The light curve inversion metrics go up with longer gaps between intra-night visits, due to the increase in the total number of visits, with a larger number of singleton images that are spread out across the observable sky (see Table \ref{tab:nightly_pairs}). Having the Rubin scheduler aim for the two visits to be separated by 33 minutes is the best compromise between optimizing the number of nightly pairs completed and heliocentric distance probed. We note that the SCOC moved the LSST baseline strategy from aiming for 22 minutes nightly pair separations to 33 minutes from the v2.0 simulations onward \citep{SCOC_Report_1}.

\begin{deluxetable}{lcccc }
\tablecaption{Diagnostics for the LSST Cadence Simulations Changing the Desired Separation Between Nightly Pairs\label{tab:nightly_pairs}}
\tablewidth{0pt}
\tablehead{
\colhead{LSST Cadence Simulation Name }  & \colhead{Ideal Separation} & \colhead{Total $\#$}  of &   \colhead{Area with} & \colhead{Mean Fraction (RMS) }\\
& \colhead{Separation} &  \colhead{of On-Sky} & \colhead{ $>$ 825  visits} &\colhead{of WFD+NES visits}   \\
 & \colhead{Between} & \colhead{Visits} &  \colhead{(degrees$^2$)}   &  \colhead{in 15-60 min}   \\
 & \colhead{Nightly Pairs} &  & & \colhead{separated pairs} \\
 & \colhead{(min)} & & &  \colhead{$g$,$r$, or $i$ filters only$^*$}
}
\startdata
\texttt{pair$\_$times$\_$11$\_$v1.7$\_$10yrs} & 11 & 1947985 & 14356.96 & 0.240 (0.061) \\
\update{\texttt{baseline$\_$nexp2$\_$v1.7$\_$10yrs}/}\texttt{pair$\_$times$\_$22$\_$v1.7$\_$10yrs}  & 22 &2045493 & 17982.71 & 0.586 (0.055) \\
\texttt{pair$\_$times$\_$33$\_$v1.7$\_$10yrs} & 33 & 2075493 &  18076.71& 0.546 (0.057) \\
\texttt{pair$\_$times$\_$44$\_$v1.7$\_$10yrs} & 44 & 2089977 & 18104.40 & 0.475 (0.061) \\
\texttt{pair$\_$times$\_$55$\_$v1.7$\_$10yrs} & 55 & 2100189 & 18108.60  & 0.398 (0.061) \\
\enddata
\tablenotetext{^*}{\update{15 minute} separations cover the full classical Kuiper belt. \update{60 minutes was chosen as the upper limit because the bulk of the nightly pairs in these runs are separated by less than this value (See Figure \ref{fig:pair_sep_hist}).}}
\tablecomments{\update{These simulations used 2 $\times$ 15 s snaps per visit.}}
\end{deluxetable}

\begin{figure}
\begin{center}
\includegraphics[width=0.91\columnwidth]{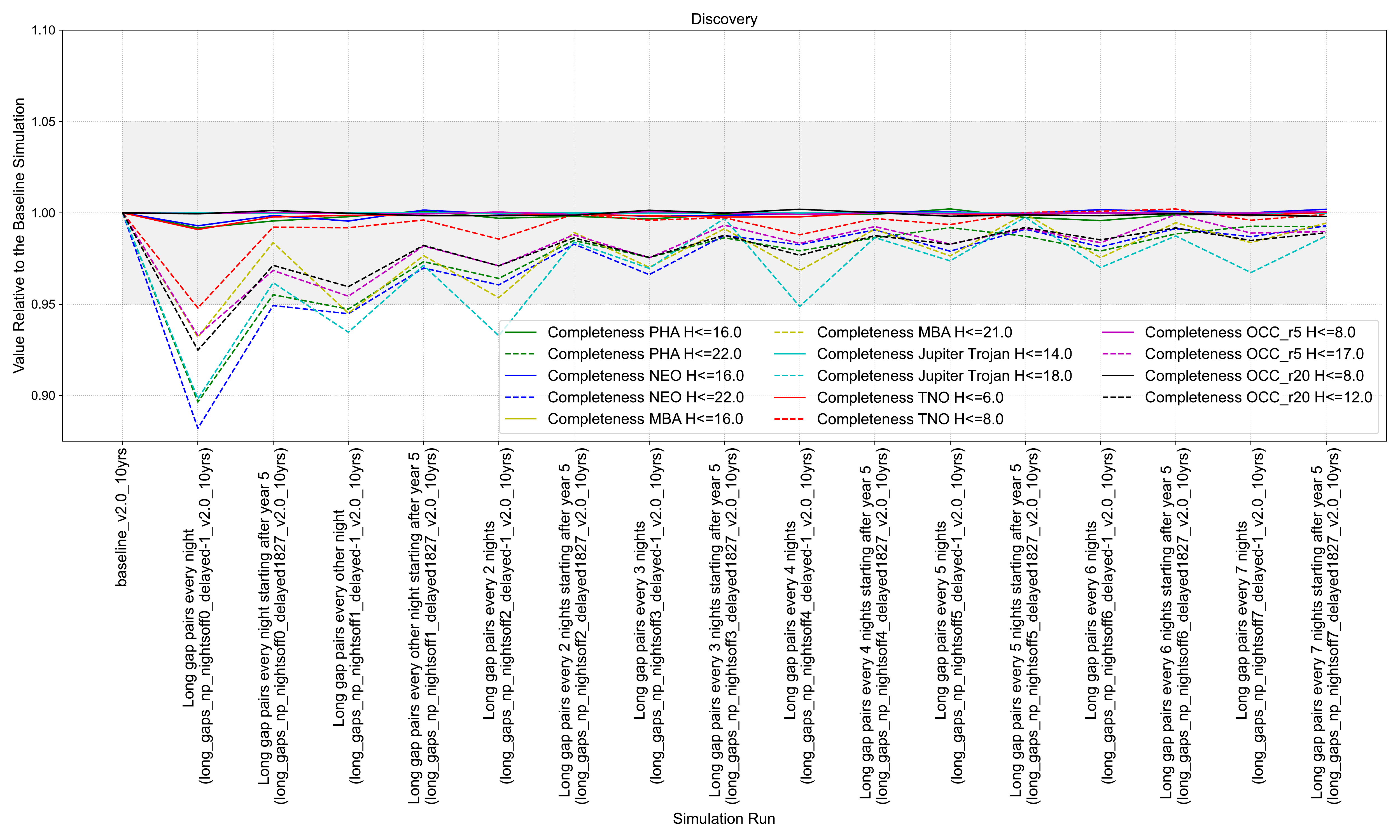}
\includegraphics[width=0.91\columnwidth]{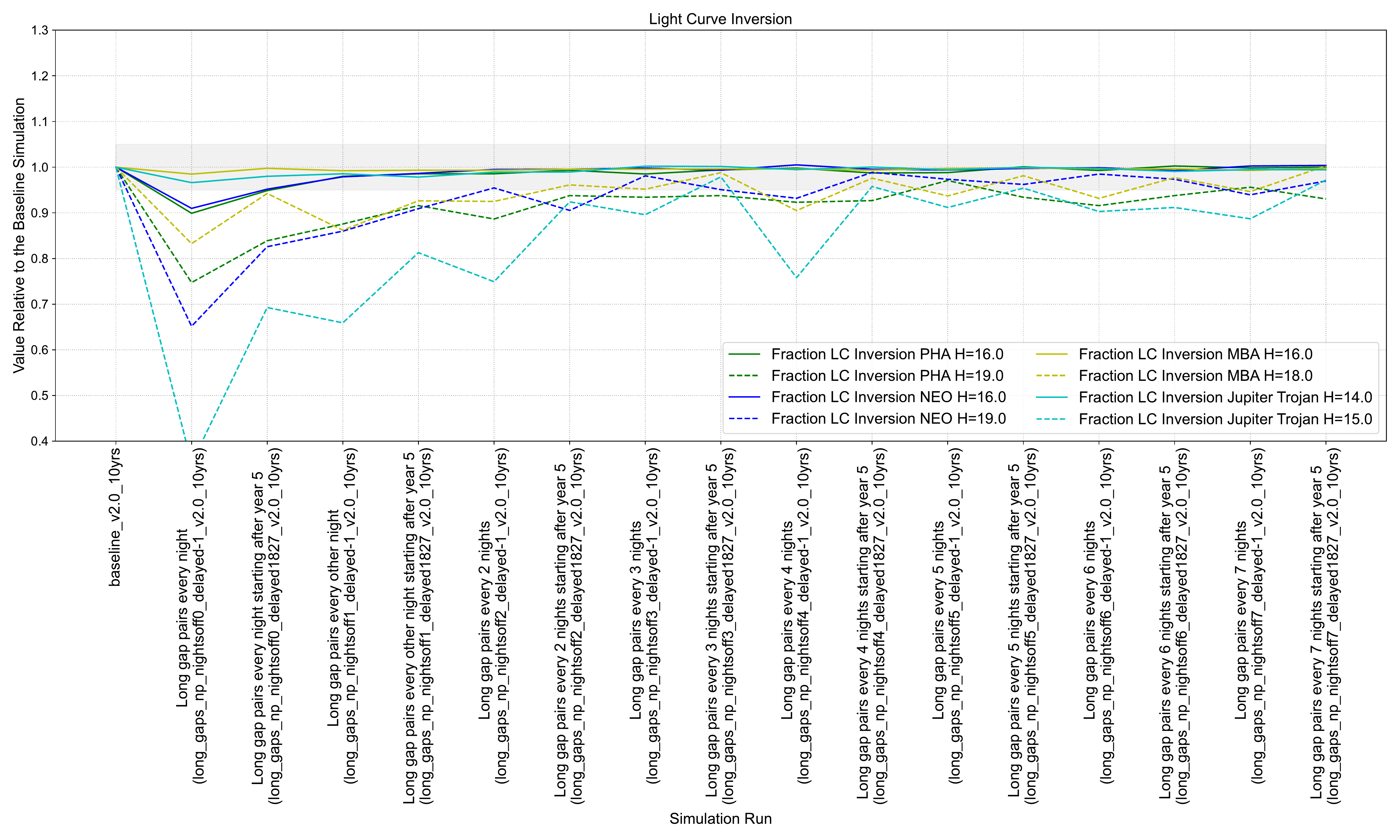}
\caption{Impact of various scenarios for lengthening the gap between pairs to be variable in the range of 2-7 hours (v2.0 simulations). \update{The baseline (reference) simulation with the default scheduler configuration} for this cadence experiment is the first entry on the left. All values have been normalized by this simulation's output. The gray shading outlines changes that are within $\pm 5\%$ of the baseline simulation. Top: Discovery Metrics. Bottom: Light Curve Inversion Metrics. The y-axis is truncated in the Light Curve Inversion plot. The Jupiter Trojans extend below the y-axis range for \texttt{long\_gaps\_np\_nightsoff0\_delayed-1\_v2.0\_10yrs}. The baseline simulation has an ideal separation of 33 minutes.\label{fig:v2.0_long_gaps_no_repeats}}
\end{center}
\end{figure}

The v2.0 \texttt{long\_gaps\_np} (long gaps no pairs) simulations extend the time between nightly repeat exposures to  2-7 hours in various combinations. These simulations either begin the extended pair separation in Year 1 (\texttt{delayed-1}) or after Year 5 (\texttt{delayed1827}). This simulation family explores the impact of executing the long gaps strategy every night and less frequently, where the \texttt{nightsoff} parameter (the number of sequential nights with no long gap sequences) is varied. On the nights when the long gap observing is not active, the scheduler aims for 33 minute nightly pair simulations like the v2.0 baseline survey. These simulations are one option explored to potentially better capture fast evolving astrophysical phenomena, as suggested by \cite{2022ApJS..258...13B}; additional strategies for addressing the temporal coverage of fast transients are explored in Section \ref{sec:third_nightly_visits}.  As seen in Table \ref{tab:nightly_pairs},  the fraction of $gri$ pairs is largest when the Rubin scheduler is tasked with 33 minute pair spacings. Therefore these hours long separations are not going to be efficient in generating nightly pairs conducive for the moving object search. This can be seen in the discovery and light curve metrics displayed in Figure \ref{fig:v2.0_long_gaps_no_repeats}. Across all populations, the light curve metrics and detections decreases. The increased sensitivity to objects beyond 150 au is not worth the tradeoff purely from a planetary astronomy perspective, but the simulations that do not have the long gaps in intra-night visits occurring every night have less impact. The less time devoted to the large time gap pair observing, the less severe the hit to the discovery and light curve metrics. Nonetheless, 33 minute pair separations are better optimized for outer solar system discoveries and the completion of repeat visits within the night. 

\subsubsection{Filter Choices for Repeat Visits in a Night}
\label{sec:repeat-filter-choice}



\begin{figure}
\begin{center}
\includegraphics[width=0.95\columnwidth]{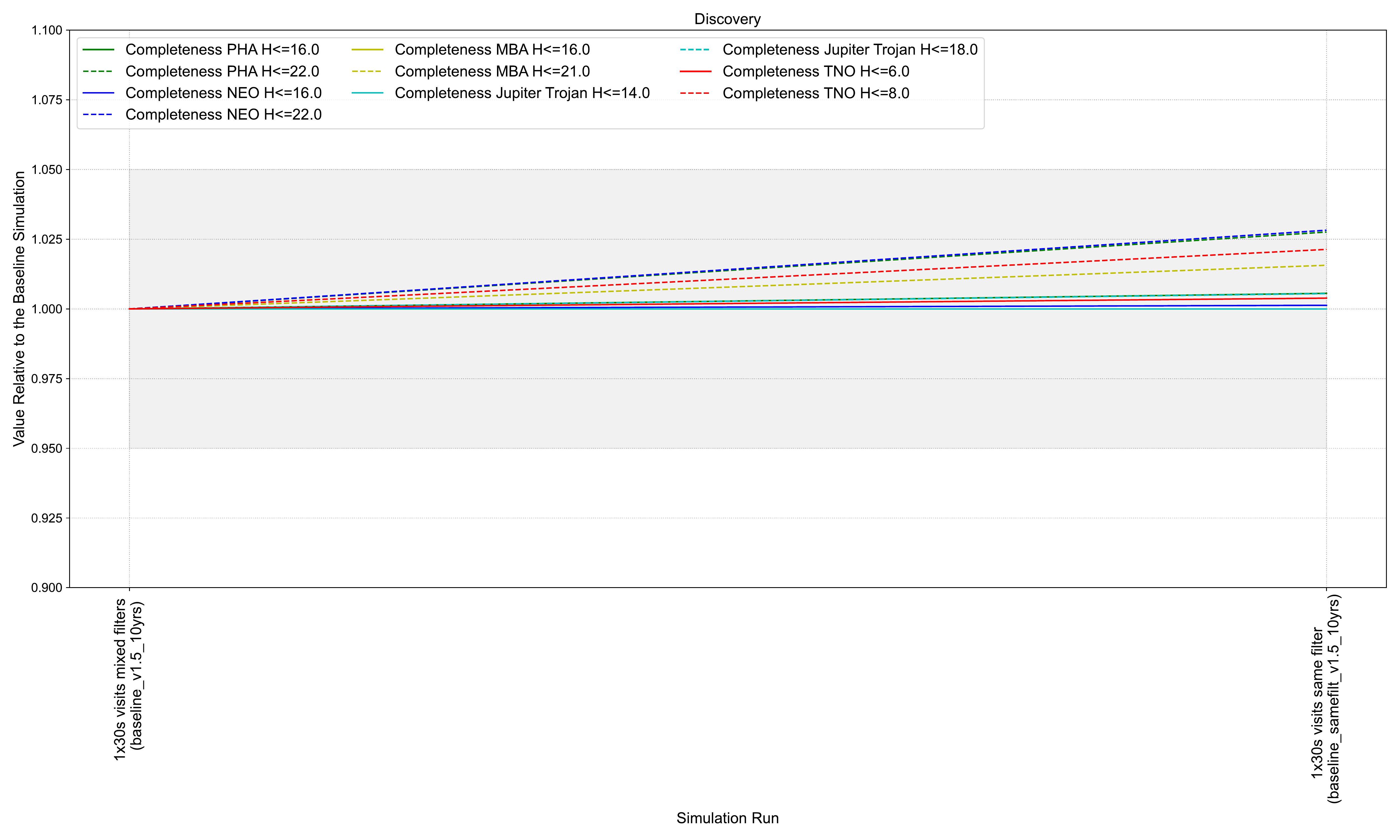}
\includegraphics[width=0.95\columnwidth]{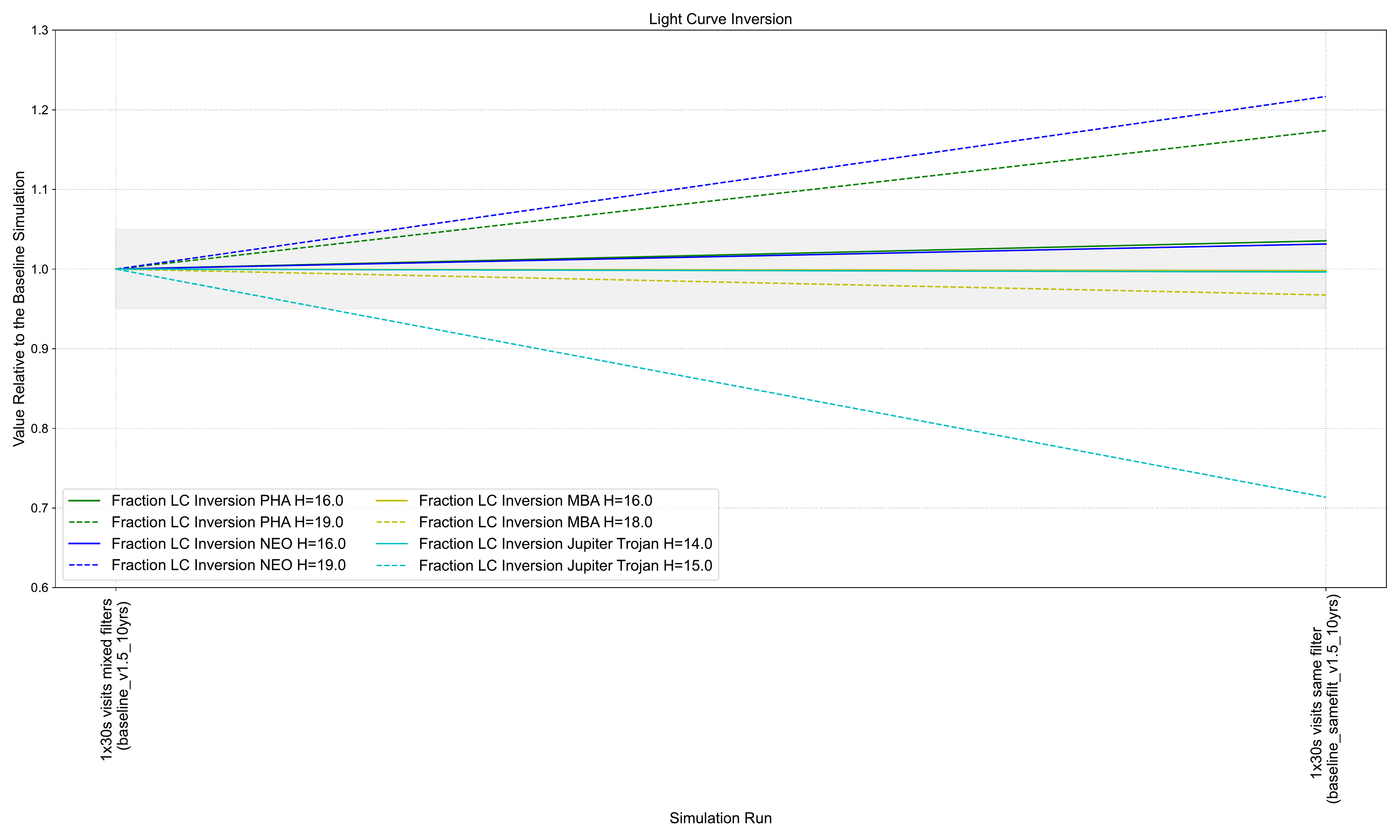}
\caption{Nightly pairs in the same versus different filters (v1.5 simulations). \update{The baseline (reference) simulation with the default scheduler configuration} for this cadence experiment is the first entry on the left. All values have been normalized by this simulation's output. The gray shading outlines changes that are within $\pm 5\%$ of the baseline simulation. Top: Discovery Metrics. Bottom: Light Curve Inversion Metrics.\label{fig:1.5_filters_in_a_night}}
\end{center}
\end{figure}
\begin{figure}
\begin{center}
\includegraphics[width=0.97\columnwidth]{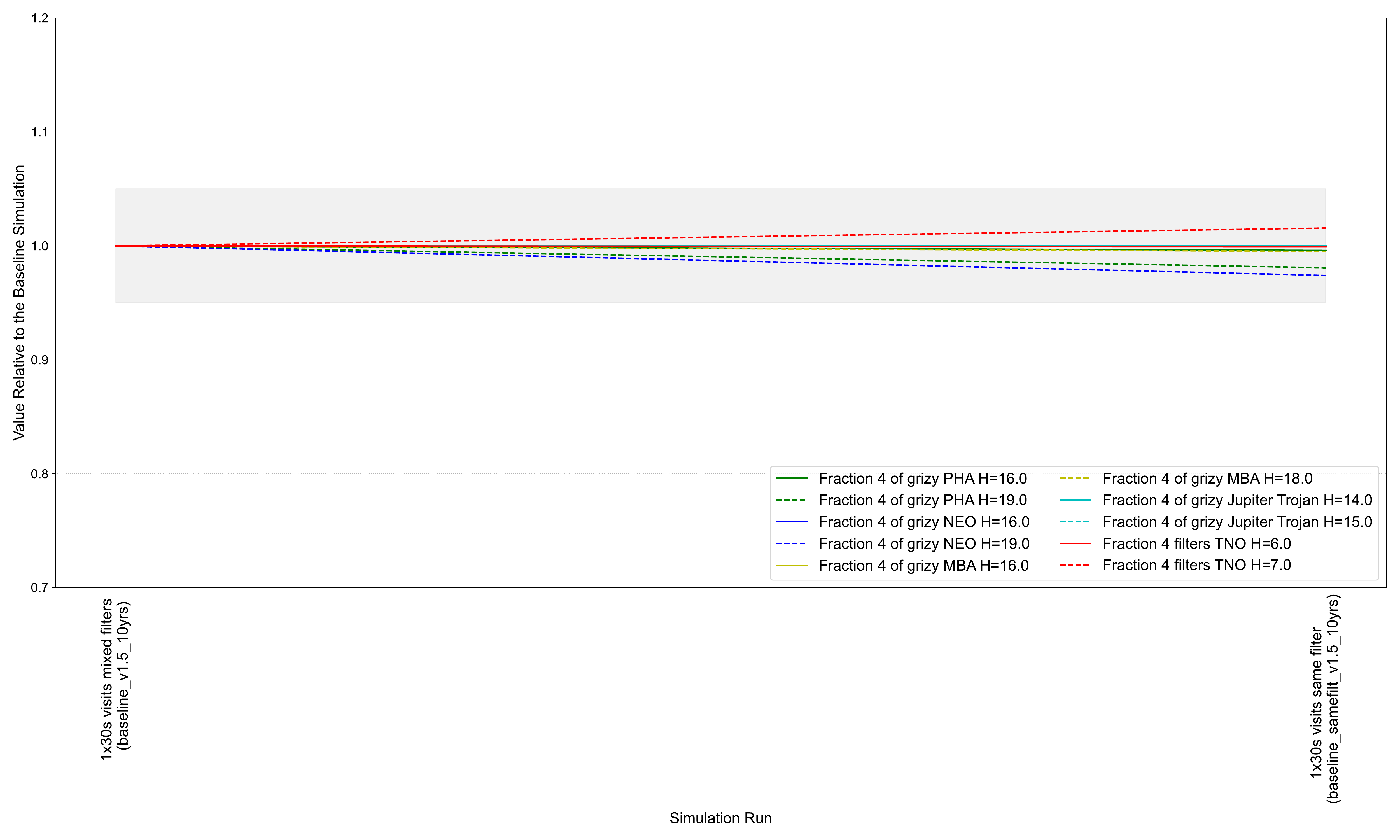}
\caption{Color-light curve metrics for observing strategies with nightly pairs in the same versus different filters (v1.5 simulations). \update{The baseline (reference) simulation with the default scheduler configuration} for this cadence experiment is the first entry on the left. All values have been normalized by this simulation's output.The gray shading outlines changes that are within $\pm 5\%$ of the baseline simulation. \label{fig:v1.5_filters_in_a_night_4filters}}
\end{center}
\end{figure}

In the v1.5 simulations, cases were run with nightly pairs of visits performed in either matching filters (\texttt{baseline\_samefilt\_v1.5\_10yrs}) or in mixed filters (\texttt{baseline\_v1.5\_10yrs}). 
The discovery metrics for solar system populations are largely unaffected by the choice (top panel of Figure~\ref{fig:1.5_filters_in_a_night}). 
This is because the mixed-filter pairs in the cadence simulations contain filter pairs such as $g$-$r$ and $r$-$i$ where the colors of solar system objects allow detections in both filters (we note that $r$-$i$ pairings are better than $g$-$r$ pairings for the reddest objects like TNOs).
Similarly as shown in Figure \ref{fig:v1.5_filters_in_a_night_4filters}, the color-light curve metrics for different populations are not significantly affected by the choice of same or mixed filters; there is likely an advantage to having mixed filters within \update{the same night} in that it could provide a \update{single-night color estimate} for objects that rotate slowly compared to the visit separation.
For faint NEOs and MBAs, nightly pairs in the same filter does boost the light curve inversion metric by 15-20\% (bottom panel of Figure~\ref{fig:1.5_filters_in_a_night}). 
This is likely due to nightly pairs of faint objects that are only detectable in a small number of filters.
However, faint Jupiter Trojans suffer a 30\% loss in the light curves metric for the same-filter pairs. 
This is likely related to the light curve metric requirement than observations in a filter span at least 90$^\circ$ in ecliptic longitude.
The Jupiter Trojans move more slowly on-sky compared to the other populations this metric is calculated for, and having same-filter nightly pairs reduces the number of different nights (and thus different longitudes) an object might be observed in that filter; for faint Jupiter Trojans, this appears to reduce the odds that successful detections in a given filter span the required range of longitudes.

\subsubsection{Suppressing Extra Visits}
\label{sec:visit_suppress}

In the baseline cadence (\texttt{baseline\_v2.1\_10yrs}), up to 20\% of the \update{pointings are visited} more than twice per night. By adding an additional basis function to suppress these repeat visits to the Rubin scheduler algorithm, the additional visits can be distributed to different nights, thus changing the inter-night cadence or season length for a given field. 
The suppress repeats (\texttt{no\_repeat\_rpw}) family of simulations explores these changes by considering six different values for the weight of the suppression factor, indicated as \texttt{rpw}, namely: 1, 2, 5, 10, 20, and 100. This number basically reflects how strongly the suppress-revisits basis function influences the scheduler: the higher the number, the lower the number of revisits per night will be. Note that some regions of the sky will still be observed more than two times within a night if they are included in overlapping pointings.

An immediate consequence of redistributing the visits over different nights is a decreased total area with more than 825 visits per pointings, from a negligible effect ($\lesssim$0.1\% at \texttt{rpw}=1) to a more significant effect of $\lesssim$5\% at \texttt{rpw}=20. However, because of the extended timeline, the discovery metrics are generally improved with respect to the WFD: a suppressing factor between 2 and 10 will increase the discovery rate for all the different families, while for \texttt{rpw}=1, 20 or 100, there is only a marginal decrease ($\lesssim$0.005\%) in the discovery rate of faint TNOs and bright comets from the Oort Cloud. A suppressing factor equal or larger than 10 will also impact the metrics of light curve inversion, reducing up to $\approx$10\% the number of faint MBAs and Jupiter Trojans for which inversion will be feasible.
Summarizing, the ``suppress visit" family cadence produce negligible effects on solar system science, with a marginal improvement on the discovery rates for \texttt{rpw}=2 and 5, and a marginal decrement of the number of faint objects for which we will be able to perform light curve inversion for \texttt{rpw}=10, 20, or 100.

\begin{figure}
\begin{center}
\includegraphics[width=0.95\columnwidth]{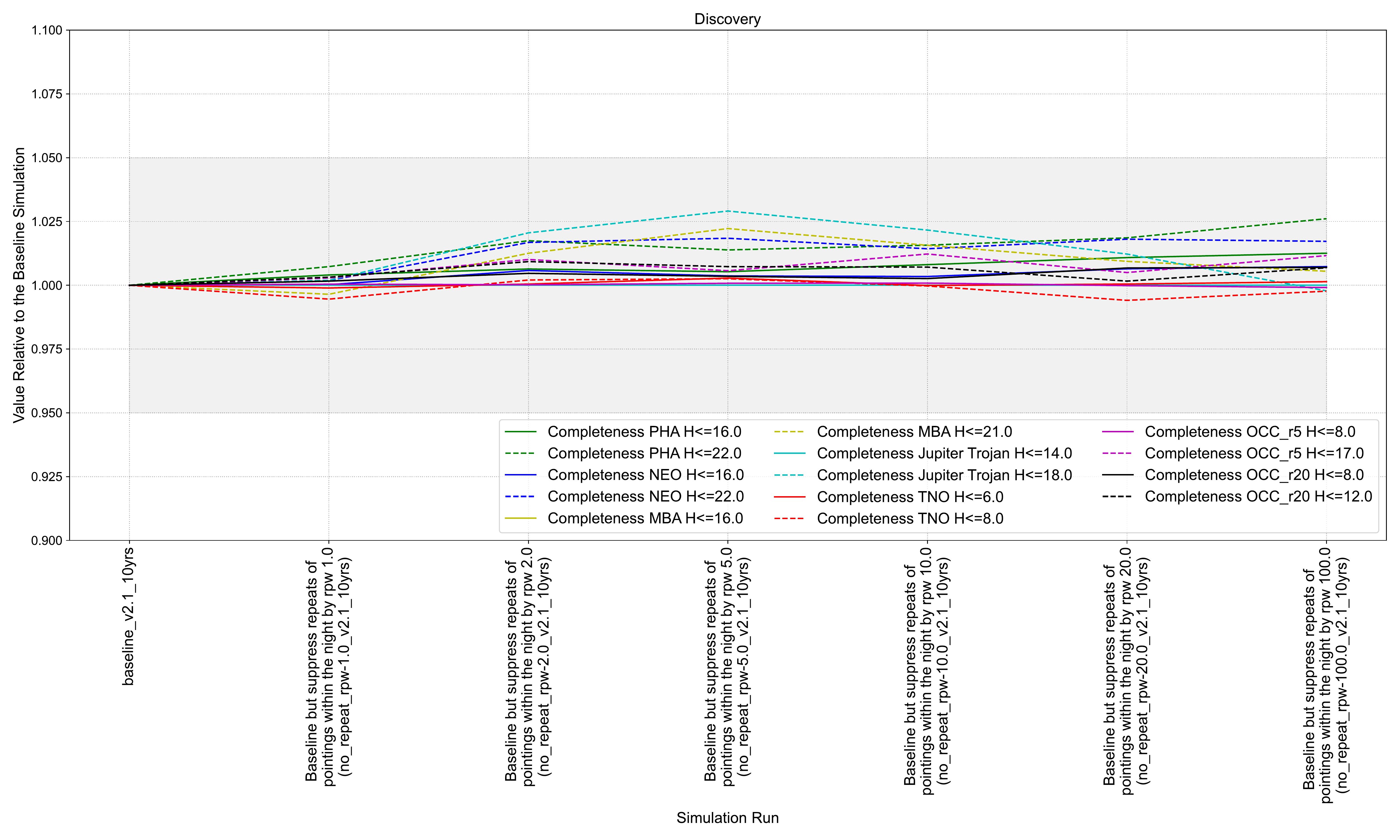}
\includegraphics[width=0.95\columnwidth]{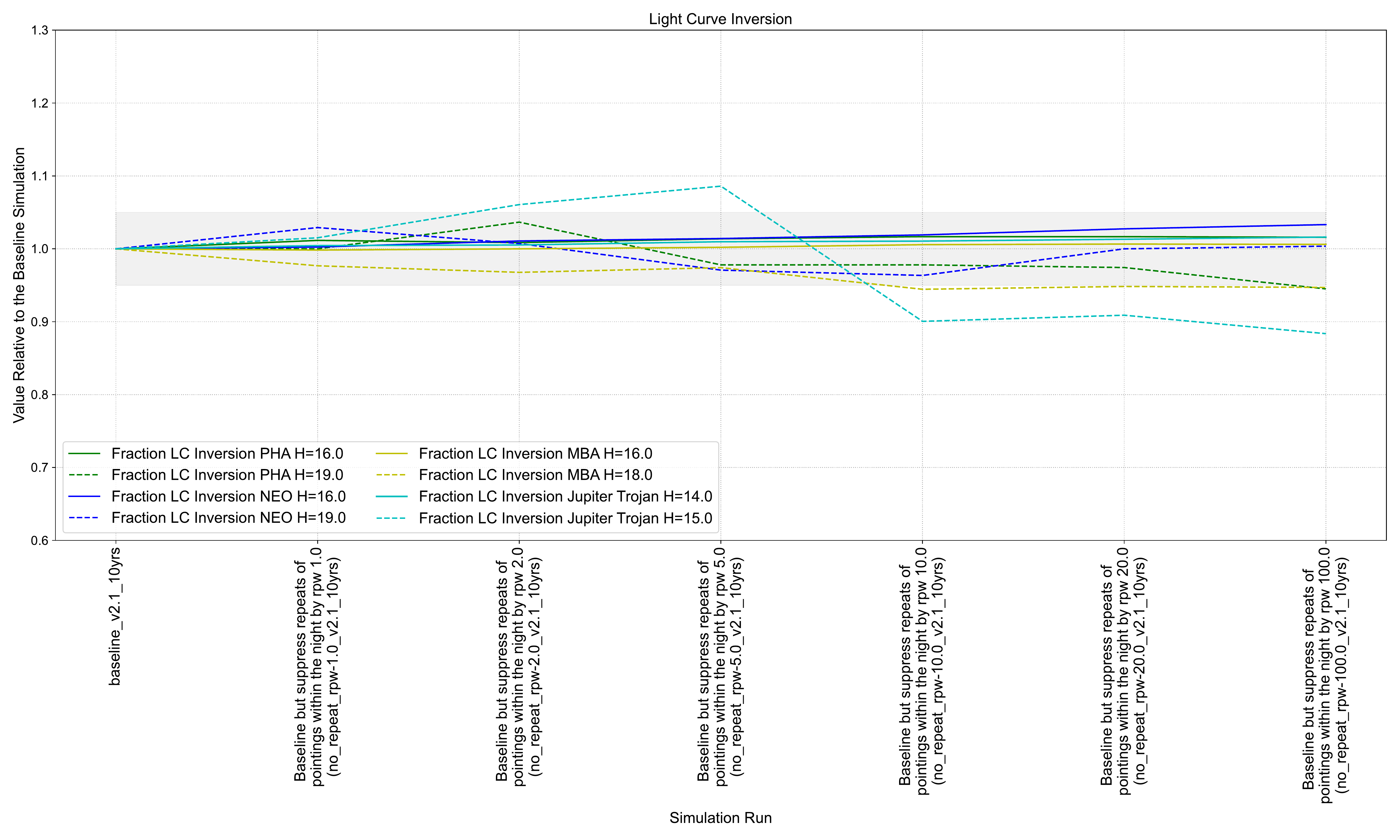}
\caption{Investigating ways of reducing extra repeat visits and redistributing (v2.1 simulations). \update{The baseline (reference) simulation with the default scheduler configuration} for this cadence experiment is the first entry on the left. All values have been normalized by this simulation's output. The gray shading outlines changes that are within $\pm 5\%$ of the baseline simulation. Top: Discovery Metrics. Bottom: Light Curve Inversion Metrics.\label{fig:v2.1no_no_repeat_rpw}}
\end{center}
\end{figure}

\subsubsection{Third Visits in a Night}
\label{sec:third_nightly_visits}

\begin{figure}
\begin{center}
\includegraphics[width=1.0\columnwidth]{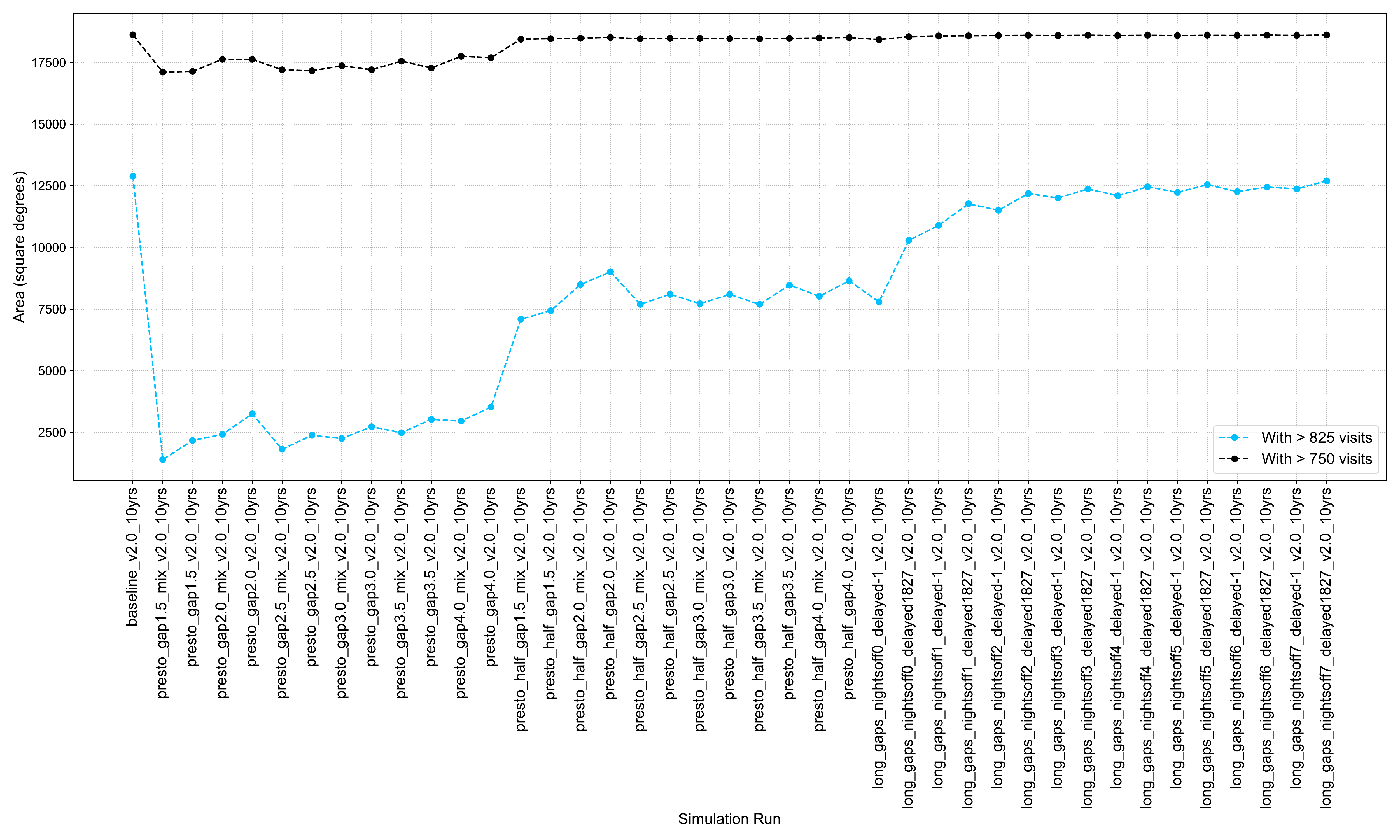}
\caption{Comparison of the sky coverage with greater than 825 deg$^2$ (in cyan) and 750 deg$^2$ in black for v2.0 cadence simulations with various options for third repeat visits. \update{The baseline (reference) simulation with the default scheduler configuration} for this cadence experiment is the first entry on the left.
\label{fig:v2.0_area_third_visit}}
\end{center}
\end{figure}

\begin{figure}
\begin{center}
\includegraphics[width=0.93\columnwidth]{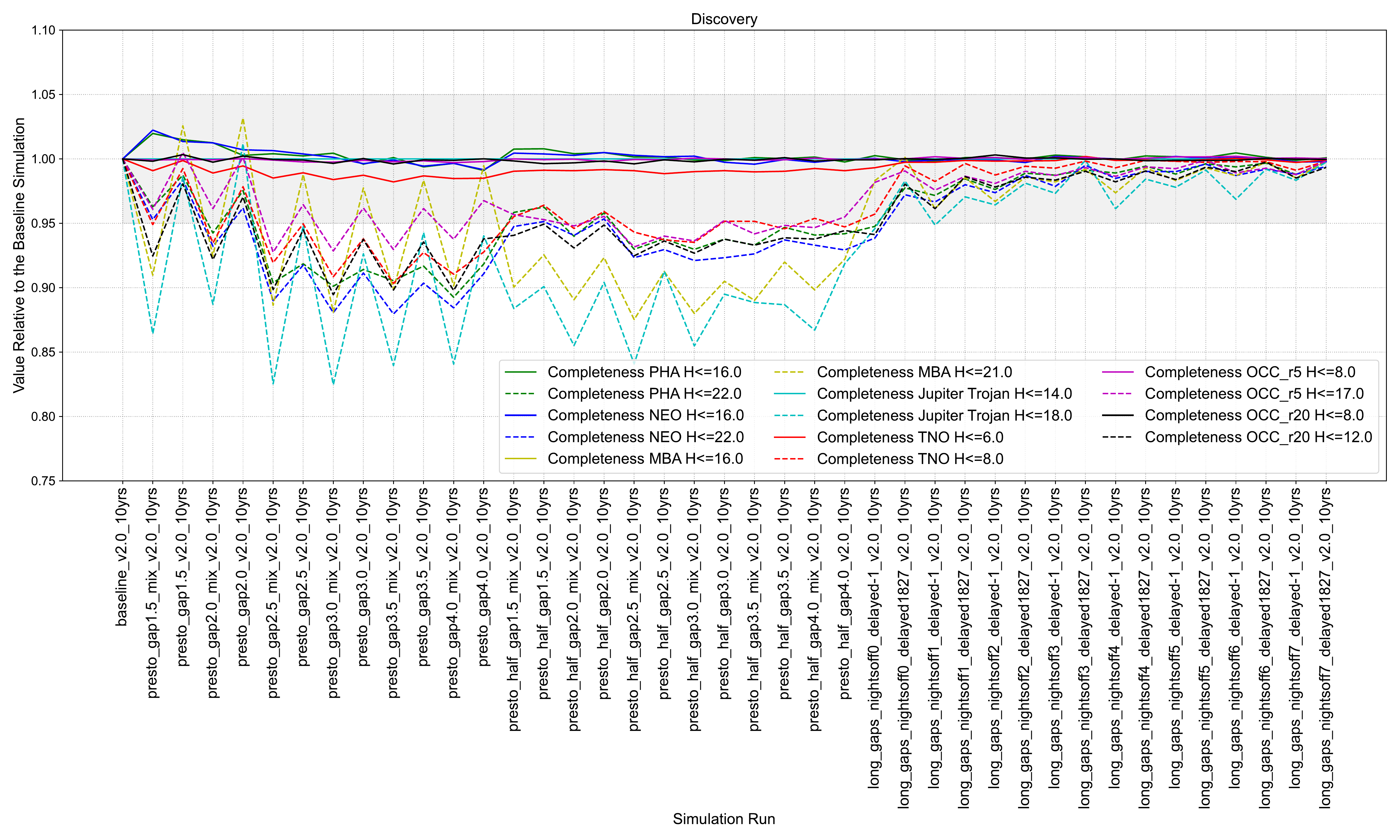}
\includegraphics[width=0.93\columnwidth]{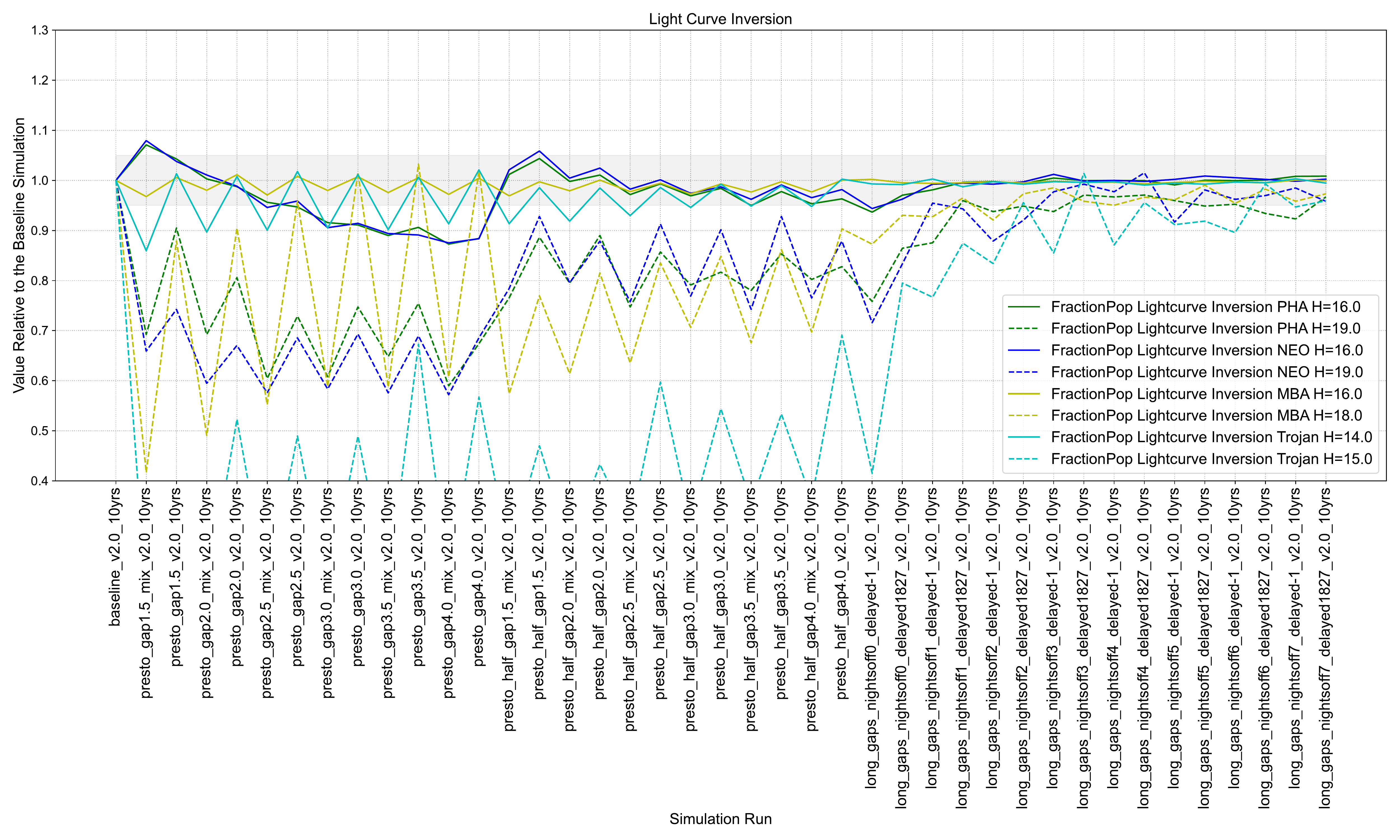}
\caption{Impact of various third visit scenarios (v2.0 simulations). \update{The baseline (reference) simulation with the default scheduler configuration} for this cadence experiment is the first entry on the left. All values have been normalized by this simulation's output. The gray shading outlines changes that are within $\pm 5\%$ of the baseline simulation. Top: Discovery Metrics. Bottom: Light Curve Inversion Metrics. We have truncated the y-axis for visibility in the Light Curve Inversion Metrics plot. The fraction of the Jupiter Trojan detections in some of these runs compared to the baseline is lower than 0.4 and off the bottom edge. \label{fig:v2.0third_visits}}
\end{center}
\end{figure}

There is a strong desire among other Rubin Observatory LSST Science Collaborations to add a third visit in a \update{different filter} to aid in capturing and identifying fast ($<1$ d) transients by adding more color information \citep[see][]{2019PASPpresto} to the base survey of nightly pairs of visits which are separated by $\sim20$--30 minutes. The \texttt{presto\_gap} family of simulations explores the effects of adding a third visit to the night's visits after a time period of 1.5--4\,hours. Within the \texttt{presto} family of simulations, there are two significant subfamilies. The \texttt{presto\_half} simulations explore the effect of adding the third image/triplet every other night rather than every night of the cadence while the \texttt{presto\_gap\_mix} has a wider separation and difference in colors between the initial pair and the third visit (e.g., $g$+\textbf{\emph{i}}, $r$+\textbf{\emph{z}}, $i$+\textbf{\emph{y}} rather than $g$+$r$, $r$+$i$, $i$+$z$ initial pairs).

 The main impact of adding this third visit is to dramatically decrease the amount of well-covered survey area (see Figure~\ref{fig:v2.0_area_third_visit}). This would have a large negative impact on science cases where the objects are sparse on the sky such as discovering rare objects (e.g. ISOs) or the onset of activity on solar system objects. 
The other large effect of adding the third visit is seen in the solar system object detection and light curve inversion metrics and illustrated in Figure~\ref{fig:v2.0third_visits}. Although there is some improvement in the detection of the brighter solar system objects at the shorter gap lengths in the 1.5--2.0\,hour regime (see e.g., the \texttt{presto\_gap1.5\_mix} simulation in Figure~\ref{fig:v2.0third_visits}), this is not a high priority for the large aperture capabilities of \RObs. For the vast majority of the other simulations and solar system populations, this family of simulations produces a 20--75\% decrease in the light curve inversion metrics, well beyond our threshold for flagging these simulation families as bad for solar system science. The impacts are less dramatic for the \texttt{presto\_half} subfamily, as might be expected, since the third visit is only carried out 50\% of the time. The impact of the \texttt{\_mix} version of a simulation (with the wider spread of observed colors in the third visit) is always worse than the corresponding ``non-mixed" simulation run.

As an alternative to the \texttt{presto} families discussed above, the \texttt{long\_gaps\_nightsoffN} family (not to be confused with the \texttt{long\_gaps\_np} family of simulations considered in Section~\ref{sec:nightly_sep}) also adds a third visit in the same filter as one of the pairs (like the \texttt{presto} family). However unlike the \texttt{presto} families, 1) the third visit forming the triplet is in one of the same filters as the earlier pair, 2) only occurs if the first pair is in the $griz$ filters and 3) occurs after a longer 2--7 hour gap from the initial pair than the standard $\sim$33\ minute gap. This is done every $N$ nights ($N=0\ldots7$) e.g., \texttt{long\_gaps\_nightsoff7} has the long gaps every 7 nights and \texttt{long\_gaps\_nightsoff0} has ``zero nights off” and the long gap third visit/triplets are done every night. These families additionally come in 2 flavors, \texttt{delayed-1} and \texttt{delayed1827} where the third visit/triplets either starts immediately before the start of the survey (night -1) or in survey year 5 (night 1827) respectively.

Overall this family of simulations has much smaller detrimental effects ($<10$\%) on the area covered (final third of Figure~\ref{fig:v2.0_area_third_visit}) and most solar system metrics, except for where this is done every or almost every night (the \texttt{\_nightsoff0} and \texttt{\_nightsoff1} simulations) which hit the area and light curve inversion metrics hard (20-60\%; see final third of Figure~\ref{fig:v2.0third_visits}). These families of survey strategy simulations with longer gaps for the potential third visit in the night, could constitute a path forward towards satisfying the desires of other science goals without unduly compromising solar system science. The impact of the loss of sky area covered in all of these third visit simulations on the detectability of rare but high value targets which are sparse on the sky such as ISOs or very distant extreme TNOs (ETNOs and IOCs), needs additional simulations with these populations added. The addition of the OCCs to the later versions of the simulations, which are more much numerous on the sky than either ISOs and Sedna-like objects, and the corresponding drop in OCC discovery when adding the third visit, shows the downside of adding the third visit on discovering the rarer solar system populations.

\subsection{Rolling Cadence}
\label{sec:rolling}

\begin{figure}
\begin{center}
\includegraphics[width=0.6\columnwidth]{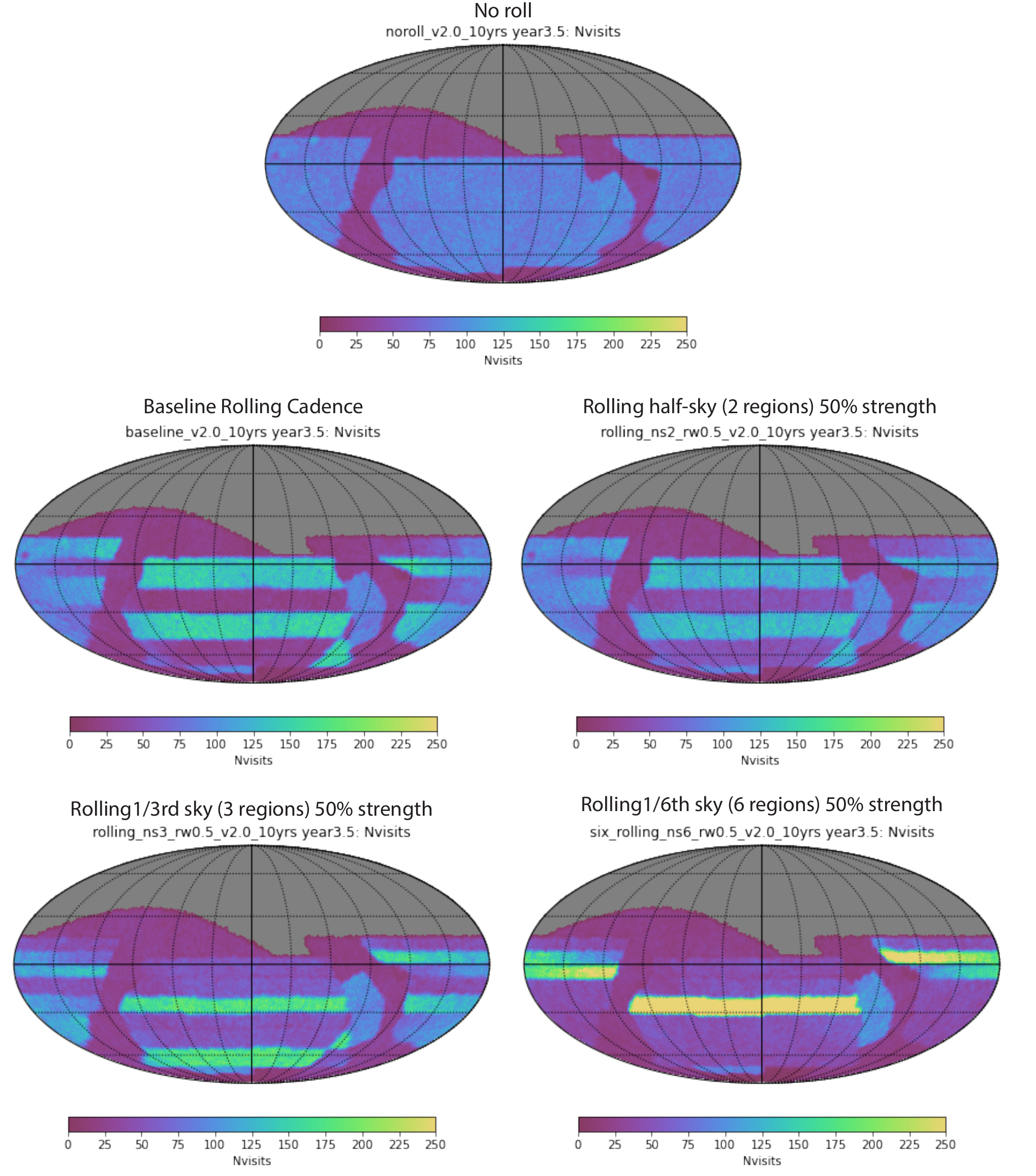}
\caption{Snapshot of the cumulative number of on-sky visits in all filters as a function of a subset of rolling cadence scenarios simulated at Year 3.5 (v2.0 simulations).  \update{The plots are centered on $\alpha$=0 and $\delta$=0. Right ascension and declination lines are marked every 30$^\circ$.} \label{fig:rolling_year_3.5}}
\end{center}
\end{figure}

\begin{figure}
\begin{center}
\begin{interactive}{animation}{baseline_v2.0_10yrs__N_Visits_mp4}
\end{interactive}
\includegraphics[width=0.7\columnwidth]{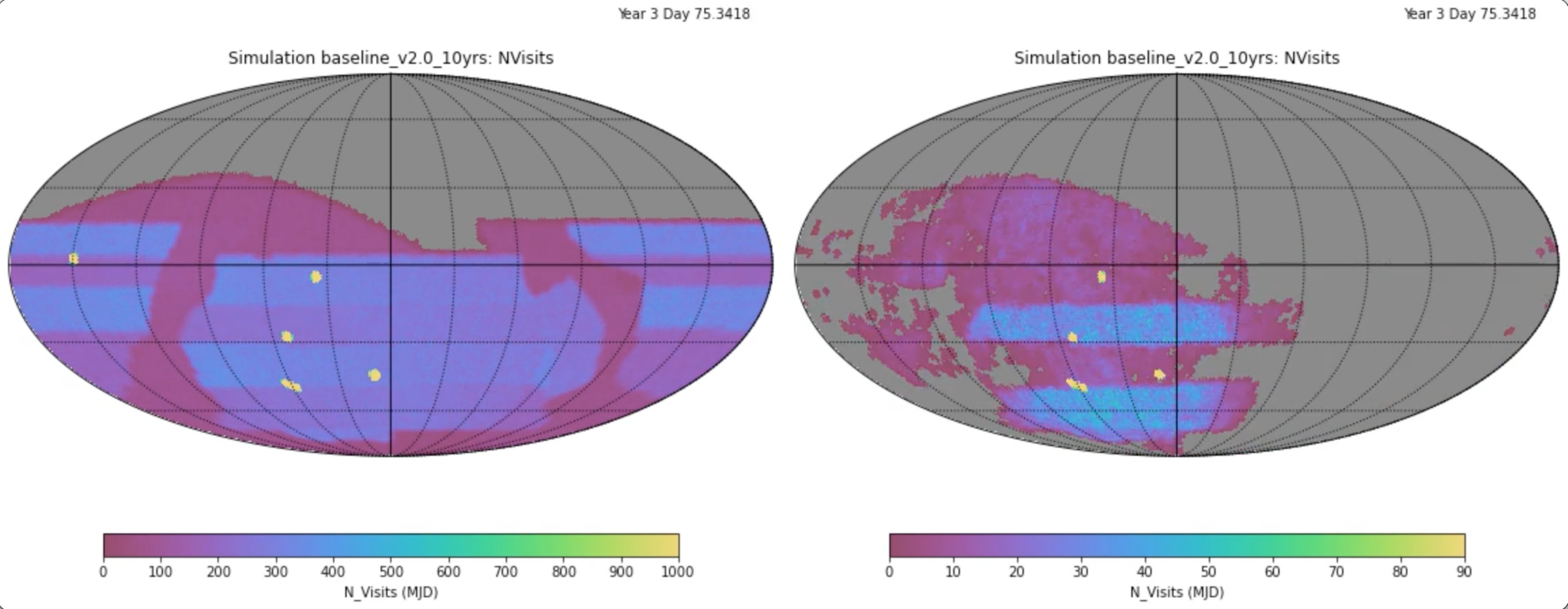}
\caption{Snapshot from a video animation of the \texttt{baseline\_v2.0\_10yrs} to demonstrate how \update{the} two band rolling cadence observing strategy is implemented over a 10-year simulated LSST survey. The animation of this figure is available in the online version of the paper. The animation steps through in 30 day intervals over 10 years displaying the cumulative number of on-sky visits in all filters (left) and presenting the total number of on-sky visits in all filters accumulated during the time step (right). The animation has a real-time duration of 25 s. \update{The plots are centered on $\alpha$=0 and $\delta$=0. Right ascension and declination lines are marked every 30$^\circ$.} \\
(An animation of this figure is available.) \label{animation:baselinev2.0}}
\end{center}
\end{figure}

\begin{figure}
\begin{center}
\begin{interactive}{animation}{rolling_ns3_rw0.9_v2.0_10yrs__N_Visits.mp4}
\end{interactive}
\includegraphics[width=0.7\columnwidth]{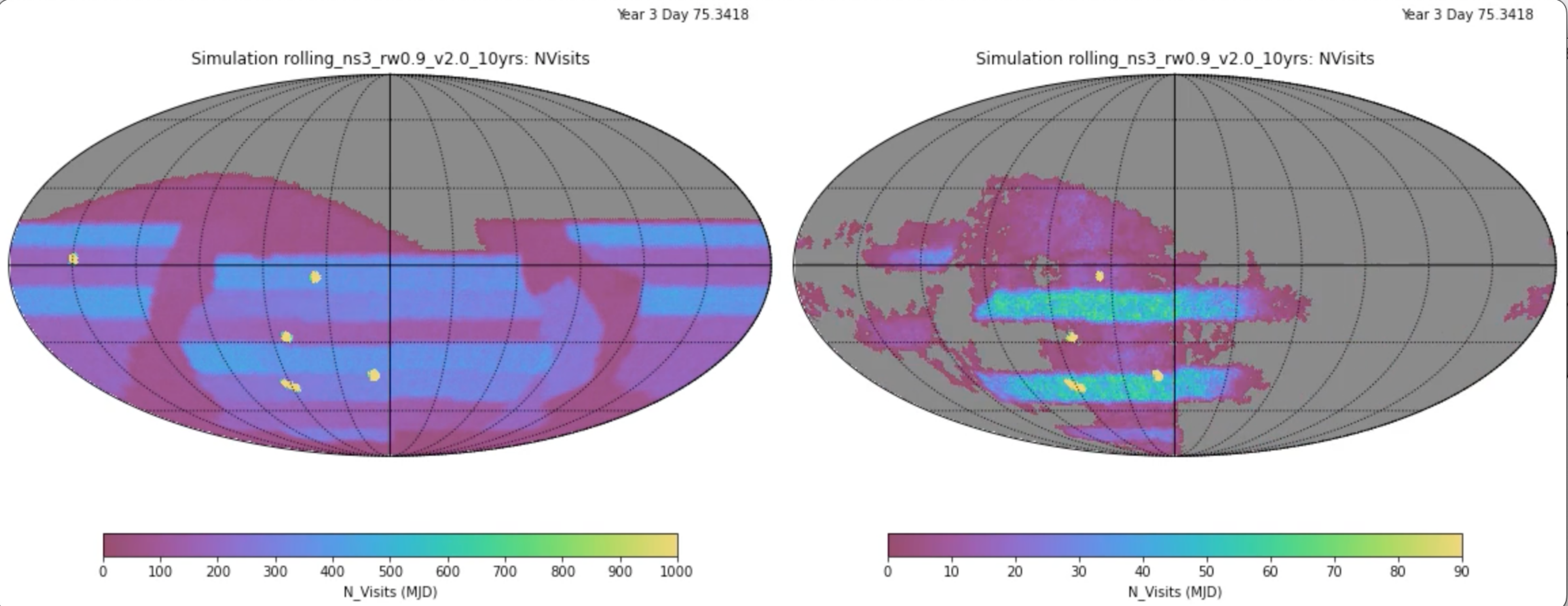}
\caption{Snapshot from a video animation of the \texttt{rolling\_ns3\_rw0.9\_v2.0\_10yrs} to demonstrate how \update{the} three band rolling cadence observing strategy is implemented over a 10-year simulated LSST survey. The animation of this figure is available in the online version of the paper. The animation steps through in 30 day intervals over 10 years displaying the cumulative number of on-sky visits in all filters (left) and presenting the total number of on-sky visits in all filters accumulated during the time step (right). The animation has a real-time duration of 25 s.  \update{The plots are centered on $\alpha$=0 and $\delta$=0. Right ascension and declination lines are marked every 30$^\circ$.} \\
(An animation of this figure is available.) \label{animation:rolling_ns3}}
\end{center}
\end{figure}

\begin{figure}
\begin{center}
\begin{interactive}{animation}{six_rolling_ns6_rw0.9_v2.0_10yrs__N_Visits.mp4}
\end{interactive}
\includegraphics[width=0.7\columnwidth]{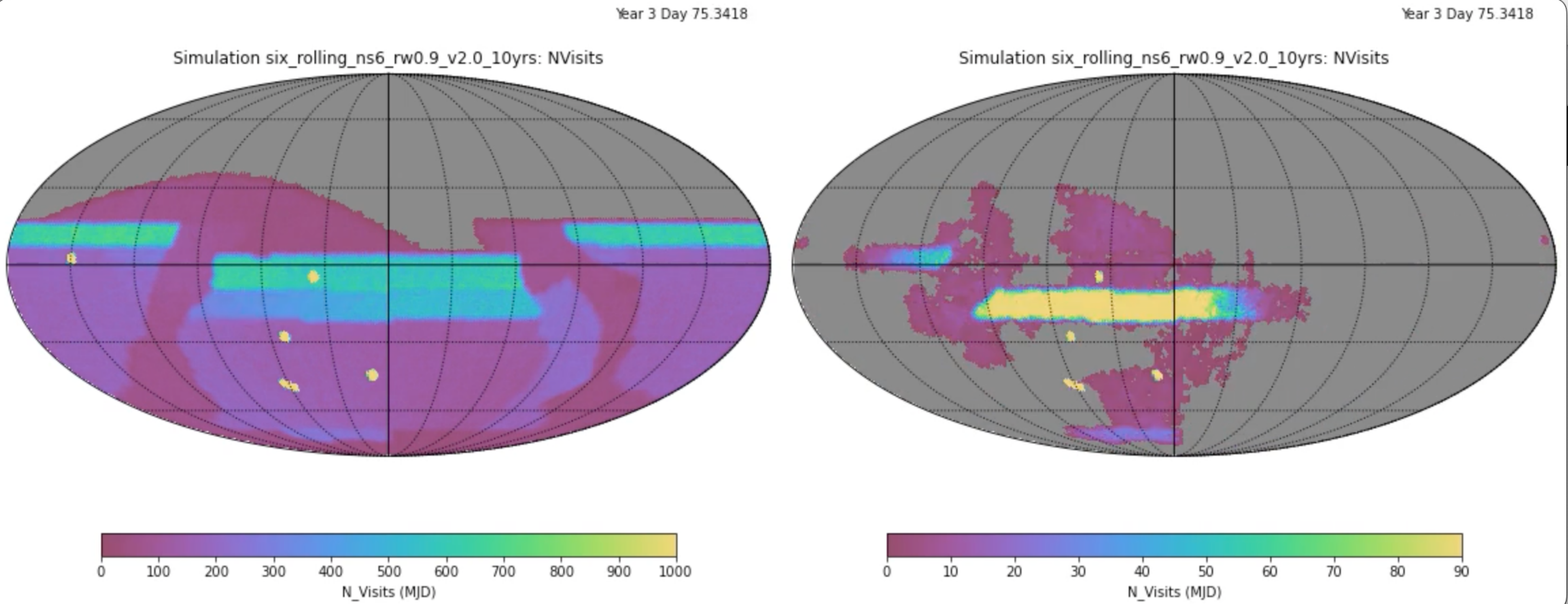}
\caption{Snapshot from a video animation of the \texttt{six\_rolling\_ns6\_rw0.9\_v2.0\_10yrs} to demonstrate how \update{the} three band rolling cadence observing strategy is implemented over a 10-year simulated LSST survey. The animation of this figure is available in the online version of the paper. The animation steps through in 30 day intervals over 10 years displaying the cumulative number of on-sky visits in all filters (left) and presenting the total number of on-sky visits in all filters accumulated during the time step (right). The animation has a real-time duration of 25 s. \update{The plots are centered on $\alpha$=0 and $\delta$=0. Right ascension and declination lines are marked every 30$^\circ$.} \\
(An animation of this figure is available.) \label{animation:six_rolling}}
\end{center}
\end{figure}

Spreading the 825 observations of each field in the WFD evenly over the periods that they are observable, over ten years, corresponds to an observation of each field every three to four nights, on average. As this is a relatively low cadence for some science topics (notably transients), a proposed pattern of observations increases the frequency in certain areas of the sky in some years, at the cost of a lower cadence elsewhere, and then reverses the pattern the following year. This is referred to as a rolling cadence. There are a variety of flavors of this approach, depending on how many stripes each half of the sky (North/South) is divided into, and the `strength' of the rolling, i.e., the fraction of the time spent in the `on' stripes compared to the `off' ones (see Figure~\ref{fig:rolling_year_3.5} for an illustration of these patterns). No rolling cadence entirely neglects the `off' stripes, but in some cases these areas see only a few observations in the entire year, to support template building. Over the 10 years of the survey, the pattern of on/off stripes balances out to give uniform coverage across the whole WFD area. For the majority of simulations rolling cadence is only applied to the WFD area (not the bulge, NES, or other `extended' survey areas) and is not used in the first and last 1.5 years of the survey. Video animations of three example rolling cadence scenarios are available (via the online version of the paper) in Figures \ref{animation:baselinev2.0}, \ref{animation:rolling_ns3}, and \ref{animation:six_rolling}.

\begin{figure}
\begin{center}
\includegraphics[width=0.93\columnwidth]{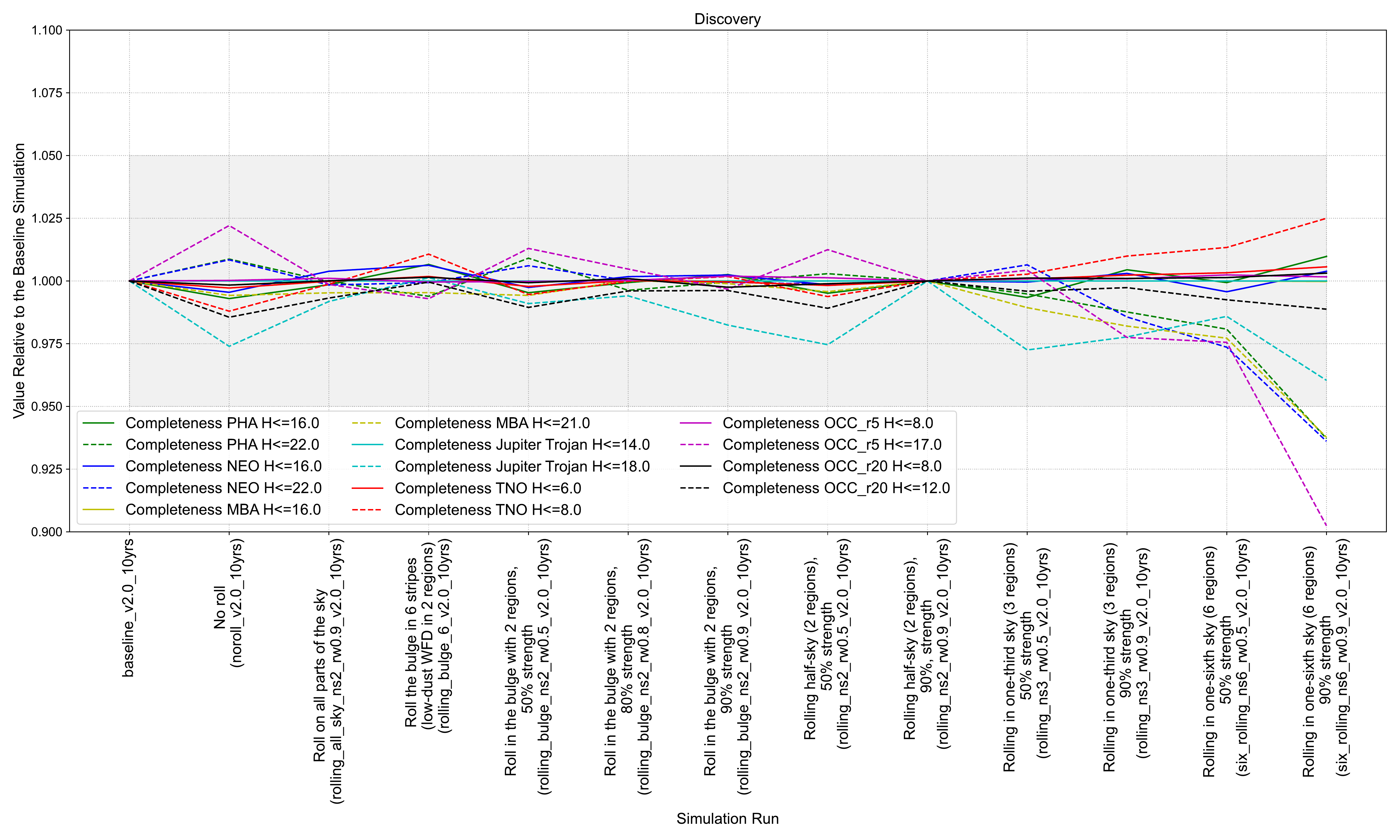}
\includegraphics[width=0.93\columnwidth]{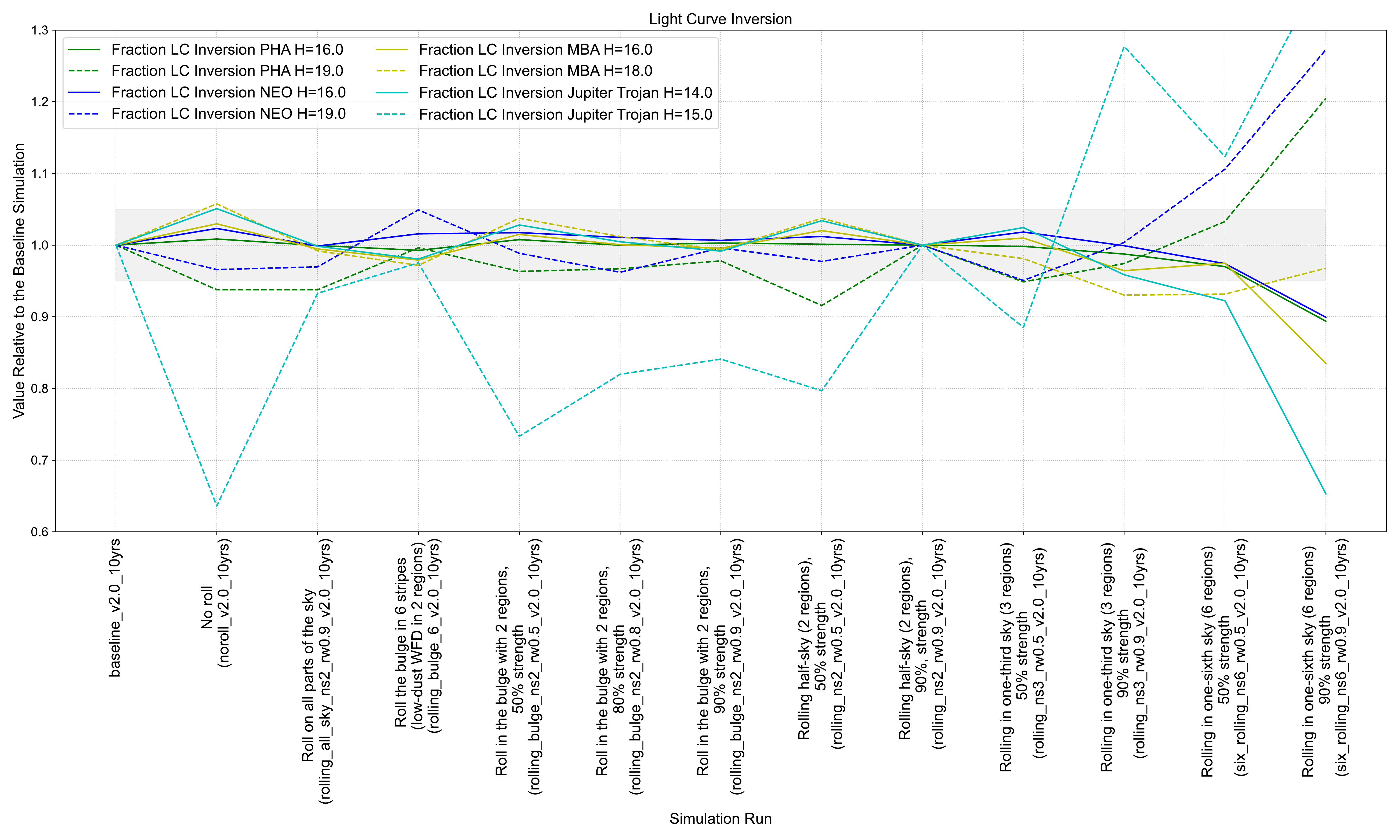}
\caption{Impact of various rolling cadence scenarios (v2.0 simulations). \update{The baseline (reference) simulation with the default scheduler configuration} for this cadence experiment is the first entry on the left. The baseline simulation has a two band rolling cadence implemented with no rolling in the Galactic plane and Northern Ecliptic Spur (NES). All values have been normalized by this simulation's output. The gray shading outlines changes that are within $\pm 5\%$ of the baseline simulation. Top: Discovery Metrics. Bottom: Light Curve Inversion Metrics. \update{Note: The light curve inversion plot has been truncated for clarity. The Jupiter Trojans extend beyond the plot for the \texttt{six\_rolling\_ns6\_rw0.9\_v2.0\_10yrs}.}\label{fig:v2.0_rolling}}
\end{center}
\end{figure}

The effect of rolling cadence is generally seen as positive for most science cases, e.g. having a denser coverage of light curves in the `on' stripes enhances transient science, and rolling cadence is included in the baseline v2.0 simulations. However, there are positive and negative effects that vary with the pattern and strength of the rolling cadence. There is little difference between patterns that split the WFD into two or three stripes North and South of Cerro Pach{\'o}n, but a more extreme 6-stripe pattern, especially at high rolling strength, has more significant effects on both discovery and light curve metrics (Figure~\ref{fig:v2.0_rolling}). Such a pattern is also vulnerable to extended periods of bad weather in one season, resulting in uneven final co-added survey depth, so is not favored for many areas of LSST science. The largest variability in the metrics shown in Figure~\ref{fig:v2.0_rolling} is for faint Jupiter Trojans, which is not surprising, as the Trojan clouds have a limited spatial extent that is in an approximately fixed direction in a given season, relative to where the corresponding planet is. As the cloud may fall into either an on or off stripe in a given season, Jupiter Trojans experience feast or famine in terms of observations, which may not even out over the years in the same way that more distant populations (TNOs) will -- this will depend on the precise timing and choice of band patterns in the final survey. 

A remaining concern with rolling cadence is the possibility that individual objects of interest may be missed, or more likely be discovered later than they could have been if they first brighten above LSST detection limits in an off-stripe. This could potentially affect follow-up of rare objects like ISOs, or impact target choice for the European Space Agency's (ESA's) \emph{Comet Interceptor} (expected to launch in 2029; \citealt{2019NatCo..10.5418S}), in the unlucky case that a suitable long-period comet is missed for a year. In general, discovery metrics for OCCs are not strongly affected by rolling cadence, so this is not seen as a major risk for the mission. Further study of the effect of rolling cadence on how early we might discover ISO and OCC targets is ongoing.

\subsection{Deep Drilling Field (DDF) Observing}
\label{sec:DDFs}

The Deep Drilling Fields (DDF) are a key component of LSST's structure, currently allocated $\sim$5\% of the total survey time in the latest survey simulation baselines. 
There are five confirmed DDF pointings (Table~\ref{tab:DDFs}), which will be observed with a completely different cadence from the WFD: a higher sampling rate, as well as a different sampling of filters \citep{jones_r_lynne_2020_4048838}. 
The locations of the DDF pointings were largely motivated by both galactic and extragalactic science goals \citep{2018arXiv181106542B, 2018arXiv181203143B, 2019arXiv190410439C, 2018arXiv181203144H, 2018arXiv181200516S}. 
However, the ability to stack the denser sampling means that these fields also provide a small, deeper dataset than the WFD \citep{ 2009arXiv0912.0201L}.

\begin{deluxetable}{rrrr}
\tablecaption{Planned LSST Deep Drilling Fields (DDFs) \label{tab:DDFs}}
\tablewidth{0pt}
\tablehead{
\colhead{Deep Drilling Field } & \colhead{Right Ascension} &   \colhead{Declination}  &   \colhead{Ecliptic} \\
& \colhead{(deg)} & \colhead{ (deg)} &   \colhead{Latitude}  \\
& \colhead{(J2000)} & \colhead{(J2000)} & \colhead{(deg)}\\
}
\tabletypesize{\scriptsize}
\startdata
ELAISS1 (European Large-Area ISO Survey-S1) Field &  9.450&  -44.00 & -43.18 \\
XMM-LSS (\emph{X-ray Multi-Mirror Mission-Newton} Large Scale Structure) Field &  35.71 &  -4.75 & -17.90 \\
ECDFS (Extended \emph{Chandra} Deep Field-South) &  53.13 & -28.10 & -45.47 \\
EDF-S (\emph{Euclid} Deep Field South) & 61.24 & -48.42 & -66.60 \\
COSMOS (Cosmological Evolution Survey) Field  &  150.10 &  2.18  & -9.40\\
\enddata
\end{deluxetable}

The DDFs provide a limited but strategic improvement to the solar system science expected from LSST (Fig.~\ref{fig:v2.0_ddf_frac}). 
The extra depth of the stacked DDF data will improve the detectability of objects that are fainter than the WFD limits \citep[e.g.][]{Smotheram_2021}, and thus either smaller or more distant.
Four out of five of the DDFs are at ecliptic latitudes $>$15$^\circ$ (Table~\ref{tab:DDFs}), which means they can only contain solar system objects on moderate to high inclination orbits. 
These objects are comparatively rare \citep[][]{2021ARA&A..59..203G, Raymond2022}, which will result in few observations of solar system objects in these four DDFs. 
The fifth field, COSMOS, is centred $\sim$9$^\circ$ from the ecliptic plane (Table~\ref{tab:DDFs}): this lower latitude makes it sensitive to the mildly dynamically excited small body populations, so it is the DDF most likely to be directly beneficial for solar system science.

\subsubsection{Fraction of Time Devoted to DDFs} \label{sec:frac_ddfs}

The time balancing of DDF and WFD observations within the LSST has a noticeable impact on overall Solar System detections and light curve inversion capability. Simulations with a larger portion of survey time for DDFs were previously trialed in the v1.6 sims ({\texttt{ddf\_heavy\_}}), but were rejected as they produced significant negative impacts on all Solar System populations and their metrics, as well as failing to meet some of the key science requirements for the WFD.

\begin{figure}
\begin{center}
\includegraphics[width=0.95\columnwidth]{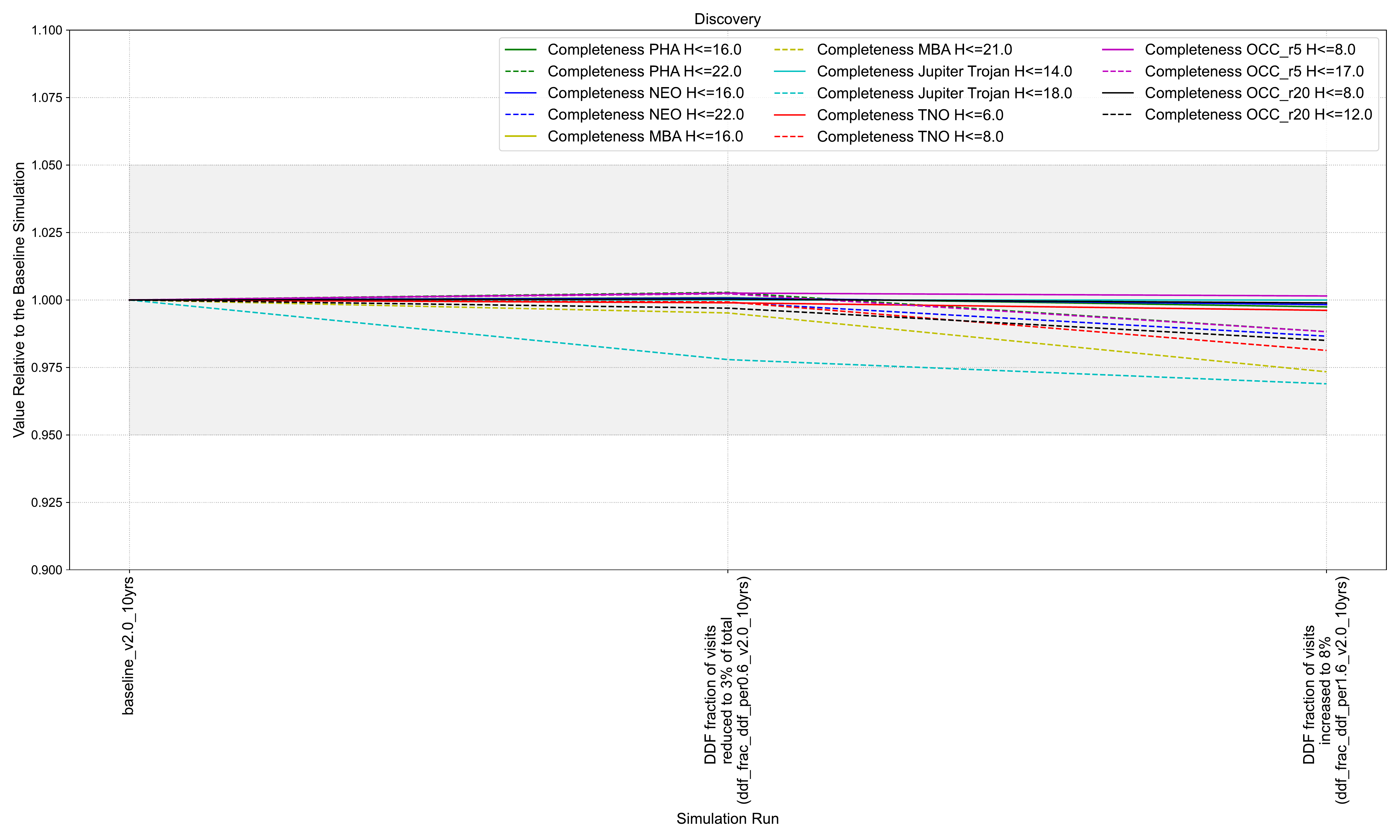}
\includegraphics[width=0.95\columnwidth]{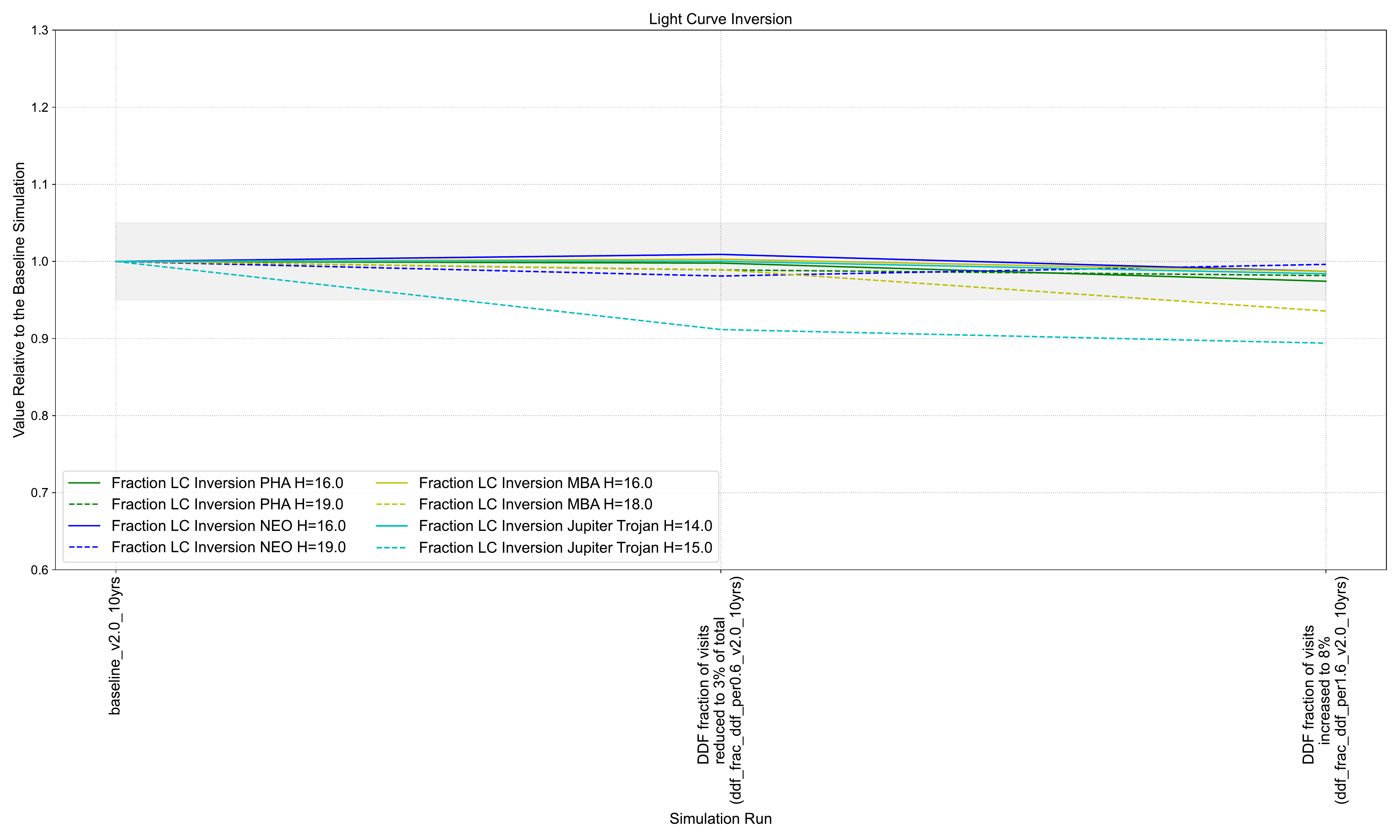}
\caption{Impact of varying the time allocated to the DDFs (v2.0 simulations). \update{The baseline (reference) simulation with the default scheduler configuration} for this cadence experiment is the first entry on the left. All values have been normalized by this simulation's output. The gray shading outlines changes that are within $\pm 5\%$ of the baseline simulation. Top: Discovery Metrics. Bottom: Light Curve Inversion Metrics. \label{fig:v2.0_ddf_frac}}
\end{center}
\end{figure}

The v2.0 simulations are the latest set of simulations which explore varying the fraction of total survey time allocated to DDFs. They test a more conservative variation of $\pm3\%$ survey time spent on DDFs from the baseline value of $5\%$. In these simulations, any extra observing time is evenly distributed across the remaining components of the LSST. Both options are satisfactory for the discovery and light curve inversion of solar system objects, for most or all populations (Figure~\ref{fig:v2.0_ddf_frac}). The simulation with 8\% of time allocated to the DDFs (\texttt{ddf\_frac\_ddf\_per1.6}\footnote{`\texttt{1.6}' indicates that the time allocated to DDFs is 1.6 times the baseline value}) provides slightly worse results for discovery and light curve inversion metrics compared to the simulation with 3\% survey time (\texttt{ddf\_frac\_ddf\_per0.6}\footnote{`\texttt{0.6}' indicates that the time allocated to DDFs is 0.6 times the baseline value}), as WFD revisits are particularly important for light curve infill and for linking the motion of solar system objects. 
However, both options are still within a negligible loss margin ($<5$\%) on both metrics when compared to the baseline.

\subsubsection{``Rolling" DDFs}\label{sec:rolling_ddfs}

\begin{figure}
\begin{center}
\includegraphics[width=0.7\columnwidth]{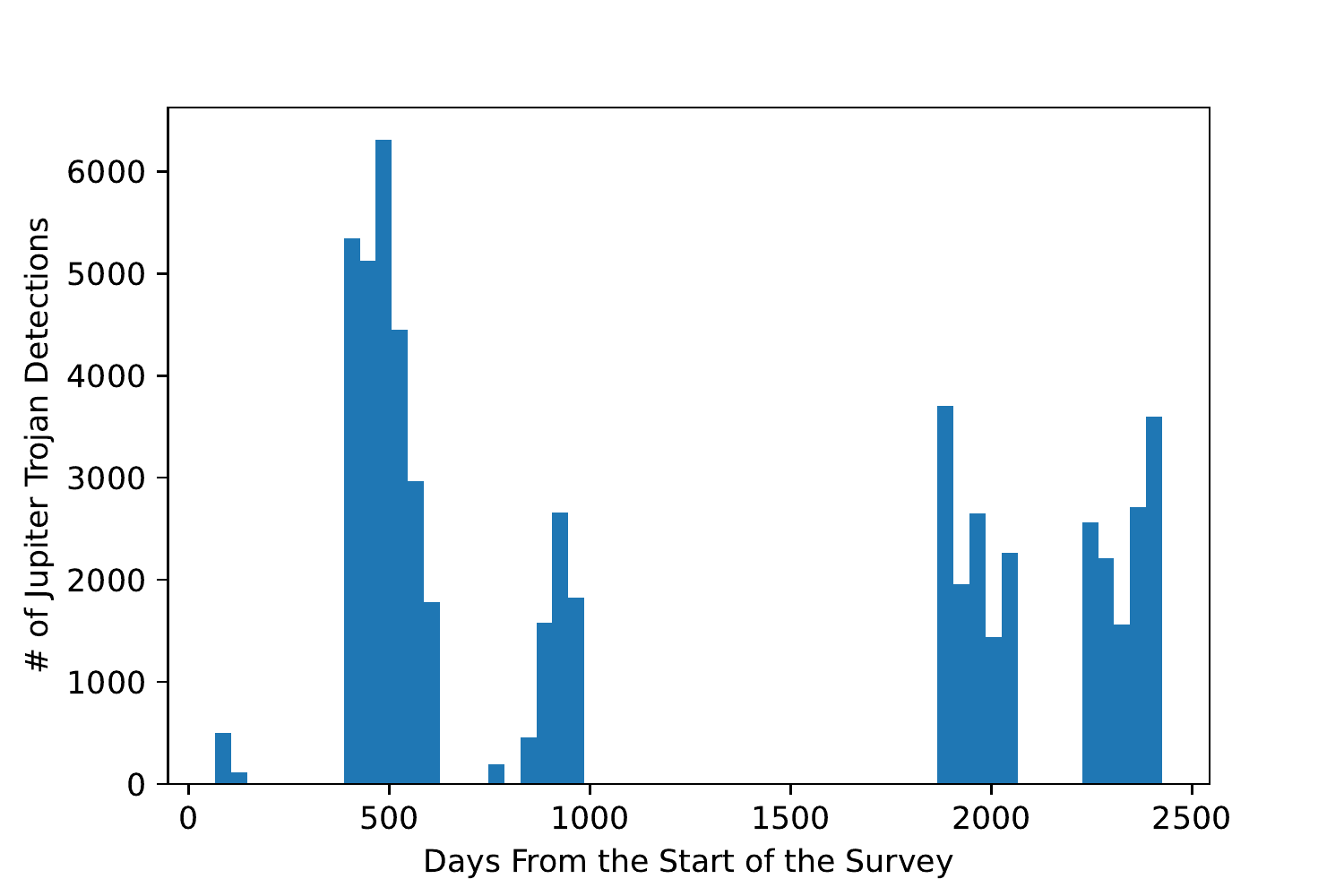}
\caption{The number of Jupiter Trojan detections \update{(of a simulated sample population of 5000 objects)} in the COSMOS DDF (Deep Drilling Field) over the course of the \texttt{baseline\_v2.2\_10yrs} simulation. This simulation starts the LSST in October 2023. By 7 years into the survey, both Jupiter Trojan clouds have traversed across the COSMOS DDF. \label{fig:ddf_n_trojans}}
\end{center}
\end{figure}

From the field population sensitivities, the key aspect of solar system science interest is the choice of rolling cadence for the COSMOS field and how it affects the small body metrics.
The suggested types of DDF rolling cadences explored in the v2.1 cadence simulations are unique to the DDFs: in contrast to the `stripes' of WFD discussed in Section \ref{sec:rolling}, these simulations instead alter the frequency of observation for each individual DDF and the relative weighting of time between DDFs.
They include cases where specific DDFs are observed only in certain years (e.g. only in the first 3 years of LSST).
Between v2.1 and v2.2, a large number of DDF strategy variations were considered, but in general solar system metrics have not been produced for these runs. The impacts of variations of DDF strategy while keeping the overall envelope of allocated DDF time and field location approximately constant are expected to produce negligible changes. Once a narrower range of DDF strategies are under consideration, solar system metrics will be produced and checked for potential impacts.   
Therefore, we consider how DDF ``rolling" cadences would probably affect solar system science, with a specific focus on the highest-yield COSMOS DDF.

The Jupiter Trojans will complete approximately one full orbit during the span of the LSST. 
The slightly asymmetric populations lead and trail the giant planet in its orbit by $\sim$60$^{\circ}$, with a mean libration amplitude of $33^{\circ}$ from the center of their respective Trojan clouds \citep{Marzari_2002}.
The more populous L4 cloud's inclination distribution is centred around the ecliptic latitude of COSMOS, while the flatter L5 inclination distribution still encompasses COSMOS \citep{Slyusarev_2014}.
The broad Jupiter Trojan libration distribution produces an on-sky distribution that has wide wings of consistent density around the libration centres. 
These orbital properties mean that Jupiter Trojans will be visible in the COSMOS DDF during distinct several-hundred-day observation periods within LSST (Figure \ref{fig:ddf_n_trojans}). 
This will permit smaller-diameter Jupiter Trojans to be discovered than can be achieved by the WFD.
Therefore, it is essential that the COSMOS DDF is observed at times when the Jupiter Trojans are passing through the field.

Throughout the LSST, the COSMOS and other DDFs will provide a constantly refreshing sample of shift-and-stack-detectable TNOs smaller than can be seen in single frames of the WFD.
As TNOs move slowly ($<5\arcsec/$hr for $r_h>30$~au; cf. Fig.~\ref{fig:on_sky_motion}), they will remain in the sidereally static DDF pointings for extended periods of time, longer than other solar system populations.
For the LSSTCam field of view, the time in years for a TNO to pass through the field is given by:
\begin{equation}
    t = \frac{3.5^\circ}{\alpha~\frac{24~hr/d\times365.25~d/yr}{3600~''/^\circ}} = \frac{3.5}{2.435\alpha} \approx \frac{1.437}{\alpha}
\end{equation}
where $\alpha$ is the opposition on-sky rate of motion in arcseconds per hour. For distances of 40--60~au, a TNO traverses the field in $\sim5$--8~months. In comparison, more distant TNOs ($r_h\geq200$~au) remain in the field for $\geq2$~yrs. This means that the population of $r_h \sim$30~au TNOs observed in a DDF is refreshed $\sim$30 times during LSST as a result of (primarily Earth's) orbital motion, compared to $r_h = 300$~au TNOs, which would take a third of the full survey to pass through the field.
COSMOS, and at lower yield, the other DDFs thus provide multi-month TNO orbital arcs that would determine parameters $r_h$ and $i$ to a precision useful for population studies.
However, these arcs are generally too short to reduce uncertainties on $a$ and $e$ to levels sufficient for Neptune resonance classification \citep[][]{Volk_2016}.
Deep revisits by LSST around the DDFs in later years to recover the DDF-sourced TNOs would be necessary for this additional improvement for outer solar system science.
Therefore, TNO science is flexible relative to the DDF ``rolling cadence" decision, as long as COSMOS and other DDFs are visited for approximately two years at some point within LSST.

\subsection{Micro-surveys}
\label{sec:all_micro-surveys}

A wide variety of special small observing programs have been proposed by the Rubin Observatory user community that have been grouped together under the micro-surveys category. Smaller than the mini-surveys that have been incorporated into the LSST footprint, each micro-survey consumes between approximately 0.3$\%$ and 3$\%$ of the total available observing time. The micro-surveys compliment the other components of the LSST  (WFD, DDFs, NES, and Galactic Plane observing) and provide unique benefits not obtained from the larger components of the LSST observing strategy.  Some of these proposed micro-surveys plan to observe new regions of sky not covered within the survey footprint, while others re-observe regions of the sky already covered in the LSST footprint with a separate observing strategy. Of all the proposed micro-surveys, the one that is most relevant to the discovery and follow-up of minor planets and ISOs is the low-SE twilight survey which aims to take short exposures closer to the Sun in order to search for small bodies in an orbital phase space that the rest of the LSST is not sensitive to.

\subsubsection{Low Solar Elongation (Low-SE) Solar System Twilight Micro-survey}
\label{sec:twilight}

\begin{figure}
\begin{center}
\includegraphics[width=0.6\columnwidth]{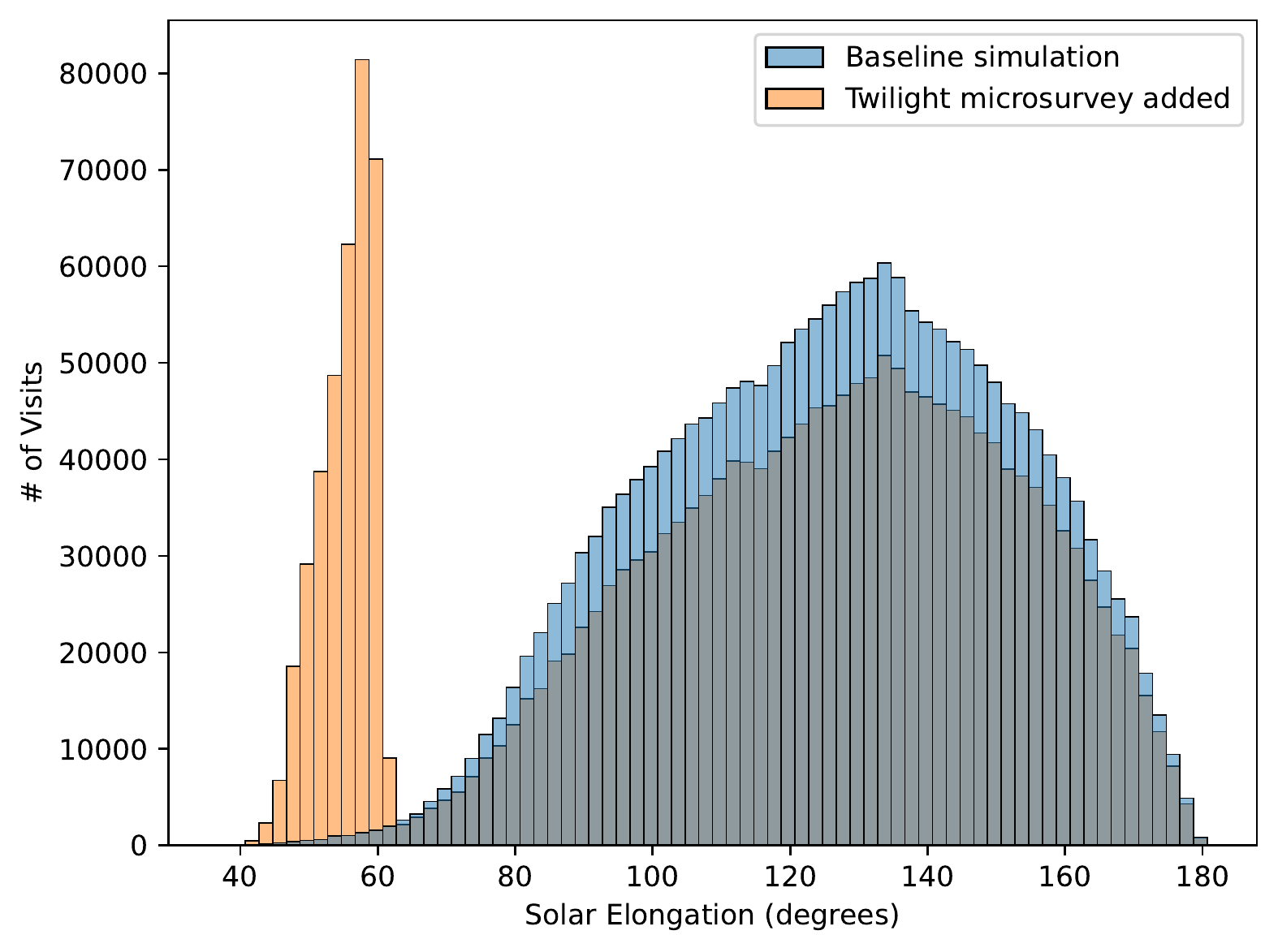}
\caption{Comparison of the number of visits as a function of solar elongation (at the center of the FOV) with and without a low solar elongation (low-SE) solar system twilight micro-survey. The binsize is \update{2$^{\circ}$}. The baseline simulation is \update{\texttt{baseline\_v2.2\_10yrs}} and the selected low solar elongation (low-SE) solar system twilight micro-survey simulation is \update{\texttt{twi\_neo\_repeat4\_iz\_np1\_v2.2\_10yrs}}. \update{We note that part of the orange histogram for the simulation that includes the  low-SE twilight micro-survey is plotted underneath the blue histogram for the baseline simulation.} \label{fig:twilight_solar_elongation}}
\end{center}
\end{figure}

\begin{figure}
\begin{center}
\includegraphics[width=0.85\columnwidth]{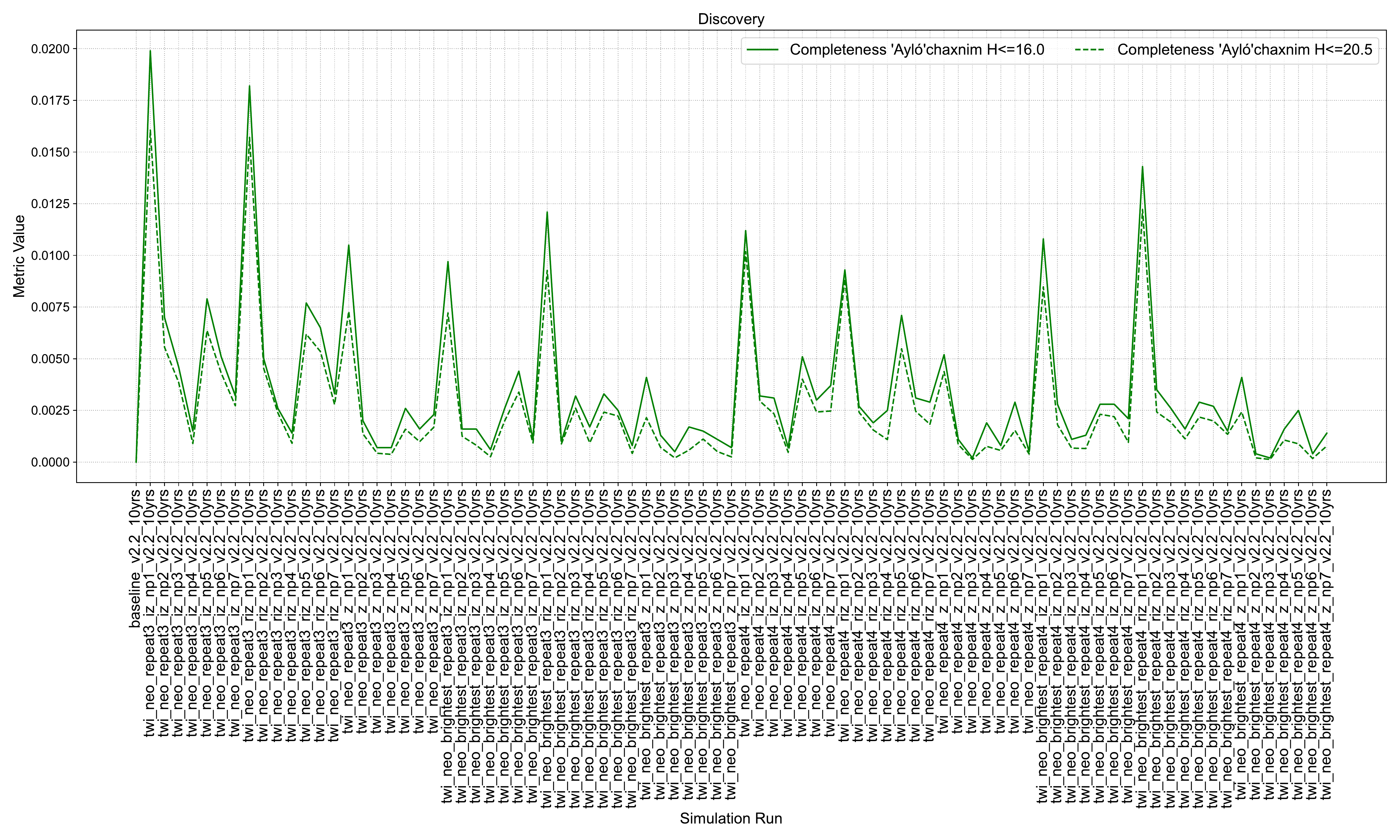}
\includegraphics[width=0.85\columnwidth]{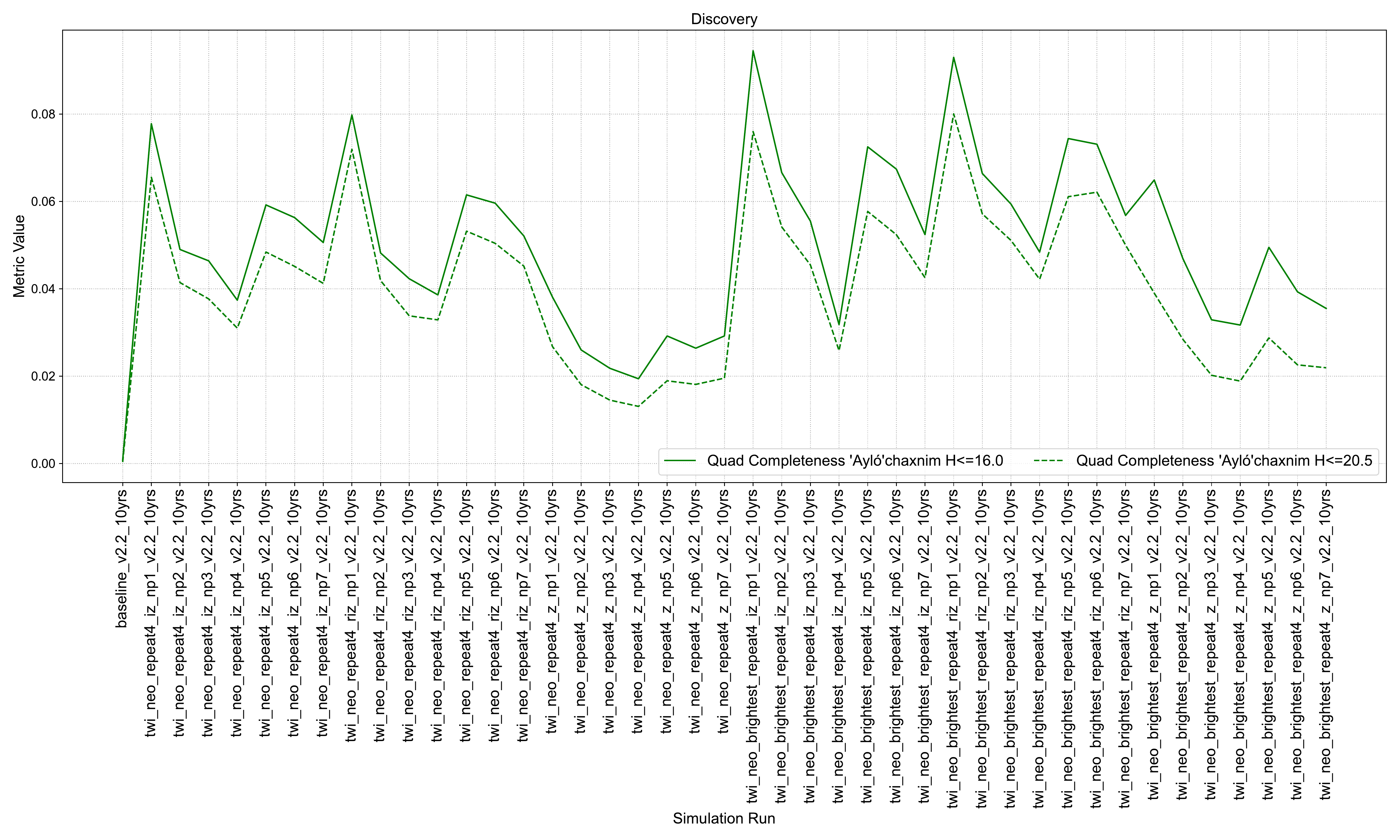}
\caption{\textquotesingle Ayl\'{o}\textquotesingle chaxnim (previously known in the literature as Vatira) population discovery completeness comparisons for cadences that include a possible low solar elongation (low-SE) solar system twilight \update{micro-survey} (v2.2 simulations). \update{The baseline (reference) simulation with the default scheduler configuration for this cadence experiment is the first entry on the left. These metric values \textbf{have not} been normalized to the baseline's simulation output since the baseline simulation, which does not include the low-SE twilight microsurvey, finds no Ayl\'{o}\textquotesingle chaxnims with 3 nightly pairs over 14 days. Top: Discovery Completeness for 3 nightly pairs in 14 days (the Rubin SSP discovery criteria). Bottom: Discovery Completeness for 4 detections in a single night for twilight simulations that take 4 visits per pointing.} Simulation legend: \texttt{twi$\_$neo$\_$repeatX$\_$Y$\_$npZ\_v2.2\_10yrs} for a micro-survey with \texttt{X} repeat visits in \texttt{Y} filter(s) per pointing per twilight observed where \texttt{Z} = 1 (on every night), 2 (1 night on/1 night off), 3 (1 night on/2 nights off), 4 (1 night on/3 nights off), 5 (4 nights on/4 nights off), 6 (3 nights on/4 nights off), 7 (2 nights on/4 nights off). \label{fig:v2.2_twilight}}
\end{center}
\end{figure}

\begin{figure}
\begin{center}
\includegraphics[width=0.88\columnwidth]{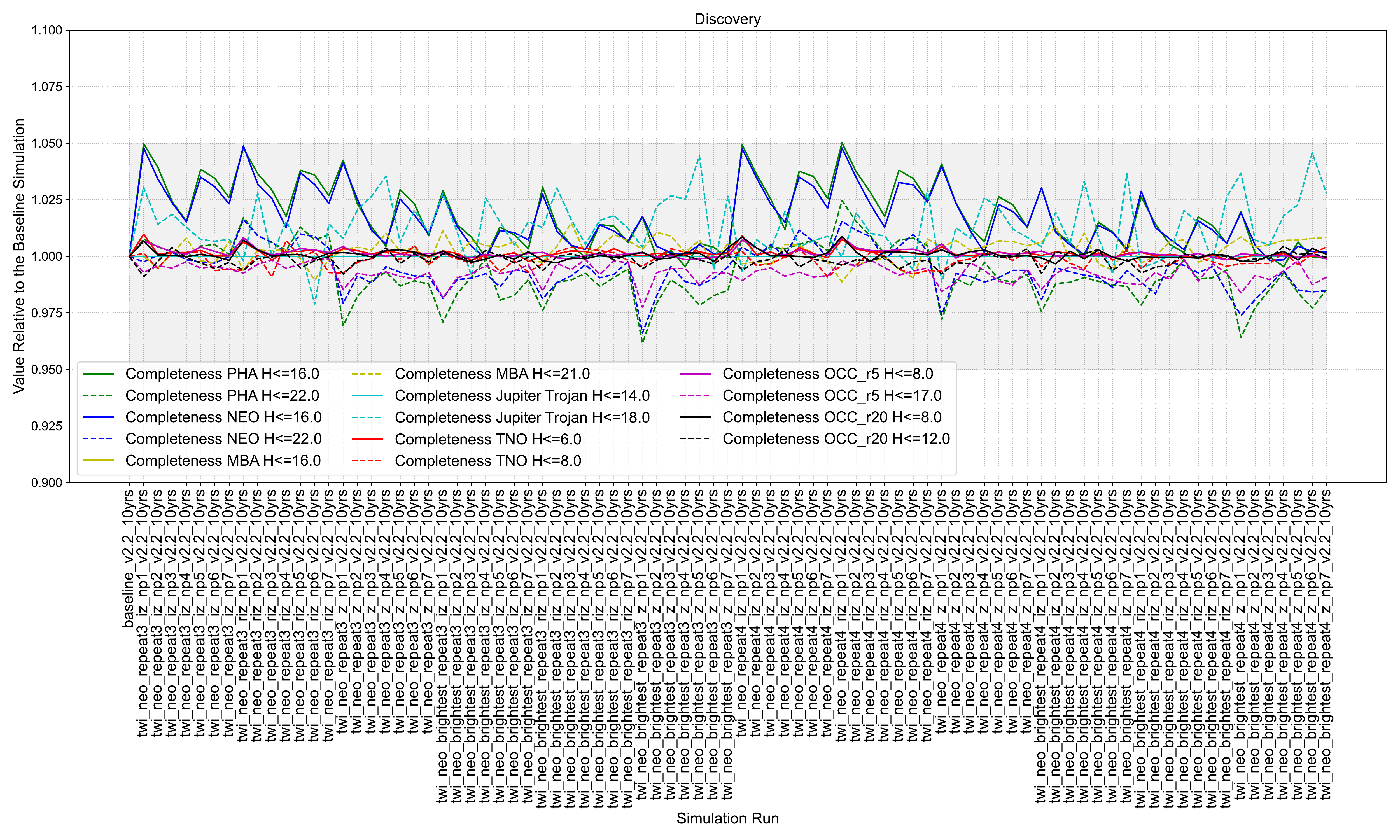}
\includegraphics[width=0.88\columnwidth]{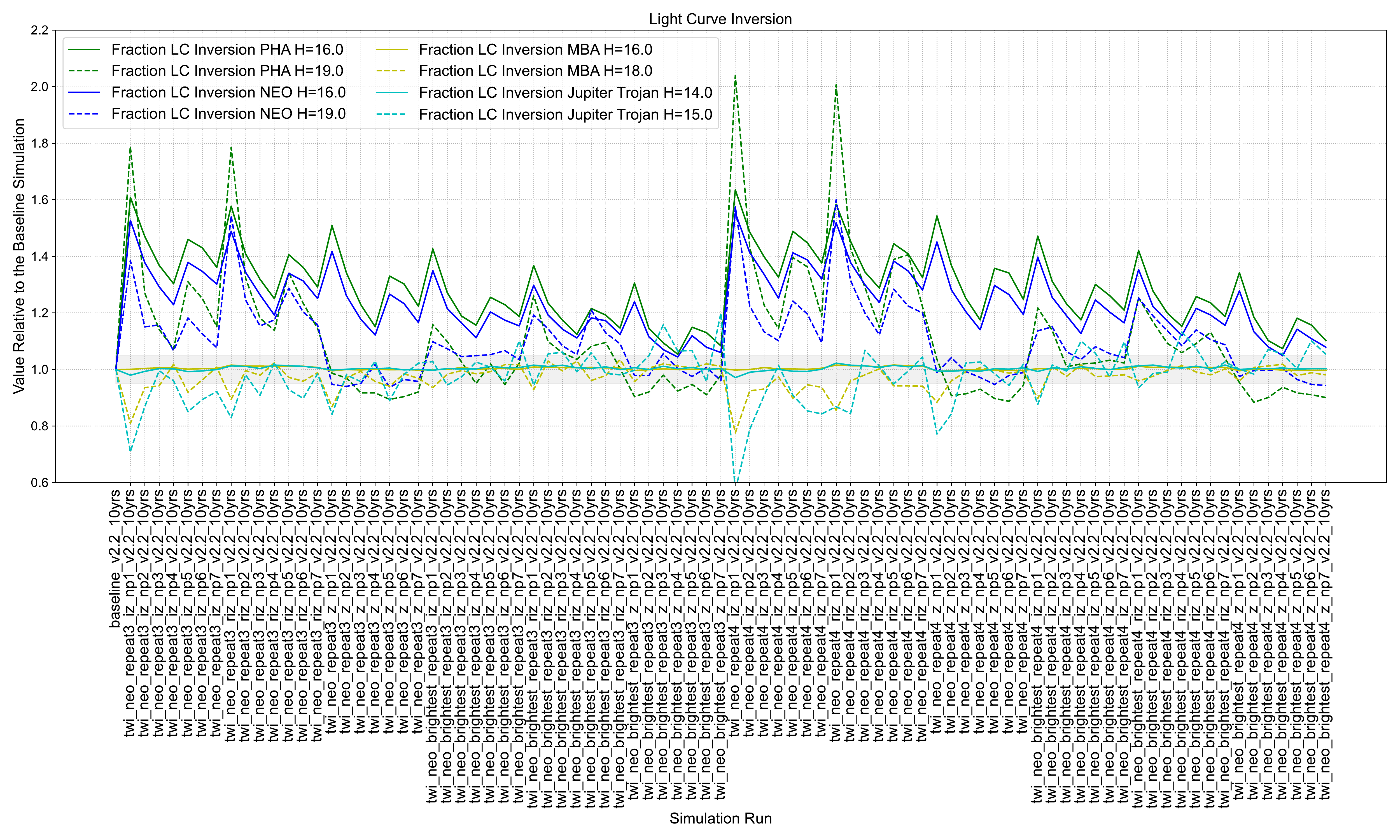}
\caption{Impact on other solar system metrics due to the inclusion of a low solar elongation \update{(low-SE)} solar system twilight \update{micro-survey}(v2.2 simulations).  \update{The baseline (reference) simulation with the default scheduler configuration} for this cadence experiment is the first entry on the left. \update{The baseline simulation does not include twilight low-SE observations.}  All values have been normalized by this simulation's output. The gray shading outlines changes that are within $\pm 5\%$ of the baseline simulation. Top: Discovery Metrics. Bottom: Light Curve Inversion Metrics. Simulation legend: \texttt{twi\_neo\_repeatX\_Y\_npZ\_v2.2\_10yrs} for micro-survey with \texttt{X} repeat visits in \texttt{Y} filter(s) per pointing per twilight observed where \texttt{Z}= 1 (on every night), 2 (1 night on/1 night off), 3 (1 night on/2 nights off), 4 (1 night/3 nights off), 5 (4 nights on/4 nights off), 6 (3 nights on/4 nights off), 7 (2 nights on/4 nights off). \label{fig:v2.2_twilight_rest_of_SS}}
\end{center}
\end{figure}

The \texttt{twi\_neo} family of simulations use $50\%$ of the available observing time during morning and evening twilight to perform a micro-survey of the low-SE ($40^{o} \lesssim \textrm{SE} \lesssim 60^{o}$) sky, which would otherwise not be observed during the WFD observing cadence (see Figure~\ref{fig:twilight_solar_elongation}). The opportunity for LSST to observe the low-SE sky during twilight is the only time when viewing solar system objects inward to Earth is possible. Although surveys similar in nature have been carried out in the past, Rubin Observatory's large aperture size would put it in a unique position to provide a more sensitive search for several populations of solar system objects such as IEOs, Earth Trojans, and sungrazing comets than has been performed previously \citep{2018arXiv181200466S}. NEOs in the region of the solar system interior to Earth's orbit (including Atiras with orbits interior to the orbit of Earth and \textquotesingle Ayl\'{o}\textquotesingle chaxnims with orbits interior to Venus' orbit\footnote{Objects on orbits entirely within the orbit of Venus have been previously referred to in the literature as Vatiras \citep{2012Icar..217..355G}. This name has acted as a placeholder until the first object in this population is discovered and named. With the recent discovery and naming of the first inner-Venus object, \textquotesingle Ayl\'{o}\textquotesingle chaxnim, \citep{2020MPEC....A...99B, Ip_2022, BolinAN2022}, we adopt the name of this population after its first known member, as is tradition.}) are the least constrained portion of currently available NEO models due to observational limitations of objects at low-SE \citep{2012Icar..217..355G,2018Icar..312..181G}. In addition, recent observational evidence and dynamical studies suggest there are possible meta-stable regions in the innermost portion of the solar system where more objects on orbits similar to that of \textquotesingle Ayl\'{o}\textquotesingle chaxnim may be lurking and awaiting discovery \update{\citep{2020MNRAS.493L.129G,2020MNRAS.494L...6D,2020MNRAS.496.3572P,Ip_2022,BolinAN2022,2022arXiv220906245S,Bolinetal_2023_accepted}}. 

In addition to the discovery of IEOs, a LSST low-SE twilight micro-survey could enhance the discovery of ISOs; ISO 2I/Borisov was discovered during twilight by an amateur astronomer in 2019 \citep{MPEC2019-R106}. Routine observations at low-SE could also provide pre-discovery images of ISOs, enabling additional astrometric measurements for improved orbit determination of the often short-lived visitors \citep{2020AJ....159...77Y,2020AJ....160...26B}. Low-SE observations of LSST-discovered ISOs would provide further opportunities beyond what the WFD observing cadence would offer for observing possible mass-shedding, outbursting, or breakup events of these interstellar interlopers, as well as extend the amount of time these short-lived visitors can be observed. Monitoring the sky in the near-Sun region could also provide the opportunity to observe cometary outbursting or breakup events as non-interstellar near-Sun comets reach heliocentric distances $<1$ au, which may otherwise not be characterized. Observing comets (either with origins from within our solar system or interstellar space) as they reach the near-Sun region will better inform us of how insolation can process cometary surfaces and connect the near-Sun comet population to comets as a whole \citep{2018arXiv181200466S}. 

Discoveries of PHAs can also be enhanced with the inclusion of a low-SE twilight micro-survey, improving our knowledge of and increasing warning times for potential asteroid impacts. In addition, the possible discovery of more Earth Trojans, which librate at the Earth-Sun L4 and L5 Lagrange points, would improve our knowledge of planetary impactor sources for both recent and ancient cratering events on the Earth and Moon \update{\citep{2018arXiv181200466S,2019NatAs...3..193M,2020MNRAS.492.6105M}}. Earth Trojans also make attractive spacecraft mission targets due to their low relative velocity with Earth. Lastly, observing asteroids in the near-Sun region with LSST can provide the opportunity to probe mechanisms responsible for the supercatastrophic disruption of asteroids with small perihelia (closest orbital distance to the Sun) \citep{2016Natur.530..303G} and test the extent to which this phenomenon occurs for asteroids that reach very small solar distances. 

Due to the large amount of science for a wide variety of small body populations that would be made possible with a low-SE twilight micro-survey, \update{a family of runs executing a variety of low-SE observing cadences during twilight} have been included in the last few rounds of cadence simulations, the most recent of which are the v2.2 simulations. \update{The v2.2 family is split into \texttt{twi\_neo} and \texttt{twi\_neo\_brightest} which execute the mini-survey when the Sun is above -17.8$^{\circ}$ and -14$^{\circ}$ elevation, respectively.} The \texttt{twi\_neo} \update{and \texttt{twi\_neo\_brightest}} simulations consist of 15 s exposures per visit and explore a variety of repeat visits, filters, and nightly twilight on-off cadences. The \texttt{twi\_neo\_repeatX\_Y\_npZ\_v2.2\_10yrs} \update{and \texttt{twi\_neo\_brightest\_repeatX\_Y\_npZ\_v2.2\_10yrs}} family consist of \texttt{X} repeat visits (i.e., triplets or quads; all separated by $\sim$3 minutes) in \texttt{Y} filter(s) per pointing per twilight observed where \texttt{Z}~= 1 (on every night), 2 (1 night on/1 night off), 3 (1 night on/2 nights off), 4 (1 night on/3 nights off), 5 (4 nights on/4 nights off), 6 (3 nights on/4 nights off), 7 (2 nights on/4 nights off). Figures~\ref{fig:v2.2_twilight} and \ref{fig:v2.2_twilight_rest_of_SS} show the impact of these low-SE twilight micro-survey cadence options on the discovery and light curve inversion solar system MAF metrics. \update{We note that Figure~\ref{fig:v2.2_twilight} provides discovery completeness values that are not normalized to the baseline simulation's output since there are no \textquotesingle Ayl\'{o}\textquotesingle chaxnims discovered with the baseline survey cadence; this is the only figure that shows the outcomes of the MAF metrics analysis that is not normalized to the baseline simulation's output.} 

The discovery completeness for the \textquotesingle Ayl\'{o}\textquotesingle chaxnim population (NEOs with orbits interior to the orbit of Venus) for a variety of cadence options for this micro-survey are compared to that for the baseline survey (with no low-SE twilight micro-survey) in Figure~\ref{fig:v2.2_twilight}. Each of the micro-survey cadence options provides a (sometimes much) higher discovery completeness than the baseline survey, which does not include any \textquotesingle Ayl\'{o}\textquotesingle chaxnim discoveries since \textquotesingle Ayl\'{o}\textquotesingle chaxnims are only visible at solar elongations smaller than the WFD cadence reaches. \update{In the top panel of Figure~\ref{fig:v2.2_twilight}, which uses the Rubin SSP discovery criteria of 3 nightly pairs in 14 days, t}he highest \textquotesingle Ayl\'{o}\textquotesingle chaxnim discovery completeness is reached when the low-SE twilight micro-survey is run every night with 3 repeat visits per pointing in either \textit{iz} or \textit{riz} filters (i.e., \texttt{twi\_neo\_repeat3\_iz\_np1\_v2.2\_10yrs} and \texttt{twi\_neo\_repeat3\_riz\_np1\_v2.2\_10yrs}). \update{These micro-survey cadences would result in the completeness of the $H\leq20.5$ and the $H\leq16.0$ \textquotesingle Ayl\'{o}\textquotesingle chaxnims increasing to $\approx$1.5\% and $\approx$2\%, respectively, providing the potential for a significant increase in the discovery of inner-Venus asteroids. Running the micro-survey with either 3 or 4 repeat visits during either the brightest twilight time (i.e., when the Sun is above -14$^{\circ}$ elevation) or full twilight time (i.e., when the Sun is above -17.8$^{\circ}$ elevation) produces similar results in \textquotesingle Ayl\'{o}\textquotesingle chaxnim discovery completeness across the various nightly `on'/`off' cadences. The largest discovery completeness increases for these options occur when the micro-survey is run every night using either \textit{iz} or \textit{riz} filters, which result in increases of $\approx$1\% to $\approx$1.5\%.} Simply using the \textit{z} filter does not get as large of a discovery boost as using either \textit{iz} or \textit{riz} filters. Unsurprisingly, the less often the micro-survey is run (fewer number of `on' nights), the lower the \textquotesingle Ayl\'{o}\textquotesingle chaxnim discovery completeness drops.

\update{If, unlike the Rubin SSP requirement of 3 nightly pairs in 14 days, 4 detections in a single night with 4 repeat visits per pointing are required, the discovery completeness improves further. This is more typical of observing cadences used for NEO discovery by current surveys \citep[e.g.,][]{1996EM&P...72..233G, 2003DPS....35.3604L, 2018PASP..130f4505T} Such a cadence would require the development and implementation of code outside SSP, which is designed only to use image pairs to make tracklets. Under this alternative cadence, running the micro-survey every night during the brightest part of twilight in either \textit{iz} or \textit{riz} (i.e., \texttt{twi\_neo\_brightest\_repeat4\_iz\_np1\_v2.2\_10yrs} and \texttt{twi\_neo\_brightest\_repeat4\_riz\_np1\_v2.2\_10yrs} in the bottom panel of Figure~\ref{fig:v2.2_twilight}), the discovery completeness increases to $\approx$8\% and $\approx$9.5\% for the $H\leq20.5$ and the $H\leq16.0$ \textquotesingle Ayl\'{o}\textquotesingle chaxnims, respectively. Given that the baseline \textquotesingle Ayl\'{o}\textquotesingle chaxnim discovery completeness is zero, a low-SE twilight micro-survey thus} has the potential for a dramatic shift in the discovery of asteroids interior to the orbit of Venus. 

In contrast, Figure~\ref{fig:v2.2_twilight_rest_of_SS} (top panel) shows that for nearly all other solar system small body populations, running the low-SE twilight micro-survey every night produces the largest drops ($\approx$3.5\%) in discovery completeness, in particular for the fainter objects. This is because the low-SE twilight micro-survey takes time away from the WFD observing that would otherwise be performed during those twilight hours. This produces a drop in the discovery completeness for faint objects that would otherwise be discovered at larger solar elongations; faint objects are also harder to see than brighter objects when looking near the Sun. This drop is also increased when the micro-survey observations are only made with the \textit{z} filter. On the other hand, bright ($H\leq16$) NEOs and PHAs get the strongest discovery boost when the micro-survey is run every night since more of the easily visible objects are picked up in the additional sky coverage. Of all the \texttt{twi\_neo} family simulations, the two best options for discovery completeness for all included small body populations are \texttt{twi\_neo\_repeat3\_riz\_np6\_v2.2\_10yrs}, which includes 3 repeat visits (i.e., triplets) per pointing in \textit{riz} filters where the low-SE twilight micro-survey is run with a 3 nights on/4 nights off cadence and \texttt{twi\_neo\_repeat4\_riz\_np4\_v2.2\_10yrs}, which includes 4 repeat visits (i.e., quads) per pointing in \textit{riz} filters with a 1 night on/3 nights off cadence. In these simulations, the only population to see a drop in discovery completeness are the $H\leq12$ OCCs with a maximum perihelia of 20 au, which get a $\sim$0.5$\%$ discovery drop; all other populations either match the baseline or gain an increased discovery completeness (up to $\sim$3.5$\%$ and $\sim$2$\%$, respectively). 

When considering the ability to perform light curve inversions for PHAs, NEOs, MBAs, and Jupiter Trojans (bottom panel of Figure~\ref{fig:v2.2_twilight_rest_of_SS}), running the low-SE twilight micro-survey every night is completely detrimental with up to \update{$40\%$} drops for $H\leq15$ Jupiter Trojans and $20\%$ drops for $H\leq18$ MBAs compared to the baseline survey that does not include any low-SE twilight micro-survey. \update{Of the micro-survey options discussed above, t}he only option that does not drop the fraction of objects for which light curve inversions can be obtained by $>5\%$ from that of the baseline survey (the level considered acceptable) is \texttt{twi\_neo\_repeat4\_riz\_np4\_v2.2\_10yrs}, which includes 4 repeat visits (i.e., quads) per pointing in \textit{riz} filters where the low-SE twilight micro-survey is run with a 1 night on/3 nights off cadence. This simulation keeps light curve inversion at baseline level or above with up to a $30\%$ increase above baseline levels for $H\leq16$ PHAs. As described above, this option also performs well for discovery completeness, which provides baseline-level performance or higher (up to $\approx2\%$) for all included small body populations except the $H\leq12$ OCCs with a maximum perihelia of 20 au, which again get a $\sim$0.5$\%$ discovery completeness drop. An additional benefit to having 4 repeat visits instead of 3 repeat visits is better resiliency to contamination by satellites streaks (for additional discussion, see Section~\ref{sec:satcons}).

In general, running the low-SE twilight micro-survey less frequently proves better for both discovery completeness and light curve inversion \update{when all solar system small body populations are considered}. Furthermore, running the low-SE twilight micro-survey at an infrequent cadence boosts discovery completeness overall, including \update{for} the \textquotesingle Ayl\'{o}\textquotesingle chaxnims, which also see a significant discovery completeness enhancement in both discussed simulations (\update{to $\approx$0.5-0.75\%} for 3 repeat visits in \textit{riz} with a 3 nights on/4 nights off cadence or \update{$\approx$0.15-0.25\%} for 4 repeat visits in \textit{riz} with 1 night on/3 nights off). Light curve inversions are also enhanced when the low-SE twilight micro-survey is run infrequently (once every 3 days with 4 repeat visits per pointing in \textit{riz} filters) compared to the baseline survey cadence. Given these enhancements and the large amount of science for a wide variety of small body populations that would be made possible with a low-SE twilight micro-survey, we thus strongly encourage an infrequently run low-SE twilight micro-survey to be included in the LSST survey cadence from the start of the survey. \update{We note for the reader, that the significant increases shown here from the discovery metrics when the twilight low-SE micro-survey is included will not translate into the exact same gains in the actual LSST \textquotesingle Ayl\'{o}\textquotesingle chaxnim, $H\leq16$  NEO, and $H\leq16$  PHA discovery yields. It depends on the size and albedo distribution of these populations which is not included in the calculation of our discovery metrics (see Section \ref{sec:discovery-metrics}). The significant increase in our metrics does show that including the micro-survey will significantly enhance LSST's chances of finding new\textquotesingle Ayl\'{o}\textquotesingle chaxnims and other IEOs, but the actual number of new discoveries may be very small.} 

\update{Given} the numerous scientific benefits and enhancements in solar system discoveries and light curve inversions that will come from running the low-SE twilight micro-survey, as described above, we recommend avoiding waiting to start this micro-survey until Year 2 or later in the 10-year survey. One additional reason for starting this micro-survey in Year 1 is the increasing number of satellite constellations being sent into low Earth orbit. These satellite constellations are most problematic for astronomic observations during twilight hours, when the numerous satellites are brightest in the sky. (For further discussion of the impact of satellite constellations on solar system science, see Section~\ref{sec:satcons}.) With the number of satellite constellations continuing to increase, and plans for that increase to continue for years to come, the problem of contamination will only get worse during the later years of the 10-year LSST survey. We thus recommend starting the low-SE twilight micro-survey in Year 1 of operations in order to reduce the level of satellite contamination as much as possible and enable the most solar system science.\update{The SCOC has recently made a recommendation for a low-SE NEO twilight micro-survey to be included in the survey strategy starting in Year 1 of LSST with further opportunities to explore the final details of the implementation \cite{SCOC_Report_2}.}

Finally, we look at possible further cadence enhancements and/or software improvements. All presently analyzed cadences assume either three or four repeat visits (i.e., triplets or quads), but the discovery criteria are the same as used for WFD observations (linking three tracklets over a \update{14}-day period). Under such assumptions, the standard observing strategy of requiring {\em pairs} should be examined as well. Hints that it may perform better are in Figure~\ref{fig:v2.2_twilight}: note how \texttt{repeat3} cadences systematically produce more discoveries than \texttt{repeat4}; a hypothetical \update{\texttt{repeat2}} cadence may even further increase our sensitivity to \textquotesingle Ayl\'{o}\textquotesingle chaxnims. We recommend such cadences are simulated and analyzed.

Should the analysis conclude 3- or 4-tracklet cadences are still preferred, we would recommend the Rubin Observatory consider reporting such tracklets to the MPC immediately (within 24 hours). This would allow for 3rd-party follow-up of such objects, which may be few enough and interesting enough that it's feasible that they may be followed within days of discovery and not require explicit self-follow-up by Rubin. It would then be interesting to examine this scenario in further detail (since it would not follow the traditional SSP detection method), and quantify the number and purity of high \update{digest 2 tracklets} (i.e., short-arc moving object detections likely to be solar system objects) Rubin would identify and report on a nightly basis, as well as the typical apparent magnitude of reported tracklets (i.e., assess whether the broader community's NEO follow-up system would be able to keep up with this modified reporting method).

\subsubsection{Other \update{M}icro-surveys}
\label{sec:other_micro-surveys}


\begin{figure}
\begin{center}
\includegraphics[width=0.93\columnwidth]{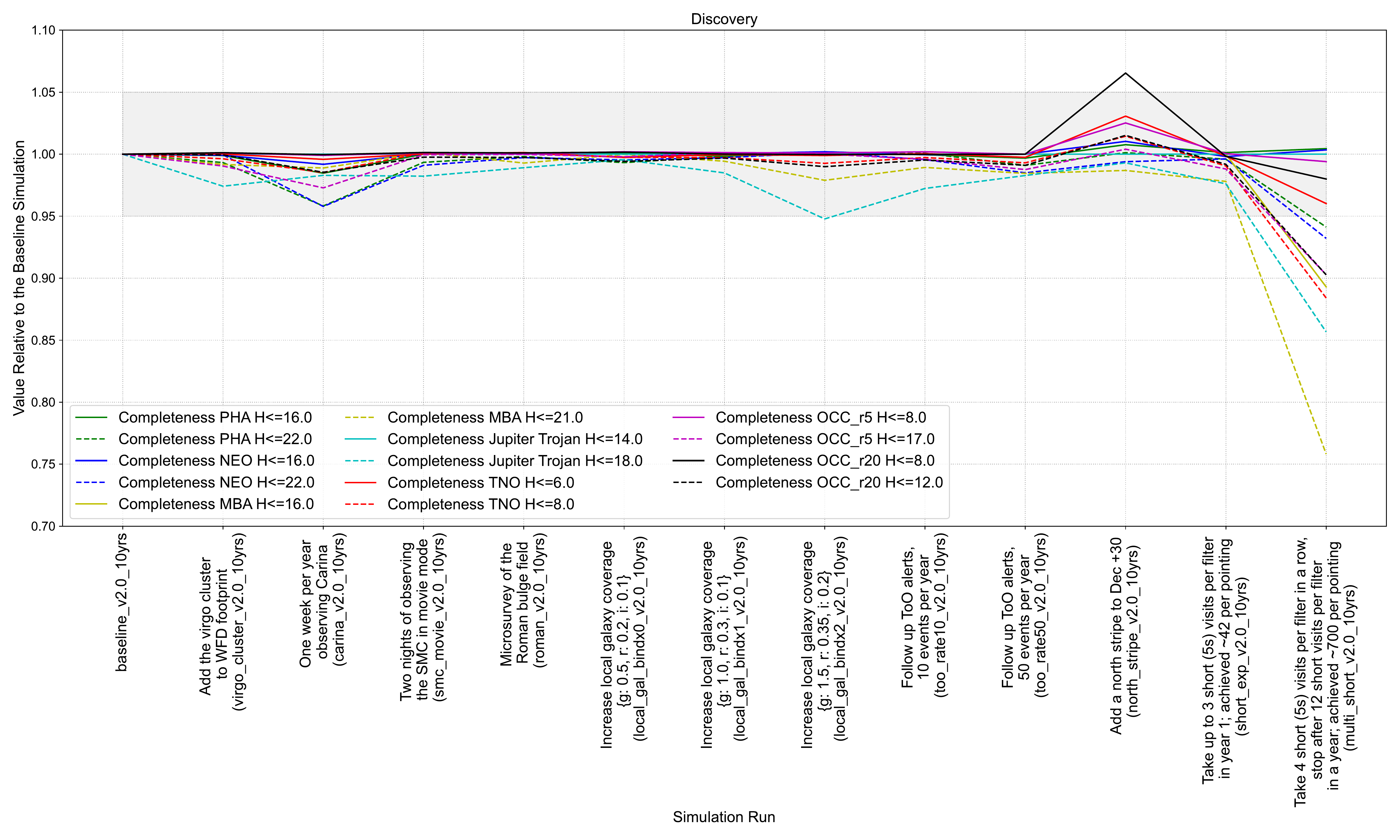}
\includegraphics[width=0.93\columnwidth]{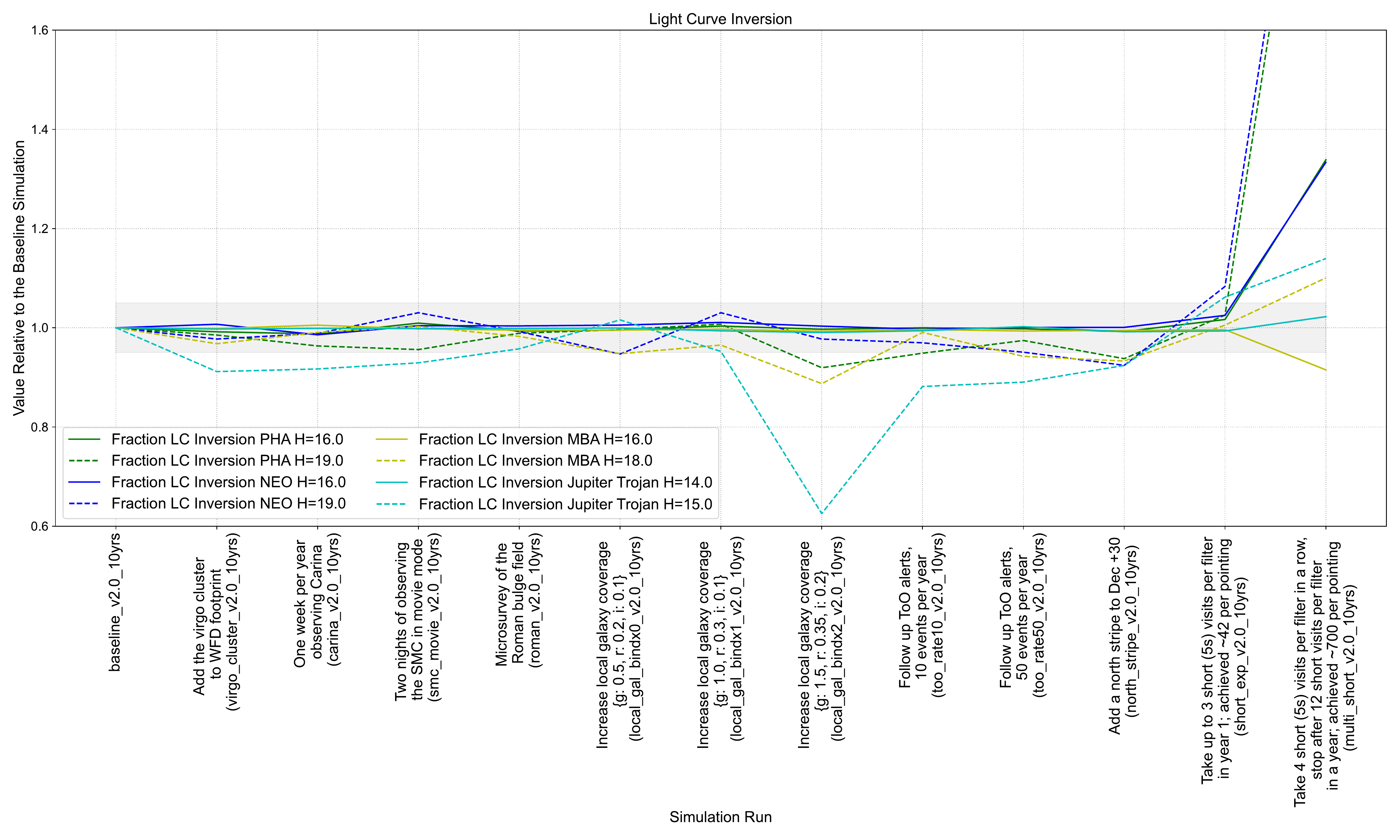}
\caption{Impact of various micro-survey scenarios. \update{The baseline (reference) simulation with the default scheduler configuration} for this cadence experiment is the first entry on the left. All values have been normalized by this simulation's output. The gray shading outlines changes that are within $\pm 5\%$ of the baseline simulation. Top: Discovery Metrics. Bottom: Light Curve Inversion Metrics. The y-axis is truncated in the bottom plot at 1.6 for readability. The light curve inversion NEO $H$=19.0 and light curve inversion PHA $H$=19.0 extend to nearly 2.5 for the \texttt{multi\_short\_v2.0\_10yrs} run.
\label{fig:v2.0_micro-survey}}
\end{center}
\end{figure}

A number of additional micro-surveys requiring 0.3--3\% of overall survey time were submitted in response to the 2018 white paper call on survey strategy\footnote{\url{https://www.lsst.org/submitted-whitepaper-2018}}. These micro-surveys include:
\begin{itemize}
\item Adding extra fields: \texttt{virgo} (adds the Virgo Cluster to WFD), \texttt{carina} (1 week/yr in Carina Nebula), \texttt{smc\_movie} (short $g$ exposures in SMC for 2 nights), \texttt{roman} (covering the \textit{Roman} bulge field twice per year),
\item \texttt{local\_gal\_bindx$<I>$}: Covers 12 Local Group galaxies with extra $gri$ exposures,
\item \texttt{too\_rate$<X>$}: Follow Targets of Opportunity, where $X$ is the number per year,
\item \texttt{north\_stripe}: Adds northern extension stripe up to Dec. +30 (illustrated in the final panel of Figure~\ref{fig:footprints}),
\item \texttt{short\_exp}: Take up to $3 \times 5$\,s exposures in year 1 of the survey,
\item \texttt{multi\_short}: Takes sequences of $4\times5$\,s exposures in each filter, with the aim of obtaining 12 total sequences of short exposures per year, achieving $\sim$700 exposures per pointing.
\end{itemize}

As shown in Figure~\ref{fig:v2.0_micro-survey}, the effects of the majority of these micro-surveys on the discovery and light curve metrics that are of most concern for Solar System science are minimal ($\lesssim5\%$) since this involves a very small fraction of the overall LSST survey time. The exception to these generally minimal effects is seen in the \texttt{multi\_short} simulation (final column in Figure~\ref{fig:v2.0_micro-survey}). This survey strategy causes a $\sim$5--25\% drop in the number of objects detected, particularly for the fainter solar system object populations (This outcome is to be expected with the switch of 12 of the $\sim$82 exposures per pointing per year to much shorter exposures). While the vast majority of the additional micro-surveys have little to no effect on the metrics of most interest to solar system researchers, there remains the possibility that the combination of several of these micro-surveys could ``constructively interfere" in such a way as to cause a large impact later. However these micro-surveys also only impact a very small amount of the survey time and the overall survey strategy, and so the decision on the details of micro-surveys may well also be delayed until later in the cadence decision process. A combined set of micro-surveys will likely be the subject of further simulation runs later when the larger and more influential parts of the cadence strategy is decided upon. 

\section{Additional Considerations Not Explored in the Cadence Simulations}
\label{sec:not_simulated}


The LSST cadence simulations are incredible tools for exploring the wide range of scenarios for how Rubin Observatory can survey the sky, but there are a few additional potential factors that could enhance or impact the LSST science returns that are not yet captured in these simulations. The survey cadence simulations do not account for the difference in Rubin Observatory operations in the first year of the survey compared to later years. The growth of low Earth orbit satellite constellations  and their future impact on LSSTCam observations is not yet quantified. Targeted small observing programs that take much less than 1$\%$ of the LSST observing time are not included in the cadence simulations, and opportunities to propose to Rubin Observatory with these very small observing requests will be considered closer to the start of the survey \citep{SCOC_Report_1}. Additionally, the combined benefits of the LSST data with future wide-field surveys cannot be derived from the cadence simulations alone.  Some of these considerations require analysis of test observations during commissioning of LSSTCam and the Simonyi Survey Telescope. 

\subsection{Incremental Template Generation in Year 1}
\label{sec:templates}

The LSST cadence simulations and the MAF metrics assume that the first year of the survey will run exactly like later years, but this is not quite the case. The data management pipelines use difference imaging to identify transient sources within the LSST exposures by subtracting off a template representing the static sky. The Rubin SSP pipelines use these catalogs nightly to identify moving sources as part of the prompt products data processing \citep{LSE-163}. In order for new solar system objects to be discovered in real time during the survey, a template for the observed field must exist in the given filter of the observation. Templates are expected to be produced during the data processing of the yearly data release. A brief overview of how templates are likely going to be made from coadded observations is available in the summary paper describing the LSST DESC DC2 (Dark Energy Science Collaboration second data challenge) simulated sky survey \citep{2021ApJS..253...31L}. Some fields will have enough observations in commissioning to generate templates at the start of the survey, but this will not be true for the vast majority of the sky. Year 1 of the LSST will have to be treated differently if solar system bodies are to be detected nightly. Otherwise these discoveries will only be made during the data release processing to make the yearly detections catalogs from LSST only data. 

Within MAF there are no solar system metrics focused on the Rubin prompt data products. Whether or not a different template generation strategy is used in Year 1 of the LSST, the total number of discoveries and the number of objects with a sufficient observations for shape inversion will remain the same. These metrics only probe what would be available in the yearly data release catalogs at the end of the 10 year span of the LSST. They do not quantify the impact on the study of transient phenomena. If no templates are produced in the LSST's first year, all time domain-related events (such as ISO apparitions, NEO/PHA close fly-bys, and cometary outbursts) present in the first year of survey observations would be discovered six months to a year \update{after they occurred. The duration of these events is on the scale of days to weeks, discovering these events at the time of the first and second LSST data release would be too late to perform any additional observational follow-up (such as obtaining spectra) with other facilities}. \cite{2021RNAAS...5..143S} highlight \update{in further detail} some of the unique opportunities for solar system science \update{enabled} in the first year of the LSST if templates are generated and implemented into Rubin Observatory's data management pipelines.

The Rubin Observatory Operations Team has committed to producing templates incrementally during the first year of the LSST \citep{RTN-011}, but the exact requirements for producing these Year 1 templates has not yet been decided. The specific strategy used will impact which observations SSP can search before the data release 1 processing and to what limiting magnitude. It will also impact the productivity of the low-SE solar system twilight micro-survey (described in Section \ref{sec:twilight}), if the SCOC recommends the micro-survey to be included in the LSST year 1 observing strategy. The micro-survey gains most benefit if astrometric follow-up observations can be performed by other observing facilities in tandem to the LSST observations. Exploring the implications of various incremental template generation strategies is beyond the scope of this paper, but \update{this analysis should} be carried out before the end of Rubin Observatory's commissioning period. 

\subsection{Solar System Deep Fields}
\label{sec:deep_fields}

\cite{2018arXiv181209705T} proposed dedicated solar system ``DDFs" in response to the 2018 LSST Cadence Optimization White Paper Call. This observing program was different in scope than the typical DDFs currently imagined as part of LSST and described in Section \ref{sec:DDFs}.  Thus we will refer to these pointings as solar system Deep Fields instead. This proposed program would consist of five different pointings at a range of ecliptic latitudes, including coverage of parts of the leading and trailing Neptune Trojan clouds. Each solar system deep field would be observed for 2.1 continuous hours in a single filter to reach the image depth ($r$ = 27.5  mag; 3 magnitudes deeper than the WFD + NES observations) required to observe TNOs as small as 25 km in diameter through shifting and stacking the images on model orbits \citep{2018arXiv181209705T}. Objects at these size ranges in the outer regions of the solar system are particularly under-observed by current surveys due to their faint apparent magnitudes, and new constraints from Rubin Observatory would provide vital information on planetary formation and collisional processes that have occurred and still occur in our solar system. An additional set of four return visits to these deep fields over a two year period is proposed that would enable dynamical classification and color measurements (if one of the later visits was taken in a different filter).
In addition to exploring the nature of the observed TNO size distribution and the physical properties of the TNOs on both sides of the broken power law, the proposed solar system deep drilling fields would also provide further characterization of other solar system objects.
Densely sampled light curves of MBAs, Centaurs, and Jupiter Trojans can reveal any temporal variability in color and brightness within the two hour observation period.
From this, physical properties such as color, size, and shape can be constrained.
 
The total program requires 40 hours of observing time over the 10 years of the LSST (totaling $\ll0.3$\% of total survey time); the equivalent of four winter observing nights. This request is well below the threshold for cadence variations to be evaluated by the SCOC, and therefore has not been included in the LSST cadence simulation runs. We highlight this proposed program here as it would deliver unique science not achieved with the TNO sample found in the WFD, NES, or DDFs. 
The SCOC has recommended to the Rubin Observatory Operations Team that a very small amount of survey time be allocated in a call for proposals for observing requests at this scale once the survey performance has been evaluated \citep{SCOC_Report_1}.

\subsection{Euclid Synergies}
\label{sec:euclid}

The \emph{Euclid} Deep Field South is the fifth DDF to be adopted into the LSST after it was proposed in a written response to the 2018 LSST Cadence Optimization White Paper Call \citep{2019arXiv190410439C}. This DDF will overlap with the southern deep field observed as part of the upcoming ESA \emph{Euclid} mission \citep{2011arXiv1110.3193L, 2012SPIE.8442E..0ZA}. \emph{Euclid} aims to map the geometry of the dark Universe during its 6 year visible and near-infrared photometric and spectroscopic survey, and will provide complementary observations to LSST's wide-field visible ground survey for a number of the LSST science goals. While only solar system objects with ecliptic latitudes beyond $\pm15^\circ$ will be observed by \emph{Euclid}, the science returns from its near-infrared capabilities, high angular resolution and densely sampled light curves will still be significant \citep{2018A&A...609A.113C}. Previous studies suggest that the entire combined Rubin-\emph{Euclid} data set would allow for the spectral classification of roughly 150,000 solar system objects largely unknown to date, and provide constraints on shape, rotation, activity and binarity for a significant number of asteroids, Centaurs, and TNOs \citep{2018arXiv181200607S,2018A&A...609A.113C,guy2022rubin}. In particular, contemporaneous observations from both \emph{Euclid} and LSST will allow rapid determination of orbits, which is vital for the recovery and follow-up of rare solar system objects \citep{2018arXiv181200607S}. For more details on proposed Rubin-\emph{Euclid} derived data products see \citet{guy2022rubin}. While LSST can and will adapt its nightly observation schedule to local weather conditions, the cadence and pointings of \emph{Euclid} are fixed and, therefore, known well in advance. As such, it would fall to LSST to optimize its nightly schedule to maximize the number of near-simultaneous pointings with \emph{Euclid}. Although an initial assessment of the simultaneous astrometry between 
LSST and \emph{Euclid} has been performed and presented at the Rubin Project and Community Workshop in 2019, no simulations currently exist that would quantify the impact of newer LSST cadences simulations with respect to joint \emph{Euclid}-Rubin solar system science.

\subsection{Unquantified Considerations: Satellite Constellations}
\label{sec:satcons}

It is now certain that the accelerating industrialisation of near-Earth space will have major adverse effects on astronomical observation \citep{satcon1:2020,dqsky1:2020,satcon2:2021,dqsky2:2021,Halferty2022}.
The future density of global satellite constellations is as yet uncertain, as it depends on commercial and regulatory decisions. 
However, regulatory approval has been granted by the United States for at least thirty thousand satellites \citep{dqsky2:2021}; $>2,500$ of these launched in the last three years, with 212 Starlinks in May 2022 alone\footnote{\url{https://www.planet4589.org/space/stats/star/starstats.html}}. 
This means Rubin Observatory will have to observe into a hyper-industrialized sky. 
The first few years of LSST will contain the effects of the iterative passes of at least \update{six} thousand low Earth orbit satellites --- and as constellation build-out continues, satellite density will only increase throughout the Survey.

For LSSTCam, there are notable streak effects when illuminated satellites cross the focal plane during an LSST exposure \citep{Tyson:2020}. 
\update{The level to which a streak could be partly or fully saturated in the LSST images depends on each satellite's orbit, morphology, reflectance properties and orientation; the severity can vary through time, such as when a single launch's `train' of co-released satellites are on its orbit raise and are brighter than when on final orbit.
In these cases, and also when the satellite is fainter and so the streaks are not saturated, the impacted pixels will not be suitable for photometry: each satellite streak decreases the effective sky coverage of the exposure.} 
Satellites at $m_g \simeq 7$, with brightnesses below saturation though at SNR $\simeq100$, are also anticipated to create substantial multi-order cross-talk. 
These highly correlated linear `ghost' streaks centre on cores surrounded by wings several hundred pixels in extent. 
The degree to which Rubin Observatory's processing pipelines will be able to mask is yet to be determined.
Even if algorithms can be developed for adequate cross-talk removal, spatially correlated noise will still generate systematics throughout the entire LSST dataset \citep{Tyson:2020}. 
Additionally, there is an increase in global sky brightness from the ensemble of the size distribution of space debris and satellites --- which has already raised the diffuse zenith luminance by 10\% as of 2021 \citep{Kocifaj:2021}. 
While the community focus to date has been on the direct solar illumination of Earth-orbiting objects, satellites also reflect moonlight. 
Moonshine, and potential additional sources of illumination (e.g., Earthshine, mutual reflectance among illuminated space objects) have yet to be modeled in published studies.
The industrially caused loss of global darkness will continue as more anthropogenic material is added to Earth's orbital environment.
For solar system science, satellites and associated debris will thus have three main effects. They obliterate, or modify in unquantified ways, the photometry of individual object detections that are blazed over by streak footprints; they introduce systematic errors at low surface brightness into sky that may later be the source of detections when images are stacked; and they reduce the dynamic range available for the lower SNR solar system detections. 

\update{A recent study by \cite{2022ApJ...941L..15H} find that $\sim$10$\%$ of all LSST images will have a streak from a launched or planned-to-be-launched Starlink (Generation 1 or 2) or OneWeb satellite, with observations in twilight more frequently impacted. If significantly more satellites are launched over the next several years, more LSST observations could have streaks.} The effects of satellite constellations have not yet been comprehensively quantified for Rubin Observatory's solar system science, and we do not attempt to do so here, as no cadence simulations yet include \update{impacts from these satellite streaks on LSSTCam}: we merely highlight a number of \update{potential} projected loss effects. First, the shallowing of LSST will decrease the detected solar system populations across any and all cadences. Across all populations, detection loss for individual objects will deplete the sparse light curves LSST generates; for instance, limiting the ability to detect small-body activity (Section~\ref{sec:lightcurve_metrics}).
Second, there will be population-dependent losses in solar system science from satellite effects. In particular, for detections of NEOs, which individually only become visible for a small subset of time within the span of LSST, the steep size distribution means the majority of detections are made toward the Rubin magnitude limit. 
NEO discovery by the SSP pipelines requires a pair of detections (cf. tracklets) and is thus fragile to satellite effects: losing single detections from a pair has a disproportionate impact on the detectability of this population, as for intra-night cadence outcomes (see Sections~\ref{sec:nightly_sep}, \ref{sec:visit_suppress} and \ref{sec:third_nightly_visits}).
Similarly, satellite-generated detection loss will also acutely affect solar system populations that are only visible for week-to-month time periods, such as ISOs and newly active comets.
The seasonality of satellite density --- more satellites are illuminated, and for longer, in summer --- will have a seasonal impact on solar system object detectability \citep{McDowell:2020,Hainaut:2020,Lawler:2022}. Seasonality detection biases adversely affect all solar system populations that cluster on parts of the sky (for instance, Trojan populations, resonant TNOs, potentially the high-q, high-a TNOs, NEOs, and ISOs); they require careful debiasing to generate accurate population models \citep[e.g.,][]{kavelaars:2020}. 
While not quantified for satellites, general outcomes of inducing this type of effect can be considered in the \texttt{vary\_NES} cadence family (Section~\ref{sec:no-wfd}).
Additionally, twilight-bright satellites will be abundant in the low-SE sky that is being targeted for detection of PHAs, IEOs, gravitationally focused ISOs, and comet comae. Illuminated satellites are most numerous near the horizon close to dawn and dusk; the low-SE twilight micro-survey runs in -12$^{\circ}$ solar elevation and darker.  For the low-SE twilight micro-survey (Section~\ref{sec:twilight}), the effects of satellite constellations will be particularly pronounced, with some 90\% of images expected to be impacted  with at least one streak per image \update{\citep{2022ApJ...941L..15H}}. 

\update{It may be possible for the Rubin Observatory scheduler to selectively observe specific pointings on the sky, which could decrease the number of WFD images with satellite streaks by a factor of two. However, this would come at the substantive cost of $\sim$10$\%$ of the LSST observing time \citep{2022ApJ...941L..15H}. The tradeoffs of implementing this algorithm would depend on the impact of the satellite streaks on LSSTCam, which as noted above has yet to be fully characterized, and the number, sky distribution, and apparent magnitude of satellites at the start of Rubin science operations.} Overall, the characterisation of the impacts of satellite constellations on the LSST cadences, given the ever-changing parameters of constellation design, replenishment, and dynamic operation, will prove challenging. It will require major effort to model and incorporate into future LSST survey simulations. 


\section{Conclusions}
\label{sec:conclusions}
By analyzing the LSST cadence simulations and the outputs for a suite of MAF metrics, we have explored the impact on solar system science for a wide range of potential LSST cadences. Our resulting analysis highlights the importance of simulating the expected small body detections for future multipurpose wide-field surveys. This allows for tensions between main science drivers to be identified in order to optimize the on-sky observing and maximize the output from next generation astronomical surveys and facilities. We hope that this paper and the entire Focus Issue that this paper contributes to may serve as resources for future SCOC reviews of the LSST cadence as well as for future wide-field survey design. 

In general, we find that a wide range of LSST survey strategies provide satisfactory temporal and spatial coverage to achieve the goals for solar system science outlined in the SSSC Science Roadmap \citep{2018arXiv180201783S}. Below, we summarize our key findings and recommendations based on the versions 1.5-2.2 LSST cadence simulations:
\begin{itemize}

\item Observing the northern regions of the ecliptic up to +10$^\circ$ ecliptic latitude (the NES) is crucial for outer solar system science and probing the solar system small body populations that are asymmetrically distributed on the sky.

\item Covering the NES to at least 25\% of the WFD level is critical for discovering and characterizing slowly moving objects (e.g., TNOs) and faint inner solar system objects. 

\item Shifting time away from the WFD to the Galactic plane can negatively impact light curve measurements of faint solar system objects.

\item Shifting the WFD footprint northward from high extinction regions to low extinction sky, such that part of the NES region obtains visits at WFD cadence, is a welcome change. The new expanded northern WFD + NES footprint used in the v2.0-v2.2 cadence simulations is conducive to solar system science.

\item We advocate for moving from 2 $\times$ 15 s snaps to a single 1 $\times$ 30 s exposure per visit due to the resulting $\sim$8$\%$ boost in on-sky visits.

\item Aiming for 33 minute separations between nightly pairs is an ideal compromise between achieving a high pair completion rate per night and for the Rubin SSP pipelines to be sensitive to moving \update{objects at distances} up to $\sim$150 au. 

\item Shorter exposure times are beneficial for the discovery of PHAs and NEOs while longer exposures are better for the discovery of more distant objects. Shorter exposures also increase the total number of on-sky visits per pointing, providing denser sampling for light curve inversion. A compromise between discovery and color/light curve characterization is to use 30 s exposures per visit when possible. 

\item Longer $u$-band exposures (50 s and above) tend to reduce discoveries at small sizes (fainter $H$) and are detrimental to light curve inversions.

\item Similarly, increasing the number of exposures in bluer filters ($u$ and $g$), decreases the number of faint objects for which light curve inversion is possible. 

\item Restricting repeat nightly visits to the same filter does not significantly improve solar system metrics over mixed filter pairs.

\item Including a third visit of a field in the same night can have a very serious effect on the coverage area and other solar system metrics, particularly if the ``Presto-Color" \citep[rapid color; ][]{2019PASPpresto} strategy is implemented. Adding the third visit in the same color as the earlier pair and increasing the gap from the initial pair is shown to have a much lower impact.

\item Having the Rubin scheduler better balance extra nightly visits beyond pairs in a given night by redistributing them across the sky has some benefit to small body discovery with typically small hits to light curve characterization. 

\item Rolling cadence strategies are generally positive for solar system metrics, although the most extreme rolling patterns (many stripes or very strong rolling) should be avoided. A rolling pattern that ensures a minimum coverage to enable discovery of rare types of \update{objects} in the `off' stripes should be considered.

\item Spending 3-8$\%$ of the survey time on DDF observations produces only minimal losses for solar system science. If some DDFs are observed only in certain years, observing the COSMOS DDF for at least two years would be the most beneficial for the detection and orbit characterization of small bodies discovered by shift-and-stack algorithms.

\item The COSMOS DDF should be observed when the Jupiter Trojans are passing through the field, which occurs in discrete windows during LSST.

\item Starting a low-SE twilight micro-survey in Year 1 of operations would make Rubin Observatory uniquely sensitive to several populations of solar system small bodies such as IEOs, Earth Trojans, and sungrazing comets and give the LSST the potential to improve asteroid models, test the theory of asteroid supercatastrophic disruption at small perihelion distances, and improve asteroid impact warning times. An infrequently run (e.g., every 3 nights) low-SE twilight micro-survey would also enhance small body discovery and light curve inversion and enable the discovery of \textquotesingle Ayl\'{o}\textquotesingle chaxnims, which are only visible during twilight.

\item The vast majority of the additional micro-surveys for specific regions of sky have little to no effect on the solar system metrics, but there remains the possibility that combining several of these micro-surveys could produce a result that causes a large impact later. This will likely need to be the subject of further simulation runs later when the larger and more influential parts of the cadence strategy is decided upon and actual operational overheads are measured from commissioning activities.

\item The production of incremental templates in the first year of the LSST is particularly important for the timely follow-up of ISO apparitions and other transient solar system phenomena. Further work is needed to explore the impact on Year 1 discovery rates for the different potential options for creating the incremental templates.

\item A 40-hour program as described in \cite{2018arXiv181209705T} to observe a set of solar system Deep Fields would probe the small size end distribution of the TNO and Neptune Trojan populations that cannot be achieved with the currently planned LSST. 

\item Creating joint data products with other contemporaneous surveys such as ESA's \emph{Euclid} would be of great scientific benefit to the solar system science community. Apart from overlapping DDFs, which are already planned, we suggest synchronizing observations of survey fields where possible when choosing LSST nightly cadences.    

\item The impact of rapidly increasing industrial activity in near-Earth space is not modeled here, but the projected adverse effects are substantial. We advocate for careful characterization of the anthropogenic impacts on the LSST.

\end{itemize}

Our analysis has focused on the individual impact of changing at the same time one or two observing constraints or parameters within the Rubin scheduler. We have not explored the impact of changing all these parameters simultaneously. We note that although tuning individual knobs for the LSST survey strategy by themselves may have little effect, the combination of several of them may not. This should be carefully considered by the SCOC when they finalize their recommendation for the initial LSST observing cadence. \update{The analysis presented here should be repeated with the finalized LSST SCOC recommended observing strategy when it becomes publicly available. Future cadence simulations should be generated and studied to examine additional options for the low-SE twilight micro-survey.} Further investigation is also needed explore the various options for incremental template generation impact the ability for real-time discovery and follow-up of our Solar System’s minor planets and ISOs in the first year of the LSST. This will be particularly important for assessing whether the low-SE twilight survey should be included as part of the Year 1 LSST observing strategy. 

\begin{acknowledgments}
The authors wish to acknowledge all of the essential workers who put their health at risk since the start of the COVID-19 global pandemic and the researchers who worked tirelessly to rapidly develop COVID-19 vaccines. Without all their efforts, we would not have been able to pursue this work. 

We thank the LSST Solar System Science Collaboration for manuscript feedback. 
The authors thank Mike Brown for useful discussions. We thank the anonymous referee for reading and reviewing this very long manuscript and providing constructive feedback. 
The authors also acknowledge the SCOC for their service to the Rubin user community. 
We thank Federica Bianco and the AAS (American Astronomical Society) Journals editorial team for facilitating the Rubin LSST Survey Strategy Optimization ApJS focus issue. 

This research has made use of NASA's Astrophysics Data System Bibliographic Services.

This work was supported in part by the LSSTC Enabling Science grants program, the B612 Foundation, the University of Washington's DiRAC (Data-intensive Research in Astrophysics and Cosmology) Institute, the Planetary Society, and Adler Planetarium through generous support of the LSST Solar System Readiness Sprints. 
MES was supported by the UK Science Technology Facilities Council (STFC) grant ST/V000691/1, and she acknowledges travel support provided by STFC for UK participation in LSST through grant ST/N002512/1. 
KV acknowledges support from the Preparing for Astrophysics with LSST Program funded by the Heising Simons Foundation (grant 2021-2975), from  NSF (grant AST-1824869), and from NASA (grants 80NSSC19K0785, 80NSSC21K0376, and 80NSSC22K0512).
MTB appreciates support by the Rutherford Discovery Fellowships from New Zealand Government funding, administered by the Royal Society Te Ap\={a}rangi.
MSK was supported by the NASA Solar System Observations program (80NSSC20K0673). 
HWL is supported by NASA grant NNX17AF21G and by NSF grant AST-2009096.
TD acknowledges support from the LSSTC Catalyst Fellowship awarded by LSST Corporation with funding from the John Templeton Foundation grant ID $\#$ 62192.
SG acknowledges support from the DIRAC Institute in the Department of Astronomy at the University of Washington. The DIRAC Institute is supported through generous gifts from the Charles and Lisa Simonyi Fund for Arts and Sciences, and the Washington Research Foundation. SG also acknowledges support from the Preparing for Astrophysics with LSST Program funded by the Heising Simons Foundation (grant 2021-2975), from  NSF (grant OAC-1934752), and from NASA (grant 80NSSC22K0978).
The work of SRC was conducted at the Jet Propulsion Laboratory, California Institute of Technology, under a contract with the National Aeronautics and Space Administration.
RCD acknowledges support from the UC Doctoral Scholarship and Canterbury Scholarship administered by the University of Canterbury, a PhD research scholarship awarded through MTB's Rutherford Discovery Fellowship grant and an LSSTC Enabling Science grant awarded by LSST Corporation. RM acknowledges support from NSF (AST-1824869) and NASA (80NSSC19K0785). 
LI acknowledges support from the Italian Space Agency (ASI) within the ASI-INAF agreements I/024/12/0 and 2020-4-HH.0. 

This material or work is supported in part by the National Science Foundation through Cooperative Agreement AST-1258333 and Cooperative Support Agreement AST1836783 managed by the Association of Universities for Research in Astronomy (AURA), and the Department of Energy under Contract No. DE-AC02-76SF00515 with the SLAC National Accelerator Laboratory managed by Stanford University.

For the purpose of open access, the author has applied a Creative Commons Attribution (CC BY) licence to any Author Accepted Manuscript version arising from this submission.

Data Access:  Data used in this paper are openly available from the Vera C. Rubin Observatory Construction Project and Operations Teams via \url{https://github.com/lsst-pst/survey_strategy/tree/main/fbs_1.7} and  \url{https://github.com/lsst-pst/survey_strategy/tree/main/fbs_2.0}. The rubin$\_$sim/OpSim LSST cadence simulation databases are available at \url{https://s3df.slac.stanford.edu/data/rubin/sim-data/}.

\end{acknowledgments}

\facility{Rubin}

\software{LSST Metrics Analysis Framework \citep[MAF,][]{2014SPIE.9149E..0BJ},
Astropy \citep{2013A&A...558A..33A,2018AJ....156..123A,2022arXiv220614220T}, 
Numpy \citep{2011CSE....13b..22V,harris2020array}, 
Matplotlib \citep{Hunter:2007}, 
Pandas \citep{reback2020pandas},
rubin$\_$sim$/$OpSim \citep{2019AJ....157..151N,jones_r_lynne_2020_4048838,peter_yoachim_2022_7374619},
sbpy \citep{2019JOSS....4.1426M}, 
JupyterHub (\url{https://jupyterhub.readthedocs.io/en/latest}), 
Jupyter Notebook \citep{soton403913}, 
python (\url{https://www.python.org}), 
OpenOrb \citep{2009MPS...44.1853G},
scipy \citep{2020NatMe..17..261V},
healpy \citep{2005ApJ...622..759G,2019JOSS....4.1298Z},
seaborn \citep{2021JOSS....6.3021W}}

\section*{Author Contributions}

MES organized and coordinated the paper writing effort as well as the review of the LSST cadence simulations and drafting of formal SSSC feedback to the SCOC that this work is derived from. She wrote the abstract, Sections \ref{sec:intro}, \ref{sec:metric_limitations}, \ref{sec:sims}, \ref{sec:NES}, \ref{sec:WFD_north}, \ref{sec:snaps}, \ref{sec:nightly_sep}, \ref{sec:templates}, and the preambles to Sections \ref{sec:analysis}, \ref{sec:footprint}, \ref{sec:exposures_and_snaps}, \ref{sec:visits}, \ref{sec:all_micro-surveys}, \ref{sec:not_simulated}. She also cowrote Sections \ref{sec:conclusions} and \ref{sec:deep_fields}. She generated the figures showing the footprints, discovery and light curve metrics, and color-light curve metrics based on software utilities and jupyter notebooks developed by RLJ and PY. She also contributed to the Planet Nine figures \update{(Figures \ref{fig:P9_orbits} and \ref{fig:P9_mags}}). She also created Tables \ref{tab:metrics}, \ref{tab:secondary_metrics} \ref{tab:snaps}, \ref{tab:nightly_pairs}, and \ref{tab:DDFs}. She also provided feedback on the entire manuscript.  
RLJ and PY provided guidance on the jupyter notebook templates used to develop the paper figures. They provided expert feedback on the performance and behavior of the Rubin scheduler and metrics. They also contributed to the discussions about the MAF metric outputs for all the cadence simulation families. They also contributed to Figure \ref{fig:pair_sep_hist}. RLJ wrote Sections \ref{sec:simulating-ssos}, \ref{sec:metrics}, and subsections within. RLJ created Tables \ref{tab:colors} and \ref{tab:BaselineMetrics} and produced the key plots for  Figures \ref{fig:baseline_metric_values}, \ref{fig:P9_orbits} and \ref{fig:P9_mags}. PY wrote Section \ref{sec:simulator} and generated Figure \ref{fig:ddf_n_trojans}. KV wrote Sections~\ref{sec:no-wfd} and~\ref{sec:repeat-filter-choice}, created Table~\ref{t:acronyms}, and provided feedback on the whole manuscript. 
RCD wrote Section~\ref{sec:DDFs} and co-wrote Sections~\ref{sec:deep_fields} and \ref{sec:euclid}. CO wrote the Section~\ref{sec:filter_dist} and Section~\ref{sec:u_band} and provided feedback on the overall manuscript. 
SG aided in reviewing the LSST cadence simulations and drafting the formal SSSC feedback to the SCOC. In particular, she led the review and formal feedback for the low-SE twilight NEO micro-survey, soliciting and organizing discussion and feedback from the NEOs and ISOs SSSC working group. She wrote Section~\ref{sec:twilight}, co-wrote Section~\ref{sec:conclusions}, and provided feedback on the overall manuscript. 
TL wrote Sections~\ref{sec:third_nightly_visits} (Third Visits in a Night) and \ref{sec:other_micro-surveys} (Other \update{M}icro-surveys), contributed to Section~\ref{sec:conclusions} (Conclusions) and provided feedback on the overall manuscript. 
CS wrote Section \ref{sec:rolling} and provided feedback on the overall manuscript.
B.T.B wrote Section \ref{sec:other_exp_variations} (Other Variations of Exposure Times) and provided feedback on Sections \ref{sec:third_nightly_visits} (Third Visits in a Night) and \ref{sec:twilight} (Low-SE Solar System Twilight micro-survey). 
LI wrote Section~\ref{sec:visit_suppress}, contributed to the discussion presented in Section~\ref{sec:u_band} and provided feedback on the overall manuscript. 
MTB wrote Section~\ref{sec:satcons} and cowrote Section~\ref{sec:DDFs}. 
SE led the writing of Section \ref{sec:euclid}\update{, provided input on Section~\ref{sec:satcons}, and contributed the description of the simulated \textquotesingle Ayl\'{o}\textquotesingle chaxnim population in Section \ref{sec:simulating-ssos}.}
MS provided feedback discussion and maintained the list of simulations across the manuscript and figures.
MSK wrote the introduction to Section \ref{sec:sim_moving_objects}, developed the cometary brightening function implemented in the OCC metric,  provided input on the OCC simulations, created the orbit OCC files, and provided feedback on the overall manuscript.  
MJ contributed text to the \ref{sec:twilight} section and provided feedback on the overall manuscript.  
HWL created Figure~\ref{fig:Neptune_Trojans}. 
AT, DR,  MMK, RM, TD, and QY provided feedback on the overall manuscript.
MG contributed to the discussions about astrometric precision and orbital characterization for Section \ref{sec:metric_limitations} and provided feedback on the overall manuscript.
CL provided feedback on Section \ref{sec:sim_moving_objects}.
PHB and WJO contributed to discussions about the Planet Nine discoverability. 
SRC, JD, DR, WCF, and AT contributed to the development of light curve metrics. WCF also provided the TNO SED.
\update{MES with contributions from RLJ, MJ, SG, PY, SE, MS, and MTB drafted the response to the referee report and revised the manuscript based on the referee's feedback. }
\\\

\appendix

\section{List of Acronyms}
\label{appendix:acronyms}

\begin{deluxetable}{ll}[!ht]
\tablecaption{List of acronyms used in this paper.\label{t:acronyms}}
\tablehead{\colhead{Acronym} & \colhead{Expansion} }
\startdata
CFEPS & Canada-France Ecliptic Plane Survey \\
COSMOS & Cosmic Evolution Survey\\
CSV & comma-separated values \\
DDF & Deep Drilling Fields\\
DES & Dark Energy Survey\\ 
DESC DC2 & Dark Energy Science Collaboration second data challenge \\
DESI & Dark Energy Spectroscopic Instrument\\
DIA  & difference imaging analysis\\
\enddata

\end{deluxetable}

\clearpage
\section{List of LSST Cadence Simulations}
\label{appendix:sims}


\startlongtable
\begin{deluxetable}{cccc}
\tablecaption{Table of all simulations referenced in this work, its family of simulations, the baseline simulation that it should be compared against, and which figures in the paper reference the simulation. A detailed summary of each of the simulations and families may be found at \url{https://github.com/lsst-pst/survey_strategy/blob/main/fbs_1.7/SummaryInfo.ipynb} (v1.5 and v1.7  simulations) or at \url{https://github.com/lsst-pst/survey_strategy/blob/main/fbs_2.0/SummaryInfo_v2.1.ipynb} (v2.0-v2.2 v simulations)}
\tablehead{\colhead{Simulation Name} & \colhead{Family} & \colhead{Comparison Baseline} & \colhead{Included in which figures} }
\startdata
\texttt{baseline\_nexp1\_v1.7\_10yrs}	&	Baseline	&	N/A	&	\ref{fig:v1.7_snaps}	\\
\texttt{baseline\_nexp2\_v1.7\_10yrs}	&	Baseline	&	N/A	&	\ref{fig:v1.7_footprint}, \ref{fig:v1.7WFDfootprints}, \ref{fig:v1.7_snaps}, \ref{fig:v1.7_long_u}, \ref{fig:v1.7_cadence_drive}, \ref{fig:v1.7_pair_times}	\\
\texttt{baseline\_retrofoot\_v2.0\_10yrs}	&	Baseline	&	N/A	&	\ref{fig:footprints}, \ref{fig:v2.1_2.0_footprint}	\\
\texttt{baseline\_samefilt\_v1.5\_10yrs}	&	intranight	&	\texttt{baseline\_v1.5\_10yrs}	&	\ref{fig:1.5_filters_in_a_night}, \ref{fig:v1.5_filters_in_a_night_4filters}	\\
\texttt{baseline\_v1.5\_10yrs}	&	Baseline	&	N/A	&	\ref{fig:v1.5_footprint}, \ref{fig:v1.5footprints},  \ref{fig:1.5_filters_in_a_night}, \ref{fig:v1.5_filters_in_a_night_4filters}	\\
\texttt{baseline\_v2.0\_10yrs}	&	Baseline	&	N/A	&	\ref{fig:footprints}, \ref{fig:v2.1_2.0_footprint}, \ref{fig:v2.1_2.0_vary_NES}, \ref{fig:v2.1_2.0_vary_NES_4filters}, \ref{fig:v2.1_2.0_vary_GP}, \ref{fig:v2.0_blue_long_u}, \ref{fig:v2.0_long_gaps_no_repeats}, \ref{fig:v2.0_area_third_visit}, \ref{fig:v2.0third_visits}, \ref{fig:rolling_year_3.5}, \ref{animation:baselinev2.0}, \ref {fig:v2.0_rolling}, \ref{fig:v2.0_ddf_frac}, \ref{fig:v2.0_micro-survey}	\\
\texttt{baseline\_v2.1\_10yrs}	&	Baseline	&	N/A	&	\update{\ref{fig:baseline_metric_values},}\ref{fig:footprints}, \ref{fig:P9_orbits}, \ref{fig:P9_mags},  \ref{fig:v2.1_GP_priority}, \ref{fig:v2.1_GP_priority_4filters},  \ref{fig:v2.1_vary_exposure_time}, \ref{fig:v2.1_good_seeing}, \ref{fig:v2.1no_no_repeat_rpw}	\\
\texttt{baseline\_v2.2\_10yrs}	&	Baseline	&	N/A	& \ref{fig:v2.2_twilight}, \ref{fig:v2.2_twilight_rest_of_SS},\ref{fig:ddf_n_trojans}	\\
\enddata

\end{deluxetable}

\section{\update{Metrics Values for the v2.1 Baseline Cadence Simulation}}
\label{appendix:BaselineMetrics}
\begin{deluxetable}{lr}[ht]
\tablecaption{Metric values for \texttt{baseline\_v2.1\_10yrs} \update{from the latest version of {\tt rubin\_sim}}. These values all represent the \update{percent} of the expected population which would ``pass" the metric requirements. \label{tab:BaselineMetrics}}
\tablewidth{0pt}
\tablehead{
\colhead{Metric} & {Value (\%)} \\
}
\startdata
Completeness PHA $H$$<=$6.0             &                 93.9 \\
Completeness PHA $H$$<=$22.0             &                 59.6 \\
Completeness NEO $H$$<=$16.0             &                 93.0 \\
Completeness NEO $H$$<=$22.0             &                 58.2 \\
Completeness MBA $H$$<=$16.0             &                100.0 \\
Completeness MBA $H$$<=$21.0             &                 54.3 \\
Completeness Jupiter Trojan $H$$<=$14.0          &                100.0 \\
Completeness Jupiter Trojan $H$$<=$18.0          &                 43.8 \\
\enddata
\tablecomments{\update{`Completeness' refers to the discovery completeness for each sample population at the indicated $H$ value, while `Fraction LC Inversion'refers to the fraction of each population which would have observations which meet the metric requirements implying that object would be a good subject for lightcurve inversion. Likewise for `Fraction 4 filters', showing the fraction of each population which would be likely to obtain colors in four filters.} Full descriptions of the metrics are listed in Tables \ref{tab:metrics} and \ref{tab:secondary_metrics}. }
\tablecomments{Values for other cadence simulations can be found at via \url{https://github.com/lsst-pst/survey_strategy/tree/main/fbs_1.7} and \url{https://github.com/lsst-pst/survey_strategy/tree/main/fbs_2.0}.}
\end{deluxetable}

\bibliography{manuscript}{}
\bibliographystyle{aasjournal}



\end{document}